\documentclass[12pt,review]{iopart}

\usepackage{amsfonts}
\usepackage{amssymb}
\usepackage{graphicx}
\usepackage{color}
\usepackage{morefloats}
\usepackage[normalem]{ulem}
\DeclareRobustCommand\msout{\bgroup\markoverwith{\color{mscolor}{\rule[0.4ex]{2pt}{0.8pt}}}\ULon}
\DeclareRobustCommand\asout{\bgroup\markoverwith{\color{ascolor}{\rule[0.4ex]{2pt}{0.8pt}}}\ULon}
\DeclareRobustCommand\vgout{\bgroup\markoverwith{\color{vgcolor}{\rule[0.4ex]{2pt}{0.8pt}}}\ULon}

\newcommand{\pom}{{I\!\!P}}
\newcommand{\Pomeron}{{I\!\!P}}
\newcommand{\regg}{{I\!\!R}}

\def\beq{\begin{equation}}
  \def\eeq{\end{equation}}

\begin{document}

%---------------------------------------------------------------------------------

\title{Selected topics in diffraction with protons and nuclei: past, present, and future}

\author{L Frankfurt$^{1,3}$, V Guzey$^2$, A Stasto$^3$, M Strikman$^3$}

\address{$^1$ Sackler School of Exact Sciences, Tel Aviv University, Tel Aviv, 69978, Israel}

\address{$^2$ National Research Center ``Kurchatov Institute'', Petersburg Nuclear Physics Institute (PNPI), Gatchina, 188300, Russia}

\address{$^3$ Department of Physics, Penn State University, University Park, PA 16802, USA}

\begin{abstract}
We review a broad range of phenomena in diffraction  in the context of hadron--hadron, hadron--nucleus collisions and deep inelastic lepton--proton/nucleus scattering focusing on the interplay between the perturbative QCD and non-perturbative models. We discuss inclusive diffraction in DIS, phenomenology of dipole models, resummation and parton saturation at low $x$, hard diffractive production of vector mesons, inelastic diffraction in hadron--hadron scattering, formalism of color fluctuations, inclusive coherent and incoherent diffraction as well as soft and hard diffraction phenomena in hadron--hadron/nucleus and photon--nucleus collisions. For each topic we review key  results from the past and present experiments including HERA and the LHC. 
Finally, we identify the remaining open questions,  which could be addressed in the continuing  experiments, in particular in photon-induced reactions at the LHC and the future Electron-Ion Collider (EIC) in the US, Large Hadron electron Collider (LHeC) and Future Circular Collider (FCC) at CERN.
\end{abstract}
 
%\keywords{}
 
\submitto{\RPP}
\maketitle

\tableofcontents
\section{Introduction}
\label{sec:intro}

One first encounters the phenomenon of diffraction in
optics, where it takes place, for example, in the scattering of light  off a completely absorptive target or a hole in an absorptive screen.
From the observed diffractive image,
one can extract information on  the properties of the target. Similarly, diffraction is also observed in quantum mechanics, for example, in the double slit setup with the variation of the intensity of  particles hitting  the screen determined by the distance between the slits and their width.
In the case of particle collisions at
high energies, elastic scattering (or elastic diffraction) provides  information on
 the dependence of the strengths of interaction on  the impact parameter of the collision 
 (the distance between the projectile and the target in the transverse plane).

Another aspect of  diffraction in hadron--proton scattering, which has no 
simple analog in quantum mechanics, is the process of $h+p\to X + p$ 
referred to as inelastic diffraction, when the minimum momentum transfer squared $t_{\rm min}\to 0$. 
In such a process, a hadron $h$ scatters off a proton $p$ at rest and transforms into a system of hadrons $X$ of a mass $M_X$, with
$-t_{\rm min}=m_N^2 M_X^4/s^2$
($m_N$ is the nucleon mass and $s$ is the invariant energy squared of the collision). 
So, in the  $s\to \infty $ limit, the interacting proton remains at rest. The idea of
    how such a process can happen has been put forward by Feinberg and Pomeranchuk~\cite{Feinberg,Feinberg2}, who pointed out that 
    only if a hadron can exist in configurations  interacting with the target with different strengths, inelastic diffraction is possible.
    Trying to understand this very difficult paper, Good and Walker\footnote{W.D.~Walker, private communication to M.~Strikman (1993).} suggested the cross section  eigenstate model~\cite{Good:1960ba}, which is  still widely used, though often beyond the range of its applicability, 
namely, away from the $t_{\rm min}\sim 0$ limit,
 see the discussion in Sec.~\ref{sec:fluctuations}. 

To a large extent theoretical modeling of inelastic diffraction remained on the sidelines of description of the strong interaction except for the calculation of the inelastic shadowing phenomenon and a related fundamental upgrade of the  Glauber approximation for  hadron--nucleus scattering~\cite{Gribov:1968jf}. 

With the advent of QCD, fluctuations of the interaction strength 
arise naturally  due to the asymptotic freedom: small-size
configurations in hadrons interact with  significantly smaller strengths than the average one.
This was supported by theoretical and experimental studies of diffraction in the 90's, which to a large extent were stimulated by 
 experiments at the electron--proton collider HERA.

In this review we will discuss results of these studies and their spin-offs for 
various phenomena involving interactions with nucleons and nuclei
emphasizing
the interplay of non-perturbative and perturbative dynamics. A special attention will be given to present and future programs to study diffraction including ultraperipheral collisions (UPCs) at the LHC \cite{Baltz:2007kq} and lepton--ion scattering at the Electron--Ion Collider (EIC) in the US \cite{Accardi:2012qut} and the Large Hadron-electron Collider (LHeC) \cite{LHeCStudyGroup:2012zhm,Bruning:2019scy} or Future Circular Collider (FCC) \cite{FCC:2018byv} at CERN.

This review is structured as follows.
In Sect.~\ref{sec:inclusive_diffraction}, we present the results of theoretical and experimental studies of diffraction in electron--proton deep inelastic scattering (DIS)
and the most recent 
analyses of the HERA data,
which demonstrated
that this process is dominated by the leading twist (LT) mechanism in a wide $x$ and $Q^2$ range and has the energy dependence consistent with expectations from models based on the soft Pomeron dynamics. Predictions for the next lepton--proton colliders, EIC, LHeC and FCC-eh are also given.

Section~\ref{sec:dipole_model} reviews the phenomenological dipole models of diffraction and the
total small-$x$ DIS inclusive cross section, which enable one
to include higher-twist effects in the inclusive and diffractive DIS.
These processes allow one not only to constrain the dipole--nucleon cross section, but also 
to determine the kinematics, where the
interaction at small impact parameters may approach the regime of complete absorption, the so-called black disk regime (BDR), which is characterized by the breakdown of perturbative QCD approximations. 

This very high-energy asymptotic behaviour and related small-$x$ methods based on resummation and leading to the notion of parton saturation are  discussed in Sec.~\ref{sec:lowx_resum_sat}.

In Sec.~\ref{sec:vm1}, we summarize results of theoretical studies of hard diffractive electro- and photoproduction of vector mesons in the framework of the dipole model, which in the limit of high $Q^2$ reproduces the leading twist factorization theorem
for exclusive processes. 
A special attention is payed to the phenomenologically important case of elastic $J/\psi$ photoproduction in UPCs in the LHC kinematics.  The rapidity gap vector meson production is discussed in various limits: at $t \sim 0$, where the gluon density fluctuations are probed; in the intermediate $-t<0.5 \; \rm GeV^2$ region, which may be described as a break-up mechanism and, finally, in the large $t$ region, which offers unique opportunities to probe 
Balitsky--Fadin--Kuraev--Lipatov
(BFKL) Pomeron with minimal interference from 
the Dokshitzer--Gribov--Lipatov--Altarelli--Parisi (DGLAP) kinematics.

A convenient formalism to treat inelastic diffraction in hadron--hadron scattering is offered by the notion of cross section fluctuations introduced above. In Sec.~\ref{sec:fluctuations}, we consider specific models of such fluctuations parameterizing the composite hadronic structure of protons, pions, and photons.

In Sec.~\ref{sec:shadowing}, we review the deep connection between diffraction off nucleons and the phenomenon of nuclear shadowing in scattering off nuclei. The application of the factorization theorems for inclusive and diffractive DIS allows one
to predict nuclear inclusive and diffractive 
parton distribution functions (PDFs) within the framework of the leading twist nuclear shadowing approximation.

A  characteristic feature  of this approach is the predictions of large nuclear gluon shadowing. 
Since for the next decade 
ultraperipheral $pp$, $pA$, and $AA$ collisions 
at the LHC would remain one of the best sources of information 
on diffraction in photon--nucleon and photon--nucleus scattering,
a comparison to the UPC data presents an important testing ground for theory. 
A good agreement of the predictions of the leading twist nuclear shadowing approximation with the UPCs data on coherent $J/\psi $ photoproduction on heavy nuclei is demonstrated in Sec.~\ref{sec:UPC}. We also make predictions for dijet photoproduction, including direct and resolved photon contributions, in UPCs at the LHC and in $ep$ scattering at the EIC.

A related subject is nuclear shadowing in coherent photoproduction of $\rho$ mesons on nuclei in heavy ion UPCs.
In Sec.~\ref{sec:UPC_rho}, we consider this process and demonstrate that the shadowing effect is much stronger than 
in the approach based on the vector meson dominance and Glauber models
and significantly stronger than in the $J/\psi$ case.

The importance of color (cross section) fluctuations in soft hadron--nucleus scattering is further demonstrated in Sec.~\ref{sec:diff_soft}, where we present
predictions for 
the diffractive dissociation
cross section 
and fluctuations in the distribution over the number of wounded nucleons in $\gamma A$ 
scattering, which could be measured at the LHC.

LHC experiments provide novel opportunities to study hard diffraction in proton--proton ($pp$) scattering. 
In particular, in Sec.~\ref{sec:pp_hard}  
we consider the diffractive process $pp\to 2 jets +X +p$ and estimate
the corresponding gap survival probability  
using information on elastic $pp$ scattering and diffractive fragmentation function. 
We provide a numerical estimate of the survival probability, which is consistent with the LHC results.
A mechanism of breakdown of soft factorization (the universal gap survival probability) is suggested as well as several new reactions involving production of four jets.

Our conclusions  and outlook are presented in Sec.~\ref{sec:summary}.

 \section{Inclusive diffraction in DIS}
\label{sec:inclusive_diffraction}
\subsection{Introduction}

During the 90's both H1 and ZEUS experiments performed 
a series of observations of events which were characterized by the presence of large rapidity gaps \cite{Derrick:1993xh,Ahmed:1994nw,Ahmed:1995ns,Breitweg:1997rg}. In these events, which constituted 
a
large fraction of about $10\%$ of all DIS events, the proton was separated from the rest of the particles by a large region in a detector which did not have any activity - a {\it rapidity gap}. Two experimental scenarios were employed, the large rapidity gap method (LRG) which relied on the observation of the large rapidity gap, and the leading proton method (LP) where the elastically scattered proton was directly measured in the forward instrumentation. There are some differences in the both methods in that the LRG method does not distinguish the case when the proton is scattered elastically from the cases when the proton dissociates into the system with small mass $M_Y$.  The LP method also allows
one to reconstruct the four-momentum squared $t$ at the proton vertex.

Given the presence of the hard scale in the DIS process, the minus virtuality of the photon $q^2=-Q^2$, it was only natural to ask if such processes are tractable within the perturbative QCD.  Veneziano and Trentadue in  \cite{Trentadue:1993ka}  postulated that  in  DIS the semi-inclusive processes, where the hadron is produced in the 
target fragmentation region,  can be described within the collinear approximation. For that purpose they introduced the notion of the fracture functions which contain the information about the structure function
of a given target hadron once it has fragmented  into another given final state hadron.
The diffractive processes, which can be classified as a special case of the processes discussed in \cite{Trentadue:1993ka}, were considered in  Refs.~\cite{Collins:1997sr, Berera:1995fj} where it was demonstrated that they can be described within the collinear approximation, in analogy to the standard non-diffractive processes in DIS.
The factorization proof, presented in \cite{Collins:1997sr}, essentially followed that of the inclusive case.
Note that, diffractive  factorization can also be applied to other semi-inclusive processes in diffractive DIS like diffractive heavy quark production or dijet production in the direct photon case (see discussion  later in this section).
Also, factorization is valid for a more general case of production of a hadron with  a fixed momentum fraction $x_F$ and a transverse momentum $p_t$ in the target fragmentation region.

\begin{figure}
\centerline{%
\includegraphics[width=0.4\textwidth]{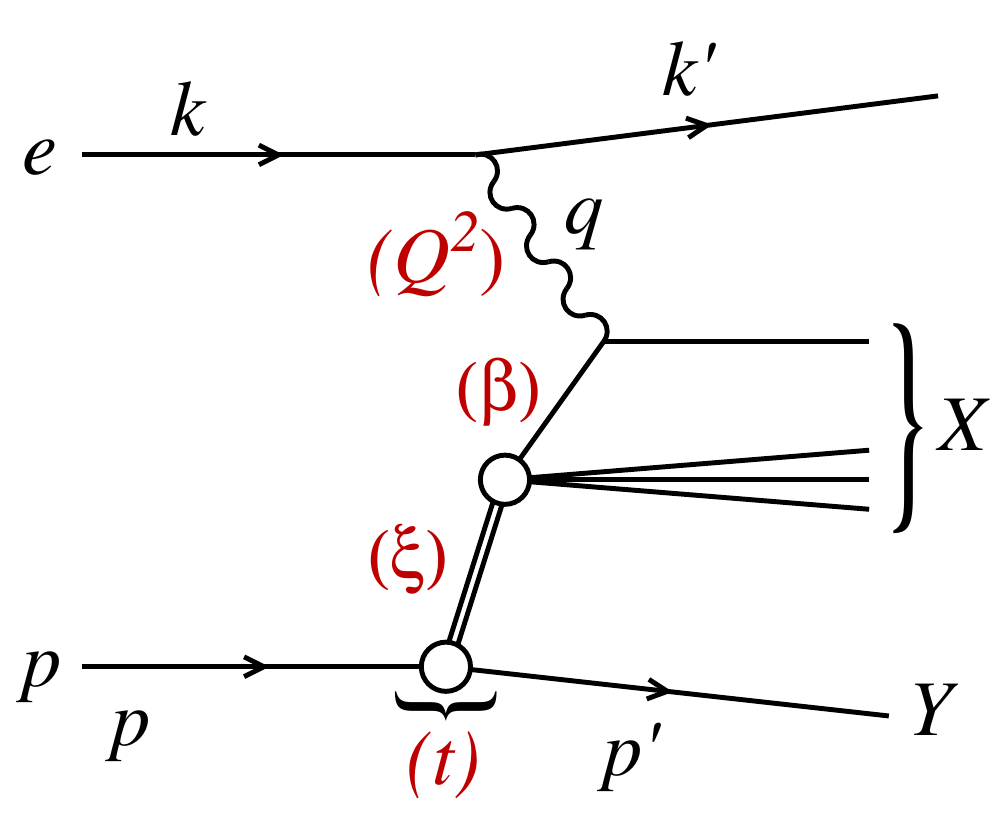}%
}
\caption{Diagram for diffractive DIS in the single photon approximation. The diffractive mass $X$ is separated from the diffractive scattered proton (or its excitation) $Y$ by a rapidity gap. See the text for the definition of the variables.
Figure from \cite{Armesto:2019gxy}, {\tt https://doi.org/10.1103/PhysRevD.100.074022}.}
\label{fig:diffractive_dis}
\end{figure}

 The typical event with a rapidity gap in DIS is depicted in a diagram shown in Fig.~\ref{fig:diffractive_dis}. 
An incoming electron or positron with four-momentum $k$
scatters off the incoming proton with four-momentum $p$. The  proton is scattered into the final state $Y$ with four-momentum $p^{\prime}$. The proton may stay intact or alternatively it can also dissociate into a low mass excitation with mass $M_Y$. The process proceeds through the exchange of 
a single photon and there is a rapidity gap between the final state $Y$ and the diffractive system $X$, see the diagram in Fig.~\ref{fig:diffractive_dis}. 

%%%%%%%%%%%%%%%%%%%%%%%%%%%%%%%%%%%%
%%%%%%%%%%%%%%%%%%%%%%%%%%%%%%%%%%%%

As any DIS process, the diffractive event  is characterized by the standard set of variables:
\begin{equation}
q^2=-Q^2\;, \qquad x=\frac{Q^2}{2p\cdot q}\;,\qquad
W^2 = (p+q)^2\;, \qquad y=\frac{p\cdot q}{p\cdot k}\;,
\label{eq:dis_def}
\end{equation}
being minus photon virtuality, Bjorken $x$, center-of-mass energy squared of the photon-proton system and inelasticity, respectively.
 In addition to these variables, there are also diffractive ones which are defined as follows
\begin{equation}
t=(p-p')^2\;, \qquad \xi=\frac{Q^2+M_X^2-t}{Q^2+W^2}\;, \qquad \beta = \frac{Q^2}{Q^2+M_X^2-t}\;,
\label{eq:diff_def}
 \end{equation}
 where $t$ is the momentum transfer squared at the proton vertex, $M_X^2$ is the mass squared of the diffractive system $X$, $\xi$ is the momentum fraction carried by the diffractive exchange, and $\beta$ is the momentum fraction carried by the struck parton with respect to the diffractive exchange. Often $\xi$ is denoted by $x_{\pom}$ in the literature. The two momentum fractions satisfy the constraint $x=\xi \beta$. The variable  $\xi$ can be related to the fraction $x_L$ of the longitudinal momentum of the initial proton carried by the final proton, i.e. $\xi = 1 - x_L$. Thus typical diffractive events are characterized by small $\xi$, or large $x_L$ meaning that the final proton carries a large fraction of the initial momentum.
The double line in diagram in Fig.~\ref{fig:diffractive_dis} depicts the diffractive exchange (often referred to as the Pomeron) between the proton and the diffractive system $X$, and is responsible for the presence of the rapidity gap.

The diffractive cross sections can be expressed by the two structure functions. In the one-photon approximation 
\begin{eqnarray}
\label{eq:sred3}
\sigma_{\rm red}^{D(3)}& = & F_2^{D(3)}(\beta,\xi,Q^2) - \frac{y^2}{Y_+} F_\mathrm{L}^{D(3)}(\beta,\xi,Q^2) \; , \\
\sigma_{\rm red}^{D(4)} & = & F_2^{D(4)}(\beta,\xi,Q^2,t) - \frac{y^2}{Y_+} F_\mathrm{L}^{D(4)}(\beta,\xi,Q^2,t) \; ,
\end{eqnarray}
where
$Y_+= 1+(1-y)^2$.
In the above equations the reduced cross sections are the rescaled differential cross sections
\begin{equation}
\frac{d^4 \sigma^{D(4)}}{d\xi d\beta dQ^2 dt} \; = \; \frac{2\pi \alpha_{\rm em}^2}{\beta Q^4} \, Y_+ \, \sigma_{\rm red}^{D(4)}\,  ,
\label{eq:sigmared4}
\end{equation}
or, upon the integration over $t$,
\begin{equation}
\label{eq:sigmared3}
\frac{d^3 \sigma^{D(3)}}{d\xi d\beta dQ^2} \; = \; \frac{2\pi \alpha_{\rm em}^2}{\beta Q^4} \, Y_+ \, \sigma_{\rm red}^{D(3)}\, .
\end{equation}

The subscripts $^{(3)}$ and $^{(4)}$ in the above formulae denote the number of variables that the diffractive cross sections or structure functions depend on. Note that the structure functions $F_{2,L}^{D(4)}$ have dimension $\rm GeV^{-2}$, whereas $F_{2,L}^{D(3)}$ are dimensionless. The contribution of the longitudinal structure function to the reduced cross sections is rather small, for the most part, except in the region of $y$ close to unity.

\subsection{Collinear factorization in diffractive DIS}

The standard perturbative QCD approach to diffractive cross sections is based on the collinear factorization \cite{Collins:1997sr,Trentadue:1993ka,Berera:1995fj}.  Similarly to the inclusive DIS cross section, the diffractive cross section can be written in a factorized form
\begin{equation}
F_{2/L}^{D(4)}(\beta,\xi,Q^2,t) \; = \; \sum_i \int_{\beta}^{1} \frac{dz}{z} \ C_{2/L,i}\left(\frac{\beta}{z},Q^2\right) \, f_i^{\rm D}(z,\xi,Q^2,t) \; ,
\label{eq:collfac}
\end{equation}
where the sum is performed over all parton flavors (gluon, $d$-quark, $u$-quark, etc.).
In the case of the lowest order parton model process, $z=\beta$. When higher order corrections are taken into account then 
 $z> \beta$.
The coefficient functions $C_{2/L,i}$  can be computed perturbatively in QCD and are the same as in 
inclusive deep inelastic scattering case. The long distance 
part $f_i^{\rm D}$ corresponds to the diffractive parton distribution functions (DPDF). Similarily to the inclusive case one can provide operator definition for the diffractive parton densities \cite{Berera:1995fj}.
The  quark diffractive distribution function is  defined as
\begin{eqnarray}
\fl f_j^{\rm D}(z,\xi,\mu,t) \; = \; \frac{1}{4\pi}\frac{1}{2} \sum_{s} \int dy^- e^{-i z p^+ y^-} \sum_{X,s'}\langle p,s|\tilde{\bar{\psi}}(0,y^-,0_T)|p',s';X\rangle \nonumber \\
\times \gamma^+ \langle p',s';X|\tilde{\psi}(0)|p,s\rangle \, ,
\label{eq:quark_diff_dist}
\end{eqnarray}
and gluon diffractive distribution 
\begin{eqnarray}
\fl G^{\rm D}(z,\xi,\mu,t) \; = \; \frac{1}{2\pi z p^+}\frac{1}{2} \sum_{s} \int dy^- e^{-i z p^+ y^-} \sum_{X,s'}\langle p,s|\tilde{F}_a^{+\mu}(0,y^-,0_T)|p',s';X\rangle \nonumber \\
\times  \langle p',s';X|\tilde{F}_{a \mu}^{+}(0)|p,s\rangle \, .
\label{eq:gluon_diff_dist}
\end{eqnarray}
In the above the quark field is defined as 
\begin{equation}
 \fl   \tilde{\psi_j}(0,y^-,0_T) \; = \; {\cal P} \, \exp\left(ig \int_{y^-}^\infty dx^- A_c^+(0,x^-,0_T) \, t_c\right) \psi_j(0,y^-,0_T) \, ,
    \label{eq:quark_field}
\end{equation}
whereas the gluon field
\begin{equation}
\fl    \tilde{F_a}^{\mu\nu} (0,y^-,0_T) \; = \; {\cal P} \, \exp\left(ig \int_{y^-}^\infty dx^- A_c^+(0,x^-,0_T) \, T_c\right)_{ab} F_b^{\mu\nu}(0,y^-,0_T) \, ,
    \label{eq:gluon_field}
\end{equation}
where ${\cal P}$ denotes the path ordering in the exponential and $t_a,T_a$ are the generators of the $SU(3)$ group
in the fundamental and adjoint representations, respectively. The definitions (\ref{eq:quark_diff_dist}) and (\ref{eq:gluon_diff_dist}) differ from the standard definitions in the inclusive case by the fact that the final state is the proton, and thus there is a sum over the spin of the proton and over any other particles that form the diffractive state $X$.
 We note that the above definitions and corresponding factorization theorem are valid for general case, which includes the fragmentation kinematics, that is for any fixed $\xi$ and $t$.

The arguments for the factorization were presented in \cite{Trentadue:1993ka,Berera:1995fj} and the proof in \cite{Collins:1997sr}. The proof basically follows the one in the
inclusive case \cite{Collins:1989gx}. The leading regions in the Feynman graphs for 
the
diffractive amplitude involve: a) the beam jet which consists of partons collinear to proton $p$ and also includes the diffractively scattered final state proton with momentum $p'$, b) one or more  final state jets, which are not in the direction of the initial proton, c) the hard interaction, which contains the lines with virtualities of the order of  $Q^2$ and connected to the virtual photon, d) soft subgraph
joined to final jets and beam jet by the soft gluons. Using gauge invariance one can demonstrate that the soft gluons do not resolve the final state jets. As a result, the gluon connections can be incorporated in the gauge link which appears in the definition of the diffractive parton densities, see  Eqs.~(\ref{eq:quark_field}) and (\ref{eq:gluon_field}). 

\begin{figure}
\centerline{%
\includegraphics[width=0.75\textwidth]{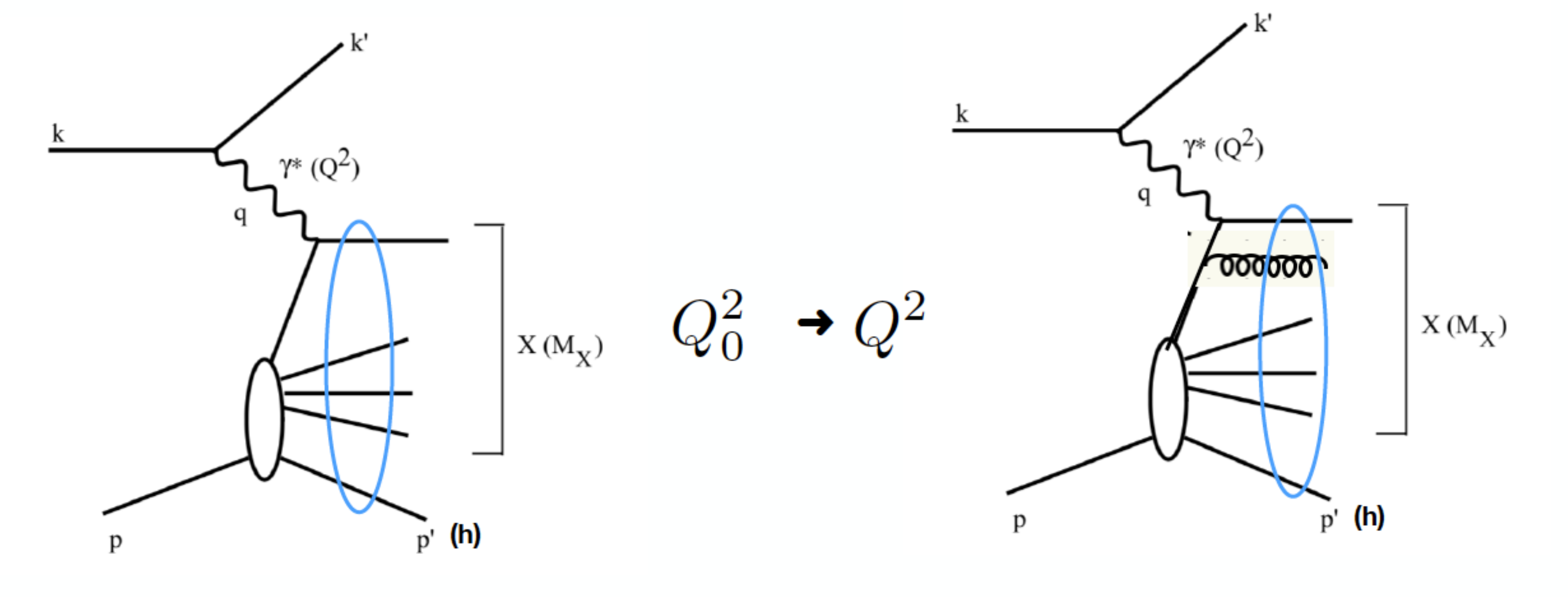}%
}
\caption{Schematic illustration of the factorization theorem for diffractive DIS. 
By changing the scale from $Q_0^2$ to $Q^2$,
an additional parton is emitted in the diffractive mass $X(M_X)$. The blue oval indicates the interaction of partons which  form the final state hadron $h$ (or its low mass excitation) with the the partons in the diffractive mass. This interaction does not change since it cannot resolve 
the $qg$ system which is localized at the transverse distance
much smaller than $1/Q_0$.}
\label{fig:fact}
\end{figure}

An alternative and intuitive way of stating the factorization may be also   illustrated schematically  in Fig.~\ref{fig:fact}. Upon changing the hard scale from $Q_0^2$ to $Q^2$,
an
additional parton (or partons) is emitted in the diffractive system $X$.  The interaction of partons which would form $h$ is not changed upon the variation of the scale since the overall interaction does not resolve the quark-gluon system  which is located at a distance $\ll 1/Q_0$. Finally, we note that the collinear factorization does not require the concept of the Pomeron
 and is distinct from the soft factorization at the proton vertex discussed below.

%%%%%%%%%%%%%%%%%%%%%%%%%%%%%%%%%%%%
%%%%%%%%%%%%%%%%%%%%%%%%%%%%%%%%%%%%

As in the inclusive case the factorization applies to other processes, which can include 
heavy quarks or jets. It also applies to 
photoproduction of jets in the direct photon case,
but is expected to fail in the resolved photon case, since in effect the latter case is like the hadron-hadron scattering where the factorization is known to be violated, for example, in the case of the diffractive dijet production at Tevatron \cite{Abe:1997rg,Affolder:2000hd,Affolder:2000vb}, see Sec.~\ref{sec:pp_hard}.

The diffractive parton densities 
 can be interpreted as conditional probabilities for finding partons in the proton, provided the proton is scattered into the final state system $Y$ with a specified four-momentum $p^{\prime}$. As mentioned above, since factorization holds for semi-inclusive processes, the DPDFs should be universal
 and hence can be used from hard process to hard process.
They are evolved using the Dokshitzer-Gribov-Lipatov-Altarelli-Parisi (DGLAP) evolution equations \cite{Gribov:1972ri,Gribov:1972rt,Altarelli:1977zs,Dokshitzer:1977sg} similarly to the inclusive case.

Note that while the definitions in Eqs.~(\ref{eq:quark_diff_dist}) and (\ref{eq:gluon_diff_dist}) do not have the form of parton number operators, 
it is still customary in the literature \cite{Berera:1995fj} to use the term diffractive parton distributions with their ensuing interpretation as conditional probabilities (see above).

%%%%%%%%%%%%%%%%%%%%%%%%%%%%%%%%%%%%%%%%%%%%%%%%%%%%%%%%%%%%%%%
\subsection{Diffractive fits to HERA data}

Both H1 \cite{Aktas:2006hy} and ZEUS \cite{Chekanov:2009aa} experiments performed fits to the diffractive structure functions based on the collinear factorization and DGLAP evolution. We shall describe the details and results of the fits below.

Since the diffractive parton distribution depends a priori on 4 variables,
a large amount
of physical information needs to be incorporated to correctly describe their shape. It was empirically found that the description of the experimental data was very good when the Ingelman-Schlein \cite{Ingelman:1984ns} proton vertex factorization is assumed. This means that the DPDF is factorized into products of two terms, one of   which depends on $\xi$ and $t$ only and another one which depends on $z$ and $Q^2$ \cite{Ingelman:1984ns}
\begin{equation}
f_i^{D(4)}(z,\xi,Q^2,t) =  f^p_{\pom}(\xi,t) \, f_i^{\pom}(z,Q^2) \;.
\label{eq:vertex_factorization}
\end{equation}
A popular
physical interpretation of this factorization is that  the diffractive exchange  can be interpreted as a colorless object called a Pomeron with a partonic structure
given by the parton distributions $f_i^{\pom}(\beta,Q^2)$, the variable $\beta$ corresponding to the fraction
of the Pomeron longitudinal momentum carried by the struck parton.

The Pomeron flux factor
$f^p_{\pom}(\xi,t)$ represents the probability that a Pomeron with particular values of $\xi$ and $t$ couples
to the proton.
It is worth to emphasize that in the diagrammatic language the  Pomeron exchange occurs over long space-time intervals,
which
is due to the fact that the exchanged gluon quanta have small plus and minus momentum components.
Also let us note that the soft factorization is highly nontrivial as  the structure of the Pomeron  could a priori depend on $Q^2$ - like it happens in the case of the Balitsky-Fadin-Kuraev-Lipatov (BFKL)  perturbative Pomeron
\cite{Balitsky:1978ic,Kuraev:1977fs, Lipatov:1985uk}. In the latter case it is known, for example, that the Pomeron intercept will depend on the value of the strong coupling which depends on the scales involved in the process, such as $Q^2$ in the case of DIS, see Sec.~\ref{sec:lowx_resum}.

There are number of diffractive fits in the literature, \cite{Aktas:2006hy,Chekanov:2009aa,Goharipour:2018yov,Khanpour:2019pzq,Zlebcik:2019tiu}. Both H1 \cite{Aktas:2006hy} and ZEUS \cite{Chekanov:2009aa} collaborations performed DGLAP fits to their own experimental data. 

Analyses of the experimental data actually show a necessity to include another component, that corresponds to the subleading term, (in a sense of the small $\xi $ behavior) to which we refer as Reggeon and which is important at large values of 
$\xi \geq 0.03$ and small $\beta$. Thus the parametrization employed in the fits is a modification of (\ref{eq:vertex_factorization}) which results in a  two-component 
model, which is a sum of two exchange contributions, $\pom$ and $\regg$, each satisfying the vertex factorization hypothesis:
\begin{equation}
f_i^{D(4)}(z,\xi,Q^2,t) =  f^p_{\pom}(\xi,t) \, f_i^{\pom}(z,Q^2)+f^p_{\regg}(\xi,t) \, f_i^{\regg}(z,Q^2) \;.
\label{eq:param_2comp}
\end{equation}

 The fluxes $f^p_{\regg,\pom}(\xi,t)$  are parametrized using the form motivated by the Regge theory,
\begin{equation}
 f^p_{\pom,\regg}(\xi,t) = A_{\pom,\regg} \frac{e^{B_{\pom,\regg}t}}{\xi^{2\alpha_{\pom,\regg}(t)-1}} \; ,
\label{eq:flux}
\end{equation}
where $A_{\pom,\regg}$ are normalization constants for the Pomeron and Reggeon, $B_{\pom,\regg}$ are $t$ slopes, and   ${\alpha_{\pom,\regg}(t)=\alpha_{\pom,\regg}(0)+\alpha_{\pom,\regg}'\,t}$ are linear trajectories .

We emphasize here that the notions of Pomeron and Reggeon in the context of diffractive deep inelastic scattering  differ from those familiar from the soft $pp$ $(p\bar{p})$ interactions. In particular, the parameters of both trajectories may be different.

Strictly speaking the parametrization in (\ref{eq:param_2comp}) is inspired by the Regge theory. One does not assume anything here about the quantum numbers of the Reggeon. Neutron production at large $z$ is strongly suppressed as compared to proton production, see \cite{Chekanov:2007tv,Chekanov:2008tn}. Thus the Reggeon  contribution in (\ref{eq:param_2comp}) most probably cannot be interpreted as solely as a  pion  exchange. This is due to the fact that in the case of the pion exchange, the ratio of the neutron to proton production is equal to two
as a consequence of the Clebsh--Gordan isospin relations between the corresponding
pion photoproduction amplitudes.

The diffractive parton distributions of the Pomeron at the initial scale $\mu_0^2$ are modeled as a  singlet quark  distribution $\Sigma=\sum_i (f_i^\pom+\bar{f}_i^\pom)$ consisting of quark and antiquark distributions and a gluon distribution $f_g^{\pom}$. The neutrality of the Pomeron implies $f_i^\pom \equiv \bar{f}_i^\pom$. In both ZEUS and H1 fits it was assumed that the light quark distributions are equal $f_u^{\pom}=f_d^{\pom}=f_s^{\pom}$ and that the charm and beauty distributions are in the variable flavour number scheme (VFNS) and are generated radiatively.
Both quarks and gluons are parametrized using similar functional forms to that used in the fits to the inclusive structure function data. H1 and ZEUS used 
the
following simple parametrization for the distributions in the Pomeron
\begin{equation}
z f_i^\pom (z,\mu_0^2)= A_i z^{B_i} (1-z)^{C_i} \; ,
\label{eq:initcond}
\end{equation}
where $i$ is a gluon or a light quark.  The parameters $C_q,C_g$ were allowed to vary
and in particular they were allowed to take both positive and negative  values. To ensure the vanishing of the  distributions at $z=1$ for the solutions to the DGLAP equation, an exponential regulating factor of  $\exp(-0.01/(1-z))$ has been included.

The parton distributions for the Reggeon component, $f_i^{\regg}$ were taken from a parametrization which was obtained from fits to the pion structure function \cite{Owens:1984zj,Gluck:1991ey}.

The fits performed by H1 \cite{Aktas:2006hy} and ZEUS \cite{Chekanov:2009aa} are very similar in general setup but differ   in the details of the choice and number of free parameters and the selection of data sets.
 The fits performed in \cite{Aktas:2006hy} were referred to as NLO H1 fit A and B. The two sets of parton densities differ mainly in the gluon density at high fractional parton momentum. This region is very poorly constrained by the inclusive diffractive scattering data at HERA kinematic range. 
ZEUS collaboration also performed  three separate fits, ZEUS C, S and SJ fits \cite{Chekanov:2009aa}. Fits C and S differ in the form of gluon parametrization at large $z$, whereas fit SJ is essentially based on fit S parametrization but in addition to inclusive data, diffractive dijet data are included in the fit.

In Table \ref{table:table1} we show values of the selected  parameters for ZEUS fits S,  C and SJ as well as H1 fits A and B. Note that the parametrizations for ZEUS were specified at $\mu_0^2 = 1.8 \, \rm GeV^2$ and for H1 for $\mu_0^2 = 1.75 \, \rm GeV^2$. We do not show the normalization parameters as they cannot be compared directly due to different conventions for both experiments.

\begin{table}
\begin{center}
 \begin{tabular}{||c| c c c c c||} 
 \hline
 Parameter & ZEUS S & ZEUS C & ZEUS SJ & H1 A & H1 B \\ [0.5ex] 
 \hline\hline
 $B_q$ & 1.34 $\pm$0.05 & 1.25 $\pm$ 0.03 & 1.23 $\pm$ 0.04 & 2.3$\pm$ 0.36 & 1.5 $\pm$ 0.12 \\
 \hline
 $C_q$ & 0.34 $\pm$ 0.043 & 0.358 $\pm$ 0.043  & 0.332 $\pm$ 0.049 & 0.57 $\pm$ 0.15 & 0.45 $\pm$ 0.09 \\
 \hline
 $B_g$ & -0.422 $\pm$ 0.066& {\bf 0} & -0.161 $\pm$ 0.051 & {\bf 0} & {\bf 0}  \\ 
  \hline
 $C_g$ & -0.725 $\pm$ 0.082 & {\bf 0} & -0.232 $\pm$ 0.058 & -0.95 $\pm$ 0.20 & {\bf 0} \\ 
  \hline
 $\alpha_{\pom}(0)$ & 1.12 $\pm$ 0.02 &  1.11 $\pm$ 0.02 & 1.11 $\pm$ 0.02 & 1.118 $\pm$ 0.008 & 1.111 $\pm$ 0.007 \\ 
  \hline
 $\alpha_{\regg}(0)$ &0.732 $\pm$ 0.031 &  0.668 $\pm$ 0.040 & 0.699 $\pm$ 0.043 & {\bf 0.5} & {\bf 0.5} \\
 \hline
 $\alpha'_{\pom}$ & {\bf 0} & {\bf 0} ${\rm GeV^{-2}}$ & {\bf 0} ${\rm GeV^{-2}}$& {\bf 0.06} ${\rm GeV^{-2}}$& {\bf 0.06} ${\rm GeV^{-2}}$ \\ 
 \hline
 $\alpha'_{\regg}$ & {\bf 0.9} ${\rm GeV^{-2}}$ & {\bf 0.9} ${\rm GeV^{-2}}$& {\bf 0.9} ${\rm GeV^{-2}}$& {\bf 0.3} ${\rm GeV^{-2}}$& {\bf 0.3} ${\rm GeV^{-2}}$ \\ 
  \hline
 $B_{\pom}$ & {\bf 7} ${\rm GeV^{-2}}$ & {\bf 7} ${\rm GeV^{-2}}$& {\bf 7} ${\rm GeV^{-2}}$& {\bf 5.5} ${\rm GeV^{-2}}$& {\bf 5.5} ${\rm GeV^{-2}}$ \\ 
  \hline
 $B_{\regg}$ & {\bf 2} ${\rm GeV^{-2}}$ & {\bf 2} ${\rm GeV^{-2}}$& {\bf 2} ${\rm GeV^{-2}}$& {\bf 1.6} ${\rm GeV^{-2}}$& {\bf 1.6} ${\rm GeV^{-2}}$ \\ 
 \hline
\end{tabular}
\end{center}
\caption{Values of the parameters for ZEUS S, C, SJ fits \cite{Chekanov:2009aa} and H1 A and B fits \cite{Aktas:2006hy}. Parameters in bold font have been fixed in the fit. }
\label{table:table1}
\end{table}

%%%%%%%%%%%%%%%%%%%%%%%%%%%%%%%%%%%%
\begin{figure}
\centerline{%
\includegraphics*[width=0.7\textwidth,trim=0 150 0 100]{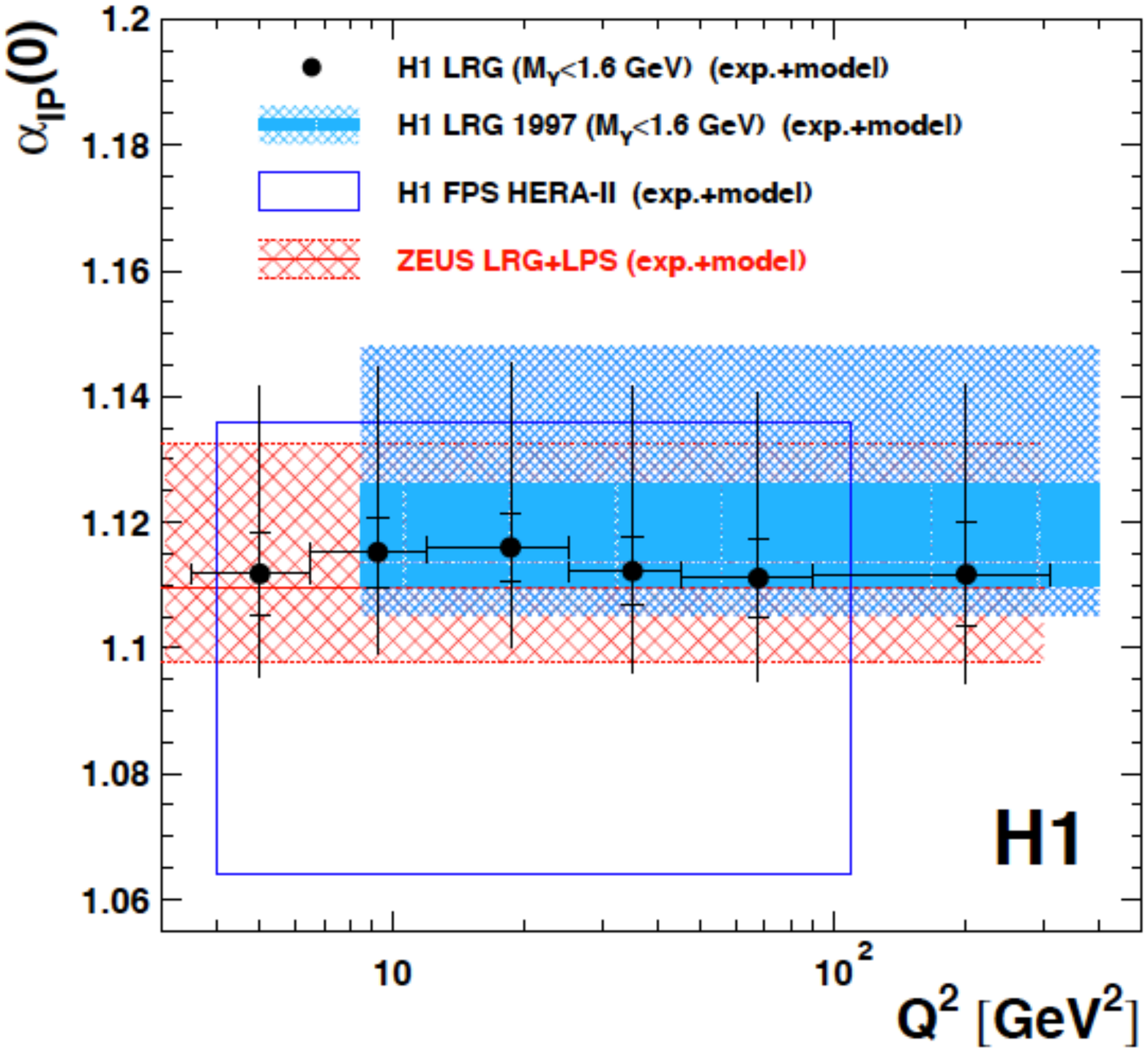}%
}
\caption{ The value of Pomeron intercept $\alpha_\pom(0)$ obtained from Regge fits in different $Q^2$ bins. The inner error bars represent the statistical and systematic errors added in
quadrature and the outer error bars include model uncertainties. Figure from \cite{Aaron:2012ad}, {\tt https://doi.org/10.1140/epjc/s10052-012-2074-2 }.}
\label{fig:alphap0}
\end{figure}

%%%%%%%%%%%%%%%%%%%%%%%%%%%%%%%%%%%%

The value of the Pomeron intercept $\alpha_{\pom}(0)$ is about $1.11-1.12$, 
which is  of
the same order as the value 
extracted from the $pp$ total cross section data.  Both collaborations tested the proton vertex factorization hypothesis, by investigating the dependence on $Q^2$. In the analysis \cite{Aaron:2012ad},
full range in $Q^2$ is divided into six intervals, and for each interval, a  Pomeron intercept was introduced. In that way the vertex factorization hypothesis was tested  by allowing for a $Q^2$ dependence of the Pomeron intercept in the fit procedure.
The results of this analysis are shown in Fig.~\ref{fig:alphap0}. As is evident from this figure no significant $Q^2$ dependence of the Pomeron intercept was observed, which supports the proton vertex factorization hypothesis.
The average value for the Pomeron intercept found 
in
this analysis was $\alpha_{\pom}(0)=1.113$, 
which is
consistent with the earlier fits.

The value of $\alpha_{\pom}(0)$
in inclusive diffraction is significantly smaller than the value extracted from the elastic diffractive production of heavy vector mesons, for example, in the $J/\psi$ photoproduction \cite{Derrick:1995ue,Aid:1996dn}, see Sec.~\ref{sec:vm1}. The $Q^2$ independence of the intercept also has to be contrasted with the increase of the effective Pomeron intercept extracted from the inclusive $F_2(x,Q^2)$ \cite{Adloff:2001rw}.

The value of $\alpha'_{\pom}$ from Table \ref{table:table1} is either fixed to zero (ZEUS fits) or to very small value $0.06 \; \rm GeV^{-2}$ in the case of H1 fits. This is much smaller than the value found from the fits to hadronic cross sections $\alpha'_{\pom}=0.25 \, \rm GeV^{-2}$, see for example \cite{Barone:2002cv}.  This parameter has been extracted by both collaborations in the measurements of the $t$ differential cross section $\sigma^{D(4)}_{\rm red}$. The H1  measurements using Forward Proton Spectrometer (FPS) gave value  $0.02-0.1 \; \rm GeV^{-2}$ \cite{Aktas:2006hx} depending on the range of $\xi$ and $0.009-0.06\, \rm GeV^{-2}$ \cite{Aaron:2010aa} depending on the range of $Q^2$. In the case of ZEUS \cite{Chekanov:2008fh}, the extracted value was $\alpha'_{\pom}=-0.01\; \rm GeV^{-2}$ with large error bars.

In the same works, the $t$ slope parameter was extracted for the Pomeron  with the value of about $7.0 \; \rm GeV^{-2}$ from ZEUS \cite{Chekanov:2008fh} and $5.17-5.78 \; \rm GeV^{-2}$  from H1 data \cite{Aaron:2010aa}.
This is  somewhat larger slope than for exclusive hard diffraction, $J/\psi$ production, which indicates that the size of the inclusive diffractive vertex is not small. 
This is consistent with $\alpha_\pom(t=0) $ being smaller for inclusive diffraction than for hard exclusive processes.

\begin{figure}
\centerline{%
	\includegraphics*[width=0.5\textwidth]{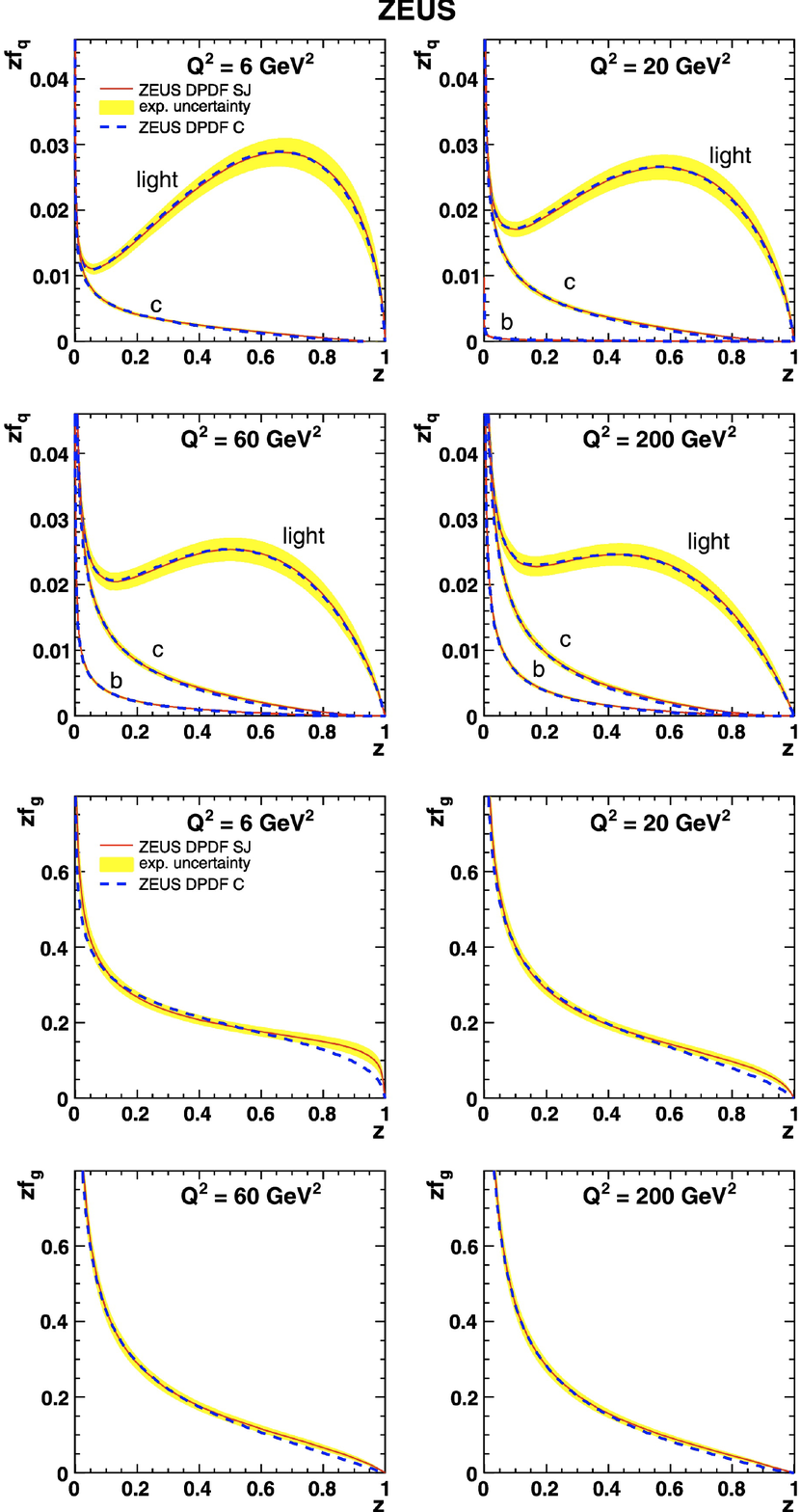}%
	\hspace*{-1cm}
		\includegraphics*[width=0.5\textwidth]{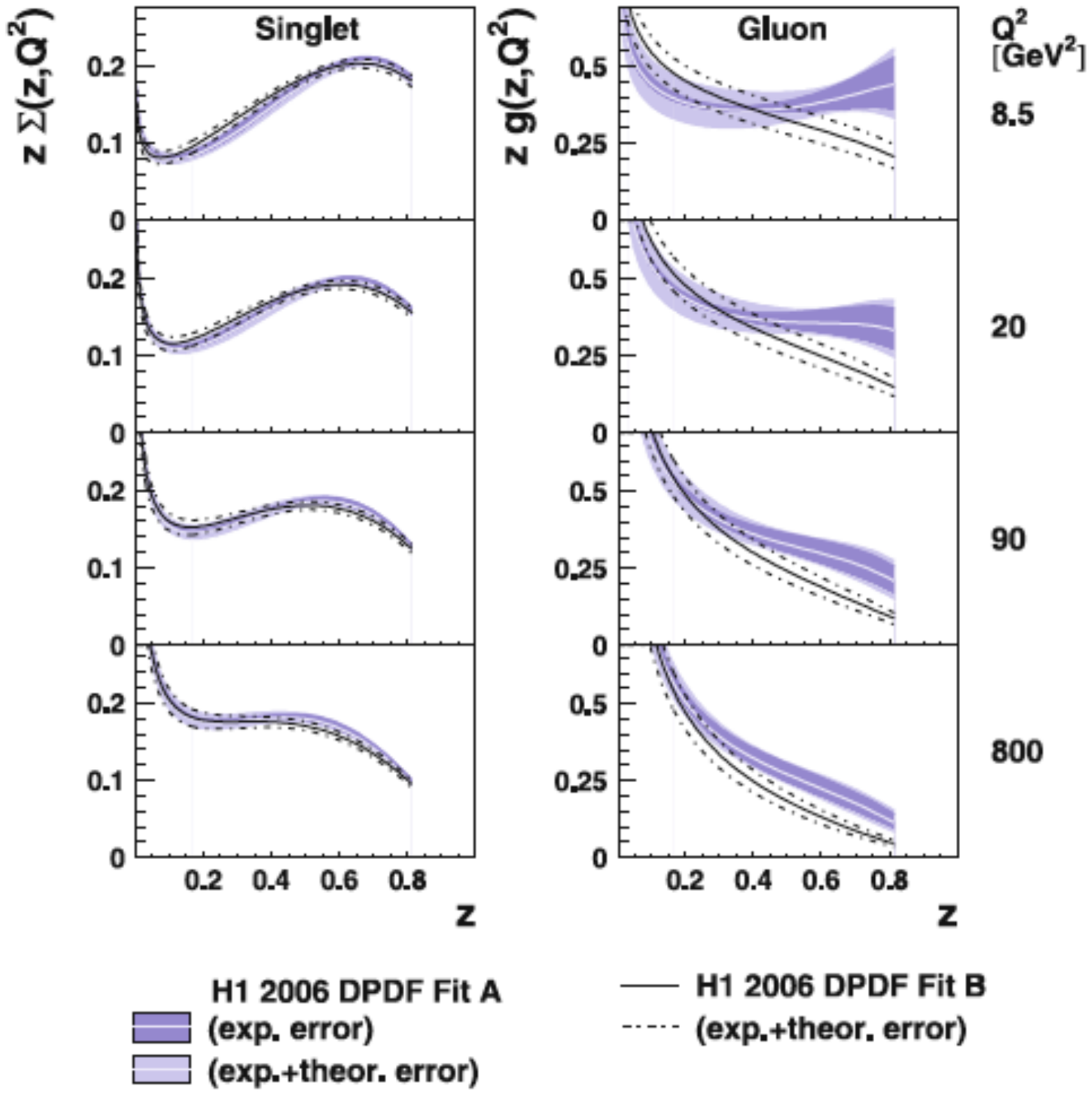}%
}
\caption{Examples of diffractive parton distribution functions from  ZEUS and H1 analysis. Figures from \cite{Chekanov:2009aa}, {\tt https://doi.org/10.1016/j.nuclphysb.2010.01.014} and  \cite{Aktas:2006hy}, {\tt https://doi.org/10.1140/epjc/s10052-006-0035-3}.}
\label{fig:dpdfs}
\end{figure}

As mentioned above the biggest difference between ZEUS fits S and C and fits H1 A and B is the behavior of the gluon at large $z$. Since both fits described the inclusive diffractive HERA data very well, it was concluded that the inclusive data are not sensitive enough to pin down the behavior of the gluon density in this region.  The
data on diffractive production of dijets turned out to be much more sensitive to the gluon distribution at large $z$ and favored ZEUS C and H1 B fits. 
The latter data on diffractive dijets were included in addition to the inclusive data in  ZEUS SJ fit.
Examples of diffractive PDFs from ZEUS and H1 are shown in Fig.~\ref{fig:dpdfs}.

%%%%%%%%%%%%%%%%%%%%%%%%%%%%%%%%%%%%

%%%%%%%%%%%%%%%%%%%%%%%%%%%%%%%%%%%%

First diffractive fits at NNLO accuracy have also been performed recently \cite{Zlebcik:2019tiu}. The fit used the combined HERA II data in addition to the HERA I data, which allowed for the increased precision. In addition, similarly to previous fits, the NNLO fit has been  constrained  by the diffractive  dijet data. The quark distribution is similar to the NLO extraction, whereas the gluon density, which is more sensitive to higher order effects is reduced by about $30\%$.  Overall, the NNLO calculation fits the data very well, above $Q^2 > 8.5 \, \rm GeV^2$, where the effects from higher twists can be neglected.

%%%%%%%%%%%%%%%%%%%%%%%%%%%%%%%%%%%%
\begin{figure}[t]
\centerline{%
	\includegraphics*[width=0.7\textwidth,trim=0 50 0 50]{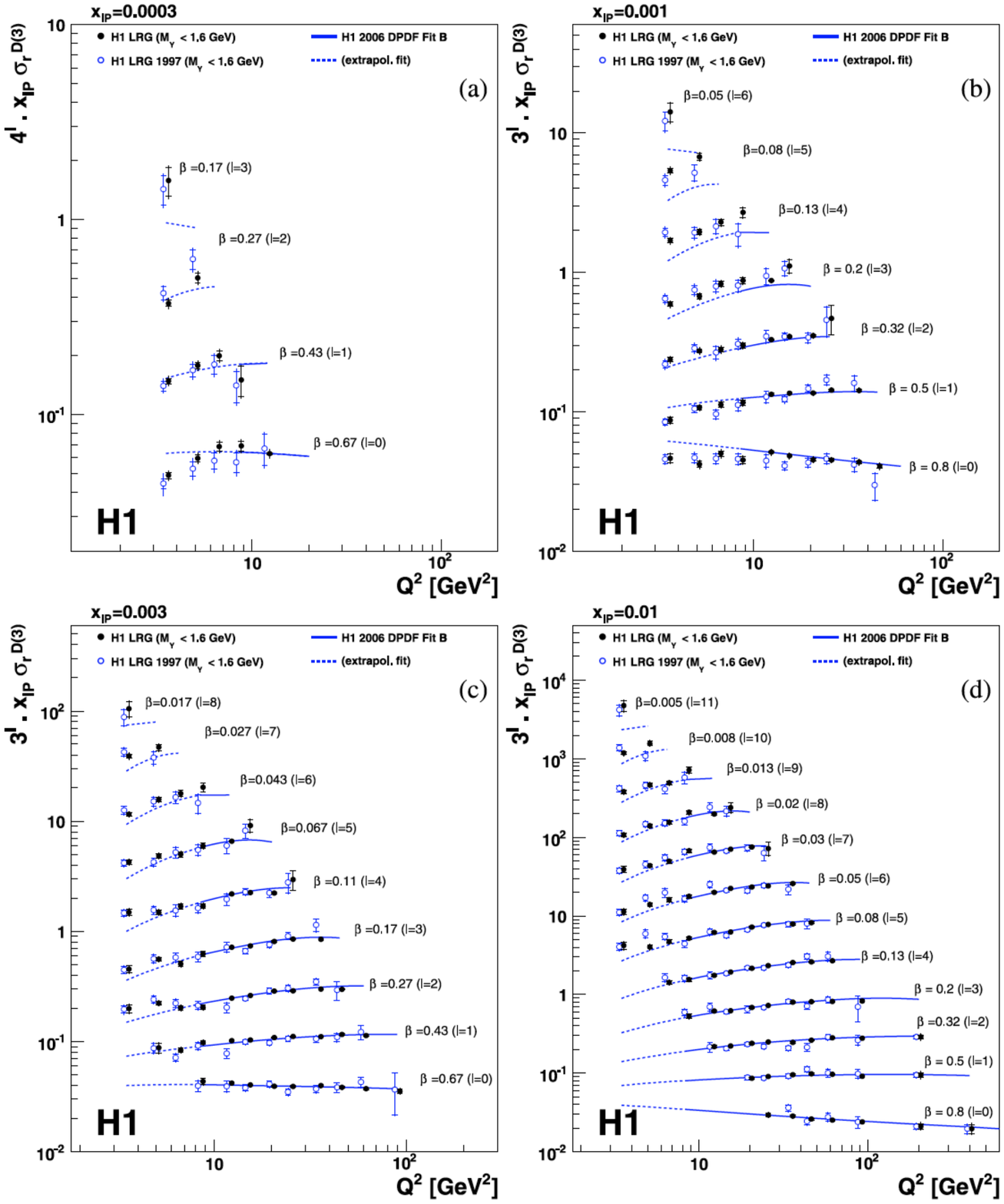}%
}
\caption{The $Q^2$ dependence of the reduced diffractive cross section, multiplied by $x_{\pom}$, at different fixed values of $x_{\pom}$=0.0003 (a), 0.001 (b), 0.003 (c) and 0.01 (d). The reduced cross section values are multiplied by a scaling factor, $4^l$ for $x_\pom$=0.0003 and $3^l$ for $x_{\pom}$=0.003, 0.001 and 0.01, with l values as indicated in parentheses.  Solid lines fit H1 B, dashed line indicates an extrapolation. 
Figure from \cite{Aaron:2012ad}, {\tt https://doi.org/10.1140/epjc/s10052-012-2074-2}.}
\label{fig:herafits}
\end{figure}

%%%%%%%%%%%%%%%%%%%%%%%%%%%%%%%%%%%%

In Fig.~\ref{fig:herafits},   
the calculations using the H1 Fit B \cite{Aktas:2006hy}  are compared
with the  more recent analysis of the  HERA data  \cite{Aaron:2012ad} on the reduced diffractive cross section. The results are shown as a function of $Q^2$ in several bins of $\beta$ and four bins bins of $\xi=x_\pom$. One observes reasonable description of the experimental data by the fits, with 
a certain deterioration 
at low values of $Q^2$. Indeed, it was noted by both ZEUS and H1
 that DGLAP fits fail in the low $Q^2$ region, the breakdown point was about $8.5 \; \rm GeV^2$  as reported by H1, and $5 \; \rm GeV^2$ for ZEUS. The solid line in Fig.~\ref{fig:herafits} indicates the fit H1 B which was only performed down to $8.5 \; \rm GeV^2$, while the dashed line is an extrapolation. From the figures we also observe that the leading twist description deteriorates at small values of $\beta$. The exact origin of this breakdown of DGLAP fits is unknown. It has been though suggested \cite{Motyka:2012ty,Maktoubian:2019ppi} that this may be an indication of the importance of the higher twists in diffraction. Both analyses  \cite{Motyka:2012ty,Maktoubian:2019ppi} demonstrate that inclusion of higher twists significantly improves the description of the inclusive diffractive data.
We briefly discuss the higher twist contributions
and their impact on the description of the experimental data
in Sec.~\ref{sec:dipole_model}.

\subsection{Diffractive dijet and charm production}
\label{sec:diff_dijets}

 The power of the factorization lies in the observation that the parton densities are universal, and once extracted from one process can be used to make predictions for the rates of other hard processes. In the case of diffraction, the diffractive parton densities can be used to calculate other processes in rapidity gap DIS where the hard scale is present. At HERA, the diffractive parton densities were used to calculate the diffractive charm production and the dijet production both in DIS and in photoproduction. 

The factorization formula for the diffractive dijet production in DIS takes the following form
\begin{equation}
 \fl   d\sigma(e + p \rightarrow e+2{\rm jets}+X'+Y) = \sum_{i} 
    \int dt \int d\xi \int dz \, d\hat{\sigma}(e + i\rightarrow {e + \rm  2jets}) f_i^{D(4)}(z,\xi,\mu^2,t)
\label{eq:fact_diff_dijet}    
\end{equation}
where the sum is over the contributing partons of type $i$,  and $d\hat{\sigma}$ is the partonic cross section. The scale $\mu^2$ is the factorization scale, which is usually identified with the jet transverse energy scale and $X'$ denotes the part of the diffractive system $X$ which does not include the two jets.

\begin{figure}
\centerline{%
\includegraphics*[width=0.85\textwidth]{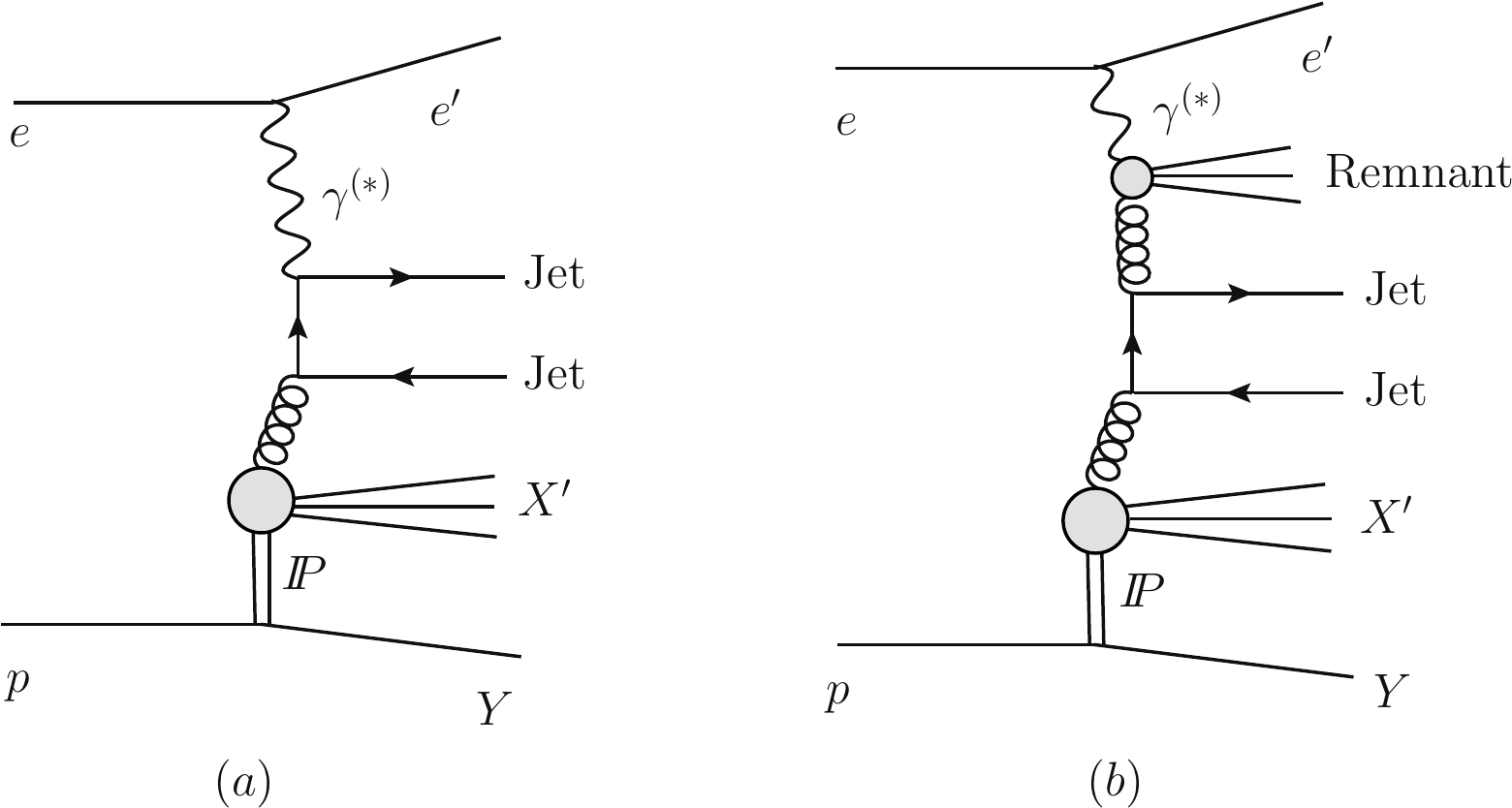}%
	}
\caption{Typical diagrams for diffractive dijet production in $ep$ scattering. 
Graphs $a$ and $b$ correspond to the direct and resolved photon contributions, respectively.}
\label{fig:photo_dijets_ep}
\end{figure}

In the case of dijet photoproduction, there are in general two ways the photon can interact with the partons in the proton, namely the {\em direct} and the {\em resolved} photon processes, see Fig.~\ref{fig:photo_dijets_ep}.
In the former case, the photon acts as a pointlike particle and  interacts with the parton from the target. An example is shown in graph ($a$) in Fig.~\ref{fig:photo_dijets_ep}, where the partonic interaction is the photon--gluon fusion which results in the quark -- antiquark pair in the final state. In this process the collinear factorization is expected to hold. 

In the second, resolved case, the photon may fluctuate into 
$q\bar{q}$ pairs and rather low mass hadronic states.  The low mass hadronic states are dominated by vector mesons. The parton from the photon will then interact with the parton from the proton,  see graph ($b$) in Fig.~\ref{fig:photo_dijets_ep}.
 In this case the interaction is more reminiscent of the hadron--hadron scattering. Therefore  the breaking of the collinear factorization is expected due to the fact that the soft interactions between the target proton and the resolved photon component are going to destroy the rapidity gap. 
 It is usually accounted for by introducing a model-dependent suppression factor (rapidity gap survival probability) either globally or only for the resolved photon contribution~\cite{Klasen:2002xb,Khoze:2000wk,Klasen:2010vk,Guzey:2016awf}. Note that the calculations at  next-to-leading
order (NLO) of QCD \cite{Klasen:1994bj} and beyond \cite{Klasen:2013cba} demonstrate that the 
direct and resolved processes are connected through the factorization of
collinear initial-state singularities. 
It is also possible that the resolved-photon suppression factor depends on
the parton flavor. 
Indeed, since the QCD factorization seems to hold for diffractive photoproduction of open charm, the suppression factor
should be introduced only for light parton flavors and should be significantly smaller compared to the situation,  when all channels are suppressed equally.
Moreover, it can also depend on the momentum fraction $x_{\gamma}$ \cite{Guzey:2016awf}. See also the discussion in the end of the current subsection.

The cross section for the dijet photoproduction in diffraction, neglecting any rapidity gap destruction effects, can be written as a convolution of the diffractive parton distribution functions, photon parton distribution functions and the hard partonic cross section~\cite{Klasen:1995ab,Klasen:1996it,Klasen:1997br,Klasen:2010vk,Klasen:2002xb}
\begin{eqnarray}
\fl    d\sigma(e + p \rightarrow e+2\,{\rm jets}+X'+Y) = \sum_{ij} \int\,  dy\,  f_{\gamma/e}(y) \int dx_{\gamma} \, f_{j/\gamma}(x_{\gamma},\mu^2) \times \nonumber \\
   \times \int dt \int d\xi \int dz \, d\hat{\sigma}(i+j\rightarrow {\rm 2\, jets}) \, f_i^{D(4)}(z,\xi,\mu^2,t)\, ,
\label{eq:fact_diff_dijet_photo}    
\end{eqnarray}
where $i$ and $j$ are parton flavors including the case when $i$ is the photon for the photon direct contribution. 
Here, $x_\gamma$ is the fraction of the longitudinal momentum of the photon carried by the parton entering the hard process and $y$ is the longitudinal momentum fraction of electron carried by the photon.
The functions $f_{j/\gamma}(x_{\gamma},\mu^2)$ are the parton distributions in the photon and $f_{\gamma/e}(y)$ is the
flux of equivalent photons calculated in the  Weizs\"aecker-Williams approximation~\cite{Budnev:1974de,Vidovic:1992ik}.
In the case of the direct photon process, the photon PDF reduces to the delta function $\delta(1-x_{\gamma})$. Finally, $d\hat{\sigma}$ is the partonic cross section for scattering of partons $i$ and $j$ and producing two jets.

Thus the diffractive dijet photoproduction offers  a unique bridge between the diffractive DIS, for large $Q^2$, where the factorization should hold,  and the hadron--hadron scattering, where the breakdown of the factorization happens due to the remnant--remnant interactions.

Both H1 and ZEUS performed measurements of diffractive dijets in DIS and photoproduction \cite{Chekanov:2007aa,Chekanov:2007rh,Aktas:2007bv,Aktas:2007hn,Chekanov:2009aa,Aaron:2010su,Aaron:2011mp,Andreev:2014yra,Andreev:2015cwa}. 
In the case of the dijets in DIS, both H1 and ZEUS data could be described very well when using H1 B and ZEUS C and SJ DPDF sets, respectively. This was not the case with H1 A and ZEUS S set, 
where
the main difference is the behavior of the gluon density at large fractional momenta. Thus the dijet data are more sensitive to this region than the inclusive data and can discriminate between different forms of the DPDFs in this region. We note that the ZEUS SJ fit was performed to the combination of the inclusive and diffractive data.

 The H1 data on dijet photoproduction \cite{Aktas:2007hn} show the suppression with respect to the theoretical predictions of about $50\%$. The overall shapes of the single differential cross sections are well reproduced when the calculations used H1 B set of DPDFs. However, the experiment did not observe the difference of the suppression of the resolved enriched ($x_{\gamma}<0.75$) and the direct enriched ($x_{\gamma}>0.75$) part of the cross section, contrary to the theoretical expectations. For the ZEUS data \cite{Chekanov:2007rh} the predictions tend to overestimate the data, but in general were consistent with no factorisation breaking
 within the large uncertainties of NLO calculations.

In \cite{Zlebcik:2011kq}  the analysis of the published H1 and ZEUS data was performed to better understand the difference between the different experiments. The conversion between the phase space of two experiments was done. It was demonstrated that the results were not very sensitive to the different photon structure functions. The different hadronization corrections implemented in H1 and ZEUS did not seem have an impact onto the results.

The factorization in diffractive DIS was also tested in the heavy quark production. 
Measurements of $D^{*}$ and open charm production in photoproduction and in DIS were performed by ZEUS and H1 \cite{Aktas:2006up,Chekanov:2007pm,H1:2017bnb}. The experimental data were well described by the NLO calculations. Thus the factorization is supported in charm production both in DIS and in $\gamma\,p$ interactions.
Note here that the leading twist production of charm jets contributes 
about  $40\%$ of the diffractive cross section for
transverse momenta 
$Q^2 \ll 4 p_t^2$ and
$p_t^2 \gg m_c^2$. So the validity of factorization for charm jets in photoproduction would require even larger breakdown of factorization for production of light quark jets than the one indicated  by the inclusive diffractive dijet data.
%%

%%%%%%%%%%%%%%%%%%%%%%%%%%%%%%%%%%%%

 \subsection{Predictions for EIC, LHeC, FCC-eh}
 \label{sec:future}

 There are several proposed future DIS machines that could explore the diffractive phenomena with much higher precision than at HERA as well as at higher energies. The prospects of the measurements of the inclusive diffraction and constraints  on the diffractive parton distribution functions have been studied in detail and presented in \cite{Armesto:2019gxy,Agostini:2020fmq,AbdulKhalek:2021gbh}, and selected results will be summarized below.
 
 At lower energy end, of about  $\sqrt{s} \sim 100\, \rm GeV $, the Electron-Ion Collider (EIC) in the US \cite{Accardi:2012qut} will be able to measure diffractive phenomena, covering a smaller kinematic region than HERA in $ep$ but a completely novel region in $e$A with respect to fixed target experiments where diffraction has been barely studied. Thanks to very high luminosity $10^{34}\, {\rm cm}^{-2} {\rm s}^{-1}$, and precise forward instrumentation EIC should be able to potentially reach higher values of $\xi$ and possibly measure the longitudinal diffractive structure function,  see \cite{Armesto:2021fws}.

 At higher energy end, there are two proposals for the electron-proton and electron-ion collisions at CERN.
 The Large Hadron-electron Collider (LHeC) 
is a project \cite{Dainton:2006wd, AbelleiraFernandez:2012cc, Klein:2018rhq, Agostini:2020fmq} 
that would
utilize the $7 \, \rm TeV$ proton beam from the High Lumi LHC and collide it 
with a $60(50)\, \rm GeV$ electron beam accelerated by an energy recovery  linac, 
thus reaching a centre-of-mass energy $\sqrt{s} = 1.3 \, \rm TeV$.  The instantaneous peak luminosity is projected  \cite{Bordry:2018gri, LHeClumi}   to reach also 
 $10^{34}\, {\rm cm}^{-2} {\rm s}^{-1}$, similar to EIC, and 
about three orders of magnitude higher than HERA. The projected running of the machine is over three periods. In the initial run period the total integrated luminosity is estimated to be 
$50 \; {\rm fb}^{-1}$. Throughout the 
entire  operation the LHeC is projected to reach $1 \,{\rm ab}^{-1}$ integrated luminosity.  
At yet higher energies, the next generation $ep$ collider would be the Future Circular 
Collider in electron-hadron mode (FCC-eh), utilizing the $50 \, \rm TeV$ proton beam 
from the FCC \cite{FCC_CDRv1,FCC_CDRv3} which would 
probe DIS at centre-of-mass energy of $\sqrt{s} = 3.5 \, \rm TeV$ 
with a total integrated luminosity of several ${\rm ab}^{-1}$. 
Both machines would also have capabilities to run in $e$A mode with center of mass energy  $\sqrt{s}=812 \, \rm GeV$ for LHeC and $\sqrt{s}=2.2 \, \rm TeV$ per nucleon for FCC-eh respectively. The projected integrated luminosity for the $e$A
 collisions would be of the order of $10 \, {\rm fb}^{-1}$.
 
%%%%%%%%%%%%%%%%%%%%%%%%%%%%%%%%%%%%
\begin{figure}
\centering{\includegraphics[width=12cm]{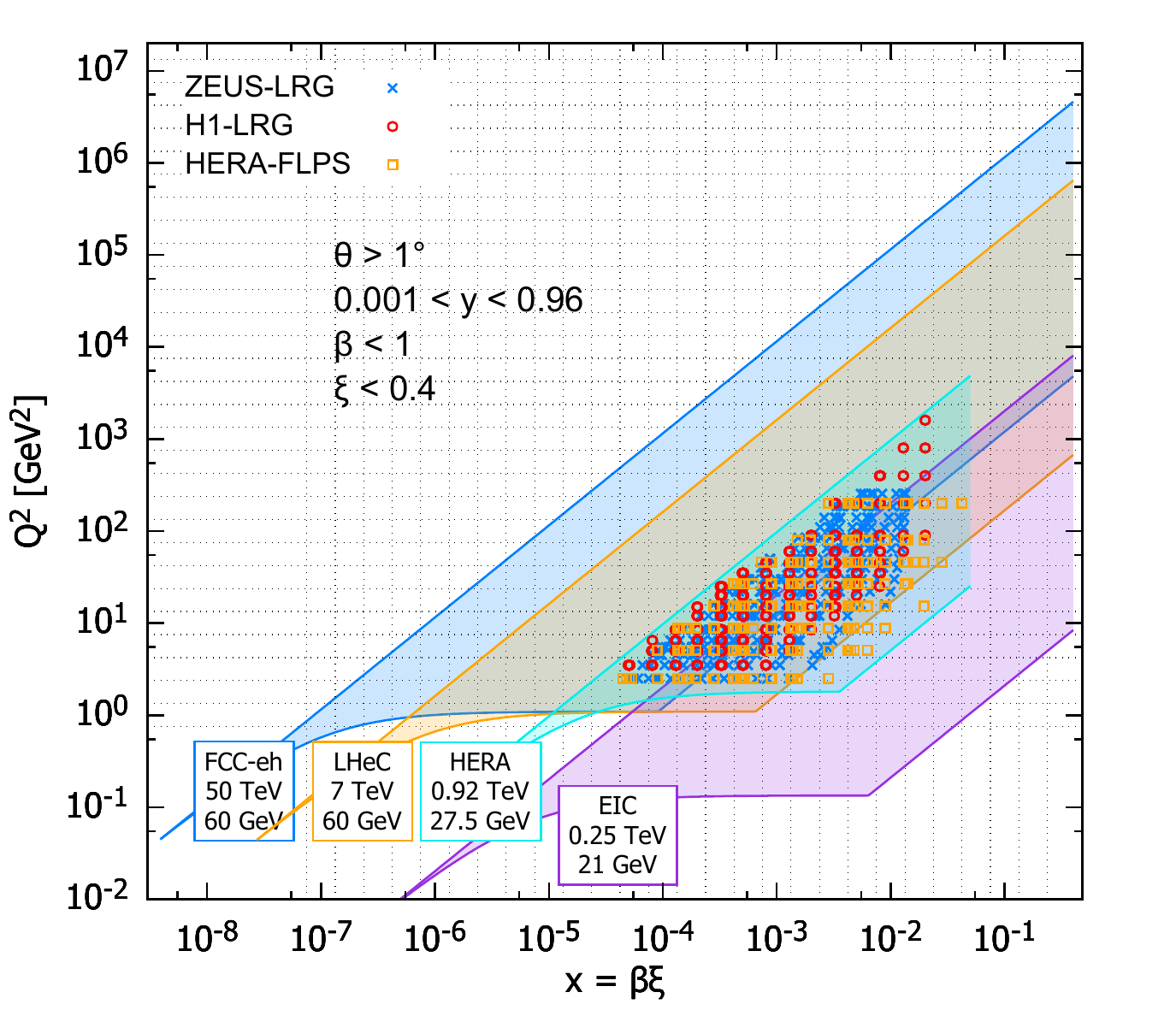}}
\caption{Kinematic phase space for inclusive diffraction in $(x,Q^2)$ for  the EIC (magenta region), the LHeC (orange region) and the FCC-eh (dark blue region) as compared with the HERA data (light blue region, ZEUS-LRG~\cite{Chekanov:2008fh}, H1-LRG~\cite{Aaron:2012ad}, HERA-FLPS~\cite{Aaron:2012hua}). The acceptance limit for the electron in the detector design has been assumed to be $ 1^{\circ}$, and we take $\xi<0.4$. Figure from \cite{Armesto:2019gxy}, {\tt https://doi.org/10.1103/PhysRevD.100.074022}.}
\label{fig:phasespace_xQ}
\end{figure}
%%%%%%%%%%%%%%%%%%%%%%%%%%%%%%%%%%%%

In Figure \ref{fig:phasespace_xQ} we show the accessible kinematic range in $(x,Q^2)$  for the four machines: HERA, EIC, LHeC and FCC-eh. The EIC region will  extend the range of HERA towards largest values of $x$ as well as lowest $Q^2$. For the LHeC design the 
range in $x$ is increased by a factor $\sim 20$ over HERA
and the maximum available $Q^2$ by a factor $\sim 100$. The FCC-eh machine would further increase this range with respect to LHeC by roughly one order of magnitude in both $x$ and $Q^2$. 

Simulations of the pseudodata were performed for these machines using extrapolations of the reduced cross sections with the ZEUS-SJ DPDFs, see Eq.~(\ref{eq:sigmared3}).
The pseudodata were generated with the errors given by the total error consisting of uncorrelated $5\%$ systematic error and the statistical error computed assuming luminosity of  $2 \, {\rm fb}^{-1}$.

\begin{figure}
\centering{\includegraphics*[width=12cm,trim=0 5 0 20]{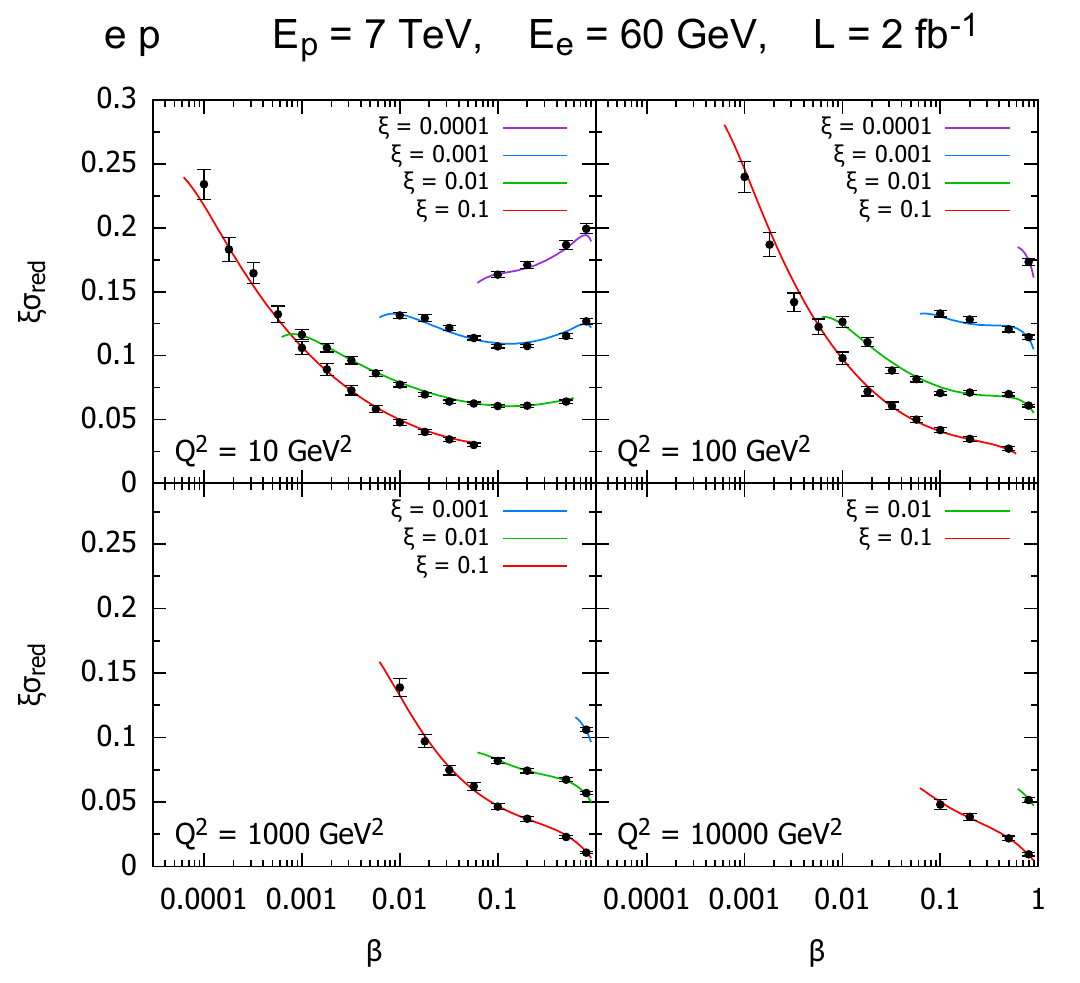}}
\caption{Selected subset of the  simulated data for the diffractive reduced cross section as a function of $\beta$ in bins of $\xi$ and $Q^2$ for $ep$ collisions at the LHeC.
The curves for $\xi = 0.01, 0.001, 0.0001$ are shifted up by 0.04, 0.08, 0.12, respectively. Figure from \cite{Armesto:2019gxy}, {\tt https://doi.org/10.1103/PhysRevD.100.074022}.
}
\label{fig:sigred_ep_lhec}
\end{figure}

In Fig.~\ref{fig:sigred_ep_lhec}  we show an example of the pseudodata simulation for the LHeC case. 
We show a subset of the simulated data for the diffractive reduced cross section 
$\xi\sigma_{\rm red}$ as a function of $\beta$ in selected bins of $\xi$ 
and $Q^2$. The errors are very small and are dominated by the systematics. 

%%%%%%%%%%%%%%%%%%%%%%%%%%%%%%%%%%%%
%%%%%%%%%%%%%%%%%%%%%%%%%%%%%%%%%%%%

Using these pseudodata one can estimate the experimental precision with which the diffractive parton densities can be extracted. In \cite{Armesto:2019gxy} detailed studies were performed, where the simulated data were used to predict possible constraints on the DPDFs.
This analysis demonstrated that the  DPDFs determination accuracy improves with respect to 
HERA by a factor of 5--7 for the LHeC and 10--15 for the FCC-eh.

\begin{figure}
\centering{\includegraphics*[width=10cm,trim=0 0 0 54]{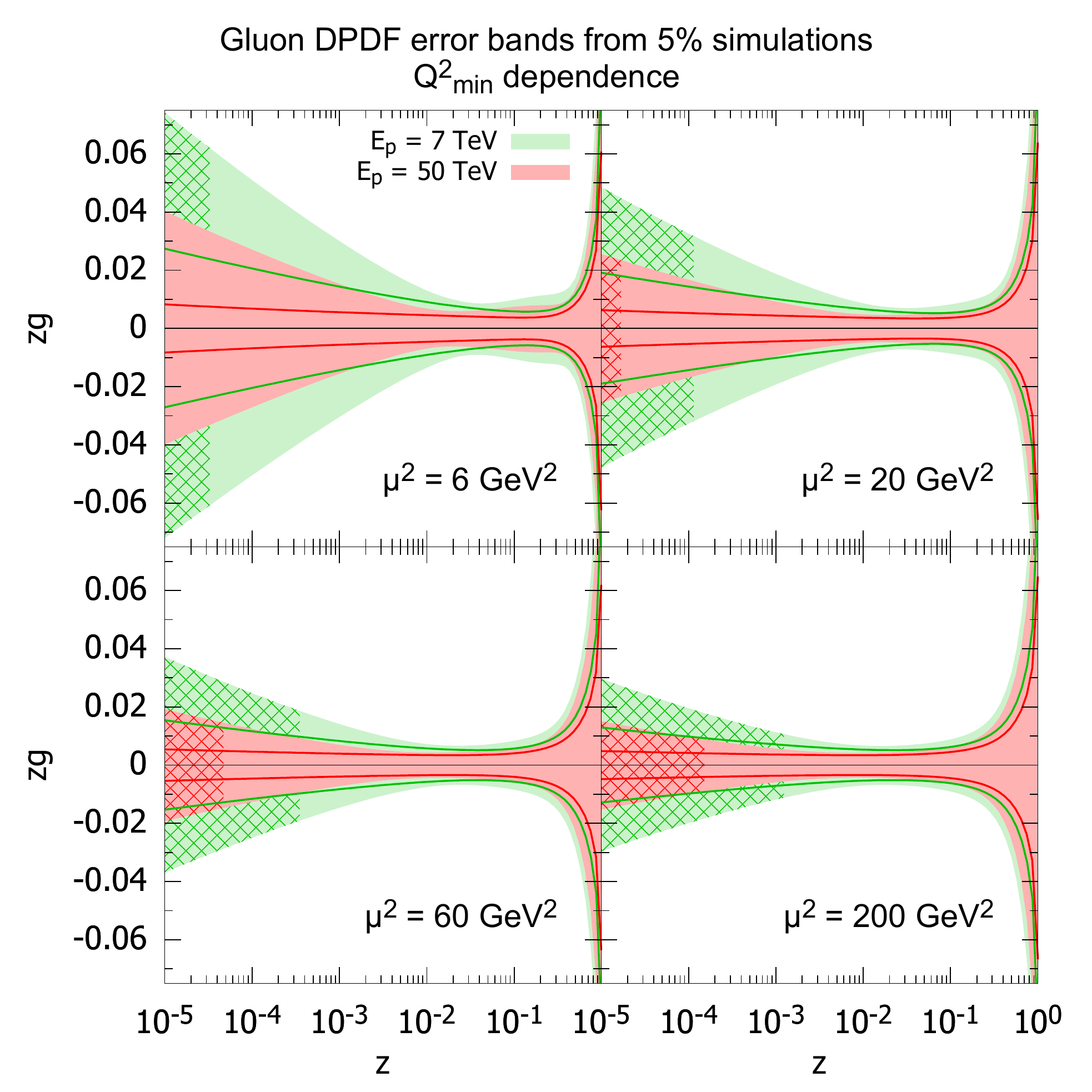}}
\caption{Relative uncertainties on the diffractive gluon PDF extraction for 
four distinct scales $\mu^2=6,20,60,200 \, \rm GeV^2$. 
The bands correspond to the choice of the high cut-off on the  
data included in the fit $Q^2_{\rm min}=5 \,\rm GeV^2$ and 
the lines correspond to the lower choice $Q^2_{\rm min}=1.8 \,\rm GeV^2$. 
The green colour corresponds to the LHeC scenario and red to the FCC-eh scenario.
The cross-hatched areas show kinematically excluded regions. The bands indicate only the experimental uncertainties, see the text. Figure from \cite{Armesto:2019gxy}, {\tt https://doi.org/10.1103/PhysRevD.100.074022}.}
\label{fig:pdf_qHL}
\end{figure}

As an example, in Fig.~\ref{fig:pdf_qHL} relative uncertainties on the diffractive PDFs are shown for the LHeC and FCC-eh  as a function of the longitudinal momentum fraction $z$ and the scale $\mu^2$. 
In particular we show the variation of the relative precision of DPDFs
with the change of the minimal value of $Q^2$ from $1.8 \ {\rm GeV^2}$ 
(curves) to $5\,\rm GeV^2$ (bands).   The LHeC scenario is indicated in green and 
FCC-eh in red. 
There is a quite substantial effect on the achieved 
precision depending on the minimal value of  $Q^2$,  and it is clear from the figure that  that both machines will be very sensitive to the low $Q^2$ region and  therefore potentially able to constrain higher twists and/or saturation effects.

%%%%%%%%%%%%%%%%%%%%%%%%%%%%%%%%%%%%
%%%%%%%%%%%%%%%%%%%%%%%%%%%%%%%%%%%%

Diffractive dijet photoproduction in $ep$ and $eA$ scattering in the EIC kinematics using next-to-leading order (NLO) of QCD was considered in Ref.~\cite{Guzey:2020gkk}. The analysis established the kinematic reaches in most important kinematic variables
for various beam types, energies and kinematic cuts.
It also showed that the EIC will provide new information on DPDFs 
of the proton, probe novel nuclear diffractive PDFs, and may illuminate 
the mechanism of factorization breaking.

Dijets can also be measured in ultraperipheral collisions (UPC) at the LHC, see discussion in Sec.~\ref{sec:UPC}.

 \section{Dipole model in diffraction}
\label{sec:dipole_model}

\subsection{Introduction}
\label{sec:dipole_intro}

The diffractive phenomena in DIS can also be described using the dipole model of QCD
 \cite{Mueller:1989st}. In that approach, when viewed from the target
rest frame,  the virtual photon fluctuates into a
quark--antiquark pair, which has large longitudinal distance to evolve, and then subsequently interacts with the target \cite{Gribov:1968gs,Bjorken:1973gc,Frankfurt:1988nt}. On average, the distance over which this pair forms is about $\sim 1/(x \, m_N)$ \cite{Ioffe:1969kf}.  The  size of the configuration close to the interaction point in impact parameter space of the dipole depends on the polarization of the photon. For the case of the longitudinaly polarized photon,
it is of the order of $\sim 1/Q^2$ for $q\bar{q}$, where $Q^2$ is the minus photon virtuality. In the case of the transversely polarized photon this size becomes larger due to the so-called `aligned jet' configurations (see discussion later in this section). In the aligned jet configuration, the photon converts into $q\bar q$ with small transverse momenta (e.g., 0.3 GeV/c) and over a distance $1/(x\,m_N )$ it evolves into a hadronic size configuration.

The $q\bar{q}$ pair size in impact parameter space can also  become larger due to the subsequent parton emissions.  
The lowest order correction to the $q\bar{q}$ dipole stems from the single gluon emission and can be computed in the perturbative QCD, provided the strong coupling is sufficiently small. Both the quark--antiquark dipole and the quark--antiquark--gluon state can be described using the lightcone wave functions. The gluon  emission is higher order in the strong coupling, but it is important for the description of the diffractive states with higher masses $M_X^2$, or equivalently in the region of small $\beta$. From the   perspective of the DGLAP evolution the gluon emissions are automatically resummed through the evolution equation in $\ln Q^2$, and thus in such a case, at large $Q^2$, there is  more than one gluon emission taken into account.  

The dipole picture has certain advantages and disadvantages over the standard collinear approach.
The advantage in that it is very convenient to include the unitarization corrections. The latter ones are important, particularly at high energy and low virtualities $Q^2$ and they are expected to play more prominent role in the diffractive cross sections than in the inclusive ones due to the fact that the diffractive cross sections are squared in the amplitudes whereas the inclusive cross sections are proportional to the imaginary part of the scattering amplitude.
Thus the use of dipole model opens up the possibility of convenient  incorporation of unitarization and some of the higher twist contributions into the  description of diffraction.

 The dipole picture has also  some limitations, namely  it is suitable when the center-of-mass energy is high, i.e., Bjorken $x$ sufficiently small, and when the ratio of $M_X^2/Q^2$ is not extremely large.
Also, it parametrically underestimates the strength of double (triple ...) scattering of a small dipole 
%vg of
with
2 (3...) nucleons leading to diffraction for the longitudinal $\sigma_L(x,Q^2)$ cross section being a higher twist effect in difference from the factorization theorem expectations.

\subsection{Diffraction in dipole model}
\label{sec:diff_dipole}

In the dipole approach the diffractive  structure function $F_2^{D}$ can be expressed as sum of the contributions from the transverse  and longitudinally polarized virtual photons.
Usually the two-component model is considered with $q\bar{q}$
and $q\bar{q}g$ contributions  \cite{Buchmuller:1996xw,Buchmuller:1998jv,Wusthoff:1997fz,Kowalski:2008sa}.
The structure function can thus be expressed as the sum of the following terms
\begin{equation}
F_2^{D(3)}(\xi,\beta,Q^2) \; = \; F_T^{q\bar{q}}+ F_L^{q\bar{q}}+F_T^{q\bar{q}g} \; .
    \label{eq:dipolef2}
\end{equation}
The longitudinal contribution from $q\bar{q}g$ state is usually neglected since it has no leading logarithm in $Q^2$  \cite{Golec-Biernat:2001gyl}.
The $q\bar{q}$ components for transversely and longitudinally polarized photons have the following expressions \cite{Wusthoff:1997fz}
\begin{eqnarray}
    \xi F_T^{q\bar{q}}(\xi,\beta,Q^2) \; = \; \frac{3Q^4}{64 \pi^4 \beta B_d} \sum_f e_f^2 \int_{z_f}^1 dz z(1-z) \times \nonumber \\
    \times \left\{ [z^2+(1-z)^2]Q_f^2 \phi_1^2+m_f^2 \phi_0^2\right\} \; ,
    \label{eq:qqbartransverse}
\end{eqnarray}
and
\begin{equation}
  \xi F_L^{q\bar{q}}(\xi,\beta,Q^2) \; = \; \frac{3Q^6}{16 \pi^4 \beta B_d} \sum_f e_f^2 \int_{z_f}^1 dz \, z^3 (1-z)^3 \, \phi_0^2 \; .
    \label{eq:qqbarlongitudinal}
\end{equation}
In the above $f$ denotes quark flavors, $m_f$ is the quark mass, $B_d$ is the diffractive slope stemming from the $t$ integration assuming an exponential form,
and $z$ is the photon momentum fraction carried by a quark (antiquark) in the dipole.
In addition, the variables used in the above are defined as
\begin{equation}
z_f = \frac{1}{2}\left(1-\sqrt{1-\frac{4 m_f^2}{M_X^2}}\right), \;\;\; Q_f^2=Q^2z(1-z)+m_f^2 \; ,
    \label{eq:zf}
\end{equation}
as well as the functions $\phi_i$ being defined as
\begin{equation}
\phi_i = \int_0^{\infty} dr r \, K_i(Q_f r) \, J_i(k_f r) \, \hat{\sigma}(\xi,r)\;\;\; ,
    \label{eq:phii}
\end{equation}
with $k_f=\sqrt{z(1-z)M_X^2-m_f^2}$, and $J_i$ and $K_i$ are the Bessel functions. We recall that $\xi=(Q^2+M_X^2) / (Q^2+W^2)$ and $\beta= Q^2 / (Q^2+M_X^2)$.
The function 
$\hat{\sigma}(\xi,r)$ in Eq.~(\ref{eq:phii})
is the {\em dipole  cross section}  which contains all the necessary information about   the interaction between the dipole and the proton. It depends on the dipole size $r$ and it has energy dependence through the dependence on $\xi$. It can be calculated from theory or modeled phenomenologically as we shall discuss below.

The formulae (\ref{eq:qqbartransverse}) and (\ref{eq:qqbarlongitudinal}) were derived in the case when the two gluons were exchanged between the color dipole  and the proton. All possible couplings were added up in order to retain the gauge invariance property. The result was obtained in the form of the $k_T$ factorization formula, with the unintegrated gluon density describing the details of the target. Finally, the  formulae were rewritten in terms of the dipole cross section which can be related to the unintegrated gluon density. 
The  $q\bar{q}g$ component is given by \cite{Buchmuller:1996xw,Wusthoff:1997fz}
\begin{eqnarray}
   \xi F_T^{q\bar{q}g}(\xi,\beta,Q^2)=\frac{81 \beta \alpha_s}{512 \pi^5 B_d} \sum_f e_f^2 \int_\beta^1  \frac{dz}{(1-z)^3} \left[ \big(1-\frac{\beta}{z}\big)^2+\big(\frac{\beta}{z}\big)^2\right] \nonumber \\
   \times \, \int_0^{(1-z) Q^2} dk^2 \, \log\bigg(\frac{(1-z)Q^2}{k^2}\bigg) \phi_2^2 \; ,
    \label{eq:qqbargtransverse}
\end{eqnarray}
where the function $\phi_2$ is defined as
\begin{equation}
\phi_2 = k^2\int_0^{\infty} dr r\,  K_2\left(\sqrt{\frac{z}{1-z}}k r\right) \, J_2(k r) \,  \hat{\sigma}(\xi,r) \; ,
    \label{eq:phi2}
\end{equation}
with $J_2$ and $K_2$ Bessel functions.

The contribution of Eq.~(\ref{eq:phi2})  is of course higher order in $\alpha_s$. The $q\bar{q}g$ term in Eq.~(\ref{eq:qqbargtransverse}) was computed in the approximation when the transverse momenta are strongly ordered 
$$
k_{Tq}\simeq k_{T\bar{q}} \gg k_{T g} \; .
$$
In the large $N_c$ approximation, it can be effectively treated as the gluonic color dipole. Thus a factor $C_A/C_F$ must be included in order to rescale  the interaction with respect to the $q\bar{q}$ dipole by changing the scattering amplitude by the factor 9/4, for details see \cite{GolecBiernat:2008gk}.

%%%%%%%%%%%%%%%%%%%%%%%%%%%%%%%%%%
\subsection{Models for the dipole cross section}
\label{sec:dipole_model_xsection}

The dipole cross section can be  obtained from theory or it can be modeled phenomenologically. One can broadly classify the dipole models into several main categories: those  based on DGLAP evolution, obtained from non-linear evolution equations at small $x$, and phenomenological parametrizations.

In the limit of small dipole size, and in the leading logarithmic approximation, accounting for the terms $\alpha_s \ln Q^2/\Lambda^2_{QCD}$, there is an important relation between the dipole cross section $\hat{\sigma}(x,r)$ and the collinear gluon density $xg(x,\mu^2)$ which was derived in \cite{Blaettel:1993rd,Frankfurt:1993it} 
\begin{equation}
\hat{\sigma}(x,r) \; = \; \frac{ \pi^2}{3} r^2 \left[\alpha_s(\mu^2) xg(x,\mu^2)\right]_{\mu^2 = C/r^2} \; \; ,
    \label{eq:dipole_xsection_gluon_density}
\end{equation}
where the coefficient $C$ can be estimated based on matching to $\sigma_L$ (cross section for longitudinally polarized photons) or phenomenologically to match large and small size contributions smoothly, see \cite{Nikolaev:1994ce,Frankfurt:1995jw}. The above formula has an important property that the dipole cross section vanishes for small dipole sizes -- the {\it color transparency} phenomenon. However, since the cross section is proportional to the gluon density, the interaction can become strong at sufficiently small $x$.  
Namely, the probability of interaction at a fixed impact parameter can reach values close to
unity.

The dipole cross section can also be derived in the limit of high energy including the unitarization corrections in the form of parton saturation effects.
Parton saturation is a phenomenon expected to occur at very high energy or equivalently at very small value of Bjorken $x$, when the parton density is very high and gluon recombination is expected to occur \cite{Gribov:1984tu,Mueller:1985wy}, for a comprehensive review of low $x$ and saturation phenomena see Ref.~\cite{Kovchegov:2012mbw}. This effect leads to the modification of the gluon evolution equations to include the nonlinear terms. There are various approaches to saturation: Color Glass Condensate (CGC) \cite{McLerran:1993ka,McLerran:1993ni,McLerran:1994vd,Iancu:2000hn,Iancu:2001ad,Ferreiro:2001qy,JalilianMarian:1996xn,JalilianMarian:1997dw,JalilianMarian:1997gr,JalilianMarian:1997jx},  Balitsky hierarchy \cite{Balitsky:1995ub,Balitsky:1998ya}, and Balitsky--Kovchegov equation \cite{Kovchegov:1999ua,Kovchegov:1999yj}; we shall describe more details of these approaches and the low $x$ physics in Sec.~\ref{sec:lowx_resum}.

\begin{figure}
\centerline{%
	\includegraphics*[width=0.6\textwidth]{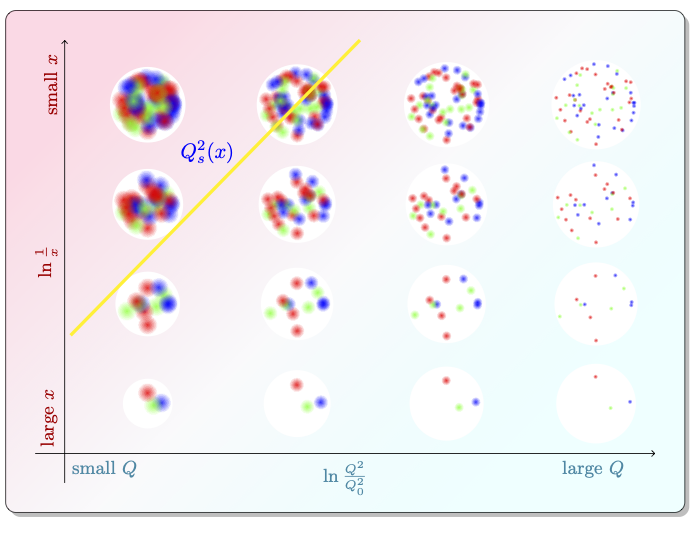}%
}
\caption{Schematic representation of parton evolution $(x,Q^2)$ plane. The dots symbolize partons, with their number density increasing towards decreasing $x$ and the resolution in transverse size, the latter one  decreasing when scale $Q^2$ is increased.  Yellow diagonal line indicates  the saturation scale $Q_s^2(x)$ which divides the dilute regime where the linear evolution is applicable (to the right of the line) and the dense regime where the nonlienar evolution needs to be taken into account (to the left of the line).}
\label{fig:satplot}
\end{figure}

  The fundamental property of the parton saturation is the presence of the dynamically generated saturation scale $Q_s(x,b)$ which depends on Bjorken $x$, impact parameter as well as the target \cite{Gribov:1984tu,Mueller:1985wy,McLerran:1993ka,McLerran:1993ni}. To be precise it can depend on the mass number $A$ of the nucleus. It divides the dilute and dense partonic regimes, i.e., the regions where the parton density is low and the evolution is linear, and the region where the parton density is high and the evolution needs to be supplemented by the nonlinear terms which tame the growth of the scattering amplitude.
The schematic view of the dilute and dense regions together with the saturation scale $Q_s$ is illustrated in Fig.~\ref{fig:satplot} (in the case averaged over impact parameter).

Using the solution to the non-linear Balitsky--Kovchegov equation, numerous fits to HERA data were performed. The fits were performed in the leading logarithmic order approximation with running coupling \cite{Albacete:2010sy}, including resummation in the form of kinematical constraints \cite{Iancu:2015joa,Ducloue:2019jmy} as well as very recently by taking into account NLO corrections to  the dipole wave function \cite{Beuf:2020dxl} for light quarks.

%%%%%%%%%%%%%%%%%%%%%%%%%%%%%%%%%%%%%%%
%%%%%%%%%%%%%%%%%%%%%%%%%%%%%%%%%%%%%%%

An alternative, phenomenological way to include the unitarization corrections is to eikonalize the formula Eq.~(\ref{eq:dipole_xsection_gluon_density}), see for example \cite{Bartels:2002cj}

\begin{equation}
    \hat{\sigma}(x,r) \; = \; \sigma_0[1-\exp(\frac{ \pi^2}{3\sigma_0} r^2 \left[\alpha_s(\mu^2) xg(x,\mu^2)\right]_{\mu^2 = C/r^2}) ] \; \; .
       \label{eq:dipole_xsection_gluon_density_eik}
\end{equation}
In the limit of small dipole sizes and when $\alpha_s(\mu^2) xg(x,\mu^2)$ is not large the eikonalized formula can be expanded and it coincides with  Eq.~(\ref{eq:dipole_xsection_gluon_density}). Otherwise in the limit of large dipole sizes and small $x$ it saturates to a constant value set by the $\sigma_0$. 

The above model can be thought of as an extension of the  Golec-Biernat--Wusthoff (GBW) model \cite{GolecBiernat:1998js,GolecBiernat:1999qd} that includes the  DGLAP evolution.
The GBW model  posseses qualitative features that can be found when solving the Balitsky--Kovchegov equation. It has  been very successful in the description of the inclusive HERA data on structure function $F_2$ as well as the diffractive structure function. The consistency of the simultaneous description of the non-diffractive and diffractive data is a very important test for the dipole models.
The GBW parametrization  \cite{GolecBiernat:1998js,GolecBiernat:1999qd} has the following form for
the $q\bar{q}$ dipole cross section
\begin{equation}
\hat{\sigma}(\xi,r) \; = \; \sigma_0 \big[ 1- \exp(-r^2 Q_s^2(\xi) /4) \big] \; ,
    \label{eq:gbw}
\end{equation}
where the normalisation constant was found to be $\sigma_0 = 29.12 \; {\rm mb}$ from the fit. The saturation scale $Q_s$ is given by
\begin{equation}
Q_s^2(\xi) = Q_0^2 \, \bigg( \frac{\xi}{x_0}\bigg)^{-\lambda} \; ,
    \label{eq:qsat}
\end{equation}
with $Q_0^2=1 \, {\rm GeV^{-2}}$ $x_0=4\cdot 10^{-5}$ and $\lambda=0.277$.

By comparing Eq.~(\ref{eq:gbw}) with Eq.~(\ref{eq:dipole_xsection_gluon_density_eik}) we observe that the saturation scale in the GBW model approximates, roughly speaking, the product of the gluon density
and the coupling constant
\begin{equation}
 \alpha_s(\mu^2) xg(x,\mu^2) \rightarrow Q_s^2(x) \; .
\end{equation}

More recently, fits based on the DGLAP improved saturation model and  Eq.~(\ref{eq:dipole_xsection_gluon_density_eik}) were performed  \cite{Golec-Biernat:2017lfv} to the  HERA data. It was demonstrated, not surprisingly, that good description of the HERA data at values of $Q^2>50 \rm \; GeV^2$ can only be achieved by including the DGLAP evolution in the form of the dipole model (\ref{eq:dipole_xsection_gluon_density_eik}).

 An example of the  model which has been successfully used to describe the inclusive $F_2(x,Q^2)$ is based on the analysis of the semi-analytical solution to the nonlinear Balitsky-Kovchegov equation.
The CGC parametrisation  of the dipole scattering amplitude (see discussion below) is given by the following  model \cite{Iancu:2003ge, Marquet:2007nf,Soyez:2007kg}
\begin{equation}
N(\xi,r,b) \; = \; S(b) \, {\cal N}(\xi,r)\; ,
    \label{eq:cgcamplitude}
\end{equation}
here $S(b)$ is the impact parameter profile $S(b) = \exp(-b^2/(2 B_d))$ with the diffractive slope $B_d$ taken from the HERA data $B_d=6 \, \rm GeV^{-2}$ \cite{Aktas:2006hx} (see discussion later in this section about the impact parameter dependence). Upon the integration of the scattering amplitude~(\ref{eq:cgcamplitude}) over the impact parameter $b$, the resulting dipole cross section reads
\begin{equation}
\hat{\sigma}(\xi,r) \; = \; 4 \pi B_d \, {\cal N}(\xi,r)\; ,
    \label{eq:cgc}
\end{equation}

With the value of the diffractive slope from HERA this gives $4\pi B_d=29 \, \rm mb$. The functional form of the CGC dipole cross section is motivated by the analysis of the approximated solution to the BK equation.
Following \cite{Soyez:2007kg} it reads
\begin{equation}
\fl
{\cal N}(\xi,r) = \left\{\begin{array}{ll}
{\cal N}_0 \bigg( \frac{rQ_s}{2}\bigg)^{2\gamma_s} \, \exp(\frac{2 \ln^2(rQ_s/2)}{\kappa \lambda \ln(\xi)}) & { \rm for} \;  rQ_s \le 2 \; ,\\
1-\exp(-4 \alpha \ln^2(\beta r Q_s)) & {\rm for} \;  rQ_s > 2 \; .
\end{array} \right.
\end{equation}    
  The fitted parameters in the saturation scale $Q_s^2(\xi)=Q_0^2 (x_0/\xi)^{\lambda}$ are  $\lambda=0.22$, $x_0=1.63 \cdot 10^{-5}$ (with $Q_0^2=1 \, {\rm GeV^2}$ being fixed). The parameters  $\alpha=0.615$, $\beta=1.006$ are chosen such that the amplitude ${\cal N}$ and its first derivative are continuous at point  ${\cal N}_0=0.7$. The  parameter $\kappa=9.9$ was taken
  from the approximated analytical solution to the non-linear BK equation. In the fit above with the heavy quarks and parameter $\gamma_c=0.7376$ was also fitted, though other fits presented in \cite{Soyez:2007kg} were using the value which can be obtained from the approximated analytical solution as well.

  %%%%%%%%%%%%%%%%%%%%%%%%%%%%%%%%%%%%%%%%
 \begin{figure}
\centerline{%
	\includegraphics*[width=0.45\textwidth]{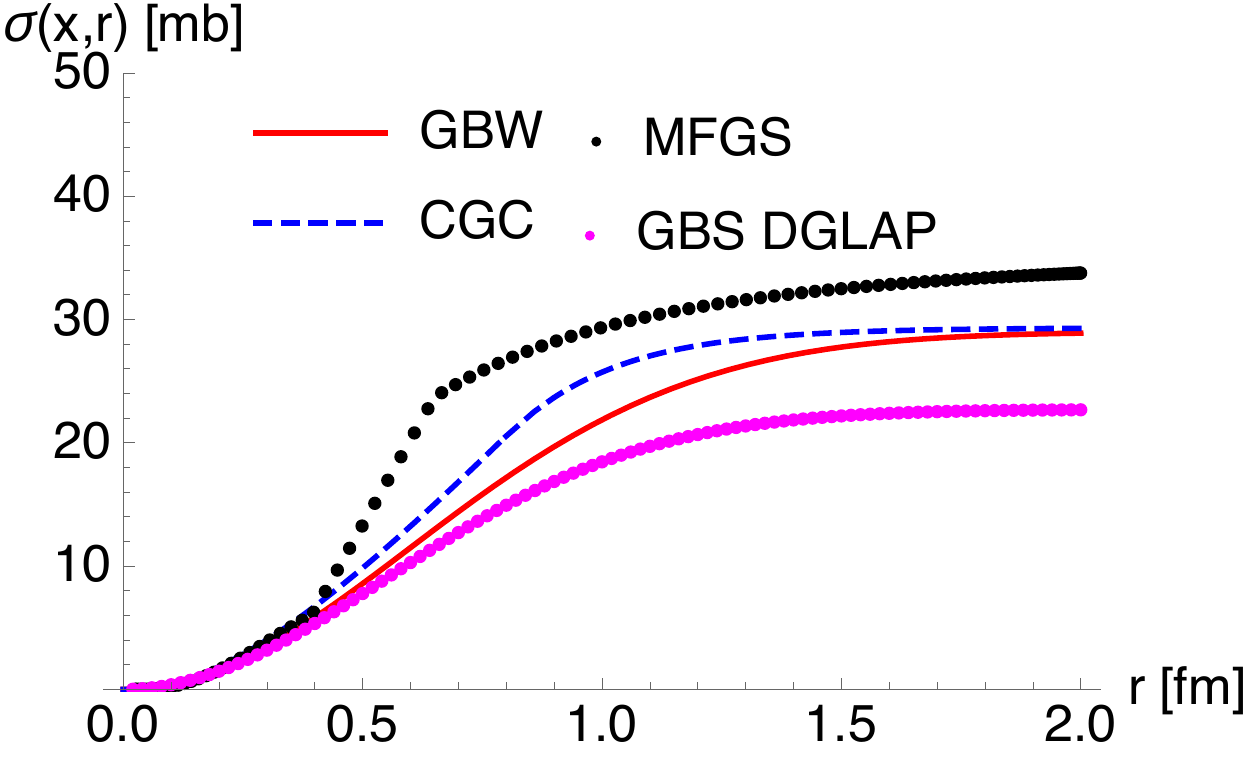}%
	\includegraphics*[width=0.45\textwidth]{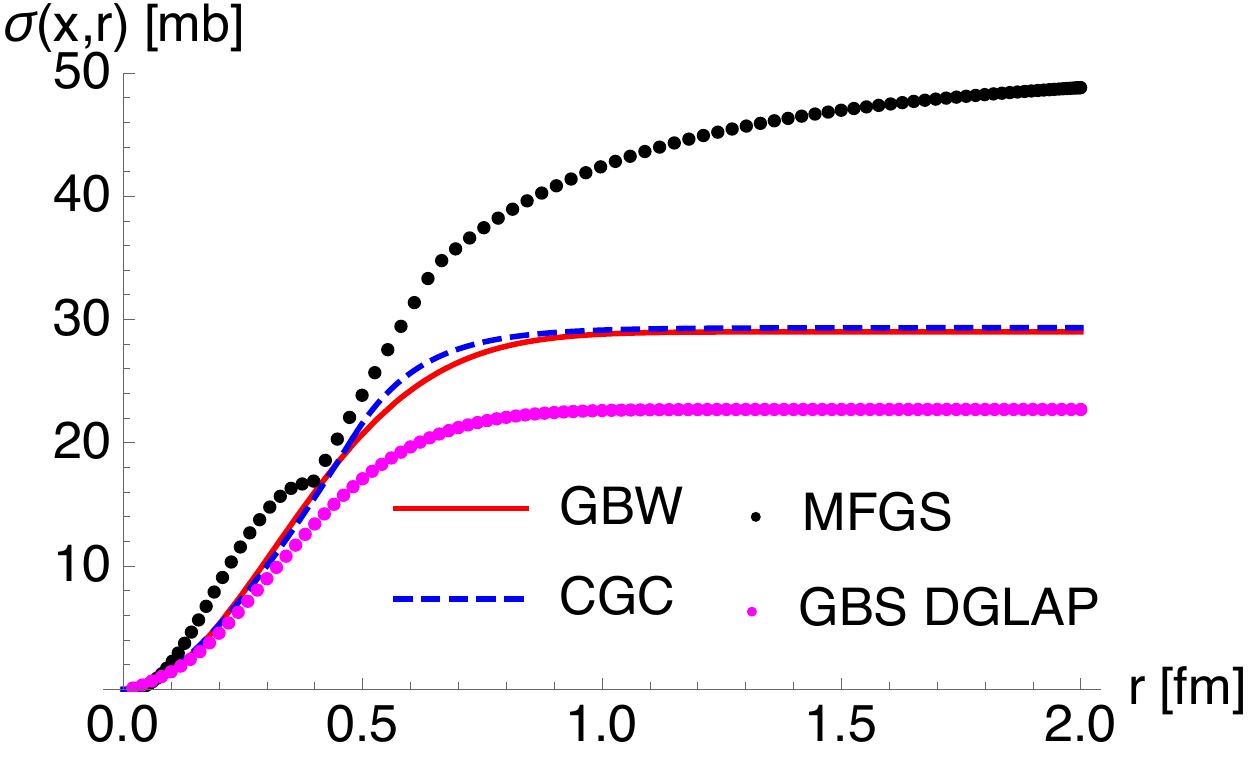}%
}
\caption{The dipole cross section in units of $\rm mb$ as a function of the dipole size in $\rm fm$ for two different values of $x=10^{-2}$ and $10^{-4}$ (left  and right panel respectively). Four models are shown: GBW, red solid,  \cite{GolecBiernat:1998js}; GBS + DGLAP, magenta points, \cite{Golec-Biernat:2017lfv}; CGC, blue dashed, \cite{Soyez:2007kg}; MFGS, black points, \cite{McDermott:1999fa}.}
\label{fig:dipole}
\end{figure}
 %%%%%%%%%%%%%%%%%%%%%%%%%%%%%%%%%%%%%%%%%%%
  In Fig.~\ref{fig:dipole} several models have been collected  for the dipole cross sections and shown for comparison.  The dipole cross sections are shown as a function of the dipole size and for two different values of $x=10^{-2}$ and $x=10^{-4}$. The models presented are GBW \cite{GolecBiernat:1998js} Eq.~(\ref{eq:gbw}), GBS with DGLAP \cite{Golec-Biernat:2017lfv} Eq.~(\ref{eq:dipole_xsection_gluon_density_eik}), MFGS \cite{McDermott:1999fa} and CGC \cite{Iancu:2003ge,Marquet:2007nf,Soyez:2007kg}.  We see that the dipole cross section grows with increasing values of the dipole size and for large values of dipole sizes the dipole cross section saturates to a constant value for GBW, GBS+DGLAP and CGC. In the case of the MFGS model there is a residual increase at large dipole sizes which is modeled by the soft Pomeron behavior $\sim x^{-0.08}$.  When $x$ is decreased the dipole cross section becomes larger overall. Characteristically it becomes saturated for lower values of the dipole size when $x$ is smaller. This is typical feature of the dipole models, and the transition between the region where the dipole cross section is small and large is given by the saturation scale, as discussed above.  
 Note that in Fig.~\ref{fig:dipole} the modelled dipole cross sections are similar in the perturbative domain, whereas there are sizeable differences in the non-perturbative regime, which could be attributed to different treatment of quark masses in the wave function.
  
  The dipole cross section depends on energy (through longitudinal momentum fraction) and on the dipole size. It is related to the imaginary part of {\em dipole scattering amplitude} through
  \begin{equation}
 \hat{\sigma}(x,{\bf r}) \; = \; 2 \int d^2 {\bf b} \, N(x,{\bf r},{\bf b}) \; ,
      \label{eq:dipole_scattering_amplitude}
  \end{equation}
  for the interaction of the dipole of size ${\bf r}$ interacting with the target at  impact parameter ${\bf b}$.
  In the above we put back explicitly dependence on two-dimensional vectors ${\bf r}$ and ${\bf b}$. Note that, in principle the amplitude depends on the relative angle between the two vectors.

  It is the dipole scattering amplitude that should satisfy the unitarity constraint
  \begin{equation}
      N \le 1 \;.
      \label{eq:unitarity}
  \end{equation}
  
  An interesting limit in the context of hadronic interactions at high energies is the {\em black disk limit}.
  It is defined as the case when the probability of inelastic interaction is equal to unity for all impact parameters less than the size of the target $R$. This corresponds to the condition 
  \begin{equation}
      N(x,{\bf r},{\bf b}) \; = \; \theta(R-b) \; .
      \label{eq:black_disk_limit}
  \end{equation}
  In this limit the inelastic cross section is equal to elastic one, and both of them equal to half of the total cross section. 
  
  In general, one can define the saturation scale from the dipole scattering amplitude for example through the condition
  \begin{equation}
  N(x,r=1/Q_s,{\bf b}) = {\rm const.} \; ,
      \label{eq:sat_scale_ampltiude}
  \end{equation}
  where the $\rm const.$ is of the order of $1/2$. The above condition defines $Q_s(x,b)$ as a function of $x$ and impact parameter. 
The solution to the nonlinear Balitsky--Kovchegov equation  \cite{Balitsky:1995ub,Balitsky:1998ya,Kovchegov:1999ua,Kovchegov:1999yj} satisfies the unitarity condition (\ref{eq:unitarity}), see Sec.~\ref{sec:lowx_resum_sat}.

Another form the dipole scattering amplitude that satisfies the unitarity constraint in QCD, has been introduced by Mueller in \cite{Mueller:1989st}
\begin{equation}
    N(x,{\bf r},{\bf b}) = 1-\exp\big[ -\frac{\pi^2}{2 N_c}r^2 \alpha_s(\mu^2) xg(x,\mu^2) T({\bf b})\big] \; ,
    \label{eq:ggm}
\end{equation}
with the scale $\mu$ defined as in Eq.~(\ref{eq:dipole_xsection_gluon_density}) and $T({}\bf b)$ is the profile function for the target. 
This expression can be viewed as generalization of Eq.~(\ref{eq:dipole_xsection_gluon_density_eik}) by including the dependence on the impact parameter ${\bf b}$.
Equation~(\ref{eq:ggm}) is usually refered to as the Gribov--Glauber--Mueller model \cite{Mueller:1989st}, which takes into account the multiple scatterings of the dipole off the target. The original idea goes back to Glauber \cite{Glauber:1955qq,Franco:1965wi,Glauber:1970jm} who treated it in a quantum-mechanical model and to Gribov  \cite{Gribov:1968jf,Gribov:1968gs} in the context of photon and hadron--nucleus scattering. Gribov expressed amplitudes for hadron--nucleus scattering through the diffractive (elastic plus inelastic) amplitudes of interaction with one, two, three, $\dots$, nucleons in the nuclei, see Sec.~\ref{sec:shadowing}.

Note that in the discussed eikonal approximation~(\ref{eq:ggm}), the rescattering terms corresponding to the interaction with $n$ nucleons are proportional to $r^{2n}$. Hence they are higher twist terms. Thus,
this approximation neglects the contribution of multiparton configurations responsible for the leading twist diffraction and leading twist nuclear shadowing, see Sec.~\ref{sec:shadowing}
and Ref.~\cite{Frankfurt:2002kd}.

  \subsection{Distribution of dipole sizes}
  \label{sec:dipole_dsitribution}
  
  For a given value of the photon (minus) virtuality $Q^2$, the photon wave function will have a distribution of the dipole configurations of different sizes $r$.  It is instructive  to investigate this distribution in the case of diffraction as compared to the inclusive cross section.
   The inclusive cross section in the dipole model is given by 
\begin{equation}
    \sigma_{T,L}^{\gamma^ * p}(x,Q^2) \; = \; \sum_f \int d^2 {\bf r} \, dz |\Psi_{T,L}^f({ r},Q,z)|^2\,\hat{\sigma}(x,{ r}) \; ,
    \label{eq:dipole_xsec_incl}
\end{equation}
  where the photon wave functions \cite{Bjorken:1970ah,Nikolaev:1990ja,Nikolaev:1991et} have the following form for the transverse case
  \begin{equation}
  |\Psi_{T}^f({r},Q,z)|^2 \; = \; \frac{3 \alpha_{\rm em}}{2 \pi^2} e_f^2 \left\{  [z^2+(1-z)^2]Q_f^2 K_1^2(Q_f r)+m_f^2 K_0^2(Q_fr) \right \} \; , 
      \label{eq:photon_T}
  \end{equation}
  and for the longitudinal photon case
   \begin{equation}
  |\Psi_{L}^f({ r},Q,z)|^2 \; = \; \frac{3 \alpha_{\rm em}}{2 \pi^2} e_f^2 \left\{  4 Q^2 z^2(1-z)^2 K_0^2(Q_f r)\right  \} \; .
      \label{eq:photon_L}
  \end{equation}

\begin{figure}
\centerline{%
	\includegraphics*[width=0.47\textwidth]{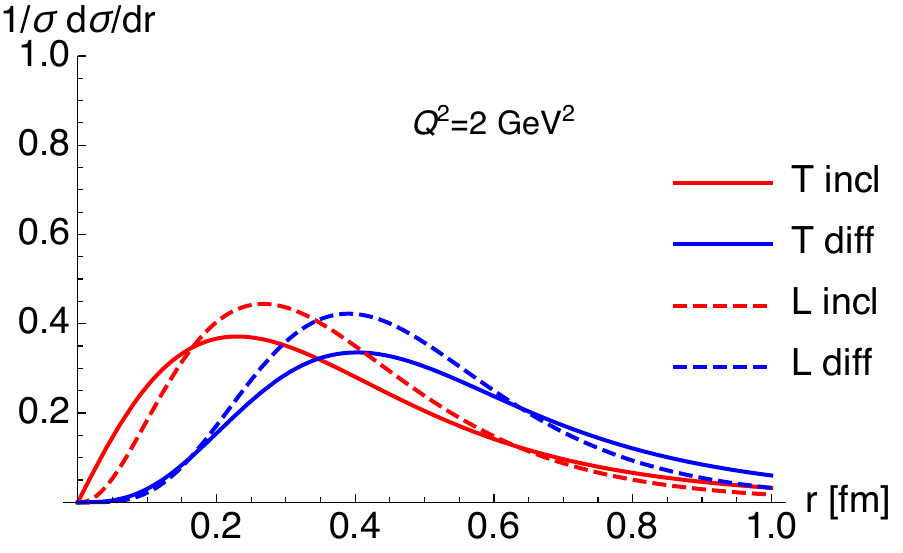}%
	\hspace*{0.5cm}
	\includegraphics*[width=0.47\textwidth]{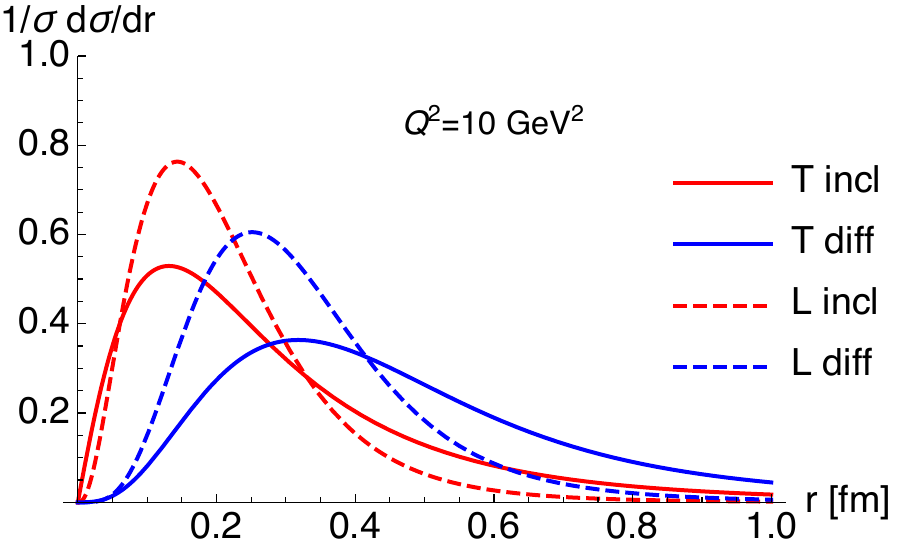}%
}
\caption{ The normalized dipole size distribution for the inclusive (red) and diffractive (blue) cross section using the GBW dipole cross section. Both transverse (solid) and longitudinal (dashed) polarizations are shown. Two values of photon virtuality are chosen : $Q^2=2 \ {\rm GeV^2}$ (left plot) and $Q^2=10 \; {\rm GeV^2}$ (right plot). $x$ was fixed so that $Q_s(x) =1 \rm GeV$ and masses  $m_u=m_d=m_s=0.14 \;  \rm GeV$. }
\label{fig:dipole_size_incl_Q2}
\end{figure}

 In Fig.~\ref{fig:dipole_size_incl_Q2}  we show the distribution of the dipole sizes for the case of the dipole cross section from the GBW model for two different values of $Q^2=2, 10 \rm \; GeV^2$ (left and right plot respectively). In particular we plot the function $1/\sigma_{T,L}^{\gamma^ * p} \;  p_{T,L}(r,x,Q^2)$ which  is defined as 
 \begin{equation}
   \sigma_{T,L}^{\gamma^ * p}(x,Q^2) \; = \; \int_0^{\infty}  dr \, p_{T,L}(r,x,Q^2) \; .
   \label{eq:p_integrand}
 \end{equation}

   \begin{figure}
 \centerline{%
 	\includegraphics*[width=0.47\textwidth]{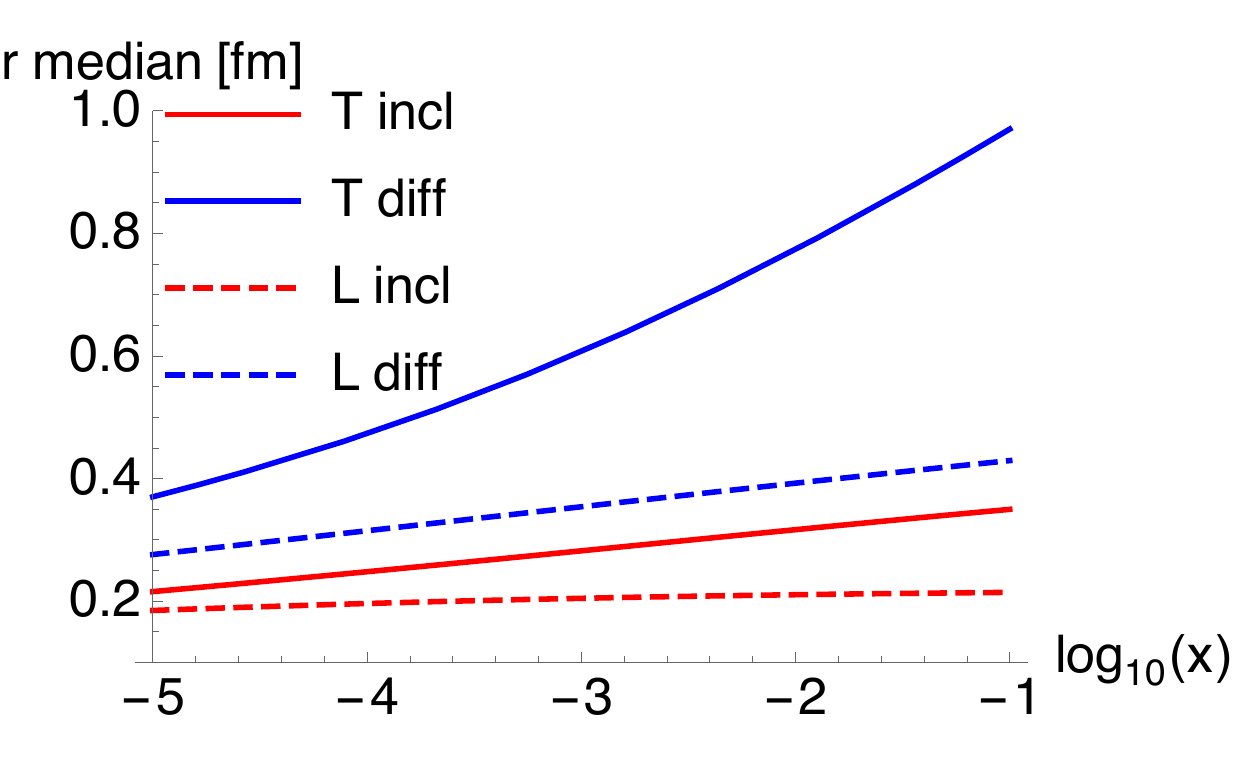}%
 	\hspace*{0.5cm}
 	\includegraphics*[width=0.47\textwidth]{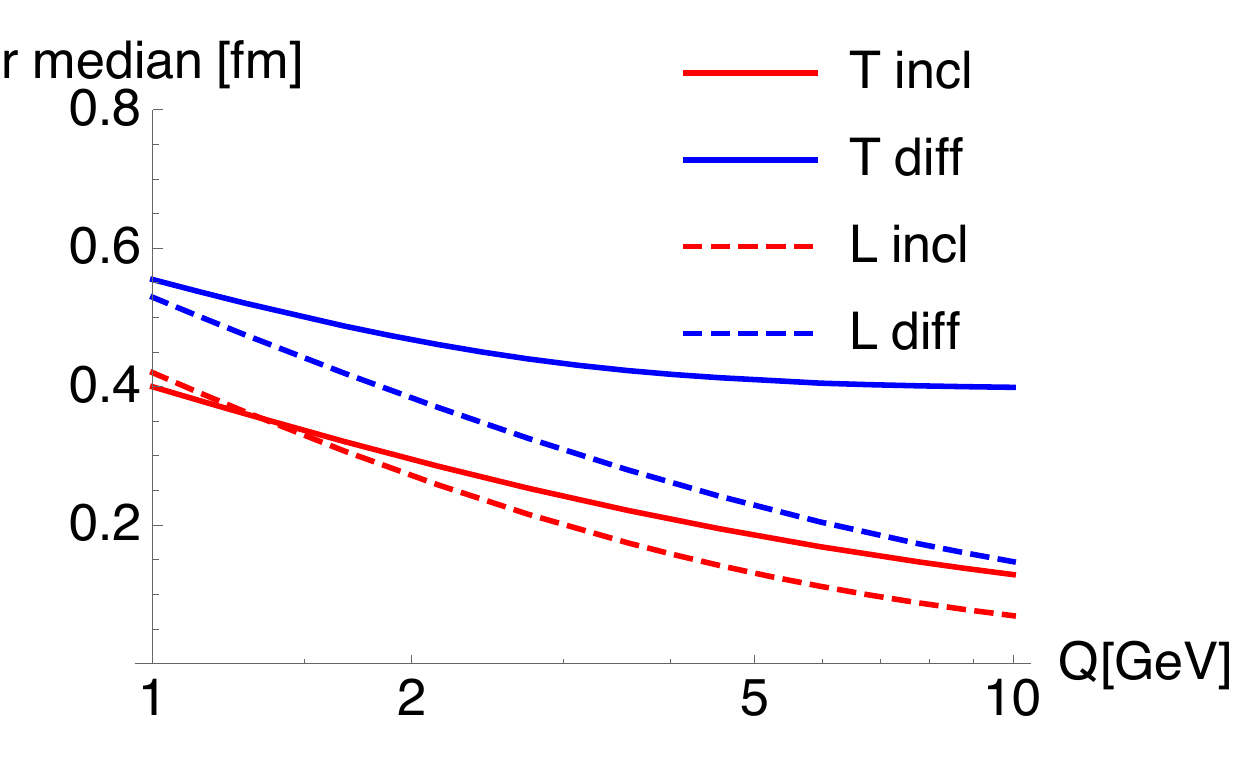}%
 }
 \caption{ The median size distribution for inclusive and diffractive cross section, based on the GBW model \cite{GolecBiernat:1998js,GolecBiernat:1999qd}, both longitudinal and transverse photon polarization cases. Left plot: function of $x$, fixed $Q^2= 10 \; {\rm GeV}^2$. Right plot: function of $Q$, fixed $x=x_0=0.4 \times 10^{-5}$ such that $Q_s(x_0)=1 \, \rm GeV$ in the GBW model. }
 \label{fig:dipole_size_median}
 \end{figure}

Solid lines denote the dipole size distribution for the transverse photon and the dashed lines for the longitudinal polarization.
Two values of $Q^2$ are chosen to illustrate the change in the distribution from larger to smaller sizes.  
  We observe that the distribution for transversely polarized photons has a longer tail, extending to larger values of dipole sizes $r$ as compared to the longitudinal one. This is due to the presence of the aligned jet configurations in the $F_T(x,Q^2)$ structure function originating from the endpoint configurations $z=0$ and $z=1$. 
  
  This effect is also illustrated in Figs.~\ref{fig:dipole_size_median}, where the median of the distribution is plotted as a function of $x$ and $Q$ (left and right plot respectively).  The median size $r_{\rm m}$, defined 
  by the following relation
  \begin{equation}
    \frac{\int_0^{r_{\rm m}} dr\, p_{T,L}(r,x,Q^2)}{\sigma_{T,L}^{\gamma^{\ast} p}}=\frac{1}{2} \,.
    \label{eq:rm}
  \end{equation}
One can see from the figure that $r_{\rm m}$
  is a slowly increasing function of $x$ due to the fact that the GBW dipole cross section becomes larger at  small $r$ when $x$ is decreased, and it always saturates to a constant at large $r$. Thus the dipole size distribution shifts to smaller values of $r$ when $x$ decreases. In the case of the $Q$ dependence, the median dipole size decreases with increasing $Q$. This is connected with the fact that the peak of the integrand in the  cross section moves to smaller values of $r$ with increasing value of $Q$ due to the effect of the photon wave function. The different dependence for the case of longitudinal and transverse photons is evident. Note that in Fig.~\ref{fig:dipole_size_median} we show the dipole sizes up to values of $x\simeq 0.1$ which is most likely  beyond the region of validity of the dipole model. This  region is shown for purely illustrative purposes to demonstrate the behavior of the model extrapolation.
  
 We also analyze this distribution for the case of the diffractive cross section. In the approximation of high energy, when $\xi$ is small, such that $\beta \simeq 1$, one can rewrite the formula for the cross section  (\ref{eq:qqbartransverse}) and (\ref{eq:qqbarlongitudinal}) using the substitution of $\xi \simeq x$ inside the dipole cross section and obtain  the following  form for the diffractive cross section, see for example \cite{Kovchegov:1999kx}
  \begin{equation}
    \sigma_{T,L}^{\gamma^ * p (D)}(x,Q^2) \; = \; \frac{1}{16 \pi B_d} \sum_f \int d^2 {\bf r} \, dz |\Psi_{T,L}^f({ r},Q,z)|^2\,\hat{\sigma}^2(x,{r}) \; .
    \label{eq:dipole_xsec_diff}
\end{equation}
 We note that only a $q\bar{q}$ component is included in this approximation.
  The above formula can be interpreted as the realization of the Good--Walker idea~\cite{Good:1960ba}, which states that the diffraction occurs due to the different absorption of 
  eigenstates of the interaction operator (scattering matrix).
  In the context of the dipole picture at small $x$, these are the $q\bar{q}$ dipoles with definite values of $r$ and $z$. 
  
 At this point, it is important to clarify the connection between the Good-Walker formalism and the dipole model. As mentioned above, the quark--antiquark wave functions corresponding to different dipole cross sections are orthogonal, which explicitly realizes the assumption of the Good--Walker model designed for soft diffractive processes, see discussion in Sec.~\ref{sec:fluctuations}. At the same time, in contrast to this model, orthogonality of scattering eigenstates does not hold for scattering at finite $t \neq 0$ since elastic scattering mixes dipoles of different sizes.
Thus, it is not clear whether one can build an orthogonal set of $q\bar q $ and $q\bar q g$, etc.~dipole states even for $t=0$. As a result, while the effects of quantum evolution and the quark--antiquark--gluon Fock states are important for the description of diffraction~\cite{Hatta:2006hs}, their connection to
the formalism of Good--Walker eigenstates in general cannot be established.

 The distributions analogous 
 to Eq.~(\ref{eq:p_integrand}) 
for the case of diffraction (using Eq.~(\ref{eq:dipole_xsec_diff}))  are shown in Figs.~\ref{fig:dipole_size_incl_Q2} and are indicated by the blue lines. As compared with the inclusive case, the distributions are wider, with the peak shifted to larger values of $r$.  This is also evident in the median plots, Fig.~\ref{fig:dipole_size_median} where the median size  shifts substantially to larger dipole sizes.  Significant difference between the cases of transverse and longitudinally polarized photons is evident.
This has physical consequence that the diffraction is sensitive more to the soft physics.
This qualitatively explains the smaller $\alpha_\pom(0)$ value for diffraction than the inclusive case.
The broader distribution for the transverse case is even more prominent in diffractive case.

  The fact that diffraction is dominated by the larger dipole sizes than the inclusive process has important implications for the saturation. As showed in \cite{GolecBiernat:1999qd} saturation of the dipole cross section is very important for diffraction, with the saturation scale $Q_s(x)$ playing a  role of a regulator which reduces the   contribution to the cross section from the infrared regime. As a result, as shown in  \cite{GolecBiernat:1999qd} the ratio of the diffractive to inclusive cross section is approximately constant (modulo logarithmic corrections) for the dipole model with saturation (when only $q\bar{q}$ contributions are considered). 
  
  Note, however, that the conclusion of Ref.~\cite{GolecBiernat:1999qd} is sensitive to modeling of the dipole cross section at large $|r|$, where it is taken to be energy independent. 
  In this approach, $N(x,r,{\bf b})$ first grows with a decrease of $x$, but then starts to decrease so that the $N(x,r,{\bf b})=1$ limit is never reached. On the other hand, in Ref.~\cite{McDermott:1999fa}, where the dipole cross section at large $|r|$ is motivated by soft 
  Pomeron exchange, the ratio of the diffractive to inclusive cross sections grows with an increase of energy.
  
It is worth mentioning that the general considerations about the distributions are valid for large values of $Q^2$. When $Q^2$ is small, like in the case of Fig.\ref{fig:dipole_size_incl_Q2} (left plot) 
 and \ref{fig:dipole_size_median} (right plot, low $Q$ region), the scales related to the non-zero quark masses have a non-negligible effect on the  distributions. Effectively, the larger masses
 tend to shift slightly the dipole size distribution towards smaller values of dipole sizes
 and increase the value of $\hat{\sigma}(x,r)$ to keep the same magnitude of $F_2(x,Q^2)$.

\begin{figure}
\centerline{%
	\includegraphics*[width=0.7\textwidth,trim=0 150 0 100]{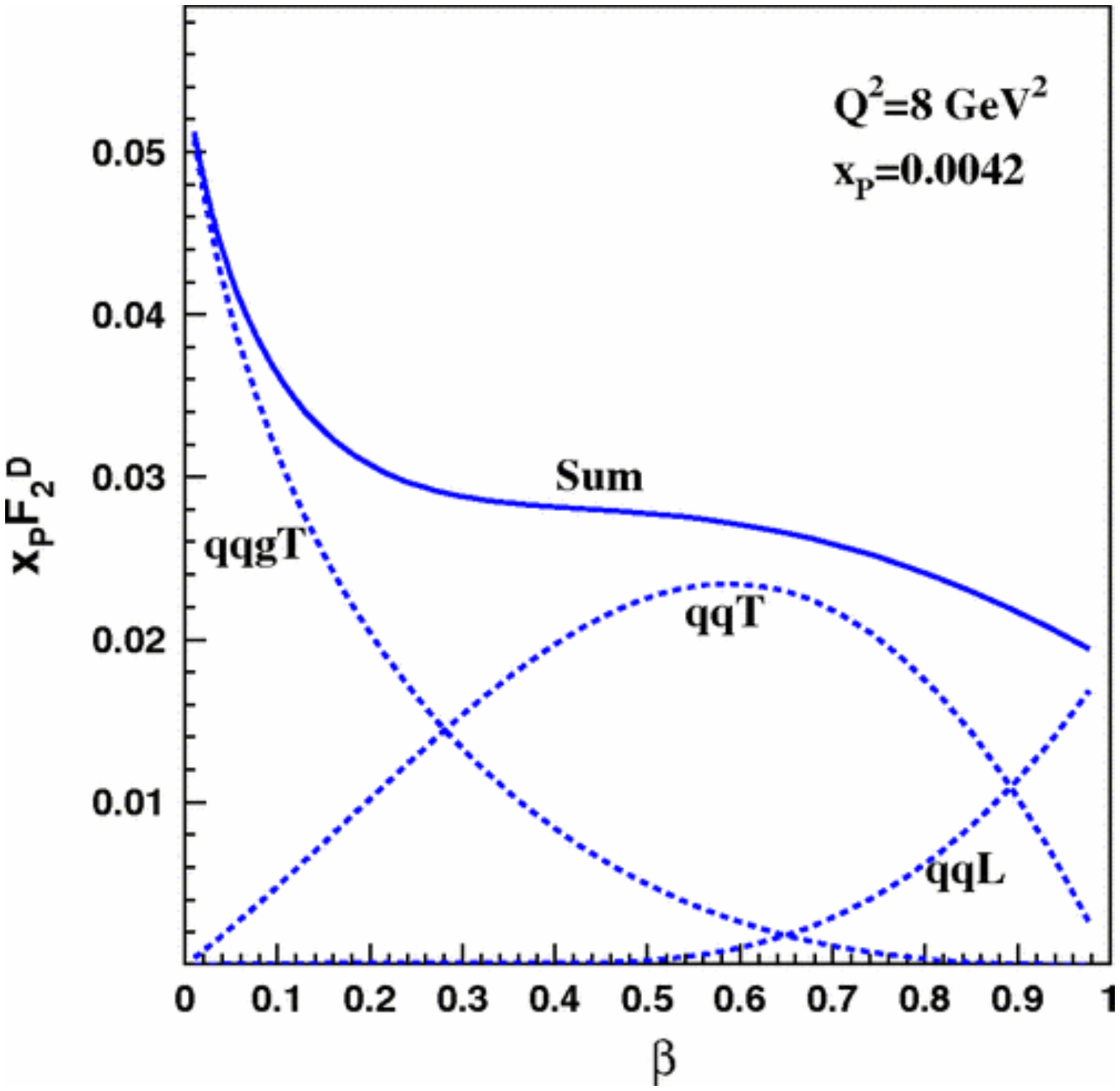}%
}
\caption{The diffractive structure function $F_2^{D(3)}$ from the dipole model based on the GBW parametrisation of the dipole cross section. Dotted lines indicate the components $q\bar{q}$, transverse and longitudinal and the $q\bar{q}g$ transverse component. Solid line indicates the sum of all three contributions. Figure from \cite{GolecBiernat:2008gk}, https://doi.org/10.1103/PhysRevD.79.114010.}
\label{fig:f2dipolecomp}
\end{figure}

%%%%%%%%%%%%%%%%%%%%%%%%%%%%%%%%%%%%%%%%%%%%%%%%%%%  
\subsection{Diffractive structure function from dipole model and higher twists}
\label{sec:dipole_diff_structure_function_ht}

  In Fig.~\ref{fig:f2dipolecomp} the diffractive structure function $\xi F_2^{D(3)}$ is shown
  as a function of $\beta$ for fixed value of $Q^2=8 \; \rm {\rm GeV^2}$ and $\xi=0.0042$. In addition to the total value, three components as given by  Eq.~(\ref{eq:dipolef2}) are separately shown. We observe that they dominate the cross section in different regions of diffractive masses, see \cite{Kowalski:2008sa,GolecBiernat:2008gk}. The $F_T^{q\bar{q}}$ dominates for $\beta \sim 1/2$ which corresponds to $M_X^2 \sim Q^2$, $F_L^{q\bar{q}}$ dominates for large $\beta \sim 1$, that is small diffractive masses $M_X^2 \ll Q^2$ , and the $F_T^{q\bar{q}g}$ is most important for large diffractive masses $M_X^2 \gg Q^2$, corresponding to  $\beta \ll 1$. By analyzing the formulae for $q\bar{q}$ contribution from transverse and longitudinal cases one can demonstrate that in the small mass limit $M^2\ll Q^2$, the longitudinal part dominates, even though it is suppressed by an additional power of $1/Q^2$. This stems from the fact that the transverse part of the cross section is dominated by the aligned jet configurations which in turn involves  large distances in the dipole size $r$ (see analysis above). As a result, in the small diffractive mass limit, this component is suppressed. On the other hand the longitudinal cross section is dominated by the symmetric configurations of $z\simeq 1/2$, and the cross section is dominated by the small dipole sizes. Thus the diffractive cross section is dominated by the contribution from the longitudinally polarized photons in the small mass $M_X^2\ll Q^2$ or large $\beta \rightarrow 1$ limit.
  Note that the same holds for inclusive vector meson production, see Sect.~\ref{sec:vm1}.

\begin{figure}
\centerline{%
	\includegraphics*[width=0.75\textwidth,trim=0 250 0 250]{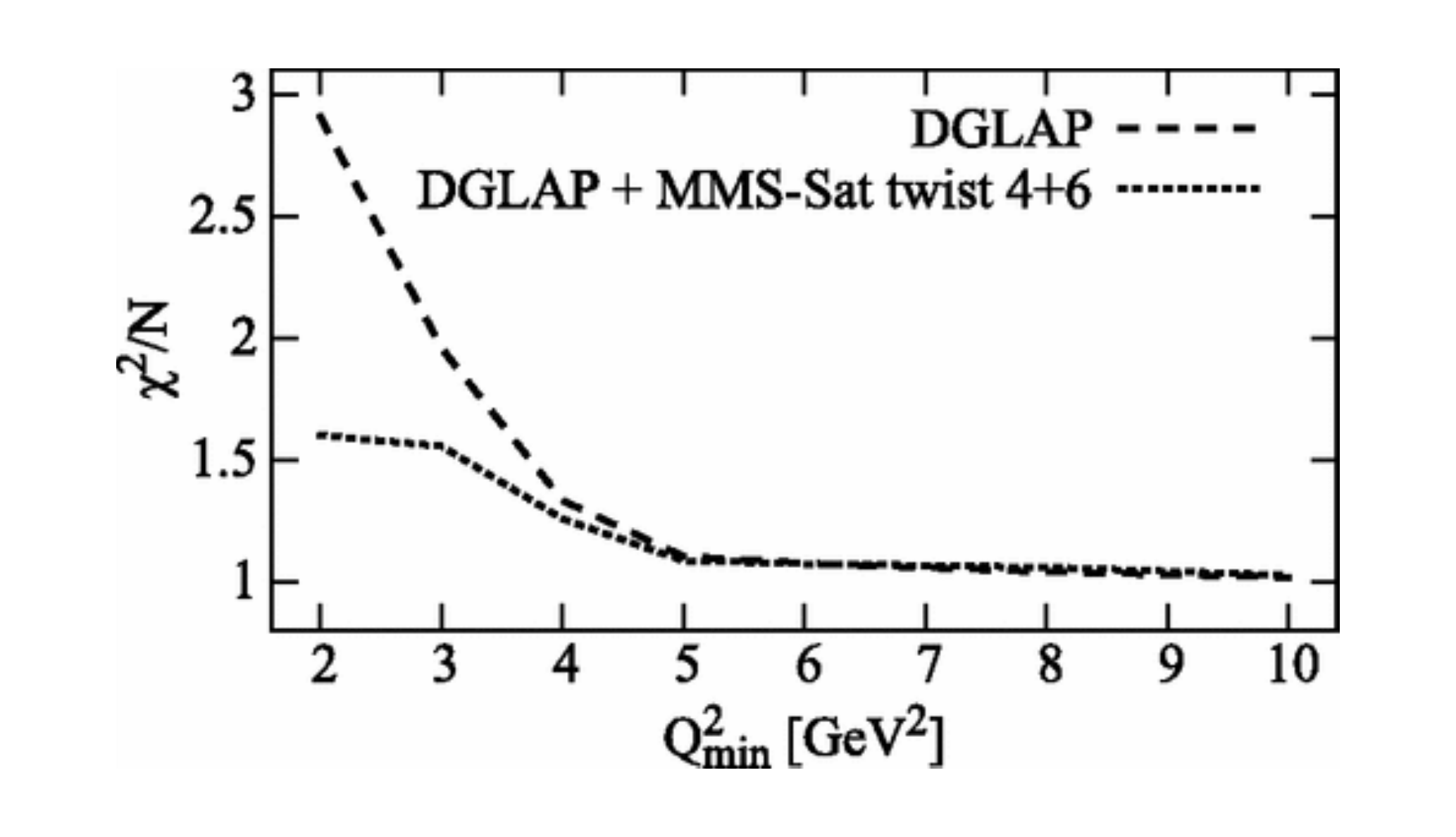}%
}
\caption{ Variation of the $\chi^2$ of the fit to the diffractive structure function as a function of the cutoff $Q^2_{\rm min}$, where only data for $Q^2 > Q^2_{\rm min}$ were included in the fit.  DGLAP fit - dhased line, DGLAP supplemented with the higher twists from the dipole model MMS \cite{Marquet:2007nf,Munier:2003zb}, solid line. Figure from \cite{Motyka:2012ty}, {\tt https://doi.org/10.1103/PhysRevD.86.111501}.}
\label{fig:chi2}
\end{figure}
 
Fits to the diffractive data using the dipole model  were performed  in \cite{GolecBiernat:2008gk}, where good description of the data was obtained. 
The dipole model with parametrization described above contains towers of higher twists effects which go beyond the leading twist DGLAP approach. As mentioned in the previous section, the DGLAP description of the diffractive inclusive data is inadequate 
for low values of $Q^2$. To be precises the fits are inadequate for  $Q^2 < 5 \; \rm GeV^2$ for ZEUS data and $Q^2 < 8.5 \; \rm GeV^2$ for H1 data. The fits deteriorate in a low $Q^2$ region where the $\chi^2$ deviates by $100\%$. The problem is illustrated in Fig.~\ref{fig:chi2}
where the value of $\chi^2$ is shown as a function of $Q^2_{\rm min}$, the latter being defined as the cutoff of the data used in DGLAP fits. To be precise, the fits were done with subset of the ZEUS LRG data with $Q^2> Q_{\rm min}^2$. The NNLO order analysis does not cure this problem, though it was limited to the same form of the initial parametrizations for the diffractive structure function.

%%%%%%%%%%%%%%%%%%%%%%%%%%%%%%%%%%%%
\begin{figure}
\centerline{%
	\includegraphics*[width=0.95\textwidth,trim=0 230 0 200]{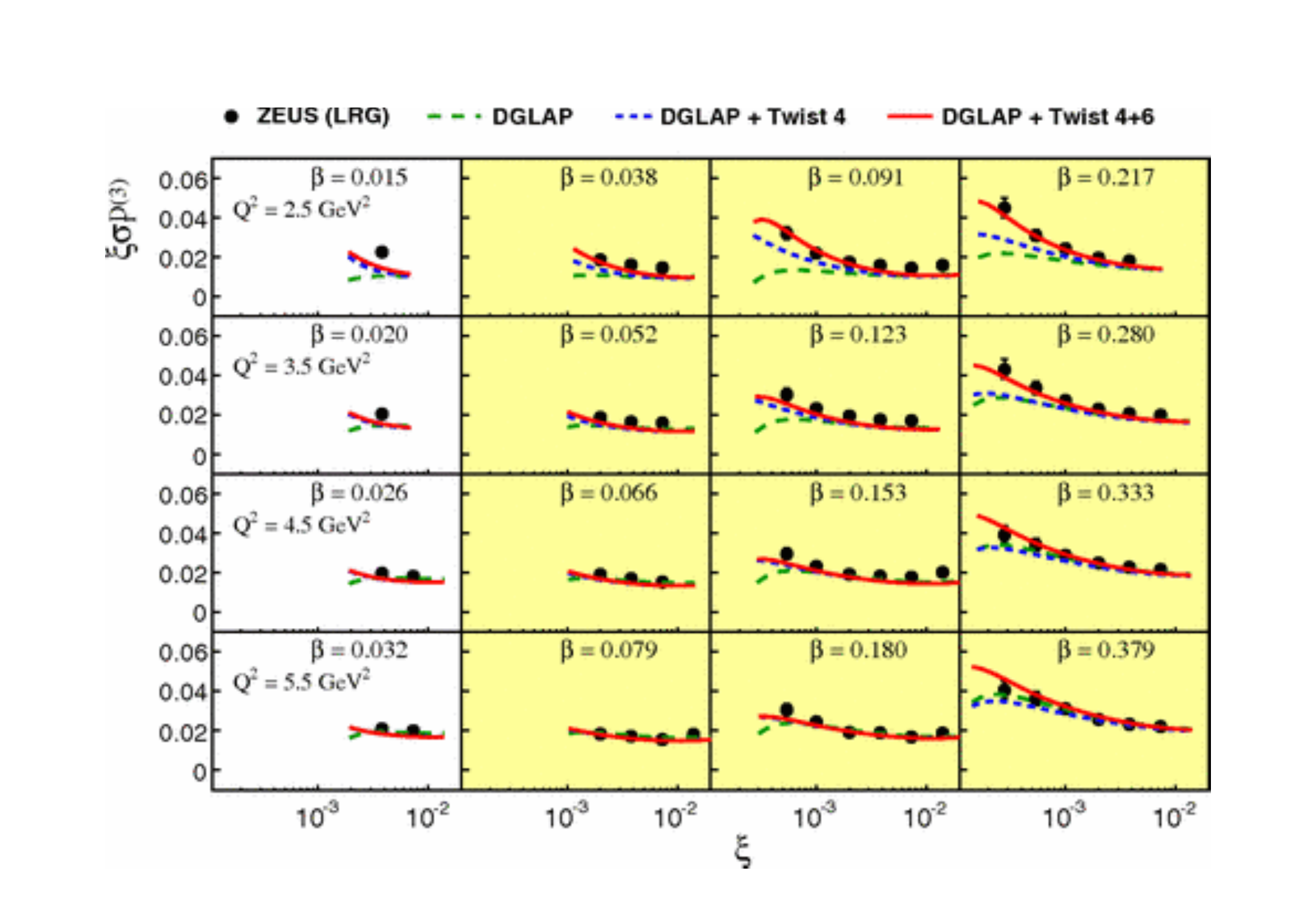}%
}
\caption{The experimental data from ZEUS \cite{Chekanov:2009aa} for $\xi \sigma^{D(3)}$ at low values of $Q^2$. Green dashed line, DGLAP ZEUS-SJ fit, blue dashed line, DGLAP+twist-4 corrections from the  MMS dipole saturation model, red solid line, DGLAP with twist-4 and 6 contributions from MMS dipole saturation model. In yellow region the contribution due to the $q\bar{q}gg$ contribution may be neglected. Figure  from \cite{Motyka:2012ty}, {\tt https://doi.org/10.1103/PhysRevD.86.111501}.}
\label{fig:twistanalysis}
\end{figure}
%%%%%%%%%%%%%%%%%%%%%%%%%%%%%%%%%%%%

In \cite{Motyka:2012ty} a twist analysis was performed, where the DGLAP fit was supplemented by the higher twist expansion from the dipole  model.
In that approach, it was shown that the dipole model formula can be systematically expanded in powers of $1/Q^2$. Several calculations were performed and compared with each other : DGLAP NLO, dipole approach with saturation models MMS \cite{Marquet:2007nf,Munier:2003zb} and GBW \cite{GolecBiernat:1998js,GolecBiernat:1999qd}, twist-2 truncation of the dipole approach
DGLAP plus twist-4 part of MMS dipole model, and DGLAP plus twist-4 and twist-6 parts for the MMS dipole model. This study concluded that the DGLAP NLO description and twist-2 parts of dipole models were consistent in the description of the HERA data at moderate and high $Q^2$ but both failed at low $Q^2$. The dipole models gave better description of the data but were still below the data. The DGLAP supplemented by the twist-4 part of the MMS saturation model describes the data much better. The best description was given by the DGLAP supplemented by the twist-4 and twist-6 parts of the dipole model.

The results are presented in Fig.~\ref{fig:twistanalysis}, where we show the comparison of the calculations to the HERA data. It is clear that the DGLAP supplemented by  twist-4 and twist-6 contributions provides best description of the data in the low $Q^2$ region. We observe however that the curves are still lower than the data in the region of low values of $\beta$. This is the region of high diffractive masses, where more resolved photon fluctuations, $q\bar{q}gg$ may become more relevant. In that analysis it was found that this correction may be neglected for values of $\beta > 0.035$.

The improvement of the description is demonstrated by the solid line in Fig.\ref{fig:chi2} where the $\chi^2$ is shown  for the case of the DGLAP supplemented by the twist-4 and twist-6 components.

In summary, the dipole model has been  successful in fitting the diffractive data on structure function \cite{GolecBiernat:2008gk}. It provides a convenient way to parametrize unitarization effects, either through the modeling or through the solution to the  nonlinear evolution equation. Models, where the additional twist contributions were added to the DGLAP evolution, lead to the improved description of the data at low values of $Q^2$. However, more studies need to be done to pin down the origin of the slight discrepancy of the collinear description based on the DGLAP evolution at low values of $Q^2$. For example, studies with more flexible parametrizations of the initial conditions for the DGLAP evolution, in particular relaxing the assumption of Regge factorization, would need to be performed in order to find out how much of the discrepancy can be accommodated in the initial conditions at very low $Q^2$.

 \section{Low $x$, resummation and parton saturation}
\label{sec:lowx_resum_sat}

\subsection{BFKL Hard Pomeron}
\label{sec:lowx_llx_nllx}

Perturbative QCD gives robust predictions for processes where hard scales are involved. This is also true in the case of the diffractive processes in deep inelastic scattering (DIS),
where the collinear factorization of hard scattering cross section and the diffractive parton distribution functions (DPDFs) apply in the case of DIS with high values of $Q^2$, see  Sec.~\ref{sec:inclusive_diffraction}. The DPDFs are evolved with DGLAP evolution which resums powers of $\alpha_s\ln Q^2/\Lambda^2$
up to the desired order of accuracy. The DGLAP framework is applicable in the case of large $Q^2$ and fixed values of Bjorken $x$. However, in the limit of high energies, or very low $x$, there are other types of logarithms, $\ln 1/x$, which are potentially very large and thus need to be resumed. This limit is usually referred to as the Regge limit, that is when $s \gg |t|$
and $t={\rm const}$.

\begin{figure}
\centerline{%
	\includegraphics*[width=0.5\textwidth,trim=0 100 0 100]{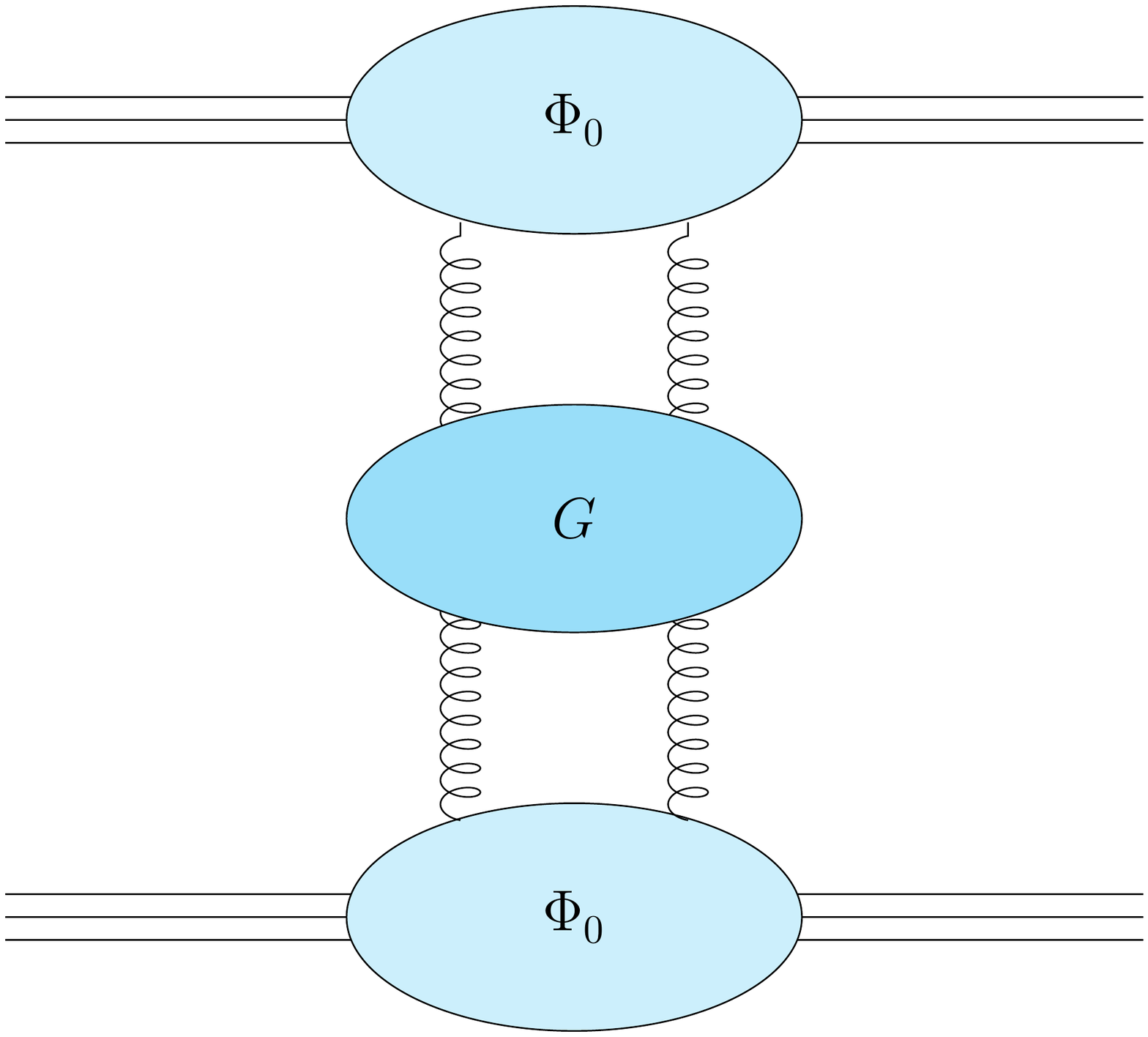}%
}
\caption{A schematic representation of the high energy scattering with the exchange of the gluon Green's function $G$. Functions $\Phi_0$ denote the impact factors which describe the coupling of the $t$-channel gluons to the incoming particles. }
\label{fig:green_gluon}
\end{figure}

The resummation of the leading logarithms of $x$, that is powers $(\alpha_s \ln 1/x)^n$, was performed in the seminal papers by Lipatov and collaborators \cite{Kuraev:1977fs,Balitsky:1978ic,Lipatov:1985uk}.
In these works, the evolution equation in variable $\ln 1/x$ or $\ln s$ was derived for the gluon Green's function, which is known as the Balitsky--Fadin--Kuraev--Lipatov (BFKL) evolution equation. 
The BFKL  evolution equation can  be written in the following form
\begin{equation}
    G(x;{\bf k},{\bf k}_0) = \delta^{(2)}({\bf k}-{\bf k}_0)+\int_x^1 \frac{dz}{z}\int d^2 {\bf k}' \, K(\alpha_s,{\bf k},{\bf k}') \, G(\frac{x}{z};{\bf k}',{\bf k}_0) \; ,
    \label{eq:bfkl}
\end{equation}
where $G(x;{\bf k},{\bf k}_0)$ is the gluon Green's function which depends on the transverse momenta of the gluons in the $t$-channel and $K$ is the BFKL kernel  which has an expansion in the strong coupling 
\begin{equation}
K(\alpha_s,{\bf k},{\bf k}') \; = \; \alpha_s K_0({\bf k},{\bf k}') \;+\; \alpha_s^2 K_1({\bf k},{\bf k}') \; + \dots\;
    \label{eq:kernel_expansion}
\end{equation}
A schematic representation of the gluon Green's function in the context of high energy scattering is shown in Fig.~\ref{fig:green_gluon} which depicts the scattering of two particles in the Regge limit $s \gg |t|$.

%%%%%%%%%%%%%%%%%%%%%%%%%%%%%%%%%%%%
%%%%%%%%%%%%%%%%%%%%%%%%%%%%%%%%%%%%

The solution to this equation can be obtained by performing  the Mellin transform of the variables, and one ends up in the following eigenvalue equation
\begin{equation}
\omega = \frac{\alpha_s N_c}{\pi} \, \chi(\gamma) \; ,
    \label{eq:bfkl_chi}
\end{equation}
where $\omega$ is the Mellin variable conjugated to the longitudinal momentum fraction $z$ and $\gamma$ is the conjugate variable to the transverse momentum ${\bf k}$. In the above, function $\chi(\gamma)$ is the BFKL eigenvalue \cite{Lipatov:1985uk} which in the leading logarithmic approximation reads
\begin{equation}
\chi_0(\gamma) = 2\psi(1)-\psi(\gamma) -\psi(1-\gamma) \;,
    \label{eq:chi0}
\end{equation}
with 
\begin{equation}
\psi(\gamma)=\frac{1}{\Gamma(\gamma)}\frac{d \Gamma(\gamma)}{d\gamma}\, ,
\label{eq:polygamma}
\end{equation}
  the polygamma function, and the index '0' denotes the leading logarithmic order in $\ln 1/x$, LLx. 
The solution to this equation exhibits powerlike behavior in $x$, or Regge-type, $\sim x^{-\lambda}$, with $\lambda = (N_c \alpha_s/\pi) \chi_0(\gamma=1/2)=(N_c \alpha_s/\pi) 4 \ln 2$ given by the saddle-point solution to Eq.~(\ref{eq:bfkl}). The important property of  QCD in high energy limit as demonstrated in \cite{Kuraev:1977fs,Balitsky:1978ic,Lipatov:1985uk} is the fact that the gluon is reggeizzed, that is its propagator can be expressed as $1/(k^2)^{\omega(t)}$, with the trajectory $\omega(t)$ which can be calculated in QCD. 
The other property of the BFKL evolution equation is the diffusion in the transverse momenta along the ladder exchanged between the scattering objects in the high energy limit. This leads to the effect, where the momenta which are initially perturbative can diffuse into the infrared region, \cite{Bartels:1993du}. 
The last property is related to the fact that the BFKL ladder is not ordered in the transverse momenta. This can be seen in the Mellin space, where the eigenvalue (\ref{eq:chi0}) can be approximated by the leading poles
\begin{equation}
\chi_0(\gamma) \; \simeq \;  \frac{1}{\gamma} \, + \, \frac{1}{1-\gamma} \;,
\label{eq:chi0_sim}
\end{equation}
where the pole  $1/\gamma$ corresponds to the collinear limit $k^2 \gg k'^2$ and the pole $1/(1-\gamma)$ to the anti-collinear limit $k^2 \ll k'^2$.

The power-like solution, $x^{-\lambda}$, in the LLx approximation,  turned out to be too steep for the experimental data \cite{Bojak:1997me}.
With moderate values of the strong coupling $\alpha_s \simeq 0.2$, this gives the intercept $\lambda \approx 0.5$, which is excluded by the HERA data on inclusive structure function $F_2(x,Q^2)$.

The next-to-leading logarithmic (NLLx) terms in $\ln 1/x$ to the BFKL evolution were calculated in \cite{Fadin:1998py,Ciafaloni:1998gs}. The NLLx terms turn out to be very large and negative, and also lead to some instabilities in the solution \cite{Salam:1998tj}, like the oscillating cross sections. The main source of the NLLx corrections were identified to be the non-singular part of the DGLAP splitting function, the choice of the energy scales and the running of the strong coupling, \cite{Salam:1998tj, Salam:1999cn}
\footnote{It is worth to mention, that the poor convergence of the series for the total cross section in terms of powers of $\ln(s/\mu^2)$, where $\mu$ is electron mass,   was first demonstrated in QED \cite{Kuraev:1973sj} by the direct calculation of the lowest order diagrams for the $e^+e^-$ pair production in electron–electron scattering.  It was explained in \cite{Kuraev:1973sj} that a fast growth of the coefficients in front of the powers of $\ln(s/\mu^2)$ reflects the highly restricted
phase space for obtaining logarithmic contributions.
}.

The kernel eigenvalue at NLL in QCD has the  following form \cite{Fadin:1998py,Ciafaloni:1998gs}
\begin{eqnarray}
 \chi_1(\gamma)& = &-\frac{b}{2} [\chi^2_0(\gamma) + \chi'_0(\gamma)]
  -\frac{1}{4} \chi_0''(\gamma) \nonumber  \\
&  &-\frac{1}{4} \left(\frac{\pi}{\sin \pi \gamma} \right)^2
  \frac{\cos \pi \gamma}{3 (1-2\gamma)}
  \left(11+\frac{\gamma (1-\gamma )}{(1+2\gamma)(3-2\gamma)}\right) \nonumber \\
 & &+\left(\frac{67}{36}-\frac{\pi^2}{12} \right) \chi_0(\gamma)
  +\frac{3}{2} \zeta(3) + \frac{\pi^3}{4\sin \pi\gamma}   - \Phi(\gamma) \; ,
\label{eq:nllorg}
\end{eqnarray}
with
\begin{equation}
\Phi(\gamma) = \sum_{n=0}^{\infty} (-1)^n
\left[ \frac{\psi(n+1+\gamma)-\psi(1)}{(n+\gamma)^2}
+\frac{\psi(n+2-\gamma)-\psi(1)}{(n+1-\gamma)^2} \right] \;,
\end{equation}
where $b=(33-2N_f)/(12\pi)$  with $N_f$ the number of (active) flavors.

The well known problem that arises at NLLx order in BFKL is due to the presence of double and triple collinear poles.  The double poles are arising due to the running coupling and the non-singular  (in $1/z$) part of the DGLAP splitting function which appear at NLLx order. To be precise, keeping most singular $\gamma \rightarrow 0$ and $\gamma \rightarrow 1$ contributions in Eq.~(\ref{eq:nllorg}) gives for the corresponding terms
\begin{equation}
\fl
-\frac{b}{2} [\chi^2_0(\gamma) + \chi'_0(\gamma)] \;\;\; \longrightarrow \;\;\; -b \frac{1}{(1-\gamma)^2} \; ,
\end{equation} 
for the running coupling, and 
\begin{equation}
\fl
 -\frac{1}{4} \left(\frac{\pi}{\sin \pi \gamma} \right)^2
\frac{\cos \pi \gamma}{3 (1-2\gamma)}
\left(11+\frac{\gamma (1-\gamma )}{(1+2\gamma)(3-2\gamma)}\right)  \;\;\; \longrightarrow \;\;\;  -\frac{11}{12} \frac{1}{\gamma^2}\, -\, \frac{11}{12} \frac{1}{(1-\gamma)^2} \; ,
\end{equation}
which is  the DGLAP contribution. There are also triple collinear poles which appear due to the kinematical constraint \cite{Ciafaloni:2003ek,Ciafaloni:2003rd}. Such constraint was discussed in the BFKL context as originating from the improved kinematics, and more precisely by the requirement that the exchanged momenta are dominated by the transverse components \cite{Andersson:1995ju,Kwiecinski:1996td}, for more recent work on  kinematical constraint see \cite{Deak:2019wms}. These  contributions, when truncated at the NLLx order, generate the double  logarithms in transverse momenta in the kernel and in the Mellin space they exhibit most singular behavior resulting in  the triple collinear poles.  The corresponding term in the NLLx eigenvalue is
\begin{equation}
-\frac{1}{4} \chi_0''(\gamma) \;\;\; \longrightarrow \;\;\; -\frac{1}{2} \frac{1}{\gamma^3}\, -\, \frac{1}{2} \frac{1}{(1-\gamma)^3}  \; .
\label{eq:triple_sym}
\end{equation}
As it has been demonstrated in \cite{Salam:1998tj,Salam:1999cn} this collinear approximation  to the NLLx eigenvalue
\begin{equation}
\chi(\gamma)_{\rm coll} = -b \frac{1}{(1-\gamma)^2} -\frac{11}{12} \frac{1}{\gamma^2}\, -\, \frac{11}{12} \frac{1}{(1-\gamma)^2}-\frac{1}{2} \frac{1}{\gamma^3}\, -\, \frac{1}{2} \frac{1}{(1-\gamma)^3} \; ,
\end{equation}
accounts for the major part of the NLLx corrections given by $\chi_1$.

\subsection{Resummation at low $x$}
\label{sec:lowx_resum}

Resummation procedures were constructed in the early 2000's to stabilize the BFKL solution \cite{Altarelli:2000mh,Altarelli:2001ji,Altarelli:2003hk,Altarelli:2008aj,Ciafaloni:1999yw,Ciafaloni:1999au,Ciafaloni:2003ek,Ciafaloni:2003kd,Ciafaloni:2003rd,Ciafaloni:2007gf,Thorne:2001nr,Vera:2005jt,Bonvini:2016wki}. 
General setup for the CCSS \cite{Ciafaloni:2003ek,Ciafaloni:2003kd,Ciafaloni:2003rd} resummation scheme was based on the analysis of poles in the Mellin space, but the final formulation and the solution to the equation was given in the momentum space.  A similar  idea for the resummation was formulated previously also in Ref.~\cite{Kwiecinski:1997ee} by combining the DGLAP and BFKL evolution with the kinematical constraint.  In the  CCSS resummation one subtracts triple and double poles and incorporates  the full DGLAP splitting function and the kinematical constraint which both resum double and triple poles respectively. 
In addition, more subtractions are needed to ensure the conservation of the momentum sum rule.

In the original  CCSS scheme \cite{Ciafaloni:2003ek,Ciafaloni:2003rd} one starts with the LLx+NLLx BFKL kernel  with LO DGLAP splitting function and puts in  kinematical constraint \cite{Kwiecinski:1996td}.
Imposing kinematical constraint means that the transverse momentum integrals in Eq.~(\ref{eq:bfkl})
are limited by
\begin{equation}
k'^2 \le \frac{k^2}{z} \;.
\label{eq:kin_constr}
\end{equation}
In the Mellin space this leads to the following modification of the LLx kernel eigenvalue
\begin{equation}
\chi_0(\gamma) \rightarrow \tilde{\chi}_0(\gamma,\omega) = 2\psi(1)-\psi(\gamma)  -\psi(1-\gamma+\omega) \;,
    \label{eq:chi_omega}
\end{equation}
with characteristic shift of the pole in $\gamma$ by $\omega$ (recall that $\omega$ is the Mellin variable conjugated to longitudinal momentum fraction $z$).
Using the above modified kernel eigenvalue in the eigenvalue equation (\ref{eq:bfkl_chi}) and expanding in $\omega$, one can see that it generates powers of $1/(1-\gamma)^{2n+1}$ poles with $n=1,2,\dots$ at NLLx, NNLLx and higher orders. It has been verified \cite{Deak:2019wms} that the kinematical constraint generates correct poles at NNLLx level for the case of the $N=4$ 
SYM
theory where the calculation at this order is available \cite{Caron-Huot:2016tzz}. The form of the kernel Eq.~(\ref{eq:chi_omega}) corresponds to the so-called asymmetric scale choice \cite{Ciafaloni:1998gs}. This is appropriate for the case of the DIS where the scale is given by $Q^2$. On the other hand, for the process like $\gamma^*\gamma^*$ scattering with two similar virtualities, the scale choice would be given by $QQ_0$ where $Q^2\simeq  Q_0^2$ are the minus virtualities of both photons. For this case one has to perform the scale change and this will result in the following modified eigenvalue
\begin{equation}
 \tilde{\chi}_0^{\rm sym}(\gamma,\omega) = 2\psi(1)-\psi(\gamma+\omega/2)  -\psi(1-\gamma+\omega/2) \;.
    \label{eq:chi_omega_sym}
\end{equation}
Expanding this form will lead to  the symmetric appearance of the triple collinear poles exactly as in Eq.~(\ref{eq:triple_sym}).

These triple poles need to be  subtracted from the NLLx expression in order to avoid the double counting. In addition, the non-singular DGLAP splitting function in leading order is added and the strong coupling is running in front of the LLx kernel. Thus the double poles also need to be subtracted from the NLLx kernel since they are already incorporated by these modifications. The expression in the Mellin space for the NLLx kernel with subtractions   is 
\begin{equation}
\fl
\chi_1^{\rm resum}(\gamma) \; = \;  \chi_1(\gamma)
+\frac{1}{2} \chi_0(\gamma)  \frac{\pi^2}{\sin^2(\pi\gamma)}-\chi_0(\gamma) \frac{A_1(0)}{\gamma(1-\gamma)}+\frac{b}{2} (\chi_0'+\chi_0^2) \;,
\label{eq:subtractions}
\end{equation}
where $A_1(0)=-11/12$.
The terms on the r.h.s of this equation are as follows: the original NLLx eigenvalue; the subtraction due to the $\omega$-shift (or kinematical constraint) giving the triple poles; the double pole DGLAP terms; the double poles from running coupling with coefficient proportional to the $\beta$ function in QCD.

The resummation removed the instability of the solutions, which remained positive, and gave a reduced value of the intercept of the order of $0.2-0.3$, more compatible with the experimental data.
In Fig.~\ref{fig:intercept} we show  the value of the BFKL intercept as a function of the  strong coupling for LLx, NLLx and the resummed case. We see that the resummed intercept value  is in between the LLx and NLLx values with the nonlinear dependence on the strong coupling.

\begin{figure}
\centerline{%
	\includegraphics*[width=0.75\textwidth]{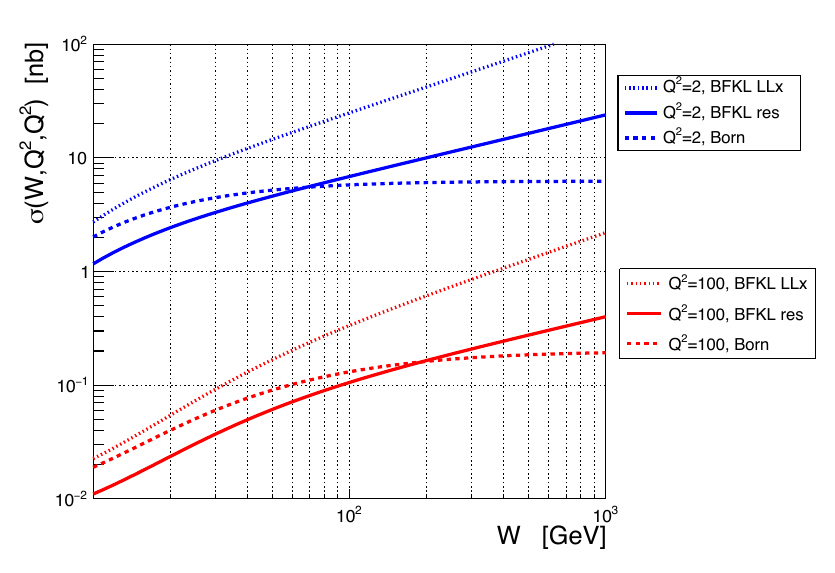}%
}
\caption{The $\gamma^*\gamma^*$ as a function of energy W for two values 
of $Q^2 = 2$ and $100\, \rm GeV$. The BFKL resummed calculation is compared to the LLx BFKL
calculation with the running coupling, and the Born case. Figure from \cite{Deak:2020zay}, {\tt https://doi.org/10.1140/epjc/s10052-019-7171-z}. }
\label{fig:born_bfkl}
\end{figure}

\begin{figure}
\centerline{%
	\includegraphics*[width=0.7\textwidth,trim=0 130 0 130]{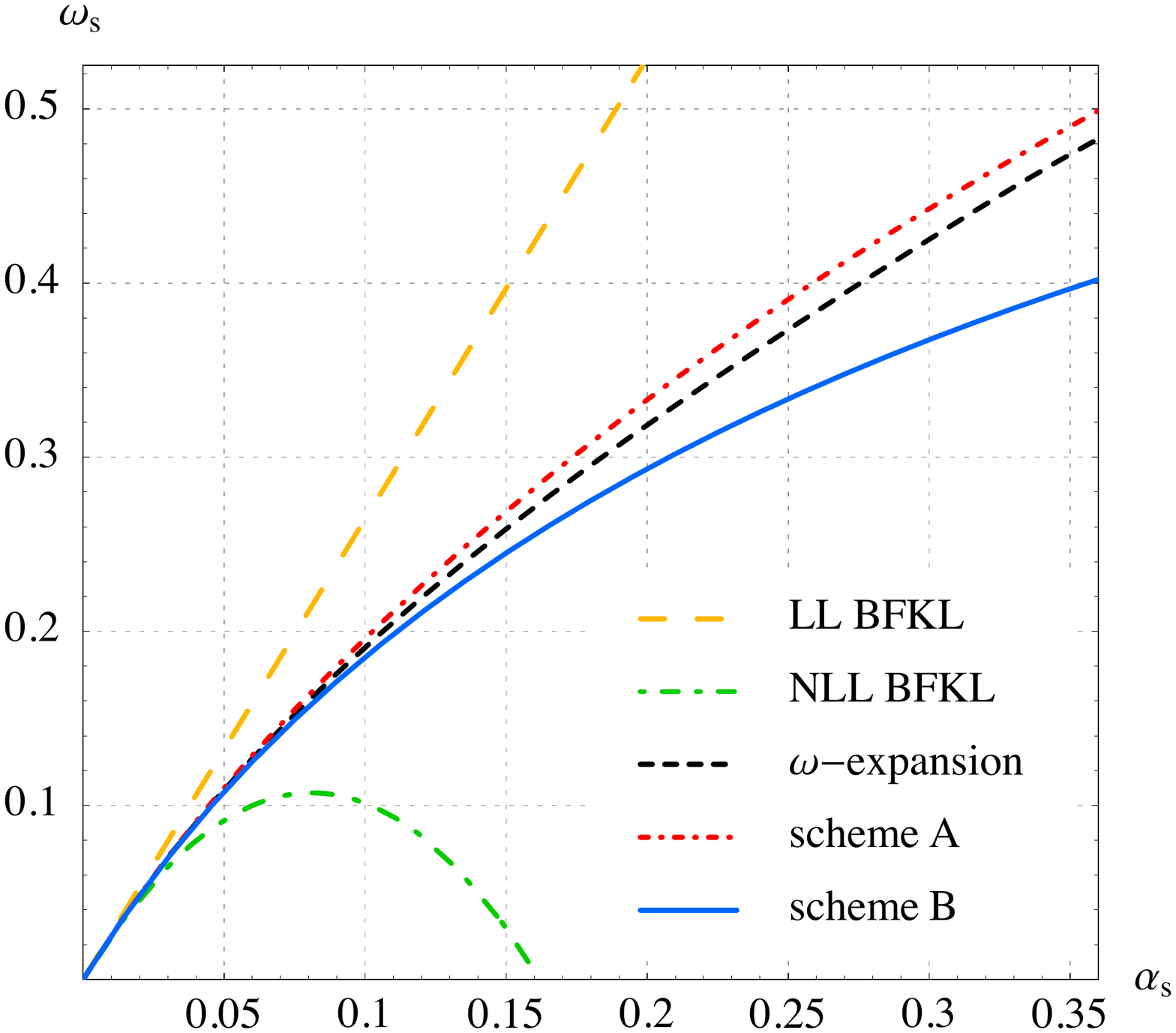}%
}
\caption{The value of the BFKL intercept as a function of the coupling constant, from the LL BFKL (yellow long dashed), NLL case (green dot- long dashed), and three resummed schemes (black dashed, red dot-short dashed, blue solid). Calculation done in the fixed coupling case. Figure from \cite{Ciafaloni:2003rd}, {\tt https://doi.org/10.1103/PhysRevD.68.114003}.} 
\label{fig:intercept}
\end{figure}

In addition to the reduced value of the intercept, the resummation  also leads to the strong preasymptotic effects. As shown through detailed analysis in \cite{Deak:2020zay} the onset of the BFKL growth may be delayed by several units of rapidity, primarly due to the effects related to the kinematical constraint. For example, for the gluon Green's function $G(x;{\bf k},{\bf k}_0)$ with  scales $k\simeq k_0$ of the order $3 - 100$ GeV the growth can be delayed by $4 - 8$ units of rapidity. In the case of the $\gamma^* \gamma^*$ scattering mediated by the BFKL Pomeron exchange, the resummed calculation leads even to {\em suppression} of the cross section for  energies  $W \lesssim 100 \, \rm GeV$ with respect to the Born level,  i.e., based on the two-gluon exchange, before the BFKL growth overcomes the lowest order at higher energies, see Fig.~\ref{fig:born_bfkl}.

%%%%%%%%%%%%%%%%%%%%%%%%%%%%%%%%%%%%

%%%%%%%%%%%%%%%%%%%%%%%%%%%%%%%%%%%%

Let us also note that, in practical applications, for example in DIS, the size of the high energy logarithms is not given by $\alpha_s \ln 1/x$ but rather by interval of rapidity available for the gluon emissions. In the context of DIS, this is usually smaller than naive expectation of $\ln 1/x$ due to the restrictions of the phase space in a  process under consideration. For example, for the case of structure function $F_2(x,Q^2)$ evaluated using the $k_T$ factorization  the relevant large logarithm resummed by BFKL would be given by $\ln 1/x_g$, where $x_g$ is the longitudinal momentum fraction of the gluon entering the photon--gluon partonic subprocess. In other words the phase space for BFKL gluon emissions is reduced by the energy needed to produce $q\bar{q}$ pair.

%%%%%%%%%%%%%%%%%%%%%%%%%%%%%%%%%%%%

%%%%%%%%%%%%%%%%%%%%%%%%%%%%%%%%%%%%

Fits based on the resummed evolution were performed, see for example \cite{Ball:2017otu,Martin:1999rn,Altarelli:2008aj,Kutak:2012rf,Hentschinski:2013id}  and the overall description of the experimental data were very good. In recent works \cite{Ball:2017otu,Abdolmaleki:2018jln}  it was shown that the fits with resummation lead to a better description of the experimental data than the fixed order calculation based on the  DGLAP in the NNLO accuracy. The resummation also suppresses strong diffusion into the infrared, characteristic for the LLx evolution, although the diffusion is still present, see for example \cite{Deak:2020zay}.

%%%%%%%%%%%%%%%%%%%%%%%%%%%%%%%%%%%
\subsection{Cross section taming}
\label{sec:lowx_sat}

The other effect, which is expected to be present in the limit of high energy, is the phenomenon of the parton saturation. The BFKL equation, even in the resummed case predicts a power-like growth of the gluon density with decreasing $x$. This leads to the untamed growth of the density and hence the cross section, and ultimately will violate the unitarity of the S-matrix. Thus, additional corrections are expected to become important and may be related to the recombination of the gluons in the region, when the gluon density is very large. This is known as the phenomenon of {\em gluon saturation} \cite{Gribov:1984tu,Mueller:1985wy}.

A heuristic argument for parton recombination can be formulated as follows. The density of gluons in the proton per unit area is roughly proportional to $\rho \sim xg(x,Q^2)/\pi R_{gN}^2$, where $R_{gN}$ is the 
proton gluon radius. 
The cross section for the gluon recombination can be estimated as $\sigma \sim \alpha_s / Q^2$. Thus the gluon saturation is expected to occur when $\rho \sigma   \ge 1$ which results in  $Q^2 \le Q_s^2(x)$, where the saturation scale $Q_s$ is  defined through the condition
\begin{equation}
Q_s^2 \sim \frac{\alpha_s(Q_s^2)}{\pi R_{gN}^2} \, xg(x,Q_s^2) \; ,
    \label{eq:satscale_arg}
\end{equation}
which leads to the qualitative behavior $Q_s^2 \sim x^{-\lambda}$
assuming that $xg(x,Q_s^2) \propto x^{-\lambda}$.

Another way to quantify the onset of  proximity to the black disk regime, see Sec.~\ref{sec:dipole_model_xsection}  and Eq.~(\ref{eq:black_disk_limit}), is to consider the cross section of dipole--nucleon scattering. Since the total cross section is proportional to the gluon density, while the elastic cross section  is proportional to the gluon density squared,  the ratio of the elastic and total cross sections grows and exceeds the black disk limit value of 1/2. 
From the requirement that the interaction at small impact parameters is completely absorptive,  one obtains using the optical theorem
\begin{equation}
    {\hat{\sigma} \over 16 \pi} \int dt F_{2g}^2(t) \le \frac{1}{2} \; ,
    \label{eq:totel1}
\end{equation}
where $\hat{\sigma}$ is the dipole--nucleon cross section\footnote{Note that the actual inequality is significantly stronger since dipole cross section $\hat{\sigma}$ corresponds to the case when elastic and diffractive intermediate state are neglected. }, see Sec.~\ref{sec:dipole_model_xsection}, and $F_{2g}(t)$ is the two-gluon form factor which we take to have an exponential form
\begin{equation}
    F_{2g}^2(t) = \exp\big( B_{2g}(x) t\big) \; ,
    \label{eq:2gluon_form_factor}
\end{equation}
where  the slope is parametrized as \cite{Frankfurt:2010ea}
\begin{equation}
    B_{2g}(x)=B_{2g}^{(0)}+2 \alpha_g' \ln (x_0/x) \;,
    \label{eq:b2g}
\end{equation}
with $x_0=0.0012$, $B_{2g}^{(0)}=4.1 \, ({}^{+0.3}_{-0.5})\, {\rm GeV}^{-2}$ and $\alpha_g'=0.140\,({}^{+0.08}_{-0.08}) \, {\rm GeV}^{-2}$.
One can rewrite Eq.~(\ref{eq:totel1}) as 
\begin{equation}
    \hat{\sigma} \le 8\pi B_{2g}(x) \approx 40  \, \mbox{mb} \; ,
    \label{eq:totel2}
\end{equation}
for $x= 10^{-3}$. 
We note that taking into account relation between the gluon density and dipole cross section Eq.~(\ref{eq:dipole_xsection_gluon_density}), the relation (\ref{eq:totel2}) is equivalent to (\ref{eq:satscale_arg}) (with fixed normalisation).

The above arguments can be extended to nuclei, in which case the saturation scale obtains the modification  due to the mass number $A$. It is  coming from the enhanced gluon density, which scales roughly like a volume, factor $A$  times reduction factors (a)  the nuclear shadowing factor and (b) smaller transverse density (nuclei are rather dilute  objects)  resulting in 
\begin{equation}
\frac{Q_{sA}^2}{Q_{sN}^2} = A \, \frac{R_{gN}^2}{R_A^2} \, \frac{g_A(x, Q^2)}{A g_N(x, Q^2)} \; .
\label{eq:satscle_nucleus}    
 \end{equation}
Taking $R_{gN}^2(x=10^{-3})=0.6 \, \mbox{fm}^2$
from analysis of the $J/\psi$ elastic production, see Sec.~\ref{sec:vm1}, 
$R_A^2 = (1.1 \, \mbox{fm} \, A^{1/3})^2$,   and nuclear shadowing factor of 0.6  for $Q^2= 3$ GeV$^2$ and $x=10^{-3}$, we 
estimate the enhancement factor for heavy nuclei ($A\sim 200$):
\begin{equation}
  \frac{Q_{sA}^2}{Q_{sN}^2} =  0.3  A^{1/3} \approx 1.75 \; .
\end{equation}

A more accurate estimate avoiding edge effects  can be done for the case of scattering at small impact parameters.
In this case we can estimate  ratio $Q_{sA}^2/Q_{sN}^2 $ for small impact parameters by comparing the product of the matter density at $b=0$, 
\begin{equation}
T_A(b=0)= \int_{-\infty}^{\infty} dz\rho_A(b=0,z)_{A=200} \approx  \mbox{2 fm}^{-2} \; ,
\end{equation}
\noindent
times the shadowing factor $S_A(x)\sim 0.5$ with the transverse gluon density in a nucleon: 
\begin{equation} {1\over \pi R^2_{gN \, tr}}
= {1\over \pi R_{gN}^2 (2/3)}\approx \frac{1}{2 R_{gN}^2} \; .
\end{equation}
Using the same value of $R_{gN}^2$ as above we find the modification factor for the saturation scale equal to 
\begin{equation}
   Q_{sA}^2(b=0)/Q_{sN}^2=T_A(b=0)\cdot S_A(x,b=0)\cdot  2 R_{gN}^2  = 1.2 \; ,
    \label{eq:sat_scale_A}
\end{equation} for heavy nuclei. The difference is mainly due to neglect of the surface effects in modeling the nuclear density.

In practice the black disk regime is difficult to reach experimentally, nevertheless it is instructive to analyze the behavior of the cross sections in this limit.
It was  first considered  by  Gribov \cite{Gribov:1968gs} for the total cross section for $\gamma^*$ - heavy nucleus scattering. In this limit for virtualities $Q^2 < Q_{sA}^2$, where $Q_{sA}^2 \gg \Lambda_{\rm QCD}^2$,  the  cross section of dipole--nucleus scattering does not depend on the dipole size for $1/r^2 < Q_{sA}^2$ and is equal to $2\pi R_A^2$.  As a result  Bjorken scaling is grossly violated: 
 $\sigma_{tot}^{\gamma^*A} (x, Q^2 )$  does not drop with $Q^2$ and grows as 
 $\ln (x_0/x)$ \cite{Gribov:1968gs}. 
 In this limit $\sigma_{\rm diff}= \pi R_A^2$ and  is dominated by the exclusive dijet production \cite{Frankfurt:2001nt}.  Also, in this limit the absolute normalization of the  vector meson coherent production cross  section is predicted. The cross section drops with $Q^2$  by a factor $1/Q^4$ slower than in the LT limit since  in the LT   cross section is proportional to   the square of the dipole cross section. To be more precise in this limit one expects
 \begin{eqnarray}
     \sigma_L(\gamma^* A\to  V \, A) \;  \propto  \; 1/Q^2,  \nonumber \\
 \sigma_T(\gamma^* A\to V \, A) \; \propto  \; M_V^2/Q^4 \; .
 \end{eqnarray}
 For a detailed discussion of the prediction in the black disk limit, see \cite{Frankfurt:2001nt,Frankfurt:2001av}.

\subsection{Color Glass Condensate and Balitsky--Kovchegov equation}
\label{sec:cgc}

The effective theory which describes parton saturation is the Color Glass Condensate (CGC) \cite{McLerran:1993ka,McLerran:1993ni,McLerran:1994vd}, with the Jalilian-Marian - Iancu - McLerran - Weigert - Leonidov - Kovner (JIMWLK) evolution equations  \cite{JalilianMarian:1996xn,JalilianMarian:1997dw,JalilianMarian:1997gr,JalilianMarian:1997jx, Iancu:2000hn,Iancu:2001ad,Ferreiro:2001qy} (see \cite{Gelis:2008zzc} for a  review together with selected phenomenological applications).  
In the  CGC effective theory the relevant degrees of freedom are the color sources $\rho({\bf x})$, which have large values of $x$ and the gauge fields $A^{\mu}$ in the region of small $x$. Here ${\bf x}$ is the transverse spatial coordinate and thus $\rho({\bf x})$ describes the distribution of these color charges in the transverse coordinate space. The color sources produce a current $J^{\mu} = \delta^{\mu +} \delta(x^-) \rho({\bf x})$ (for a target moving in the positive $z$ direction, or $x^+$ direction). Due to the time dilation the color sources are effectively frozen at the time scales relevant for the strong interaction. The fast and slow degrees of freedom are then coupled through the gauge field and the current, i.e. $A^{\mu} J_{\mu}$. The distribution of the fast partons is a stochastic quantity which is different in every collision and thus the central object in CGC is the statistical distribution $W_y[\rho]$ of the color sources. From this distribution one can calculate various operators  through the averaging procedure
\begin{equation}
\langle {\cal O} \rangle_y \; = \; \int [D \rho] \, W_y[\rho] \, {\cal O}[\rho] \; ,
    \label{eq:cgc_weight}
\end{equation}
where ${\cal O}[\rho]$ is the expectation value of the operator for the particular configuration $\rho$ of the color sources.
The statistical distribution  $W_y[\rho]$ encodes all the correlations of the color charge density and it  depends on the cutoff $y$, which divides the fast and slow partons. Variable $y$ may be related to the rapidity, which in the leading logarithmic approximation is given by $y=\ln 1/x$, where $x$ would be Bjorken $x$ in the DIS case. The evolution of 
 the distribution $W_y[\rho]$ of the color sources is provided by  renormalization group equation 
 \begin{equation}
 \frac{\partial W_y[\rho]}{\partial y} \; = \;  {\cal H}[\rho,\frac{\delta }{\delta \rho}] \; W_y[\rho] \; ,
     \label{eq:renorm_group}
 \end{equation}
 with ${\cal H}$ being 
 the JIMWLK Hamiltonian. This operator contains up to two derivatives $\delta/\delta \rho$ and arbitrary powers of $\rho$. 
 From Eq.~(\ref{eq:cgc_weight}) and (\ref{eq:renorm_group}) one can derive evolution equations for the different operators. 
 
 More specifically, the scattering between the dipole and the fields generated by the target can be described using the product of two Wilson lines, one for the quark with transverse coordinate ${\bf x}$ and the antiquark with transverse coordinate ${\bf y}$. The relevant operator is given by
 \begin{equation}
 1 - \frac{1}{N_c} {\rm Tr} (U^{\dagger}({\bf x}) \, U({\bf y})) \; ,
     \label{eq:dipole_wilson_lines}
 \end{equation}
 where 
\begin{equation}
U^\dagger({\bf x}) \; = \; P \, \exp\left[ ig \, \int dx^- A_a^+(x^+\simeq0,x^-,{\bf x}) t^a \right] \; ,
    \label{eq:wilson_line_cgc}
\end{equation}
is the Wilson line operator, corresponding to the left moving quark. Here, $t^a$ are the generators of $SU(N_c)$ in the fundamental representation, $P$ denotes the $x^-$ ordering of color matrices. This Wilson line describes the quark propagating through the target, preserving the straight trajectory in the eikonal approximation and undergoing color precession.  
 
 The physical dipole scattering amplitude is defined in this framework through the average over all the color configurations of the target given by the functional $ W_y[\rho]$
 \begin{eqnarray}
     N_y({\bf x},{\bf y}) &  =  &
     \langle 1 - \frac{1}{N_c} {\rm Tr} (U^\dagger({\bf x}) \, U({\bf y})))\rangle_y \\
     & =  & \int [D\rho] \, W_y[\rho] \, 
     \left[ 1 - \frac{1}{N_c} {\rm Tr} (U^\dagger({\bf x}) \, U({\bf y}))\right] \; .
     \label{eq:dipole_cgc}
 \end{eqnarray}
By inserting the above definitions into the evolution equation for the weight function Eq.~(\ref{eq:renorm_group}) one can derive the following evolution equation
\begin{eqnarray}
    \frac{\partial}{\partial y} \langle {\rm Tr}(U^{\dagger}({\bf x}) \, U({\bf y})) \rangle_y \;  = \; & -\int d^2 {\bf z} \; {\cal K}({\bf x},{\bf y},{\bf z}) \; \langle N_c {\rm Tr}((U^{\dagger}({\bf x}) \, U^{\dagger}({\bf y})) \nonumber \\
    & - {\rm Tr} (U^{\dagger}({\bf x}) \, U^{\dagger}({\bf z}))\, {\rm Tr}(U^{\dagger}({\bf z}) \, U({\bf y}))  \rangle_y \; .
    \label{eq:balitsky_hierarchy}
\end{eqnarray}
Here ${\cal K}$ is the dipole kernel, which  in the leading order case has the form
\begin{equation}
{\cal K}({\bf x},{\bf y},{\bf z}) \; = \;  \frac{ \alpha_s }{ 2\pi^2 } \frac{({\bf x}-{\bf y})^2}{({\bf x}-{\bf z})^2({\bf z}-{\bf y})^2} \; .
\label{eq:dipole_kernel}
\end{equation}
The above equation  for the S-matrix is not closed since it relates the 2-Wilson line correlator to the 4-Wilson line correlator. For the latter, one can derive the evolution equation which in turn will involve 6-Wilson line correlator.  Thus Eq.~(\ref{eq:balitsky_hierarchy}) is just one of the  set of equations which form an  infinite hierarchy. This hierarchy was derived in 
an alternative approach  by Balitsky \cite{Balitsky:1995ub,Balitsky:1998ya} using the operator product expansion for high energies.  One can make however some simplifications to  Eq.~(\ref{eq:balitsky_hierarchy}),  that is in the large multicolor limit the line correlation  function factorizes
\begin{equation}
\fl
    \langle  {\rm Tr} (U({\bf x}) \, U^{\dagger}({\bf z}))\, {\rm Tr}(U({\bf z}) \, U^{\dagger}({\bf y}))\rangle_y \rightarrow \langle {\rm Tr} (U({\bf x}) \, U^{\dagger}({\bf z}))\rangle_y\, \langle {\rm Tr}(U({\bf z}) \, U^{\dagger}({\bf y})) \rangle_y \; .
\end{equation}

As a result the first equation of the hierarchy decouples, and has the following form
\begin{equation}
\fl 
\frac{\partial  N_y({\bf x},{\bf y})}{\partial y}  = \int d^2 {\bf z} \; {\cal K}({\bf x},{\bf y},{\bf z}) \left[ N_y({\bf x},{\bf z})+N_y({\bf z},{\bf y})-N_y({\bf x},{\bf y})-N_y({\bf x},{\bf z})N_y({\bf z},{\bf y})  \right] \; ,
\label{eq:bkequation}
\end{equation}
where we used definition Eq.~(\ref{eq:dipole_cgc}) for the dipole scattering ampliutude.

The above equation is the Balitsky--Kovchegov (BK) equation for the dipole scattering amplitude, derived independently by Kovchegov \cite{Kovchegov:1999ua,Kovchegov:1999yj} from the description of the soft gluons in the dipole wave function at high energy  \cite{Mueller:1993rr}. The linearlized version of this equation, without the $N_y({\bf x},{\bf z})N_y({\bf z},{\bf y})$ term is equivalent to the famous BFKL evolution equation \cite{Balitsky:1978ic,Kuraev:1977fs,Lipatov:1985uk} which is an equation allowing for the resummation of the large logarithms in $\ln 1/x$. The Balitsky-Kovchegov equation contains an extra negative and non-linear contribution which is relevant when the parton density is large. It is evident that the solution $N=1$ satisfies this equation, and one can demonstrate that this is the stable point of the solution. 
The solution to this equation generates dynamically saturation scale $Q_s^2(x)$ which has a power like behavior in $x$.

The BK equation is most often analyzed in the approximation where $N_y({\bf x},{\bf y})=N_y(|{\bf x}-{\bf y}|)$, that is the amplitude depends on the  dipole size (absolute value) $r=|{\bf x}-{\bf y}|$. This simplifies greatly the numerical solution to the BK equation. It however corresponds to the crude approximation of the infinite nucleus. The full solution to the BK equation with impact parameter was presented in \cite{GolecBiernat:2003ym,Berger:2010sh}. In that case the solution for the amplitude as a function of the dipole size, for fixed values of the impact parameter,  drops down for very large dipole sizes. This is due to the fact that very large dipoles will simply miss the target. Of course in reality, there is confinement, which will cut off the large dipole sizes which extend  beyond the confinement scale. In \cite{GolecBiernat:2003ym,Berger:2010sh} it was also shown that even after including the saturation effects, the BK solution will violate the Froissart bound, i.e. the dipole cross section will increase as a power of $1/x$, rather than $\ln^2 1/x$. This is related with the fact that the BK equation in the leading order approximation is scale invariant and the kernel has power like tails, which need to be regulated by confinement.

The JIMWLK and Balitsky equations have been  derived at leading and next-to-leading logarithmic order \cite{Balitsky:2008zza,Kovner:2014lca,Lublinsky:2016meo}.  Resummation schemes were also applied to the non-linear evolution equations \cite{Motyka:2009gi,Beuf:2014uia,Iancu:2015vea,Hatta:2016ujq}, and  a good description of the experimental data on the structure function $F_2$ at HERA was achieved based on the resummed nonlinear evolution \cite{Ducloue:2019jmy}. It would be interesting to investigate whether the energy-momentum conservation effects, which are discussed for 
one-Pomeron exchange (see discussion in the previous section) are amplified for the multi-Pomeron exchanges which are underpinning the non-linear evolution equation Eq.~(\ref{eq:bkequation}).

A non-linear evolution equation in the small $x$ limit was also derived for the case of the high-mass diffraction by Kovchegov and Levin \cite{Kovchegov:1999ji}, see also \cite{Kovchegov:2012mbw}.  The equation derived was for the %vg dipole-target amplitude
$N^{D}({\bf x},{\bf y},Y,Y_0)$ cross section per impact parameter ${\bf b}$
for diffractive interaction with rapidity gap greater than or equal to  $Y_0$, and where $Y$ is the total rapidity interval, so that $Y > Y_0$.
The corresponding  diffractive cross section with rapidity gap greater than or equal to $Y_0$ in the dipole--nucleus scattering is 
\begin{equation}
\hat{\sigma}_{\rm diff} = \int d^2 {\bf b} \;  N^{D}({\bf x},{\bf y},Y,Y_0) \; .
\label{eq:kovchegov_levin_2}
\end{equation}
In order to find the diffraction cross section for a given fixed rapidity gap $Y_0$ one needs to differentiate the $N^D$ with respect to $Y_0$
\begin{equation}
  M_X^2 \frac{d \sigma_{\rm diff}}{dM_X^2} \; = \; -   \int d^2 {\bf b} \;  \frac{\partial N^{D}({\bf x},{\bf y},Y,Y_0)}{\partial Y_0} \; ,
    \label{eq:eq:kovchegov_levin_3}
\end{equation}
where the diffractive mass $M_X$ is related to the rapidity gap $Y_0$ through
\begin{equation}
Y_0 = \Delta Y_{\rm gap} \simeq \ln \frac{W^2}{M_X^2+Q^2} \; .
    \label{eq:rapidity_gap}
\end{equation}

The evolution equation for the diffractive amplitude in the LLx approximation reads
\begin{eqnarray}
\fl
    \frac{\partial}{\partial y} N^{D}({\bf x},{\bf y},Y,Y_0)  & = 
   \int d^2 {\bf z} \; {\cal K}({\bf x},{\bf y},{\bf z}) \nonumber\\
   & \times \left [ N^D({\bf x},{\bf z},Y,Y_0) + N^D({\bf z},{\bf y},Y,Y_0) -
    N^D({\bf x},{\bf y},Y,Y_0) \right. \nonumber\\
    & + N^D({\bf x},{\bf z},Y,Y_0)N^D({\bf z},{\bf y},Y,Y_0)- 2  N({\bf x},{\bf z},Y) N^D({\bf z},{\bf y},Y,Y_0) \nonumber\\
    & \left. - 2 N^D({\bf x},{\bf z},Y,Y_0) N({\bf z},{\bf y},Y)+ 2  N({\bf x},{\bf z},Y) N({\bf z},{\bf y},Y)
    \right] \; .
    \label{eq:kovchegov_levin_equation}
\end{eqnarray}

Numerical solutions were studied in \cite{Levin:2001pr,Levin:2002fj}, where it was shown that the saturation scale has the same dependence on $x$ as the saturation scale for the total cross section. It was also shown that the ratio of $\sigma_{\rm diff}/\sigma_{\rm tot}$ has only mild dependence on the energy and $M_X^2$.

We note that this equation was derived in LLx approximation, and similarly to the inclusive case described previously,  it is expected  that the higher orders will be very important (e.g., see Fig.~\ref{fig:intercept}). It will be interesting to see whether the predictions from Eq.~(\ref{eq:kovchegov_levin_equation}) reproduce the value of the $\alpha_{\pom}(t=0) \simeq 1.11$.

 \section{Hard diffractive production of vector mesons}
\label{sec:vm1}

Studies of vector meson production in hard diffractive processes off nucleons and nuclear targets, 
$\gamma^*+T\to (V, V_{Q\bar Q}, \gamma)+T$, 
where $V=\rho,\omega,\phi$ or a  heavy quarkonium  $V_{Q\bar Q}=J/\psi, \Upsilon$,  is one of major areas of the theoretical and experimental studies of QCD.
A number of measurements has been performed using muon and electron beams with fixed targets, electron--proton colliding beams at HERA, and more recently using ultraperipheral heavy ion collisions at the LHC (see Sec.~\ref{sec:UPC_Jpsi}), for reviews, see Refs.~\cite{Levy:2007fb,Newman:2013ada}.

Hard exclusive diffractive processes are the source of valuable information on the origin of small $x$ processes, i.e., on the origin of the phenomenon known as  Pomeron. 
It enables one to probe the evolution of high energy processes with a decrease of $x$ and to compare 
the data with the competing LLx and NLLx  $\ln(x/x_0)$ approximations and with the resummation of these terms, see Sec.~\ref{sec:lowx_resum_sat}.
Also, 
studies of these processes allow one to access generalized parton distributions (GPDs), which can be rigorously defined in the framework of the QCD factorization theorem~\cite{Collins:1996fb} and the DGLAP approximation. 
These studies should allow one to establish spatial distributions of quarks and gluons in the hadronic targets and to investigate the role of color in high energy processes. We will begin our consideration from the formulation of basic properties of hard diffractive production of vector mesons.

\subsection{Space--time evolution and factorization of high energy processes }

Vector meson production at small $x$ in the target rest frame can be described as a three-stage process~\cite{Brodsky:1994kf}:

(i) The virtual photon $\gamma^{\ast}$ with the large longitudinal momentum $q$ converts into a $\bar q q$ pair, where the partons carry the longitudinal momenta $zq$ and $(1-z)q$ and the transverse momenta ${\bf k}_t$ and $-{\bf k}_t$, respectively; $z$ is the momentum fraction describing the longitudinal momentum sharing between the quarks. The lifetime of such fluctuations is also called the coherence length $l_{\rm coh}$ and is given by
 \begin{equation}
\tau_i=l_{\rm coh}/c= {2q\over Q^2 + {k_\perp^2+ m_q^2 \over z(1-z)}}
\approx {1\over m_Nx} \,,
\end{equation}
which follows from the energy--time uncertainty principle.
The values of $l_{\rm coh}$, which could be reached at colliders, are
$l_{\rm coh} \sim  10^3$ fm at HERA and $l_{\rm coh}  \sim 10^2$ fm at the EIC.
%%%%%%%%%%

As noted earlier in Sec.~\ref{sec:dipole_model}, in the case of longitudinally polarized photons, the transverse size of the $q\bar q$ pair is $\propto 1/Q$, making it possible to justify the applicability of perturbative QCD.  In the case of transversely polarized photons, the low $k_\perp$ contribution is not parametrically suppressed leading to significant nonperturbative effects. 
 
(ii)  Then the $q \bar q$ pair scatters off the target with the dipole cross section discussed in Sec.~\ref{sec:dipole_model}.
Note that the $q\bar q$ dipole lives for the time 
\begin{equation}
\tau_f=l_f/c= {2q \over {k_\perp^2+m_{q}^2\over z(1-z)}} \,,
\end{equation}
so that $\tau_f \geq \tau_i$.

(iii)
Finally, the vector meson in the final state is formed.  

\begin{figure}[t]
\centering
 \includegraphics[width=0.65\textwidth]{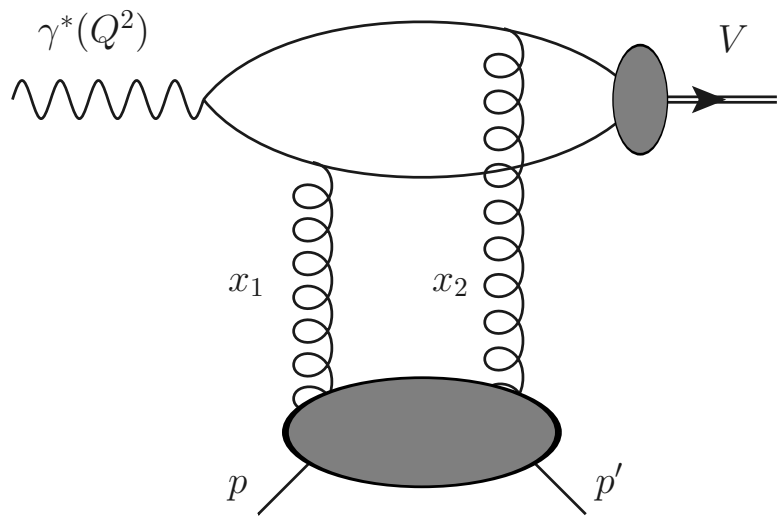}
\caption{Electroproduction of vector mesons in the dipole model.
The three stages of the process involve the transition of the photon into a quark-antiquark pair (dipole), its scattering off the target depicted as a two-gluon exchange here, and formation of the final-state vector meson. The longitudinal momentum fractions of the exchanged gluons are denoted $x_1$ and $x_2$.}
\label{fig:Jpsi_dipole}
\end{figure}

 As a result, the amplitude of vector meson electroduction by longitudinal photons ${\cal A}(\gamma^{\ast}_L+p\to V+p)$ can  be written as a convolution of the light-cone wave function of  the photon,   
$\Psi_{L}^f({\bf r},Q,z)$,
the dipole cross section,
and the wave function of the vector meson  $\Psi_{V}^f$  in a $q\bar q$ configuration,
\begin{equation}
\fl
{\cal A}(q_t,Q)= \sum_f \int d^2 {\bf b}  \int d^2 {\bf r} \int dz \,  e^{i {\bf q_t}\cdot  {\bf b}} \,\Psi_{L}^f({\bf r},Q,z) \, N(x,{\bf r},{\bf b}) \, \Psi_V^f({\bf r},z) \,,
\label{bb}
\end{equation}
where $N(x,{\bf r},{\bf b})$ is the dipole amplitude, ${\bf r}$ is the  transverse separation between $q$ and $\bar q$ (the dipole size); ${\bf b}$ is the impact parameter defined as the distance between the center of mass of the dipole and the center of the target; ${\bf q_t}$ is the momentum transfer to the target, which is assumed to be purely transverse
here (the effect of the non-zero longitudinal momentum transfer will be discussed below); $\sum_f$ is a sum over quark flavors depending on the flavor structure of the produced vector meson $V$. 
The corresponding graph is shown in Fig.~\ref{fig:Jpsi_dipole}. 
The dipole amplitude $N(x,{\bf r},{\bf b})$ when integrated over the impact parameter ${\bf b}$ is reduced to the dipole cross section $\hat{\sigma}$, see Eqs.~(\ref{eq:dipole_scattering_amplitude}) and  (\ref{eq:dipole_xsection_gluon_density}).

Note that at large $Q^2$,
QCD factorization into the three blocks in Eq.~(\ref{bb})  is valid for all $x$ and any two-body final states at fixed $t$ \cite{Collins:1996fb}. However, for small $x$ the space-time picture of the process is greatly simplified.
Note also that the same three-block description is valid for hard diffractive production of heavy quarkonia, where the heavy quark mass $m_{c,b}$ defines the hard scale of process~\cite{Ryskin:1992ui,Frankfurt:2000ez}.

The $\gamma_L^{\ast}+p\to V +p$ differential cross section is given by the usual expression
\begin{equation}
\frac{d\sigma(\gamma_L^{\ast}+p\to V +p)}{dt} = \frac{1}{16 \pi} | {\cal A}(q_t,Q)|^2 \,,
    \label{eq:diffdt}
\end{equation}
where $t \approx -{\bf q_t}^2$.

In the leading $\alpha_s \ln Q^2$ approximation and
 in the leading twist limit, one can use the explicit relation of the dipole cross section to the gluon density, see Eq.~(\ref{eq:dipole_xsection_gluon_density}), to obtain 
 the following expression for
 the cross section of diffractive vector meson electroproduction~\cite{Ryskin:1992ui,Frankfurt:1995jw,Brodsky:1994kf,Frankfurt:1997fj}
\begin{equation}
\fl
\left. {d\sigma^L_{\gamma^*N\rightarrow VN}\over dt}\right|_{t=0} =
{12\pi^3\Gamma_{V \rightarrow e^{+}e^-} M_{V}\alpha_s^2(Q)\eta^2_V
\left|\left(1 + i{\pi\over2}{d \over d\ln x}\right)xg_T(x,Q^2)\right|^2
\over \alpha_{EM}Q^6N_c^2} \,,
\label{master}
\end{equation}
where $M_V$ is the vector meson mass; $\alpha_s(Q^2)$ is the running coupling constant; $xg_T(x,Q^2)$ is the gluon density of the target; $\alpha_{EM}$ is the fine-structure constant; $N_c=3$ is the number of colors; $\Gamma_{V \rightarrow e^{+}e^{-}}$ is the $V\to e^+e^-$ decay width. The parameter $\eta_V$ 
is defined as follows,
\begin{equation}
\eta_V={1\over 2}\frac{\int\frac{dz}{z(1-z)}\, \Psi_V(r=0,z)}
{\int dz \,\Psi_V(r=0,z)} ={1\over 2}{\int{dz\,d^2{\bf k_t}\over z(1-z)}\,\Phi_V(k_t,z)\over
\int dz\,d^2{\bf k_t}\,\Phi_V(k_t,z)} \,,
\label{phi}
\end{equation}
where $\Phi_V(k_t,z)$ is the Fourier transform of the vector meson wave function $\Psi_V(r,z)$. 
In perturbative QCD, $\Phi_V(k_t,z)$ has the $Q$ dependence given by the ERBL equations and becomes proportional to $z(1-z)$ in the $Q^2 \to \infty$ limit. Hence, $\eta_V$ quantifies the deviation of the $z$ dependence of the vector meson wave function $\Psi_V(r,z)$ from its asymptotic form, which leads to $\eta_V \to 3$ at high $Q^2$.

For the first time the expression for the cross section of $J/\psi$ electroproduction and photoproduction 
neglecting the Fermi motion of quarks and the real part of the scattering amplitude was derived in Ref.~\cite{Ryskin:1992ui}.

 Note that in the leading twist approximation, the integration over quark transverse momenta in the graph in Fig.~\ref{fig:Jpsi_dipole} extends up to $k_t \sim Q$.
 As a result, the expression for the cross section of vector meson electroproduction involves $\Psi_V$ at $|{\bf r}| \propto  1/Q$, i.e., the vector meson wave function at the origin (the vanishing separation between the quark and antiquark), see
 Eqs.~(\ref{master}) and (\ref{phi}).

On the phenomenology side,  the rapid onset of the leading twist behavior for the $\sigma(e^{+}e^{-} \to hadrons)$
electron--positron annihilation cross section
suggests that for $\rho$ and $\phi$ mesons, $\Phi_V(z,k_t)$ and hence $\eta$ should already be close to the asymptotic value  at $Q^2 \sim $ a few GeV$^2$.

\subsection{The role of GPDs in hard diffractive production of vector mesons}

\begin{figure}[h]
\centering
\includegraphics{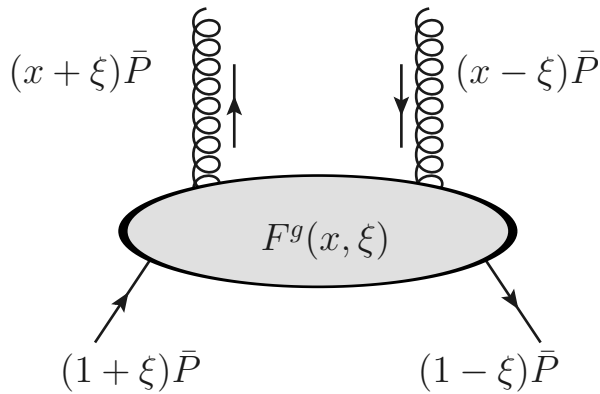}
\caption{Momentum fractions in the definition of GPDs in the symmetric notation.}
\label{fig:gpd_def}
\end{figure}

In the framework of collinear factorization for exclusive processes~\cite{Collins:1996fb}, cross sections of hard exclusive electroproduction of vector mesons are expressed in terms generalized parton distributions (GPDs) of the target~\cite{Mueller:1998fv,Radyushkin:1997ki,Ji:1996nm,Goeke:2001tz,Belitsky:2001ns,Diehl:2003ny,Belitsky:2005qn}. GPDs are defined as matrix elements of quark and gluon QCD operators between states with non-equal momenta. For instance, 
the gluon GPD of the proton target reads in the symmetric notation~\cite{Ji:1996nm}
\begin{equation}
\fl
F^g(x,\xi,t) = \frac{1}{(\bar{p} \cdot n)} \int \frac{d \lambda}{2 \pi} e^{i x (\bar{p} \cdot z)}n_{\rho} n_{\sigma} \langle p^{\prime}|F^{\rho \mu}\left(-\frac{z}{2}\right)F^{\ \sigma}_{\mu}\left(\frac{z}{2}\right) 
|p \rangle |_{z^{+}=z_{\perp}=0,z=\lambda n}
\,,
\label{eq:gpd_def_1}
\end{equation}
where $F_{\mu \nu}$ is the gluon field strength tensor; 
$n$ is a light-like vector in the direction of the initial photon momentum; 
$x_1=x+\xi$ and $x_2=x-\xi$ are the light-cone momentum fractions of the gluons along 
the $\bar{p}=(p+p^{\prime})/2$ direction, see Fig.~\ref{fig:gpd_def}.
The averaging and summation over colors is assumed.

The momentum fraction $\xi$ is associated with the longitudinal momentum transfer and determined by the kinematics of the process. For electroduction of vector mesons,
\begin{equation}
\xi=\frac{x}{2}=\frac{1}{2}\frac{Q^2+M_V^2}{W^2} \,.
\label{eq:xi}
\end{equation}

In the forward limit, i.e.,  in the $\xi=0$ and $t=0$ limit, one obtains from Eq.~(\ref{eq:gpd_def_1}):
\begin{eqnarray}
\fl
F^g(x,\xi=0,t=0) & =&  \frac{1}{(p \cdot n)} \int \frac{d \lambda}{2 \pi} e^{i x (p \cdot z)}n_{\rho} n_{\sigma} \langle p|F^{\rho \mu}\left(-\frac{z}{2}\right)F^{\ \sigma}_{\mu}\left(\frac{z}{2}\right) 
|p \rangle |_{z^{+}=z_{\perp}=0,z=\lambda n} \nonumber\\
& =& xg(x) \,,
\label{eq:gpd_forward}
\end{eqnarray}
where $g(x)$ is the usual gluon distribution (we suppress the scale dependence for brevity), see, e.g.~Ref.~\cite{Brock:1993sz}.

Note that in Eq.~(\ref{master}) the difference between the light-cone fractions of the gluons attached to the $q\bar q$ pair, $x_1-x_2=2 \xi$, see Fig.~\ref{fig:Jpsi_dipole}, was neglected, and the cross section was expressed in terms of the usual gluon distribution of the target.
One can correct for this effect phenomenologically by examining the $Q^2$ evolution of GPDs~\cite{Frankfurt:1997ha}. In particular, making a natural assumption that at  
the input evolution scale the difference between $x_1$ and $x_2$ can be neglected, the effect of $x_1 \neq x_2$ at higher $Q^2$ scales is generated by the evolution, see also Refs.~\cite{Shuvaev:1999fm,Shuvaev:1999ce}. The natural qualitative feature of this method is that the effect of $x_1 \neq x_2$ increases with an increase of $Q^2$ and the mass of the produced vector meson.

In the phenomenologically important case of $J/\psi$ production, the gluon GPD of the target can be approximated well by the usual gluon density evaluated at $x_{\rm eff}=(x_1 + x_2)/2=\xi=x/2$
multiplied by
the factor of 
$R_g$~\cite{Shuvaev:1999ce,Martin:2007sb,Harland-Lang:2013xba}
\begin{equation}
R_g=\frac{2^{3+2\lambda}}{\sqrt{\pi}}
\frac{\Gamma(\frac{5}{2}+\lambda)}{\Gamma(4 +\lambda)} \,,
    \label{eq:Rg}
\end{equation}
where the parameter $\lambda$ parametrizes the small-$x$ behavior of the gluon density, $xg(x) \sim 1/x^{\lambda}$. For the realistic value of $\lambda=0.2$, $R_g=1.2$.
Note however that the result of Eq.~(\ref{eq:Rg}) was derived in the limit when $x_1 \gg x_2$, which may not be the case for $J/\psi$ production in the dipole model \cite{Frankfurt:1997fj}. This topic thus deserves further investigation.
In the case of $\Upsilon$ production, the discussed affect is more pronounced~\cite{Frankfurt:1998yf}.

%%%%%%%%%%%%%%%%%%%%%%%%%%%%%%%%%%%
\subsection{Modeling finite-$Q^2$  effects}

In general, the $Q^2$ dependence of the differential cross section of vector meson electroduction
is contained in the amplitude in Eq.~(\ref{bb}). 
It comes from the intrinsic $Q^2$ dependence of the photon wave functuon  $\Psi_{L}^f(r,Q,z)$
and the action of the scattering operator on the energy denominator of $\Psi_{L}^f({\bf r},Q,z)$ in momentum space,
see discussion in Refs.~\cite{Frankfurt:1995jw,Frankfurt:1997fj}.
This leads to the following overall $Q^2$ dependence of the cross section
\begin{equation}
{Q^2 \over \left(Q^2+ {k_\perp^2+m^2\over z(1-z)}\right)^4} \to {1\over Q^6} \,,
\label{factoren}
\end{equation}
where $M^2_{q {\bar q}}=(k_\perp^2+m_q^2)/[z(1-z)]$ is the invariant mass squared of the $q{\bar q}$ dipole.
The asymptotic behavior of $1/Q^6$ corresponds to the leading twist approximation in the $Q^2 \to \infty$ limit. Alternatively, one can obtain this result by working in coordinate space and noticing that in the leading twist approximation, the integral
over the dipole size in Eq.~(\ref{bb})
is dominated by $|{\bf r}|\propto 1/Q$.

\begin{figure}[h]  
   \centering
   \includegraphics[width=0.8\textwidth]{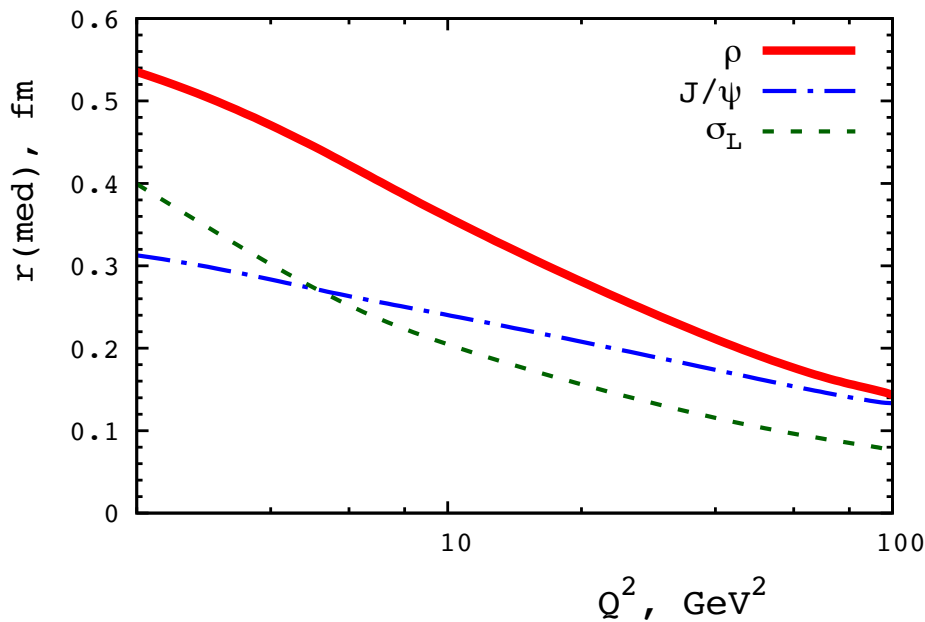} 
 \caption{The dependence of the median dipole size $r(\rm med)$  
 on the photon virtuality $Q^2$
for electroproduction of light and vector mesons and also the total photoabsorption cross secion $\sigma_L(x,Q^2)$.} 
\label{fig:bsize}
 \end{figure}

To obtain a quantitative picture of the dipole sizes characteristic for different processes, it is instructive to introduce the median dipole size $r(\rm med)$, which is defined as the value of the upper limit in the integration in Eq.~(\ref{bb}) corresponding to half of the full answer. 
Figure~\ref{fig:bsize} shows $r(\rm med)$ as a function of the photon virtuality $Q^2$ for electroproduction of light ($\rho$) and heavy ($J/\psi$) vector mesons. For comparison,
we also give $r(\rm med)$ for the total photoabsorption cross section $\sigma_L(x,Q^2)$. The calculations were carried out using the MFGS dipole model~\cite{McDermott:1999fa}
and the Gauss-LC vector meson wave function~\cite{Kowalski:2006hc}.

 One can see from the figure that, as expected, the median dipole size for $J/\psi$ is smaller than that for $\rho$ and $\sigma_L(x,Q^2)$ at low $Q^2$. As $Q^2$ is increased, the median sizes for light and heavy vector mesons converge since the details of their wave functions become unimportant.

The results presented in Fig.~\ref{fig:bsize} agree with those of Refs.~\cite{Frankfurt:1995jw,Frankfurt:1997fj}.

A related effect is that at pre-asymptotic energies, one cannot substitute $\Psi_V(r,z)$ by $\Psi_V(r=0,z)$  as has been done in Eq.~(\ref{master}).
The higher-twist correction associated with this 
substitution can be quantified by the following suppression factor
\begin{equation}
T(Q^2)=
{\left|\int d^2 {\bf r}\,dz\, \Psi_{L}^f(r,Q,z) \hat{\sigma}(x,r)\phi_V(r,z)\right|^2
\over 
\left|\int d^2{\bf r}\,dz \, \Psi_{L}^f(r,Q,z)\hat{\sigma}(x,r)
\phi_V(r=0,z)\right|^2} \,.
\label{tfac}
\end{equation}
This suppression factor is closely related to the  term $M^2_{q\bar q}=(k_\perp^2+m^2)/[z(1-z)]$ in Eq.~(\ref{factoren}). Indeed, even if one took the minimal value of  this term for light vector mesons $\sim 1$ GeV$^2$, one would still find a reduction of the cross section by 30\% at $Q^2 = 10$ GeV$^2$.
In fact,  the transverse momentum distribution appears to be  rather broad due to a singular structure of the $q\bar q$ component of the meson wave function leading to an even larger value of $M^2_{q\bar q}$ and, hence, to a larger suppression. The suppression factor gradually disappears with an increase of $Q^2$ leading to a slower decrease of the cross section with an increase of $Q^2$ than in the leading twist approximation. Note that the suppression effect is stronger for electroproduction of heavy vector mesons than for light ones. 

\begin{figure}[h]  %  figure placement: here, top, bottom, or page
   \centering
   \includegraphics[width=0.7\textwidth]{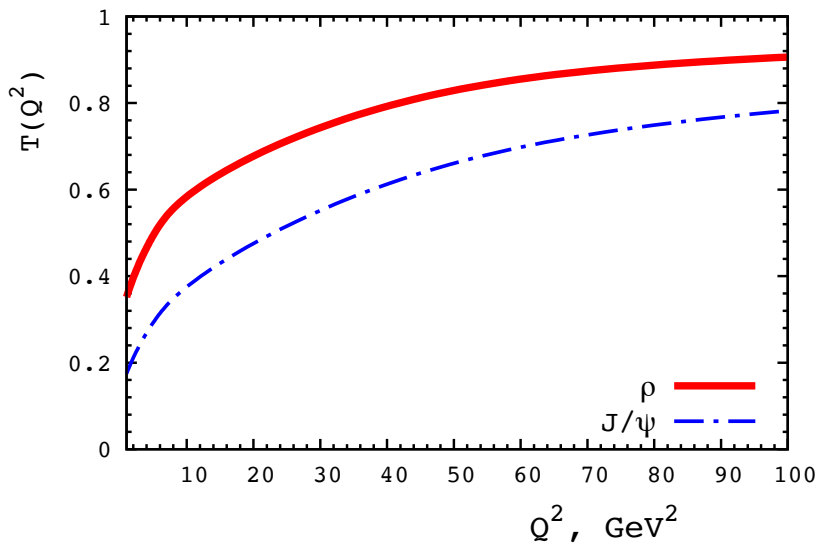} 
 \caption{The suppression factor of $T(Q^2)$, see Eq.~(\ref{tfac}), for electroproducion of light and heavy vector mesons.} 
\label{fig:T_sup}
 \end{figure}

The suppression factor of $T(Q^2)$ as a function of $Q^2$ and the trends of its behavior discussed above are presented in Fig.~\ref{fig:T_sup}.
It also allows one to explain the $Q^2$ dependence of $J/\psi$ electroproduction~\cite{Frankfurt:1997fj}, 
which is too steep in the model of~\cite{Ryskin:1992ui}, where the Fermi motion of quarks is neglected.

The finite-$Q^2$ effects considered in this section are nothing but higher-twist effects. As we mentioned above, they can be effectively modeled and estimated in the dipole formalism, for a review, 
see~\cite{Kowalski:2006hc}. This is also intertwined with possible gluon saturation effects in exclusive 
vector meson production at HERA~\cite{Kowalski:2006hc,Munier:2001nr}, see also the discussion below.

\subsection{Elastic photoproduction of $J/\psi$: from HERA to LHC}

The phenomenologically important case of vector meson production 
is elastic photoproduction of $J/\psi$, where the hard scale is provided by the mass
of $J/\psi$ (mass of the charm quark).
 The $\gamma+p \to J/\psi+p$ differential cross
section reads~\cite{Ryskin:1992ui,Frankfurt:1997fj,Frankfurt:2000ez} [compare to Eq.~(\ref{master})]
\begin{equation}
\frac{d \sigma^{\gamma p \to J/\psi p}(t=0)}{dt}= \frac{12 \pi^3}{\alpha_{\rm e.m.}} \frac{\Gamma_V M_V^3}{(4 m_c^2)^4}
\left[\alpha_s(Q^2_{\rm eff}) xg(x,Q^2_{\rm eff}) \right]^2 C(Q^2=0)\,,
\label{eq:cs_photo}
\end{equation}
where $Q_{\rm eff}$ is the effective hard scale of the process (see the discussion below).
The factor of $C(Q^2=0)$ depends on the details of the vector meson wave function and takes into account
the intrinsic motion (transverse momentum) of charm quarks in the diagram in Fig.~\ref{fig:Jpsi_dipole}. Hence, $C(Q^2=0)$ describes the effect of 
higher-twist effects in the $\gamma+p \to J/\psi+p$ cross section. It is given by the following expression,
\begin{equation}
C(Q^2=0) = \left(\frac{\eta_V}{3} m_{c}^4\right)^2 T(0)R(0) \,,
\label{eq:C}
\end{equation}
where 
\begin{equation}
T(0)R(0) = \frac{1}{M_V^4} \left[\frac{\int \frac{dz}{z(1-z)} \int dr r^3 m_{c,r}^2 \phi_V(z,r) \phi_{\gamma}(z,r)}
{\int \frac{dz}{z(1-z)} \phi_V(z,r=0)} \right]^2\,.
\label{eq:TR}
\end{equation}
Here $\phi_{\gamma}(z,r)=K_0(m_c r)$; $m_{c,r}$ is the running mass of the charm quark.

To compare predictions of Eq.~(\ref{eq:cs_photo}) to those available in the literature, it is convenient to cast it 
in the following form
\begin{equation}
\fl
\frac{d \sigma^{\gamma p \to J/\psi p}(t=0)}{dt}= \frac{\pi^3}{48 \alpha_{\rm e.m.}} \frac{\Gamma_V M_V^3}{Q_{\rm eff}^8}
\left[\alpha_s(Q^2_{\rm eff}) xg(x,Q^2_{\rm eff}) \right]^2 \tilde{C}(Q^2=0)\,,
\label{eq:cs_photo_2}
\end{equation}
where 
\begin{equation}
\fl
\tilde{C}(Q^2=0) = \frac{9}{4}\left(\frac{\eta_V}{3}\right)^2 \frac{Q_{\rm eff}^8}{\left(\frac{M_V}{2}\right)^2 m_c^4}
\left[\frac{\int \frac{dz}{z(1-z)} \int dr r^3 (m_{c,r} m_c)^2 \phi_V(z,r) \phi_{\gamma}(z,r)}
{\int \frac{dz}{z(1-z)} \phi_V(z,r=0)} \right]^2 \,.
\label{eq:C_2}
\end{equation}

Using the LC-Gauss wave function of $J/\psi$~\cite{Kowalski:2006hc}, one can explicitly calculate the overlap of the 
photon and vector meson wave functions and obtain
 \begin{equation}
\tilde{C}(Q^2=0) \approx \frac{1}{2} \frac{Q_{\rm eff}^8}{\left(\frac{M_V}{2}\right)^2 m_c^4} \,.
\label{eq:C_3}
\end{equation}
As one can see from Eq.~(\ref{eq:C_3}), the coefficient $\tilde{C}(Q^2=0)$ very strongly depends
on the effective scale $Q_{\rm eff}$ and the value of $m_c$, which affects the normalization of the $\gamma+p \to J/\psi+p$.

Using as an example the MFGS dipole model with the LC-Gauss wave function of $J/\psi$, one can estimate that
$Q_{\rm eff}^2=\lambda/d_{t}^2(\rm med) =2.8$ GeV$^2$, where $d_{t}(\rm med)$ is the corresponding median dipole size, see the discussion in the previous section. This value should be compared to $Q_{\rm eff}^2=M_V^2/4=2.4$ GeV$^{2}$ obtained in
Ref.~\cite{Ryskin:1992ui} and $Q_{\rm eff}^2=3$ GeV$^{2}$, which was determined phenomenologically~\cite{Guzey:2013qza} 
by requiring that 
Eq.~(\ref{eq:cs_photo_2}) correctly reproduces the measurements of $W$ dependence of the $\gamma+p \to J/\psi +p$ cross section at high energies. For the latter, one usually assumes the exponential $t$ dependence with the slope $B_{J/\psi}(W)$ giving
\begin{equation}
\sigma^{\gamma p \to J/\psi p}=\frac{1}{B_{J/\psi}(W)} \frac{d \sigma^{\gamma p \to J/\psi p}(t=0)}{dt} \,.
\label{eq:cs_tint}
\end{equation}
The slope $B_{J/\psi}(W)$ can be parameterized in the form $B_{J/\psi}(W)=4.5 + 0.4 \ln(W/90 \ {\rm GeV})$, which is consistent with the HERA measurements~\cite{Adloff:2000vm,Alexa:2013xxa,Chekanov:2002xi}.

\begin{figure}[h] 
\centering
\includegraphics[width=0.9\textwidth]{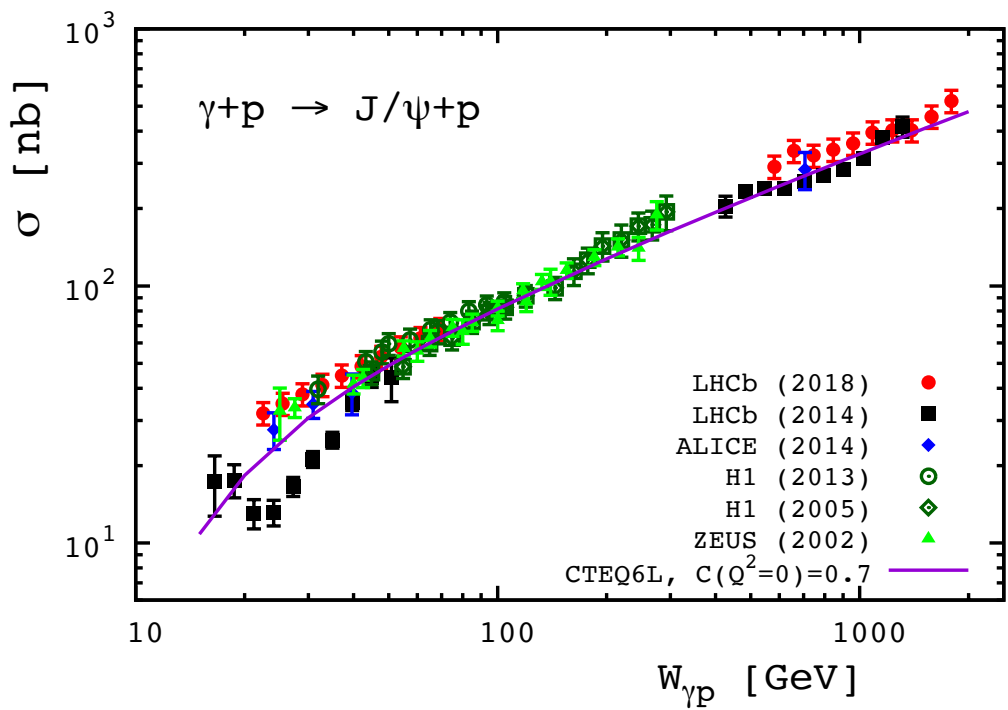} 
 \caption{The cross section of elastic $J/\psi$ photoproduction on the proton as a function of $W$:
predictions of Eq.~(\ref{eq:cs_photo_2}) vs. the H1~\cite{Aktas:2005xu,Alexa:2013xxa}, 
ZEUS~\cite{Chekanov:2002xi}, LHCb~\cite{Aaij:2014iea,Aaij:2018arx}, and ALICE~\cite{TheALICE:2014dwa} data.}
\label{fig:Sigma_proton_pQCD_Q2_3_2021}
 \end{figure}

Predictions of Eq.~(\ref{eq:cs_photo_2}) for the $\gamma+p \to J/\psi +p$ cross section as a function of $W$ and comparison to
the available H1~\cite{Aktas:2005xu,Alexa:2013xxa}, 
ZEUS~\cite{Chekanov:2002xi}, LHCb~\cite{Aaij:2014iea,Aaij:2018arx}, and ALICE~\cite{TheALICE:2014dwa} data are shown in
Fig.~\ref{fig:Sigma_proton_pQCD_Q2_3_2021}. For the calculation, we used 
$Q_{\rm eff}^2=3$ GeV$^{2}$, the CTEQ6L gluon density~\cite{Pumplin:2002vw}, and $\tilde{C}(Q^2=0)=0.7$, see details in~\cite{Guzey:2013qza}. One can see from the figure that Eq.~(\ref{eq:cs_photo_2}) provides a good description of 
the $W$ dependence and normalization of the cross section of elastic $J/\psi$ photoproduction on the proton covering
 a very wide range of energies extending into a TeV-range.

The data on $J/\Psi$ exclusive photoproduction can be described using collinear approach with a modeling of some of the  NLO effects \cite{Martin:2007sb,Jones:2015nna,Flett:2020duk}. The resulting gluon density from the fit has the $x$ dependence of the form $\sim x^{-\lambda}$ with $\lambda\simeq 0.13$ which is not far from $\alpha_{\pom}$ in the soft regime.

Alternatively, a good description of these data can also be achieved within the color dipole framework employing different models of the dipole cross sections \cite{Kowalski:2006hc} or the unintegrated gluon density~\cite{Garcia:2019tne,Hentschinski:2020yfm}.
In work \cite{Kowalski:2006hc} it was shown that the saturation based models provide very good description of the experimental data.
In ~\cite{Garcia:2019tne,Hentschinski:2020yfm} it was found that while the data can be described by both linear and non-linear low-$x$ QCD evolution, the former requires unnaturally large perturbative corrections. The authors consider it as a strong hint of the gluon saturation in exclusive $J/\psi$
photoproduction.
Finally, we mention that similar conclusions were reached in  \cite{Munier:2001nr} and \cite{Rogers:2003vi} by extracting the S-matrix from the $t$ dependence of this process. For example, for $Q^2=2 \, \rm GeV^2$ and $x\simeq 10^{-4}$ the interaction of the dipole at small impact parameters becomes predominantly absorptive, see Ref.~\cite{Rogers:2003vi}.

At the same time, one should mention that this range of impact parameters gives a rather small contribution to the cross section integrated over the impact parameter $b$.

Further, combing the dipole model with fluctuations of the gluon density, which is taken into account in the spirit of the Good-Walker formalism applied to the proton or nuclear targets, one can 
provide simultaneous description of coherent and incoherent vector meson production~\cite{Mantysaari:2016jaz,Mantysaari:2016ykx,Mantysaari:2019jhh,Krelina:2019gee}.

%%%%%%%%%%%%%%%%%%%%%%%%%%%%%%%%%%%%%%%%%%%%%%
\subsection{Lessons and open problems}
 
The HERA data, for reviews, see~\cite{Levy:2007fb,Newman:2013ada}, and the LHC data on production of $J/\psi$ in exclusive photoproduction  in ultraperipheral collisions (UPCs), see Sec.~\ref{sec:UPC},
 have  confirmed the following basic predictions of perturbative QCD.

(i) 
The rapid increase with energy of cross sections of vector meson production, which are proportional to $(xG_N(x,Q^2_{\rm eff}))^2 \propto W^{0.8}$ for $Q^2_{\rm eff} \sim 4$ GeV$^2$ in the case of
$\rho$ production for $Q^2=10-20$ GeV$^2$ and in the case  of $J/\psi$ production for $Q^2\le 10$ GeV$^2$. 
For $\Upsilon$ production, $Q^2_{\rm eff} \approx 40$ GeV$^2$, which leads to 
$\sigma(W) \propto W^{1.7}$. This prediction can be tested in the 
 ultraperipheral collisions at the LHC, but so far the statistics is not high enough.

(ii)
The data on elastic photoproduction of $J/\psi$ can be described rather well within the collinear approximation, see the discussion in the previous section. This can be interpreted as an example of the DGLAP dominance.
At the same time, the strong discrepancy between the LLx approximation and the data can be accommodated within the resummation approach,
see Sec.~\ref{sec:lowx_resum_sat}.
The restriction on the region of applicability
of the leading $\log(x_0/x)$ 
approximation follows from the necessity to take into account energy--momentum conservation. Indeed, in multi-Regge kinematics, the interval in rapidity  between adjacent radiations within the ladder is $\Delta y\gg 2$. This number is comparable with the interval in rapidities achieved (to be achieved) in DIS: 
\begin{equation}
\Delta y=\ln(1/x)+2 \ln(Q/m_N) \,.
\end{equation}
For the edge of the kinematics achieved at HERA,  $\Delta y\approx 10$. Since four units of rapidity are occupied by the two fragmentation regions,
only
two-to-three gluons are allowed to be radiated in this kinematics.  

The energy dependence of the $\rho$ production at small $-t$ is consistent with expectations of soft dynamics, 
$\alpha(t) \sim 1.1$,
however for large negative $t\sim 1.5$ GeV$^2$ the trajectory appears to be flattening out at $\alpha(t) < 1$, \cite{Aaron:2009xp}. In the perturbative regime we would expect these values to be above $1$, see Fig.\ref{fig:intercept}. In this region we expect weak dependence on $t$, see discussion in Sec.~\ref{sec:vm2}.

(iii) 
The absolute values of the cross sections of vector meson production are well  reproduced,  provided 
that  the factor $T(Q^2)$ (Eq.~(\ref{tfac})) is taken into account. In the case   of  $\Upsilon$ photoproduction, the
skewedness effects due to a large difference between $x_1 $ 
 and $x_2$ as well as the large value of the real part of the amplitude  are also important.  
 Together they  increase the predicted cross section by a factor of about four~\cite{Frankfurt:1998yf,Martin:1999rn}.
 Note that the  decrease of the $\sigma^L_{\gamma^{\ast}p \to Vp}$ cross section with an increase of $Q^2$ is slower than ${1/Q^6}$ because of the 
$\left(\alpha_SG_N\right)^2$ and $T(Q^2)$ factors. 
The cross section ratio $\sigma_L/\sigma_T \gg 1$ for $Q^2\gg m_V^2$.

\begin{figure}
   \centering
\includegraphics[width=.6\textwidth]{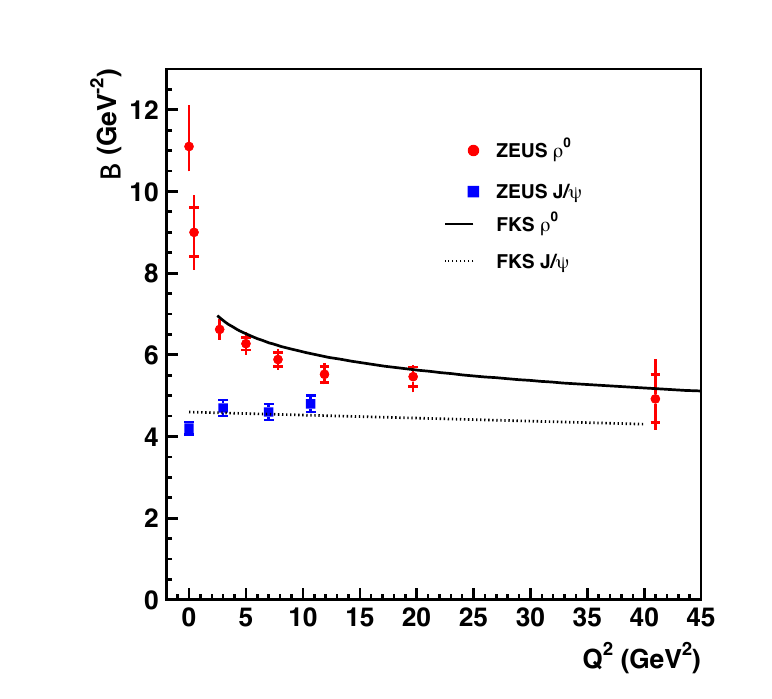}
\caption[]{The convergence of the $t$-slopes, $B$,  of $\rho$ and $J/\psi$
 electroproduction 
 off a nucleon 
  at high $Q^2$. The data are from~\cite{Chekanov:2004mw,Chekanov:2007zr}; 
 the curves are the predictions 
 of~\cite{Frankfurt:1995jw}.
 The figure is from section ``Diffractive phenomena in high energy processes'', L.~Frankfurt and M.~Strikman,
 part of book ``100 Years of Subatomic Physics'', Eds. E.M. Henley and S.D. Ellis, 2013, World Scientific.
 }
 \label{slope}
\end{figure}

(iv)
 There is a universal $t$ dependence for large $Q^2$ originating solely from the two-gluon form factor of the nucleon.
The model of Ref.~\cite{Frankfurt:1997fj}, which takes into account transverse squeezing of $\gamma_L^{\ast}$ (decrease of the average $r$)  with $Q^2$, provides a reasonable description 
of the convergence of the $t$-slopes of light mesons and $J/\psi$ production and makes the observation that the slope 
of $J/\psi$  production is practically $Q^2$ independent (Fig.~\ref{slope}).
The observed  dependence of the $t$-slopes on $Q^2$ indicates that in the case of light vector mesons,  the LT dominance (universality  regime)  is reached only at $Q^2 \ge 15 \; \mbox{GeV}^2$. 

A confirmation of a late onset of the regime of the universality follows also from the measurements of the $\phi$ and $\rho$ cross sections \cite{Aaron:2009xp}  up  to $Q^2 \simeq  \mbox{30 GeV}^2$, where the measured cross section ratio is $\sigma_{\phi}/\sigma_{\rho} = 0.191 \pm 0.007 ({\rm stat.}) ^{+0.008}_{-0.006} ({\rm syst.}) \pm 0.008 ({\rm norm.})$. The measurements are close to the value expected from quark charge counting $ 2 : 9$, but they tend to be slightly lower.  On the other hand  the LT prediction of Eq.~(\ref{master}) for $\sigma_{\phi}/\sigma_{\rho} \times 9/2$ is expected to be $1.06\pm 0.02$ (here we use the values of $\Gamma_V$ and $M_V$ from \cite{Patrignani:2016xqp}) and the theoretical expectation is that at this value of  $Q^2$, the wave functions of $\phi$ and $\rho$ should be very close. 

(v) The
observed $t$ dependence of the slopes of hard diffractive processes in the LT regime provides unique information on the distribution of quarks and gluons within hadronic targets. 
The interior of the nucleon of the radius $\approx$ 0.5 fm is filled with quarks and gluons, but not mesons (hadrons) as was
popular before the onset of QCD and currently in the low energy nuclear physics. A small number of antiquarks in the nucleon (nucleus) 
indicates that the observation of hard diffractive processes off nuclear target will allow one to demonstrate the absence of short-range meson currents in nuclei.

(vi) 
In the case of light vector mesons, one 
can estimate the effective size of a $q\bar q$ dipole as 
\begin{equation}
{B(Q^2)-B_{2g} \over  B(Q^2=0)-B_{2g}} \sim {R^2({\rm dipole}) \over R^2_{\rho}} \,,
\end{equation}
where $B_{2g} $ is the slope of the square of the two-gluon form factor.
Based on the HERA data \cite{Chekanov:2007zr},  we  conclude that
\begin{equation}
 {R^2({\rm dipole}) (Q^2\ge 3 \ {\rm GeV}^2)/ R^2_{\rho}} \le {1/2  - 1/3} 
\end{equation}
for collider energies.  
Accordingly, it appears that the soft  energy dependence of the 
cross section persists over a significant range of the dipole sizes. This is consistent with the observed similarity of the energy dependence of $\rho$ and $\phi$ photoproduction.

(vii) 
In the pQCD regime, the $t$-slope the dipole--nucleon amplitude should be a weak  function 
 of $s=W^2$, $B(s)=B(s_0)+2\alpha^{\prime}_{\rm eff}\ln(s/s_0)$, since the Gribov diffusion in the hard regime is small 
 (see the discussion in Sect.~\ref{sec:lowx_resum}.  Hence, a significant contribution to $\alpha^{\prime}$  comes 
 from the variation of the $t$ dependence of the gluon GPD with a decrease in  $x $ at $Q_0^2$.  There are indications that $\alpha_{\rm eff}(t)$ for $\rho$-meson exclusive electroproduction  drops below unity at $t \lesssim  - 1$ GeV$^2$  though in the limit $-t \sim Q^2$, one expects $\alpha_{\rm eff} > 1$.
 This may indicate that the soft physics dominates in GPDs at such $t$.

(viii) 
The contribution of soft QCD physics in the overlapping integral between the wave functions of the virtual photon and 
the transversely polarized vector meson is suppressed by the Sudakov form factor (see, e.g., the discussion 
in~\cite{Mankiewicz:1999tt}), which is absent in the case of  the processes initiated by longitudinally polarized photons. 
This is relevant for 
the understanding of 
 similar dependence of $\sigma_{L,T}$ on $x$ and on $t$ that were observed at HERA. Note, however, the indication of a slightly larger $t$ slope for $\sigma_T$ for $Q^2 = 8.6 \, \rm GeV^2$ as reported by the H1 collaboration \cite{Aaron:2009xp}.  

(ix) 
The account of NLO effects was so far concerned with 
NLO effects in the scattering amplitude, while the contribution of the $q\bar q g$ components was not analyzed. For the charmonium case, such effects appear to be important, see the discussion below.

(x) 
There exists no formal proof of factorization for photo/electroproduction of quarkonia in the 
$Q^2 \lesssim M_V^2$ limit.
However,  such factorization seems natural due to  small sizes of the lowest mass quarkonium states.
Still there are substantial difference  between low $Q^2$ limit and the limit,  where the factorization theorem was derived.
Below we list only several of the important  effects, for an extensive discussion, see \cite{Frankfurt:1997fj,McDermott:1999fa} .

(a) In the case of $J/\psi$ photoproduction, the $c\bar c g$ component of the wave function is comparable to the $c\bar c$ component since the $ \alpha_s$ factor is 
compensated by a large numerical coefficient.

(b) The nonrelativisitic approximation is not valid for $J/\psi $ production because in the overlap integrals, the region $k\sim m_c$ gives a significant contribution. Overall the quark Fermi motion results in the strong suppression of the cross section as compared to the approximation, when the Fermi motion is neglected~\cite{Ryskin:1992ui}, see also the discussion above.
The relativistic corrections to the charmonoium wave function and their effect on the cross section of elastic $J/\psi$ photoproduction
were recently considered in Refs.~\cite{Escobedo:2019bxn,Lappi:2020ufv}.

(c) The predicted cross section strongly depends on $m_c$ so one in principle needs to take into account the 
$Q^2$ dependence of $m_c$ in such a way that the condition that the intermediate mass $M^2_{c\bar c} > m^2_{J/\psi}$ is satisfied.

(d) In the case of $\psi^{\prime}$ production, the application of the dipole model appears problematic due to the proximity of the $D\bar D$ threshold. Hence, one may expect that the $\psi^{\prime}$ wave function at large distances $r$ effectively describes $D\bar D$ configurations, which  have a small overlapping integral with the photon wave function.

Since  production of $J/\psi $ serves as one of probes of the $x$ dependence of gluon density at moderate $Q^2$ and small $x$, it is important to investigate to what accuracy the effects mentioned above could be factorized so that they do not result in significant corrections to the ratio of gluon densities of nuclei and proton, see discussion in 
Sec.~\ref{sec:UPC_Jpsi}.

 \subsection{Hard diffraction in DIS and rapidity gap processes}
\label{sec:vm2}

 Another process of interest is the production of heavy vector mesons with a rapidity gap, which constitutes 
a rich source of information on high energy QCD dynamics, in particular, on the gluonic structure of the nucleon and the BFKL  dynamics in small dipole--parton scattering.

\begin{figure}[h]
\centering
\includegraphics[width=.35\textwidth]{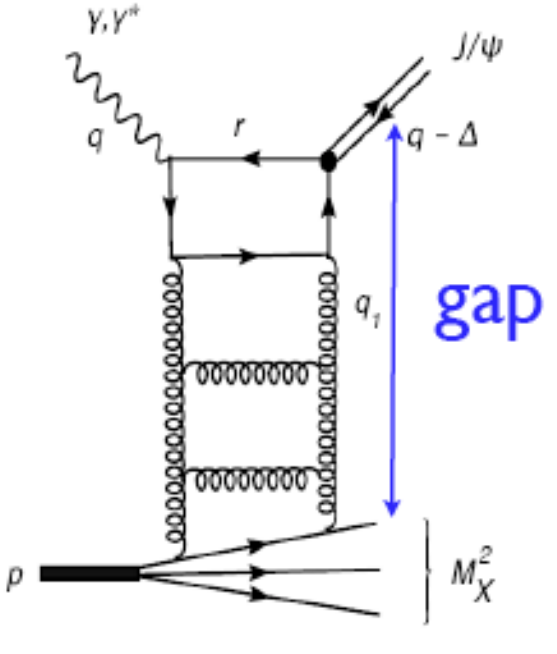}
\caption{A typical QCD diagram for the rapidity gap process (\ref{eq:gap}).} 
\label{fig:gap}
\end{figure}

 The exclusive process $\gamma^{\ast}  p \rightarrow V  p$, which is discussed in the previous section, dominates at low values of the momentum transfer, i.e., $|t| < 1 \; { \rm GeV^2}$. The requirement of the heavy vector meson in the final state allows for the selection of the small  size $q\bar{q}$ dipole pair, which scatters elastically of a nucleon. As discussed previously, this process has a steep dependence on the momentum transfer $t$.
In the diffractive vector meson production with a rapidity gap, see  Fig.~\ref{fig:gap},  
\begin{equation} \gamma^{\ast}+ p \to V + { \rm rapidity~gap} + X  \; ,
\label{eq:gap}
\end{equation}
  one can regulate hardness of the interaction by changing the virtuality of the photon, quark content of the meson, and considering a large momentum transfer $t= (p_\gamma - p_{V})^2$. 
  
 %%%%%%%%%%%%%%%%%%%%%%%%%%%%%%%%%%%%%%%%%%%%%
%%%%%%%%%%%%%%%%%%%%%%%%%%%%%%%%%%%%%%%%%%%%%

Of particular interest is the momentum transfer dependence of this process.
Below we shall discuss separately different regimes of $t$ and the underlying physical mechanisms.

\subsubsection{Probing fluctuations of the gluon field at $t\sim 0$.}

Let us first consider the reaction given by Eq.~(\ref{eq:gap}) in the limit $t\to 0$
and large $Q^2$ or heavy onium production.
In this limit, 
the two-gluon ladder couples only to one parton in the target in the leading twist approximation.
 If the strength of the coupling  to all configurations in the nucleon  containing a parton with a given $x$ was the same,
  it would be  impossible to produce an inelastic final system $X$ at $t=0$. As a result, 
  the discussed process measures the quantum fluctuations  (variance)  of the gluon field. 
  It is given by the ratio of the diffraction dissociation and elastic cross sections for 
  vector meson production at $t=0$, see \cite{Frankfurt:2008vi},
\begin{equation}
\omega_{g}={d\sigma^{\gamma^{\ast} + p \to V +{\rm rapidity \ gap} +X}(x,Q^2)/dt \over d\sigma^{\gamma^{\ast} + p \to V +p}(x,Q^2)/dt}\bigg|_{t=0} \, .
\end{equation}

To see that this ratio is proportional to the fluctuations of the gluon density, we shall follow the arguments presented in \cite{Frankfurt:2008vi}.  Let us assume that
the initial proton state can be expanded in a set of
partonic states characterized by the number of partons and their 
transverse positions. The states can be  schematically labeled as $|n\rangle$, and the expansion can thus be written as
\begin{equation}
    |p \rangle = \sum_n a_n |n\rangle \; .
    \label{eq:proton_expansion}
\end{equation}
Each configuration given by $n$ has a 
definite gluon density $G(x, Q^2| n)$, given by the expectation
value of the twist--2 gluon operator in the state $|n\rangle$. The average  gluon density in the proton is 
\begin{equation}
G(x, Q^2) \;\; = \;\; {\textstyle \sum_n}\,  |a_n|^2 \, G(x, Q^2| n) \; \equiv \;
\langle G \rangle .
\end{equation}

The assumption of factorization implies that the  the partonic states appear ``frozen'' on the typical
timescale of the hard scattering process. Therefore the 
 QCD factorization allows one to  calculate
the amplitude for vector meson production configuration
by configuration. It is (up to small calculable corrections) 
proportional to the gluon density in that 
particular configuration \cite{Brodsky:1994kf}. An essential
point is now that in the leading-twist approximation the hard 
scattering process attaches to a single parton, and, moreover,
does not transfer momentum to that parton (in the low $t$ limit).
Therefore the  partonic state $|n\rangle$ will not change. 
Making use of 
the completeness of partonic states, we find that the elastic
($X = p$) and total diffractive ($X$ is arbitrary) cross sections are
proportional to
\begin{eqnarray}
\frac{d\sigma_{\rm el}}{dt}\bigg|_{t=0} &\propto& 
\left[ {\textstyle\sum_n} |a_n|^2 G(x, Q^2| n) \right]^2  \equiv \; 
\langle G \rangle^2 \; , 
\\
\frac{d\sigma_{\rm diff}}{dt}\bigg|_{t=0} &\propto& 
{\textstyle\sum_n} |a_n|^2 \left[ G(x, Q^2| n) \right]^2  \equiv \; 
\langle G^2 \rangle \; . 
\end{eqnarray}
Therefore for the cross section of the diffraction with a rapidity gap, we have
\begin{equation}
{\sigma^{\gamma^* p \rightarrow V+{\rm rap gap}+X}} =
{\sigma_{\rm diff}} - {\sigma_{{\rm el}}} \;,
\end{equation}
and thus obtain
\begin{equation}
\omega_{g} \;\; \equiv \;\; {d\sigma^{\gamma^{\ast} + p \to V +{\rm rap \; gap} +X}/dt \over d\sigma^{\gamma^{\ast} + p \to V +p}/dt}\bigg|_{t=0}=
\frac{\langle G^2 \rangle - \langle G \rangle^2}{\langle G \rangle^2} \; .
\label{omega_g}
\end{equation}

The $x$ and  $Q^2$ evolution of $\omega_{g}$ originates from the DGLAP evolution of the gluon distribution \cite{Frankfurt:2008vi} and from multiparton correlations induced by the parton splitting (the effect similar to the one discussed for the double parton scattering, see review in \cite{Blok:2017alw}).

The value of $\omega_{g} \approx 0.15 $ for electroproduction of $\rho$-mesons in  reaction (\ref{eq:gap}) was reported in \cite{Janssen:2010zz}. A very similar value of $\omega_{g}$   for production of $J/\psi $ mesons  was derived using the data
 \cite{Alexa:2013xxa}. The  
$J/\psi$ data also indicate that 
 $\omega_{g} $ tends to decrease with increase of $W$.
 Indeed,
 the $J/\psi$ data were also fitted in the Regge phenomenology spirit using an effective Pomeron trajectory for both elastic and rapidity gap channels with $\alpha_{el}=1.2\pm 0.01$ and $\alpha_{pd}= 1.09 \pm 0.02$ leading to $\omega_g$ decreasing with an increase of $W$ as
\begin{equation}
\omega_g \propto (W/W_0)^{\Delta_R} \,, \quad \Delta_R=-0.44\pm 0.08 \,, \end{equation}
where  $\Delta_R = 4(\alpha(0)- \alpha_{el}(0))$.

In Sec.~\ref{sec:fluctuations} the fluctuations of soft interactions are discussed. There are indications that fluctuations of the strength of soft interactions given 
by $\omega_{\sigma}$ may be related to the fluctuation of the overall  size of configurations in the nucleon. Since the gluon field in small configurations should be screened and hence reduced, this picture leads to fluctuations of the strength of the gluon field in nucleons.  A model \cite{Frankfurt:2008vi} based on this picture allows one to reproduce  the magnitude of $\omega_{g}$. However, it does  not include the effect of the increase of fluctuations with $Q^2$ due to parton splitting and resulting pQCD parton--parton correlations (these issues require further theoretical studies).

\subsubsection{Competition of gluon field fluctuations and break up mechanisms of diffraction   at finite $-t\leq 0.5 \; {\rm GeV}^2$.} 

It was argued above that for $t=0$, inelastic hard diffraction is possible only due to quantum fluctuations of the gluon field. However, for finite $t$, a two-gluon ladder, which is attached to one parton of the nucleon, transfers to the parton the  transverse momentum 
$\sqrt{-t}$,
 which would most likely lead to inelastic diffractive states, if    
$\sqrt{-t} \ge \left <k_t\right>$ ($\left<k_t\right>$ is the parton
average transverse momentum).
At the same time, at moderate $t$ the final state could remain intact leading to a contribution to the elastic scattering.

The fraction of the elastic and inelastic channels can be estimated as follows, 
\begin{equation}
 \sigma_{el} \propto F_N^2(t),  \;\;\;  \sigma_{break} \propto  1 - F_N^2(t) \,,
 \end{equation}  
 where $F_N(t)$ is the two-gluon nucleon form factor,  which has approximately the same $t$ dependence as the elastic process $\gamma^* +p\to V+ p$.   
Note here that  $\sigma_{break}$ goes to zero at $t=0$ in line with the general arguments presented above since it does not include the gluon field fluctuations.

In this intermediate $t$ regime, the cross section can be expressed as a  sum of the fluctuation and break up contributions 
\begin{equation}
\fl
\frac{d\sigma (\gamma^* +p \to V + X)}{dt} \;  =  \; \omega_{g} \;
\frac{d\sigma (\gamma^*+ p\to J/\psi+ p)}{dt}\bigg|_{t=0} \exp (\tilde Bt)  +  \bigg(1 - F_N^2(t)\bigg) \,  \phi(t) \; ,
\label{break} 
\end{equation}
where $\phi(t)$ contains the residual $t$ dependence due to the coupling of two gluons to the parton of the nucleon.
Here we used the observation that fluctuations, which contribute to the cross section, are close to the average size since $\omega_{g}$ is rather small, but nondiagonal transitions should lead to  a somewhat slower  $t$ dependence of this term than for 
for elastic $J/\psi $  production (similarly to the case of hadron--nucleon diffractive scattering).

Equation~(\ref{break}) gives a qualitative explanation of the  trend observed in \cite{Alexa:2013xxa}, where the fit to the diffractive cross section systematically exceeds the data at small  $t$.

 For the function $\phi(t)$ in the second term in Eq.~(\ref{break}) in the $t$-range $-t < m_V^2$, we expect that the main contribution to the $t$  dependence originates from the cross section of gluon ladder--parton elastic scattering.
(The other sources of the $t$-dependence involve the interaction at the $M_V$ or $Q$ scales.) Therefore the function $\phi(t)$
 should have the $t$-dependence similar to that of  
 $\gamma^{\ast}$--quark (or gluon) scattering, i.e., 
 \begin{equation}
\phi(t) \propto 1/(1-t/m_0^2) \,, 
%vg\; \; \;\;m_0^2 \sim 0.5 - 0.8 \; \mbox{GeV}^2, \;
 \label{br}
 \end{equation}
 where $m_0^2\sim 0.5 - 0.8$ GeV$^2$ is the scale describing the soft-hard transition.
 
 We find that indeed Eq.~(\ref{br}) with $m_0^2= 0.7$ GeV$^2$ gives a good description of the diffractive data at $-t\leq 0.8$ GeV$^2$ 
 providing a strong support to the two-component picture of diffraction described above.

Note that in the model based on the notion of ``hot spots'' in the nucleon \cite{Mantysaari:2016ykx,Mantysaari:2016jaz},
it is assumed that the dominant contribution originated from multiple rescatterings  of small dipoles and the smallness of the rescattering amplitude is compensated by introducing strong positive short-range correlations. It is not clear how to reconcile such an assumption with  the observed strength of double parton interactions, see the discussion in Sec.~\ref{sec:pp_hard}. Also, phenomenology based on the dominance of a single gluon ladder in the $t$-channel (Fig.~\ref{fig:gap}) provides a good description of the data.

 \subsubsection{Rapidity gaps at large $t$.} 

Finally, let us consider the limit of large $-t$. In this case, on the microscopic level this process  is elastic scattering of a small dipole off a parton -- a quark or a gluon coming from the proton.
The proton dissociates, but its fragmentation remnants are still well separated from the vector meson by the large rapidity gap.
Thanks to the presence of hard scales, $Q^2$ and/or $m_{J/\psi}$ and momentum transfer $t$,
the DGLAP evolution  is strongly suppressed as well as the diffusion in the transverse momenta in the gluon ladder which is exchanged between the parton and the vector meson, see Fig.~\ref{fig:gap}. 
Hence, in the high energy limit, this process can be suitably described in terms of the exchange of the perturbative BFKL Pomeron \cite{Bartels:1996fs}. 
The latter one can be obtained from the solution to the BFKL equation in the non-forward case, that is at $t \neq 0$ \cite{Lipatov:1985uk}.  Due to the presence of large scales, that is a  large momentum transfer scale $-t$, a  heavy vector meson scale and the fact that the dissociated proton has a larger mass than the proton mass, the Pomeron ladder is `squeezed' at both ends. To be more precise, this means that the typical scales at both ends of the Pomeron are perturbative and large. This results in the suppression of the diffusion of the transverse momenta along the Pomeron ladder to the infrared regime. In addition, the multi-Pomeron exchanges which can fill the rapidity gap, are suppressed in this kinematics. The presence of the two hard scales allows for the study of the energy dependence of the Pomeron, and also to separate the effects stemming from the summation of different types of large logarithms. As discussed in  Sec.~\ref{sec:lowx_resum}, the $\ln 1/x$ logarithms are relevant at low $x$, and need to be resummed. This kinematics allows one to separate $\ln 1/x$ effects from $\ln Q^2$ logarithms. In this way, by studying the  dependence of the hard diffractive vector meson cross section as a function of the rapidity gap, one can learn about the energy dependence of BFKL Pomeron.

The process discussed here  is schematically represented in Fig.~\ref{fig:gap}. The electron--proton cross section may be factorized into a universal photon flux and the $\gamma p$ subprocess $\gamma p \rightarrow V Y$.  Since we are considering the photoproduction, the photon is quasi-real. It interacts with the incoming proton, and produces state $Y$ as well as the vector meson $V(p)$, separated by the rapidity gap of size $\Delta y$.  At large $-t$, such that $|t| \gg \Lambda_{QCD}^2$ the process is tractable within the perturbative QCD by means of the perturbative Pomeron given by the non-forward BFKL equation \cite{Balitsky:1978ic,Kuraev:1977fs,Lipatov:1985uk}. The cross section can be written in the factorized form, which is the convolution of the partonic cross section describing  the interaction of the photon with the quark or a gluon from the proton and the collinear parton distribution function
\begin{equation}
d\sigma \; = \; \sum_{i=g,q,\bar{q}} \, \int dx \, f_i (x,\mu) \, d\hat{\sigma}_{\gamma i} (\hat{s},t,\mu) \; .
    \label{eq:vm_larget}
\end{equation}
Here, $\hat{s} = x s$ is the photon--proton invariant mass squared and $f_i$ is the collinear parton distribution for the parton of type $i$, i.e., gluon, quark or anti-quark. 
 Also, $x$ is the fraction of the longitudinal momentum of the proton carried by the parton, a quark or a  gluon that participates in the sub-process, and $\mu$ is the factorization scale that is given by the transverse momentum in the process. 

\begin{figure}[h]
\hspace{-0.5cm}
\includegraphics[width=6cm, height=4.7cm]{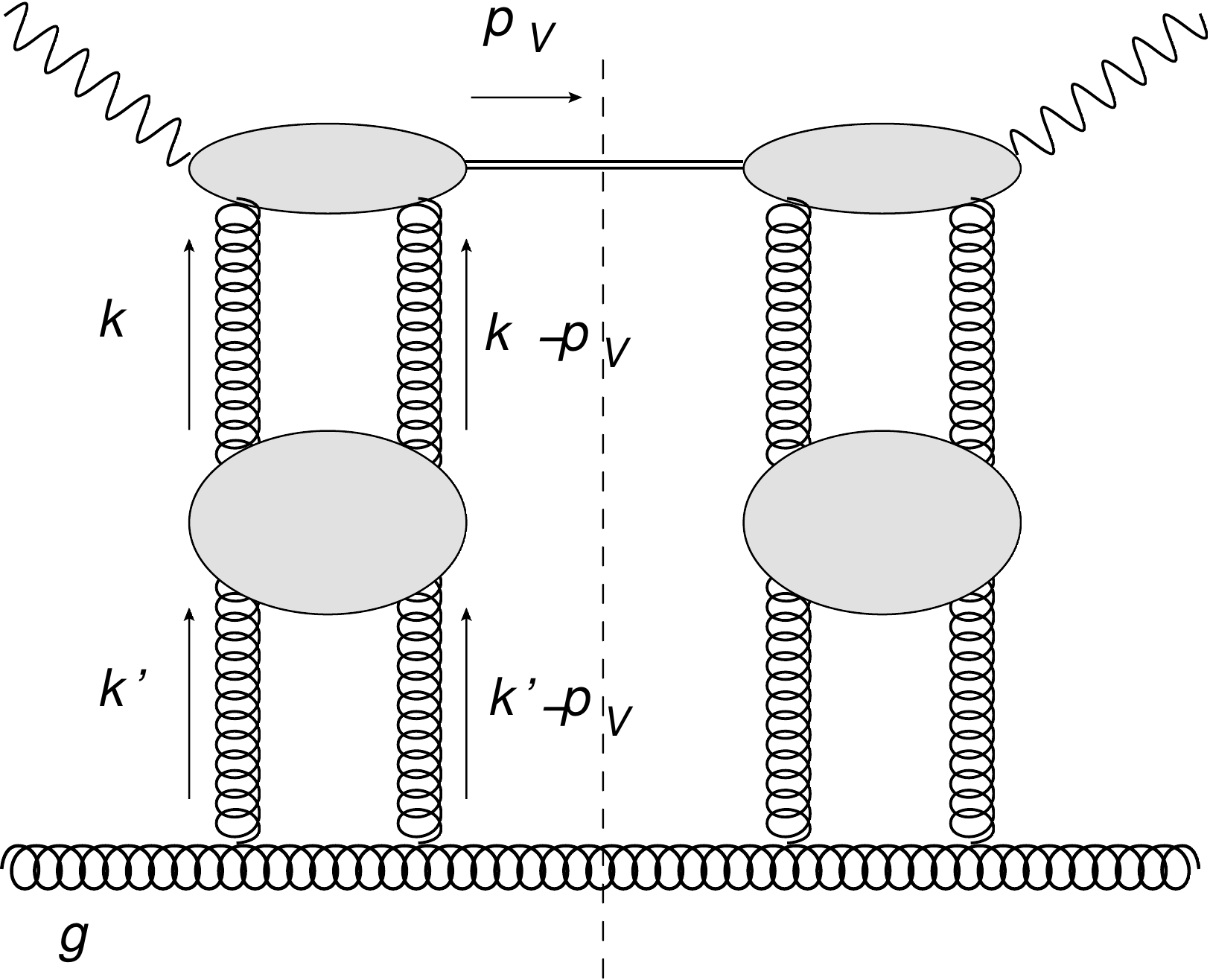}
\hspace{1.0cm}
\includegraphics[width=6cm, height=4.7cm]{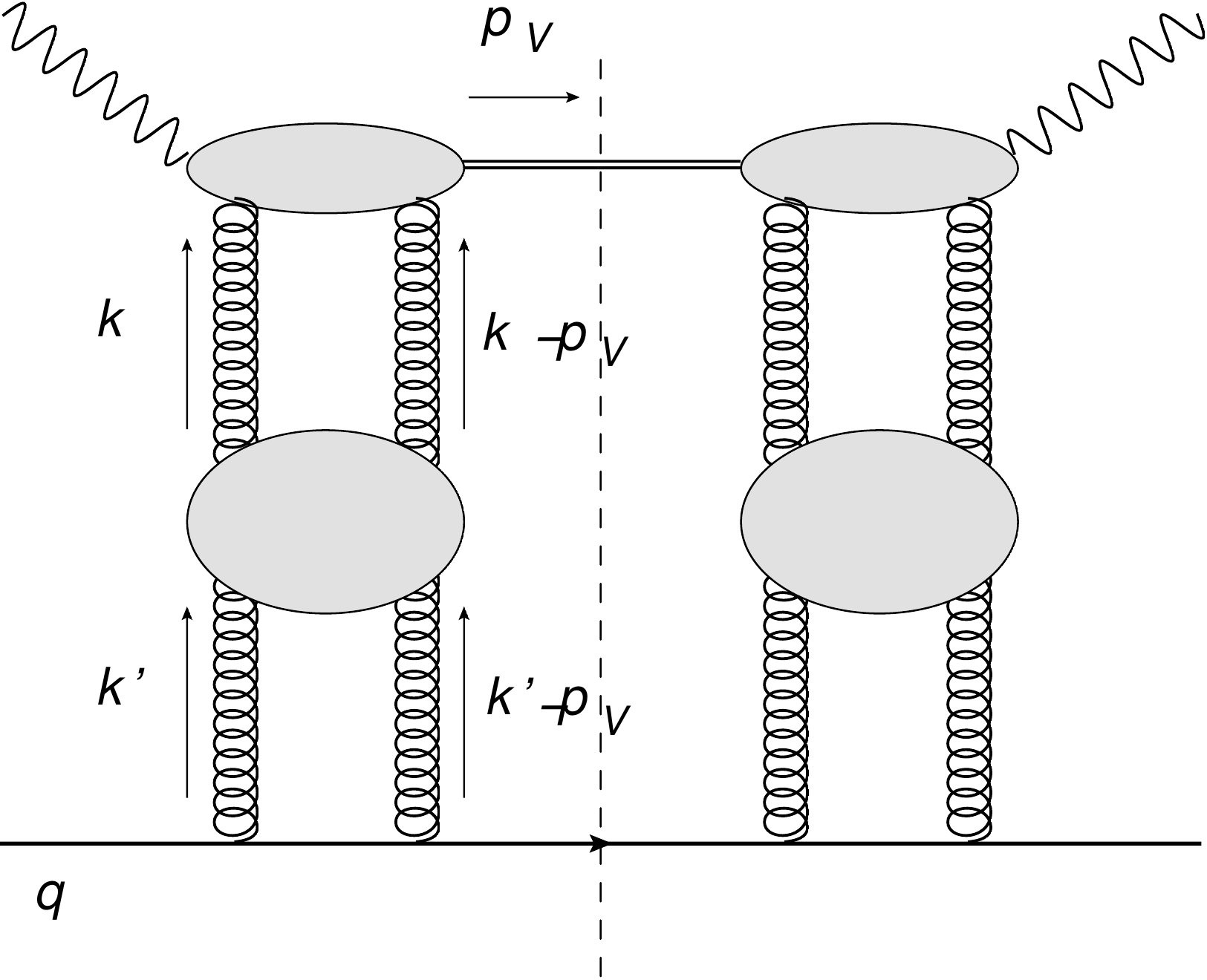}
\centering\caption{Diagrams for the partonic sub-process of the  diffractive $J/\psi$ production in scattering of the photon off the parton (gluon and quark). Vertical dashed line indicates the diffractive cut, the upper blob denotes the photon-meson impact factor and the lower blob indicates the gluon Green's function of the BFKL hard Pomeron. Figure from \cite{Deak:2020zay}, {\tt https://doi.org/10.1103/PhysRevD.103.014022}.}
\label{fig:pomeronex}
\end{figure}

The partonic cross section can be written as 
\begin{equation}
d\hat{\sigma}_{\gamma i} \; = \; C_{\gamma i} \, \frac{1}{16 \pi \hat{s}^2} \, |{\cal A}(\hat{s},t)|^2 \frac{d^2 {\bf p}}{\pi} \; ,
    \label{eq:partonic_xsection_vm_larget}
\end{equation}
where $C_{\gamma i}$ is the color factor for $i=g,q$ gluon or a quark and ${\cal A}$ is the amplitude to produce a vector meson through the Pomeron exchange which can be written as
\begin{equation}
{\rm Im} \, {\cal A}(\hat{s},t=-p^2) \; = \; \hat{s} \int \frac{d^2 {\bf k}}{2 \pi} \frac{\Phi_V({\bf k},{\bf p}) \, \Phi_q(\Delta y,{\bf k}, {\bf p})}{({\bf k}^2+s_0)[({\bf p}-{\bf k})^2+s_0]} \; ,
    \label{eq:amplitude_vm_larget}
\end{equation}
The diagrams which are taken into account in the partonic subprocess are shown  in Figs.~\ref{fig:pomeronex}. 
The  ${\bf k}$ and ${\bf k}-{\bf p}$ are the transverse momenta of the exchanged gluons in the $t$ channel and ${\bf p}$ is the transverse momentum of the produced vector meson. In this kinematical approximation, $t = -{\bf p}^2$. Also, $\Phi_V$ and $\Phi_q$ are the impact factors of the vector meson and quark, respectively. Parameter $s_0$ is the infrared cutoff, also set in the solution of the BFKL equation. 

The expression for the vector meson impact factor was calculated in Ref.~\cite{Ginzburg:1996vq} using a non-relativistic  
model for the wave function, which differs significantly from the light-cone $q\bar q $ wave function discussed in the beginning of Sec.~\ref{sec:vm1}.

The parton impact factor depends on the rapidity interval  $\Delta y$, and in the calculation can be obtained from the BFKL equation in the following form
\begin{equation}
 \Phi_q(\Delta y,{\bf k},{\bf p}) \; = \; \int d^2 {\bf k}' \, \Phi_q({\bf k}',{\bf p}) \, {\cal G}(\Delta y, {\bf k},{\bf k}',{\bf p}) \; ,
    \label{eq:parton_impfactor}
\end{equation}
where the function ${\cal G}$ is the  gluon Green's function for the non-forward BFKL equation, compare Eq.~(\ref{eq:bfkl}) with the Dirac delta $\delta^{(2)}({\bf k}-{\bf k}')$ as the initial condition. The un-evolved impact factor in the lowest order is just $\Phi_{q,0} = \alpha_s$.

The setup described above was used to compute the cross section and describe the data for diffractive production of $J/\psi$ at HERA at large $-t$.  Both ZEUS \cite{Chekanov:2002rm} and H1 collaborations \cite{Aktas:2003zi} have performed the measurements of this process, which are consistent with each other.  The ZEUS \cite{Chekanov:2002rm} data were performed in the $\gamma p$ energy range of $80 - 120 \; \rm GeV$ and H1 \cite{Aktas:2003zi} at $50 - 200 \; \rm GeV$. The range of the momentum transfer was $|t| < 12 \; \rm GeV^2$ and  $2<|t|<30 \; \rm GeV^2$ for ZEUS and H1 respectively.

\begin{figure}[ht]
\centering
\includegraphics[width=10cm,angle=0]{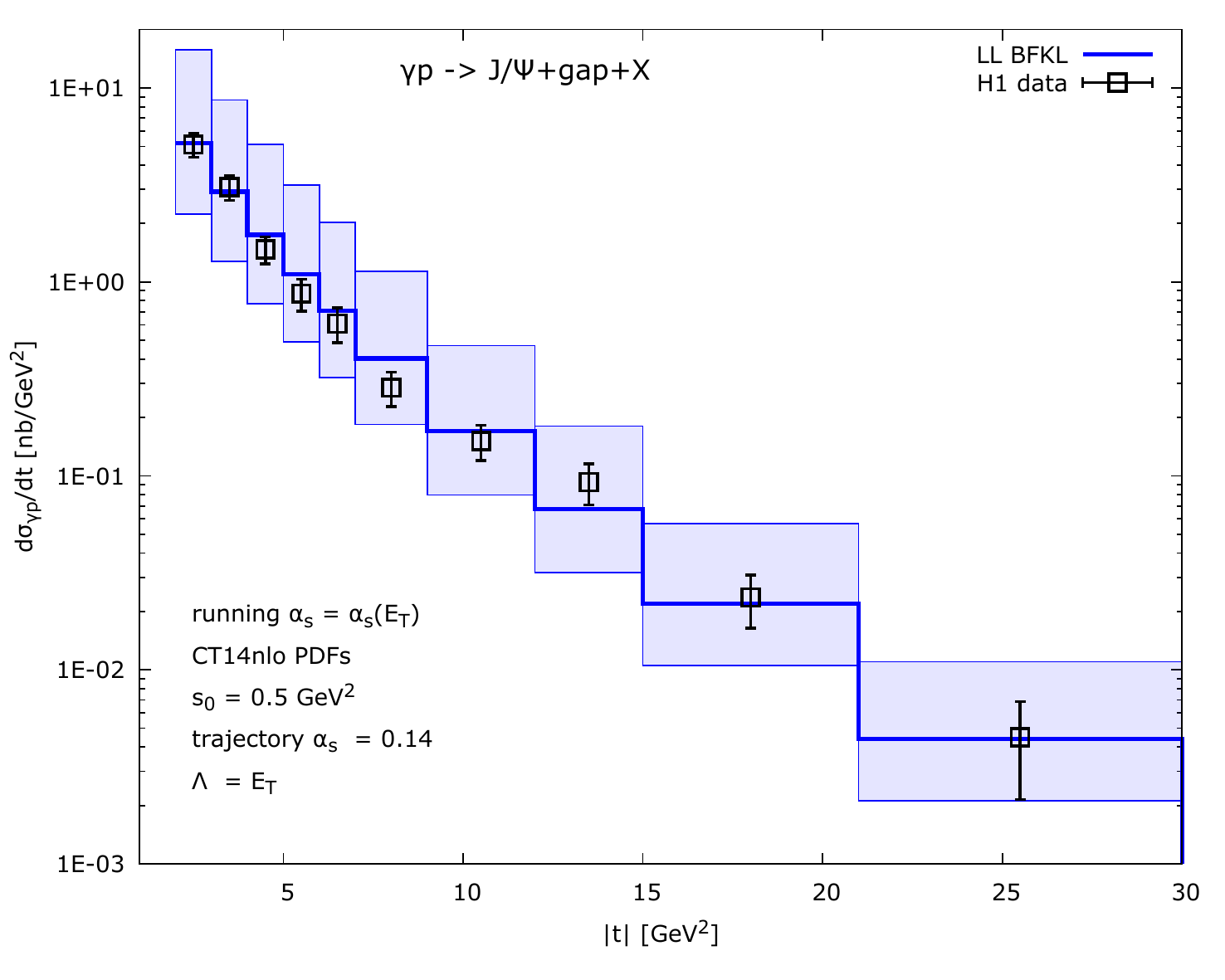}
\caption{The differential cross section $d\sigma/dt$ for the diffractive $J/\psi$ production vs H1 data \cite{Aktas:2003zi}. 
The calculation was done with LL BFKL using the coupling constant as a fit parameter equal to 
$\bar\alpha_s=0.14$. Figure from \cite{Kotko:2019kma}, {\tt https://doi.org/10.1007/JHEP07(2019)129}.}
\label{fig:vm_large_t_calc}
\end{figure}

In Fig.~\ref{fig:vm_large_t_calc} a comparison of the calculation following \cite{Kotko:2019kma} with the experimental data from H1 \cite{Aktas:2003zi} is shown.  We note that, the calculation presented here is applicable for large values of $t$. There are data from H1  \cite{Alexa:2013xxa} which extend to $|t|<5.7 \; {\rm GeV^2}$, in the region where both sets of data have common range of $t$, they are consistent with each other. We show in Fig.~\ref{fig:vm_large_t_calc} only data from \cite{Aktas:2003zi}.  The mass of the diffractively produced state is restricted to be $M_Y^2< |t|/x$ with $M_Y = 30 \; \rm GeV$. For this calculation the rapidity gap interval was defined as 
$$\Delta y = \log \left( \frac{\hat{s}}{E_T^2}\right) \; ,$$
with $E_T=\sqrt{M^2_V+p_T^2}$. 
The more exact definition of the rapidity gap is  \cite{Frankfurt:2008er} 
$$
\Delta y = \log \left( \frac{\hat{s}}{\sqrt{-t(-t+M_V^2)}}\right) \; ,
$$
which coincides with the one above when $|t| \gg M_V^2$.
The choice of the scale $\mu$ appearing in Eq.~(\ref{eq:vm_larget}) both in PDFs and in the coupling constant was taken to be $\mu^2 = E_T^2 = p_T^2 + M_V^2$, with CT14nlo PDF set. The parameter $s_0$ was set to be equal to $0.5 \;\rm GeV^2$. It is important to note that, the setup is finite for $s_0 \; \rightarrow 0$, including the BFKL equation which is infrared finite. However this parameter was introduced, following \cite{Motyka:2001zh}, to mimic the effects of the confinement.

As evident from Fig.~\ref{fig:vm_large_t_calc} the calculation describes the experimental data very well, which was also found in earlier calculations based on the BFKL equation \cite{Forshaw:2001pf}. However, the HERA data and in particular the experimental setup of the detectors at HERA did not allow for the precise measurement of the energy dependence on the rapidity gap interval. The energy dependence of the cross section in Eq.~(\ref{eq:vm_larget}) stems from the two sources: the energy dependence of the BFKL Pomeron and the $x$ dependence of the collinear distribution function. In order to be able to extract the Pomeron energy dependence it is important to perform a measurement in which the $x$ of the parton is fixed, or in other words to measure for fixed rapidity gap intervals, and not integrate over the entire range. 

In \cite{Deak:2020zay}  analysis was performed of the different experimental scenarios for the measurement of the rapidity gaps in this process at the Electron Ion Collider (EIC). 
Two cases for detection were considered:

\begin{enumerate}  
\item Request the  detection of the $J/\psi$ vector meson, a rapidity gap  and an activity in the detector in the direction of the proton beam. The latter, formed by the fragments of the dissociated target, is separated from the vector meson by the rapidity gap.
\item Request detection of the $J/\psi$ meson and a rapidity gap. No other activity in the central  detector is registered.
\end{enumerate}

In the second scenario, 
 given the limitations in the coverage of the central part of the detector, one does not have a knowledge of the exact size of the rapidity gap $\Delta y$.
 Therefore, one needs to integrate over the longitudinal momentum fraction $x$ of the parton 
 in the range of rapidity not accessible by the detector.  Since the maximum rapidity covered is related to the minimum polar angle, in the calculation the   cut was applied on the polar angle of the last particle at the edge of the rapidity gap, i.e. $\theta_j > \theta_{\rm min}$. 
 
\begin{figure}[h]
\hspace*{-2cm}
\begin{center}
\includegraphics[width=7cm, height=5cm]{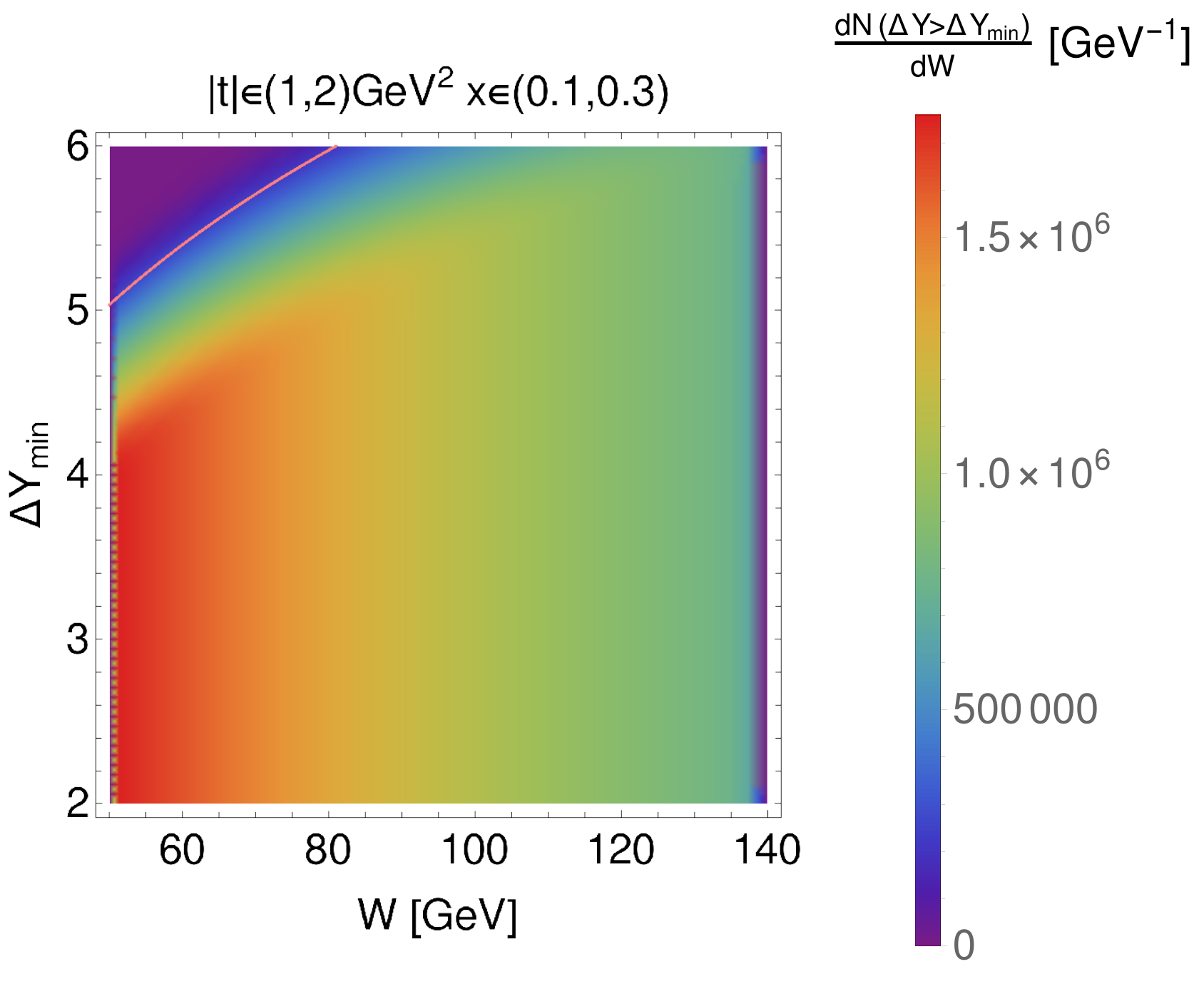}
\includegraphics[width=7cm, height=5cm]{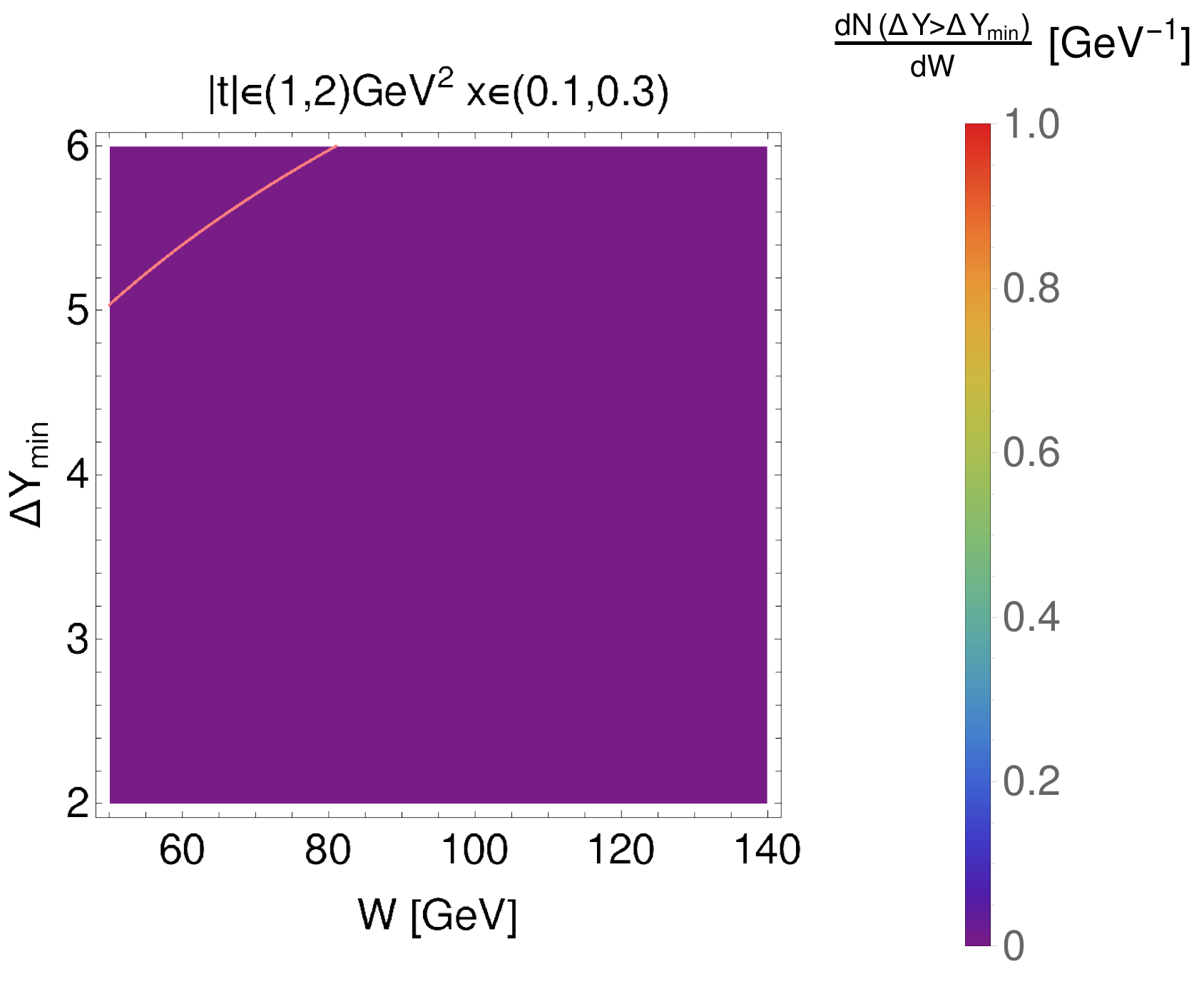}
\includegraphics[width=7cm, height=5cm]{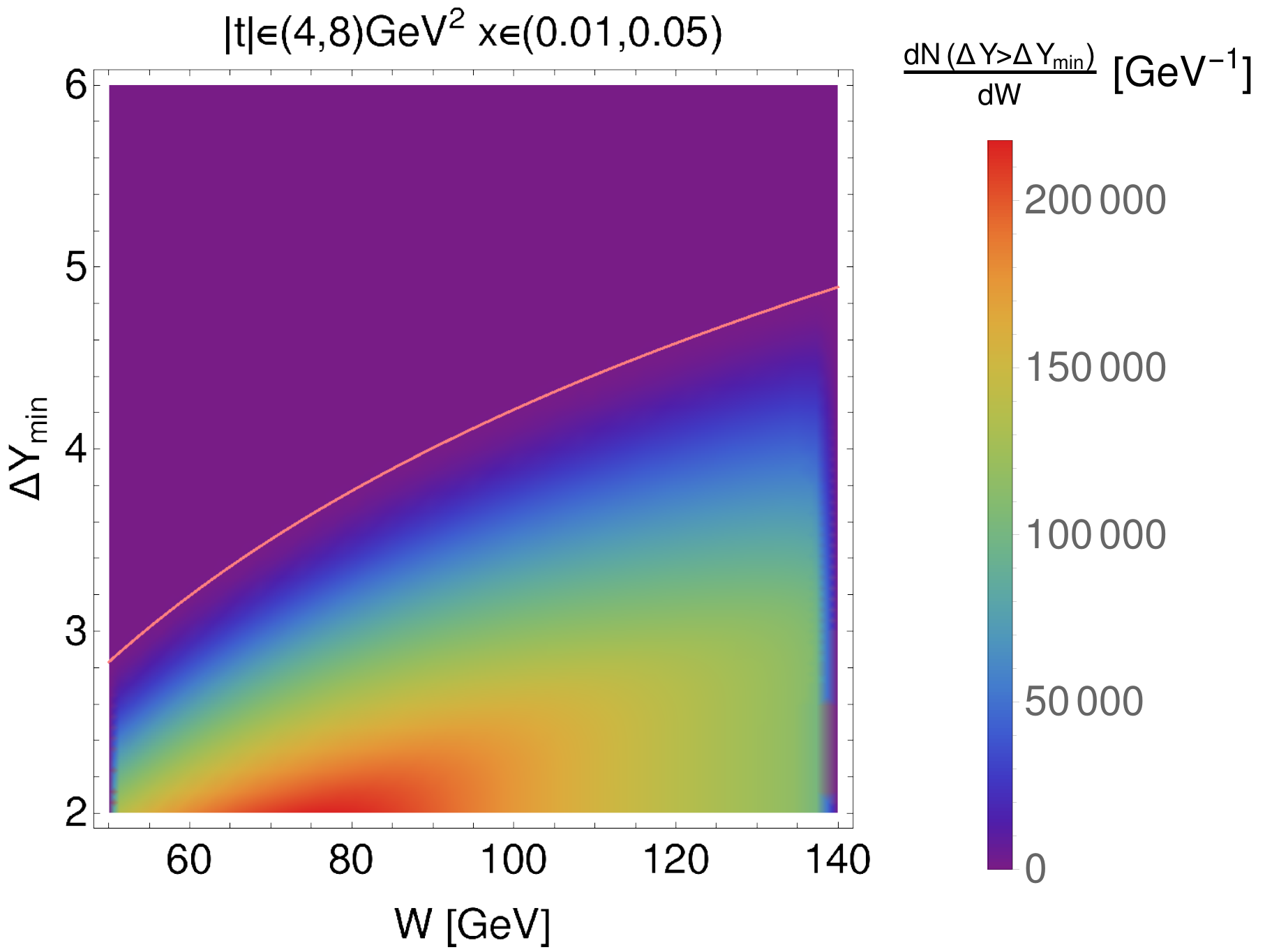}
\includegraphics[width=7cm, height=5cm]{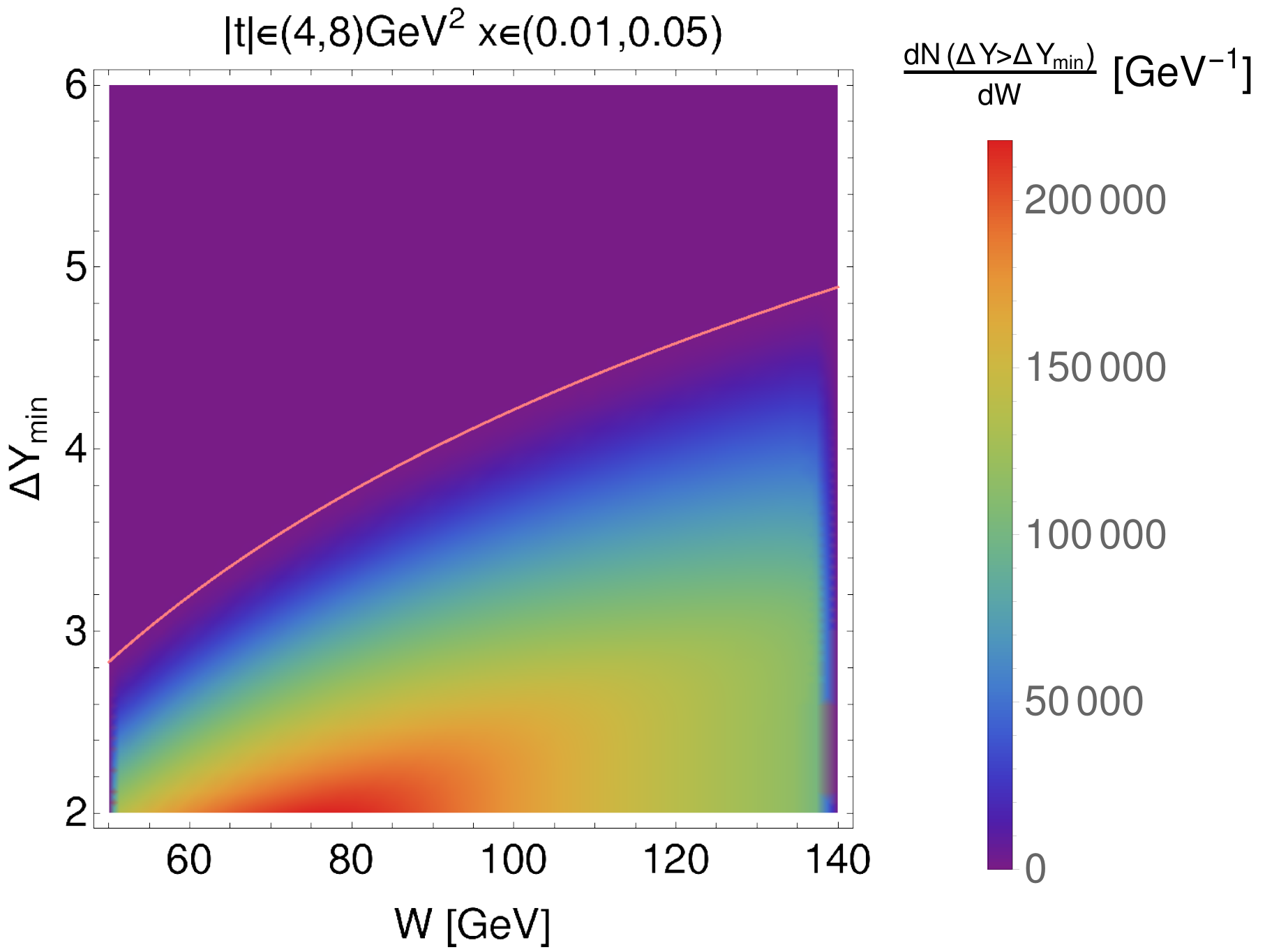}
\end{center}
\caption{Differential number of events in $W$ in bins in $t$ and $x$ as a two-dimensional function of $W$ and $\Delta Y_{\rm min}$ for the EIC kinematics. Left column: no cuts on angle, right column: restriction on angles $4^{\circ}$.
  Upper row: bin in $|t|\in (1,2) { \rm GeV^2}$ and $x \in (0.1,0.3)$, lower row:   bin in $|t|\in (4,8) { \rm GeV^2}$ and $x \in (0.01,0.05)$.
 Integrated luminosity ${\cal L }=10 \;\rm fb^{-1}$.  Figure from \cite{Deak:2020zay}, {\tt https://doi.org/10.1103/PhysRevD.103.014022}.}
\label{fig:2dim_vm_t_deltay}
\end{figure}

 In Fig.~\ref{fig:2dim_vm_t_deltay} the structure of the phase space  of vector meson production in diffraction is shown. The two-dimensional plots indicate the number of events differential in the energy $W$ as a function of $W$ and the   $\Delta Y_{\rm min}$, which is the minimum rapidity gap. The multiplicity is also integrated over $t$ and $x$ over the bins. The smaller the value of the momentum transfer $|t|$ and the larger the $x$, the larger accessible rapidity gap. For example, we see that for $|t| \in (1,2)$ and $x\in (0.1,0.3)$ the range in $\Delta y$ is well above $4$  and even up to $6$ for largest energies, left top panel in Fig.~\ref{fig:2dim_vm_t_deltay}.  On the other hand for larger $|t| \in (4,8)$ and smaller $x\in (0.01,0.06)$ the rapidity range is at most $4$ for largest energies at the EIC, left lower panel in Fig.~\ref{fig:2dim_vm_t_deltay}. In the right panels, the calculations is shown when the experimental cut on the angle is applied. We observed that for certain bins in $x$ and $t$ this cut acts as a veto on the vector meson
production in this process.  This is particularly striking in the upper right panel of Fig.~\ref{fig:2dim_vm_t_deltay}  since large $x$ and small $t$ mean
lower angle at which the vector meson is produced.

Thus it is seen that the EIC, with the proper detector setup will be able to measure range in the rapidity sizes and extract some limited range of the dependence from this measurement. Obviously, the higher energy machine, like the LHeC would have a much wider range of the rapidity, thus allowing for the longer lever arm and possibility of the more precise extraction of the energy behavior of the Pomeron. Finally, studies of rapidity gap
in the $J/\psi$ and $\psi^{\prime}$ process are feasible also in 
ultraperipheral $pA$ and $AA$ collisions at the
LHC \cite{Frankfurt:2006tp,Baltz:2007kq}.

 \section{Inelastic diffraction in hadron--hadron scattering and formalism of cross section fluctuations}
\label{sec:fluctuations}

\subsection{Inelastic diffraction and orthogonality of wave functions}

Inelastic diffraction in hadron--hadron scattering at high energies has been attracting attention of theorists since the 50's.
In a sense it appears as a counterintuitive process: a high energy hadron $h$ can scatter off a stationary nucleon and transform into a hadronic system of
the mass $M$ leaving the target nucleon practically at rest with the longitudinal momentum 
$p_{L}=(M^2 -m_h^2)/ (2 p_h) \to 0$, when the projectile momentum $p_h\to \infty$. The corresponding minimum invariant momentum transfer squared is
\begin{equation}
-t_{\rm min}=p_L^2=\frac{(M^2 -m_h^2)^2}{(2 p_h)^2} \approx 0 \, .
\label{eq:t_min}
\end{equation}

To see why the existence of such processes is non-trivial, it is instructive to consider an example of scattering of a bound state of two nucleons (e.g.~a deuteron) off a stationary hadron 
in the impulse approximation. Assuming for simplicity that the cross sections of proton--hadron and neutron--hadron interactions are equal, the cross section of
deuteron--hadron inelastic diffraction can be  written in the following form
\begin{equation} 
\fl
{d \sigma(D+ h \to M +h)\over dt } = 
2{d \sigma(N+ h \to M +h)\over dt } \left|\langle \psi_D\left(k,-k\right)| \psi_M\left(k+\frac{q}{2},- k-\frac{q}{2}\right) \rangle\right|^2 ,
\label{eq:Dh}
\end{equation}
 where $\psi_D$ and $\psi_M$ are the wave functions of the incoming deuteron and the produced state $M$, respectively, which depend on the momenta
 in the two-nucleon center of mass system. In particular, $k$ is the initial proton momentum and $k+q/2$ is the proton momentum after the scattering, 
 where $q$ is the momentum transfer.
 
 If $M$ is a two-nucleon system, the condition of orthogonality of the wave functions of the continuum and bound states leads to
 \begin{equation} 
 {d \sigma(D+ h \to "pn"+h)\over dt } \bigg|_{t=0}\, =\,0 \,,
 \end{equation}
 i.e., to the absence of inelastic diffraction at $t=0$. 
 
 In a general case of $t \neq 0$, summing over the states $M$ and using their completeness in Eq.~(\ref{eq:Dh}), one can write
   \begin{eqnarray}
   \fl
& {d \sigma(D+ h \to "pn"+h)\over dt } \bigg|_{\rm incoh} \,=\, 2 {d \sigma(N+ h \to N+h)\over dt }\big(1-F^2_D(4t)\big) \,, \nonumber\\
 &{d \sigma(D+ h \to D+h)\over dt }\,= \, 4{d \sigma(N+ h \to N+h)\over dt }F_D^2(t) \,,
 \label{eq:Dh2}
 \end{eqnarray}
 where $F_D(t)$ is the deuteron elastic form factor. Equation~(\ref{eq:Dh2})
 indicates that inelastic diffraction is impossible at $t=0$, if $h$ interacts  with just one nucleon.  At the same time  at $-t > 0$, inelastic diffraction is present and dominates for $t$ satisfying the condition $-t \ge 1/R^2_D$
 ($R_D$ is the effective deuteron radius).

 Note that no cancellation of inelastic diffraction at $t=0$ takes place, if $h$ can interact (at the amplitude level) with both nucleons of the  deuteron.
 This is the case of the nuclear shadowing correction to the total pion--deuteron (hadron--nucleus) cross section discussed in Sec.~\ref{sec:shadowing}.
 
 \subsection{Cross section fluctuations}
 
 The idea that inelastic diffraction can take place at small $t$ due to the  presence of configurations in the nucleon, which can interact with different strengths, was first suggested in the paper of Feinberg and Pomeranchuk \cite{Feinberg,Feinberg2}. In this work, as an example of fluctuation of the interaction strength, the authors considered
 fluctuations of the nucleon into the nucleon and the pion, where the latter could originate from the pion field of the nucleon.
 
 A model illustrating  this idea was suggested by Good and Walker \cite{Good:1960ba}. They assumed that the projectile can interact 
 with different interaction strengths in contributing configurations,
 which do not change while the wave packet passes through  the target.
 The corresponding coherence length 
 (time) $l_{c}$ denotes the
distance, over which the incoming hadron remains in the state with the mass $M^{\ast}$,  
\begin{equation}
\fl
l_{c} = \frac{1}{\Delta E}=
\left(\sqrt{M^{\ast\,2}+p_{\rm lab}^2}-\sqrt{m_h^2+p_{\rm lab}^2}\right)^{-1}
\simeq \frac{2p_{\rm lab}}{M^{\ast 2}-m_h^2} \ \gg R_{\rm target} \,,
\label{cls}
\end{equation}
where $R_{\rm target}$ is the radius of the target.
From Eq.\ (\ref{cls}), one can immediately see that the coherence length linearly grows with an increase of the energy of the incoming hadron. Therefore, the range of masses $M^{\ast}$, which contribute to the fluctuations and which can be considered ``frozen'', increases.

Good and Walker \cite{Good:1960ba} assumed that the state of an energetic incident hadron $|\Psi \rangle$ can be represented as a coherent superposition of eigenstates $|\Psi_{k} \rangle$  of the scattering matrix
\begin{equation}
|\Psi \rangle=\sum_{k}c_{k}|\Psi_{k} \rangle \ ,
\label{gw1}
\end{equation}
where
\begin{eqnarray}
{\rm Im} T|\Psi_{k} \rangle&=&t_{k}  |\Psi_{k} \rangle \ , \nonumber\\
\sum_{k}|c_{k}|^2&=&1 \ .
\label{gw2}
\end{eqnarray}
Here, $T$ is the scattering operator, and $t_{k}$ is the imaginary part of the  eigenvalue corresponding to the eigenstate $|\Psi_{k} \rangle$.

Various states $|\Psi_{k} \rangle$ interact with the target with different cross sections $\sigma_{k}$. By the optical theorem, $\sigma_k$ is related to the imaginary part of the scattering amplitude $t_{k}$,
\begin{equation}
\sigma_{k}=t_{k} \,.
\label{gw3}
\end{equation}
Thus, the coherent superposition of the eigenstates, which form  the final state 
emerging after the scattering, could be 
different from the initial state. 
Note that introduction of states, which interact with different 
cross sections, is natural in QCD, see the discussion in Sec.~\ref{sec:dipole_model},  where (at least in the perturbative regime) the strength of interaction is related to the area occupied by color. 

 Thus, the formalism of eigenstates of the scattering matrix is natural for
 describing diffractive dissociation of hadrons.   
However, the model is valid  only for small $t$, which is the kinematics considered in the original paper \cite{Good:1960ba},  since the authors discussed diffractive dissociation for 
scattering off nuclei. Thus, they effectively assumed that $t$ is very small, $-t \le 2/R_A^2$, where $R_A$ is the effective nucleus size.  
Later on in a number of papers it was  assumed  that Eq.~(\ref{gw2}) can be applied in a wide range of the momentum transfer, which seems problematic.  Indeed, elastic scattering of one of the constituents of the diffracting hadron can break it at finite $t$ even in the absence of fluctuations, see the case of the deuteron--proton  scattering considered above and the rapidity gap process discussed in Sec.~\ref{sec:vm2}.  The assumption that cross section eigenstates   
are orthogonal at $t \neq 0$ is in contradiction with calculations in the dipole model, where one obtains for 
two states with different transverse sizes $r_t$ and $r_t^{\prime}$ 
\begin{equation}
\langle \psi(r_t)| T(t \neq 0)|\psi(r_t^{\prime}) \rangle \neq 0 \;.
\end{equation}
Also, additional evidence comes from the analysis of soft diffraction in $pp$ scattering \cite{Alberi:1981af},  which shows that spin-flip amplitudes become important and dominate at large $-t \ge 0.2 - 0.3$ GeV$^2$. 

 Thus, the formalism
of eigenstates of the scattering matrix is suitable for
 describing diffractive dissociation of hadrons for $t\sim 0$.

Using Eqs.~(\ref{gw1})--(\ref{gw3}), diffractive dissociation can be presented as follows. 
Diffractive scattering occurs when the final state carries the same quantum numbers as the initial state, i.e., whenever the initial state overlaps with any $|\Psi_{k} \rangle$. Then,
the total diffractive differential cross section at $t=0$ can be presented  as 
\begin{equation}
\Big(\frac{d \sigma}{dt} \Big)^{\rm diff}_{t=0}=\frac{1}{16\pi}\sum_{k}|\langle\Psi_{k}| {\rm Im } T |\Psi \rangle|^2=\frac{1}{16\pi}\sum_{k}|c_{k}|^2 t_{k}^2 \equiv \frac{1}{16\pi}\langle \sigma^2 \rangle \ .
\label{o1}
\end{equation}
In Eq.\ (\ref{o1}),  we have used the completeness of the set of states $|\Psi_{k}\rangle$ and the optical theorem\ (\ref{gw3}). 
Similarly, the elastic 
differential cross section at $t=0$ reads
\begin{equation}
\Big(\frac{d \sigma}{dt} \Big)^{\rm el}_{t=0}=\frac{1}{16\pi}|\langle\Psi| {\rm Im} T |\Psi \rangle|^2=\frac{1}{16\pi}\Big(\sum_{k}|c_{k}|^2 t_{k}\Big)^2 \equiv \frac{1}{16\pi}\langle \sigma \rangle  ^2 \,.
\label{o2}
\end{equation}
In these equations, we introduced the first and second moments 
of the distribution over cross sections
\begin{eqnarray}
\langle \sigma \rangle &=& \sum_{k}|c_{k}|^2 t_{k} \,, \nonumber\\
\langle \sigma^2 \rangle &=& \sum_{k}|c_{k}|^2 t_{k}^2 \,.
\end{eqnarray}
Subtracting the elastic cross section 
from the total diffractive cross section, one obtains 
the diffractive  dissociation cross section  (inelastic diffractive cross section)

\begin{equation}
\Big(\frac{d \sigma}{dt} \Big)^{\rm diss}_{t=0}=\Big(\frac{d \sigma}{dt} \Big)^{\rm diff}_{t=0}-\Big(\frac{d \sigma}{dt} \Big)^{\rm el}_{t=0}=\frac{1}{16\pi}\Big(\langle \sigma^2 \rangle-\langle \sigma \rangle ^2\Big) \, .
\label{ineldiff}
\end{equation}

Equation~(\ref{ineldiff}) was first derived in \cite{Miettinen:1978jb} to describe diffractive dissociation within the framework of cross section fluctuations. It
explicitly demonstrates
 that diffractive dissociation occurs only if different components $|\Psi_{k} \rangle$ of the incident hadron interact with the target with different strengths $t_{k}$, i.e., when cross section fluctuations take place in the wave function of the incoming hadron.

\subsection{Properties of distribution over interaction strength $P(\sigma)$}

Within the eigenstate scattering formalism it is useful to  introduce 
the distribution over cross sections $P(\sigma)$,  which gives 
the probability to find in an energetic projectile (proton, pion, real or virtual photon) a hadronic (quark--gluon) configuration interacting with the target with the cross section $\sigma$. This distribution is a continuum version of the descrete formalism discussed in the previous section,
\begin{eqnarray}
\sum_k |c_{k}|^2 &\to & \int d\sigma P(\sigma)   \,, \nonumber\\
\langle \sigma \rangle &=& \int d\sigma P(\sigma)   \sigma \,, \nonumber\\
\langle \sigma^2 \rangle &=& \int d\sigma P(\sigma)   \sigma^2 \,. 
\label{eq:Psigma}
\end{eqnarray}
Note that the $P(\sigma)$ distribution was introduced by Miettinen and Pumplin~\cite{Miettinen:1978jb} in the context of a  parton model, where the strength of interaction is proportional to the number of slow partons. It was revived in Ref.~\cite{Blaettel:1993rd}  within the color screening framework.

The distribution of $P(\sigma)$   
 satisfies the normalization and the average cross section sum rules
 \begin{eqnarray}
 \int d\sigma P(\sigma) &=&1 \,, \nonumber\\
 \int d\sigma P(\sigma) \sigma &=&\sigma_{\rm tot} \,,
 \label{norm}
 \end{eqnarray}
 where $\sigma_{\rm tot}$ is the total projectile--target cross section.
 
 The dispersion of $P(\sigma)$ around $\langle \sigma \rangle$ is naturally related to the cross section of diffractive dissociation, see Eq.~(\ref{ineldiff}), 
 \begin{equation}
\int d\sigma P(\sigma) (\sigma^2/\sigma_{tot}^2 -1) \, = \,
\Big(\frac{d \sigma}{dt} \Big)^{\rm diss}_{t=0}\Big/\Big(\frac{d \sigma}{dt} \Big)^{\rm el}_{t=0} \, \equiv \, \omega_\sigma \,.
\label{ineldiffsum}
\end{equation}

The shape of the distribution 
$P(\sigma)$  can be modeled using Eqs.~(\ref{norm})
and (\ref{ineldiffsum}) and making a natural assumption that  fluctuations around $\langle \sigma \rangle$ should be approximately Gaussian. 
    In addition one can use the analysis \cite{Blaettel:1993ah} of the coherent diffraction off the deuteron, which indicates  that $\left< (1 - \sigma/\sigma_{\rm tot} )^3\right> \approx 0$.

 The  behavior of $P(\sigma \to 0)$ is determined by the interplay of the probability for a hadron to be in a small-size configuration and the cross section for such a configuration leading to 
 \begin{equation}
P_{h}(\sigma) \propto \sigma^{n_{q}-2} \ ,
\label{o3}
\end{equation}
where $n_{q}$ is the number of valence quarks in the hadron. Thus, for protons and pions, when $\sigma$ is much smaller than the average value of $\sigma$, i.e., when $\sigma \ll \langle \sigma \rangle$, one obtains that 
\begin{eqnarray}
P_{p}(\sigma) &\sim&  \sigma  \,, \nonumber\\ 
P_{\pi}(\sigma) &\sim &  {\rm const} \,.
\label{pproton}
\end{eqnarray}

The parameterizations of the proton and pion distribution 
functions $P(\sigma)$  suggested in the literature~\cite{Blaettel:1993rd,Blaettel:1993ah} describe correctly the behaviour at small and large $\sigma$ discussed above,
\begin{eqnarray}
P_{p}(\sigma)&=&N_p\,\frac{\sigma / \sigma_{0}}{\sigma / 
\sigma_{0}+1}e^{-(\sigma-\sigma_{0})^{2}/(\Omega \sigma_{0})^{2}} \ , \nonumber\\
P_{\pi}(\sigma)&=& N_{\pi}\,e^{-(\sigma-\sigma_{0})^{2}/
(\Omega \sigma_{0})^{2}} \,,
\label{param}
\end{eqnarray} 
where $\sigma_0 \sim \sigma_{\rm tot}$ and $\Omega$ determine the peak and the width of $P(\sigma)$, respectively. These parameters are found using the sum rules of Eqs.~(\ref{norm}) and (\ref{ineldiffsum}), and,
therefore, they depend on energy.

In  the pion  case one can go further and use pQCD to calculate the probability of small-size configurations, i.e., the constant entering Eq.\ (\ref{pproton}).
Indeed, using the QCD factorization theorem \cite{Frankfurt:1993it,Blaettel:1993rd}  for the interaction of a small-size dipole with hadrons, one finds for $P_{\pi}(\sigma)$ at $\sigma \ll \langle \sigma \rangle$,
\begin{equation}
P_{\pi}(\sigma \ll \langle \sigma \rangle)\, =\,\frac{6f^2_{\pi}}{5 \alpha_s(4k^{2}_{t})\bar x
  G_{N}(\bar x, 4k^{2}_{t})} \; .
\label{ppion}
\end{equation}
In this equation, $\alpha_{s}(4k^{2}_{t})$ is the QCD running coupling constant;
$G_N(\bar x,4k^{2}_{t})$ is the gluon distribution in the nucleon;
 $\bar x= 4k^2_t/s_{\pi N}$, where  $s_{\pi N}$ is the center of mass energy squared; $k^2_t \propto 1/d_t^2$, 
where $d_t$ is the transverse size of the $q \bar{q}$ pair in the pion wave function; $f_{\pi}$ is the constant of the $\pi \to \mu \nu$ decay. 

This estimate 
gives a correct magnitude of $P_{\pi}(\sigma \ll \sigma_{tot}(\pi N) $, see Fig.~2 of Ref.~\cite{Frankfurt:1996ri}.

To present explicit forms of $P_p(\sigma)$ and $P_{\pi}(\sigma)$,
we used the COMPETE parametrization of the total proton--proton and pion--proton cross sections~\cite{Patrignani:2016xqp}.

The variance of the $P_p(\sigma)$ distribution for protons, $\omega_\sigma$ in Eq.~(\ref{ineldiffsum}), can be extracted 
from the data at fixed-target and collider energies.
It was found~\cite{Blaettel:1993ah,Guzey:2005tk} that $\omega_{\sigma}$  initially grows with energy reaching a broad maximum value $\omega_{\sigma} \sim 0.30 \pm 0.05$ around  $ \sqrt{s} \sim \mbox{60 GeV}$ and 
then starts to decrease for $\sqrt{s} > 200$ GeV toward the Tevatron and LHC energies reaching the value of $\omega_{\sigma} \sim 0.1$ at 
$\sqrt{s} \sim \mbox{8 TeV}$. Thus, one can parameterize the energy dependence of $\omega_\sigma$ for the proton in the following simple form
\begin{equation}
\fl
 \omega_{\sigma}(s)=\left\{\begin{array}{ll}
\beta \,\sqrt{s}/(24 {\rm \, GeV}) \,, & \sqrt{s} < 24 \ {\rm GeV} \,, \\
\beta  \,, &  24 < \sqrt{s} < 200 \  {\rm GeV} \,, \\
\beta  -0.056 \ln (\sqrt{s}/200{\rm \, GeV}) \,, &  \sqrt{s} > 200 \ {\rm GeV} \,,
\end{array} \right.
\label{eq:omega_fit}   
\end{equation}
where $\beta=0.30 \pm 0.05$.

\begin{figure}
\centering
 \includegraphics[width=0.75\textwidth]{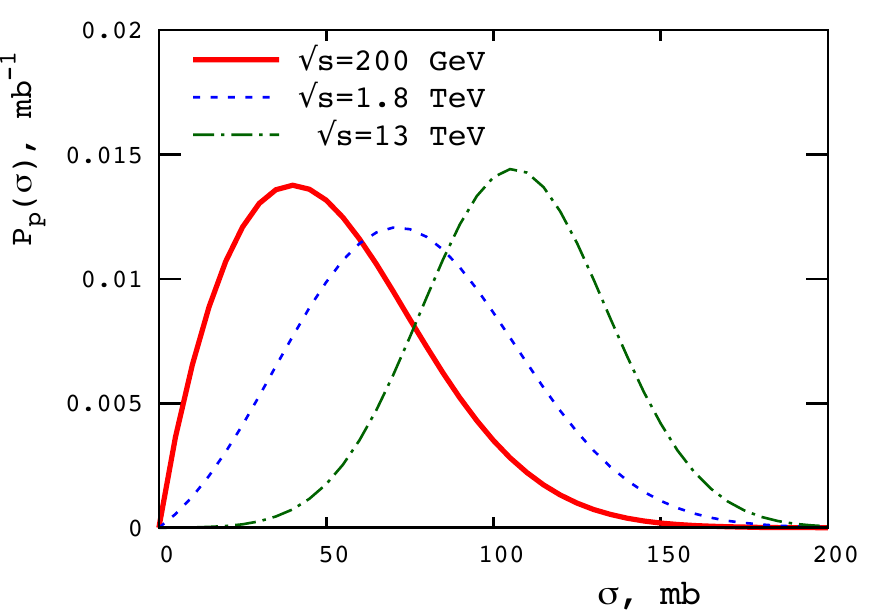}
\caption{The distribution $P_{p}(\sigma)$ as a function of $\sigma$
for $\sqrt{s}=200$ GeV (solid red), $\sqrt{s}=1.8$ TeV (blue short-dashed), and $\sqrt{s}=13$ TeV (green dot-dashed).} 
\label{fig:P_p}
\end{figure}

The resulting $P_{p}(\sigma)$ as a function of $\sigma$
for three typical values of energies ($\sqrt{s}=200$ GeV, $\sqrt{s}=1.8$ TeV, and $\sqrt{s}=13$ TeV) is presented in Fig.~\ref{fig:P_p}.
One can see from the figure that the distribution $P_{p}(\sigma)$ remains rather broad for all studied  energies since a decrease of $\omega_{\sigma}$ is compensated by an increase of $\sigma_{\rm tot}$~\cite{Patrignani:2016xqp}.

\begin{figure}[h]
\centering
 \includegraphics[width=0.75\textwidth]{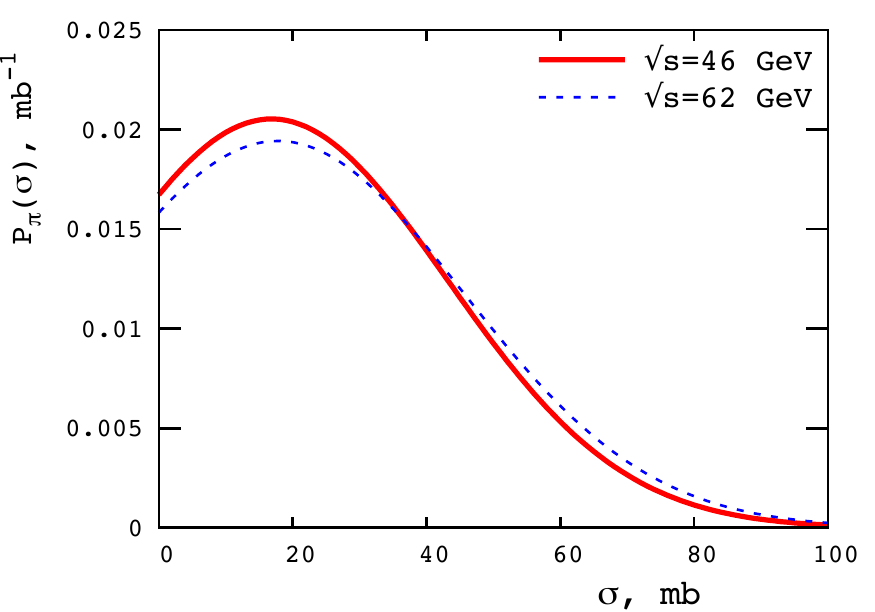}
\caption{The distribution $P_{\pi}(\sigma)$ as a function of $\sigma$
for $\sqrt{s}=46$ GeV (solid red) and $\sqrt{s}=62$ GeV (blue dashed).} 
\label{fig:P_pi}
\end{figure}

For the pion projectile, one can use the constituent quark counting rule for the ratio of the nucleon–nucleon and the pion--nucleon total cross sections~\cite{Levin:1965mi} to obtain the following simple estimate for $\omega_{\sigma}$ for pions
\begin{equation}
 \omega_{\sigma}^{\pi}=\frac{3}{2} \omega_{\sigma} \,.   
\end{equation}
The resulting $P_{\pi}(\sigma)$ distribution for pions as a function of $\sigma$ for $\sqrt{s}=46$ GeV and $\sqrt{s}=62$ GeV is shown in Fig.~\ref{fig:P_pi}.
These values correspond to the invariant photon--nucleon energies 
accessed in photoproduction of $\rho$ mesons in heavy-ion ultraperipheral collisions at central rapidities at the LHC, see next section.

\subsection{$P_{\rho}(\sigma)$ distribution for $\rho$ mesons}

Color fluctuation phenomena can also be studied in $\rho$ meson photoproduction, which was explored at HERA and   
 in heavy-ion ultraperipheral collisions at the LHC.
We note that $P_{\rho}(\sigma)$ corresponds actually to the $\gamma \to \rho$ transition (we shall use $P_{\rho}$ as a shorthand notation).

Based on the constituent quark counting rule, it is generally expected that the $P_{\rho}(\sigma)$ distribution for $\rho$ mesons should be similar to that for pions. However, this does not seem to be supported by the HERA data on $\rho$ photoproduction. Indeed, the model based on the combination of the assumption that $\sigma_{\rho N}=\sigma_{\pi N}$
($\sigma_{\rho N}$ and $\sigma_{\pi N}$ are the total $\rho$ meson--nucleon and pion--nucleon total cross sections, respectively) with the vector meson meson dominance (VMD) model somewhat overestimates the HERA data on the $\sigma_{\gamma p \to \rho p}$ cross section of elastic 
$\rho$ photoproduction on the proton~
\cite{Derrick:1995vq,Derrick:1996vw,Breitweg:1997ed,Weber:2006di}.
This calls for modifications of $P_{\rho}(\sigma)$  compared to 
$P_{\pi}(\sigma)$.
First, a natural mechanism of reduction of the 
 $\sigma_{\gamma p \to \rho p}$ cross section is offered by the color dipole model, where due to the point-like coupling of the photon to quarks,  the overlap between the real photon and $\rho$ meson light-cone wave functions selects on average dipoles with a smaller transverse sizes than  those characteristic for the pion ($\rho$ meson) wave function. In the language of $P_{\rho}(\sigma)$, it leads to an enhanced contribution of small $\sigma$, which can be modeled in the following form~\cite{Frankfurt:2015cwa}
\begin{equation}
P_{\rho}(\sigma)= N_{\rho} \frac{1}{(\sigma/\sigma_{0})^2+1}e^{-(\sigma-\sigma_{0})^{2}/
(\Omega \sigma_{0})^{2}} \,. 
\label{eq:P_rho}
\end{equation} 

Second, small-size quark--antiquark dipoles are characterized by the 
large relative transverse momentum and the large invariant mass. To take this into account, one should model the variance of the $P_{\rho}(\sigma)$ distribution, $\omega_{\sigma}^{\rho}$, using information on photon diffractive dissociation on the proton. This can be done as follows~\cite{Frankfurt:2015cwa}.
Using the formalism of cross section fluctuations, the cross section of photon diffractive dissociation on the proton can be written in the following form [compare to Eq.~(\ref{ineldiffsum})]
\begin{equation}
\fl
\frac{d\sigma_{\gamma p \to X p}(t=0)}{dt}=\frac{1}{16 \pi} 
\left(\frac{e}{f_{\rho}} \right)^2 \int d \sigma  P_{\rho}(\sigma)(\sigma^2 -\sigma_{\rho N}^2)
=\frac{1}{16 \pi} 
\left(\frac{e}{f_{\rho}} \right)^2 \omega_{\sigma}^{\rho}\sigma_{\rho N}^2 \,,
\label{eq:photon_DD}
\end{equation}
where $f_{\rho}$ is the $\gamma-\rho$ coupling constant fixed 
by the $\Gamma(\rho \to e^{+}e^{-})$ width of the 
$\rho \to e^{+}e^{-}$ decay,
$f_{\rho}^2/( 4 \pi)=2.01 \pm 0.1$. In Eq.~(\ref{eq:photon_DD}), 
$\sigma_{\rho N}$ is the total $\rho$--nucleon cross section,
which is determined by fitting the available fixed-target
and HERA experimental data on the elasic $d\sigma_{\gamma p \to \rho p}(t=0)/dt$ cross section,
\begin{equation}
\sigma_{\rho N}=\int d \sigma P_{\rho}(\sigma) \sigma =\frac{f_{\rho}}{e} \sqrt{16 \pi \frac{d\sigma_{\gamma p \to \rho p}(t=0)}{dt}} \,.
\label{eq:photon_DD_2}
\end{equation}

To proceed with the determination of $\omega_{\sigma}^{\rho}$, one invokes the result of the analysis in Ref.~\cite{Chapin:1985mf}, which demonstrated that the cross sections of photon and pion 
diffractive dissociation can be related as follows
\begin{equation}
\frac{d\sigma_{\gamma p \to X p}(t=0)/dt}{\sigma_{\gamma p}} \approx
\frac{d\sigma_{\pi p \to X p}(t=0)/dt}{\sigma_{\pi p}}=\frac{\omega_{\sigma}^{\pi}}{16 \pi} \sigma_{\pi p} \,,
\label{eq:photon_DD_3}
\end{equation}
where $\sigma_{\gamma p}$ is the total photoabsorption cross section. Combing Eqs.~(\ref{eq:photon_DD}) and (\ref{eq:photon_DD_3}), 
one obtains
\begin{equation}
\omega_{\sigma}^{\rho}=\left(\frac{f_{\rho}}{e}\right)^2 \frac{\sigma_{\gamma p} \sigma_{\pi p} }{\sigma_{\rho N}^2} \,
\omega_{\sigma}^{\pi}\,.
\label{eq:photon_DD_4}
\end{equation}

\begin{figure}[h]
\centering
 \includegraphics[width=0.75\textwidth]{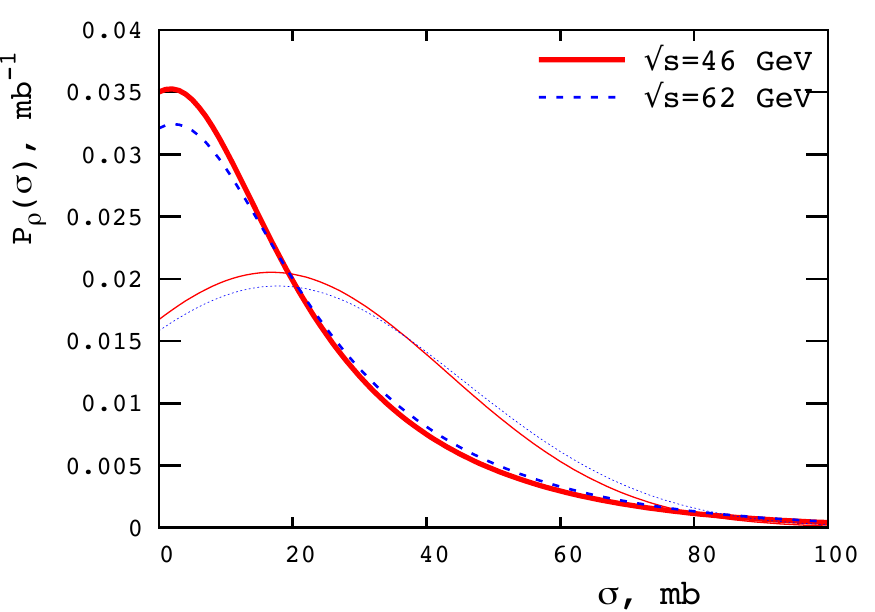}
\caption{The distribution $P_{\rho}(\sigma)$ for $\rho$ mesons as a function of $\sigma$
for $\sqrt{s}=46$ GeV (solid red) and $\sqrt{s}=62$ GeV (blue dashed).
For comparison, the thin curves show  $P_{\pi}(\sigma)$ for pions.} 
\label{fig:P_rho}
\end{figure}

Figure~\ref{fig:P_rho} shows the distribution $P_{\rho}(\sigma)$ for $\rho$ mesons as a function of $\sigma$ at $\sqrt{s}=46$ GeV and $\sqrt{s}=62$ GeV. For comparison with the pion case, we also give the corresponding $P_{\pi}(\sigma)$ by thin curves.

\subsection{$P_{\gamma^{(\ast)}}(\sigma)$ distribution for real and virtual photons}

It is well known that real and virtual photons also reveal their hadron-like nature in strong interactions. For example, in the vector dominance (VMD) model, approximately 70\% of the total photoabsorption cross section comes from the contribution of $\rho$, $\omega$ and $\phi$ mesons~\cite{Bauer:1977iq}.

In QCD, it is instructive to discuss the hadronic structure of 
real and virtual photons in the language of the color dipole model, see Sec.~\ref{sec:dipole_model}.
In general, the photon at high energies can be viewed as superposition of the following two types of components. First, 
the photon can fluctuate into aligned quark--antiquark pairs with the invariant mass $M$,  where the quarks share asymmetrically the photon 
longitudinal momentum and have small transverse momenta $p_t$.
Such configurations are characterized by large cross sections of the order of $\sigma_{\rho N}$ and small probabilities of the order of
$\mu^2/M^2$, where $\mu$ is a soft QCD scale. The latter is required to comply with the approximate Bjorken scaling of the total virtual photon--nucleon cross section~\cite{Bjorken:1972uk}.
Second, in addition to the aligned pairs,  there are also configurations with large $p_t$, which are characterized by small cross sections of the order of $\alpha_s(p_t^2)/p_t^2$  and large probabilities to find in the photon wave function~\cite{Frankfurt:1988nt}.
The relative importance of these two types of contributions depends
on the photon virtuality $Q^2$, the longitudinal momentum, and the invariant mass of the produced diffractive state.

As in the cases of protons, pions, and $\rho$ mesons considered above, it is convenient to quantify the hadronic structure of photons in terms of the distribution $P_{\gamma^{(\ast)}}(\sigma)$. Starting with the real photon case, the corresponding distribution $P_{\gamma}(\sigma)$ satisfies the following constraints,
\begin{eqnarray}
\int d\sigma P_{\gamma}(\sigma) \sigma & = & \sigma_{\gamma p}(W)\,, \nonumber\\ 
\int d\sigma P_{\gamma}(\sigma) \sigma^2 &=& 16 \pi 
\frac{d \sigma_{\gamma p \to Xp}(t=0)}{dt} \,, 
\label{eq:P_gamma_const}
\end{eqnarray}
where $\sigma_{\gamma p}$ is the total photon--nucleon cross section; $d \sigma_{\gamma p \to Xp}(t=0)/dt$ is the 
cross section of photon diffractive dissociation on the proton including the $\rho$ meson peak.
Note that, the distribution $P_{\gamma}(\sigma)$ is not 
normalizable, i.e., the integral  $\int d\sigma P_{\gamma}(\sigma)$ is divergent at the lower integration limit due to the infinite renormalization of the photon Green's function
(the vacuum polarization). 

A model for $P_{\gamma}(\sigma)$ should interpolate between 
the regimes of small and large $\sigma$; the following presentation in based on Ref.~\cite{Alvioli:2016gfo}.
For small $\sigma \ll \sigma_{\pi N}$, we use the color dipole model, which allows one to readily present $P_{\gamma}(\sigma)$  in the following form, (see \cite{Alvioli:2016gfo} for details)

\begin{equation}
\fl
P_{\gamma}^{\rm dipole}(\sigma)=\sum_q e_q^2 \left|\frac{\pi d {\bf r}^2}{d\sigma_{q \bar{q}}(r,m_q)}\right| 
\int dz |\Psi_{\gamma}(z,r(\sigma_{q \bar{q}}),m_q)|^{2}_{\ \big|\sigma_{q \bar{q}}(r,m_q)=\sigma}  \; ,
\label{eq:P_gamma}
\end{equation}
where $\Psi_{\gamma}(z,r(\sigma_{q \bar{q}}),m_q)$ is the photon wave function, which depends on the the quark longitudinal momentum fraction $z$, the dipole transverse size $r$, and the quark mass $m_q$. In Eq.~(\ref{eq:P_gamma}), $r(\sigma_{q\bar{q}})$ is related to the dipole cross section using the following implementation (see Sec.~\ref{sec:dipole_model}) and Eq.~(\ref{eq:dipole_xsection_gluon_density}) of the dipole formalism~\cite{McDermott:1999fa}
\begin{equation}
\sigma_{q \bar{q}}(r,m_q)=\frac{\pi^2}{3} r^2 \alpha_s(Q_{\rm eff}^2) x_{\rm eff} g(x_{\rm eff},Q_{\rm eff}^2) \,,
\label{eq:P_gamma_2}
\end{equation}
where $Q_{\rm eff}^2=\lambda/r^2$ for light quarks and 
$Q_{\rm eff}^2=m_q^2+\lambda/r^2$ for heavy quarks (note that heavy quarks give a negligible contribution to the discussed quantities); $x_{\rm eff}=4 m_q^2/W^2+0.75 \lambda/(W^2 r^2)$;
$m_q=300$ MeV for light $u$, $d$ and $s$ quarks and $m_c=1.5$ GeV.
This choice of the quark masses ensures that the average transverse size of $q\bar q$ configurations in the photon 
wave function is close to that of the pion, $d_{\pi}=0.65$ fm, and also leads to a smoother interpolation between small 
and large $\sigma$ regimes.  The parameter $\lambda=4$ is chosen to best reproduce the HERA data on diffractive $J/\psi$ 
photoproduction~\cite{Frankfurt:2000ez}.

For large $\sigma \geq \sigma_{\rho N}$, the distribution $P_{\gamma}(\sigma)$ can be approximated by $P_{\rho}(\sigma)$ for $\rho$ mesons.  
Taking the sum of the $\rho$, $\omega$ and $\phi$ meson  contributions, the resulting distribution reads:
\begin{equation}
P_{(\rho+\omega+\phi)/\gamma}(\sigma)=\frac{11}{9}\left(\frac{e}{f_{\rho}}\right)^2 P_{\rho}(\sigma) \,,
\label{eq:P_gamma_3}
\end{equation} 
where the coefficient of $11/9$ takes into account the $\omega$ and 
$\phi$ contributions in the flavor SU(3) approximation.

Interpolating between the regimes of small and large $\sigma$, one 
arrives at the following hybrid model for $P_{\gamma}(\sigma)$ 
\begin{equation}
P_{\gamma}(\sigma,W)=\left\{\begin{array}{ll}
P_{\gamma}^{\rm dipole}(\sigma,W) \,, & \sigma \leq 10 \ {\rm mb} \,, \\
P_{\rm int}(\sigma,W) \,, & 10 \ {\rm mb} \leq \sigma \leq 20 \ {\rm mb} \,, \\
P_{(\rho+\omega+\phi)/\gamma}(\sigma,W) \,, &  \quad \sigma \geq 20 \ {\rm mb} \,.\end{array} \right. 
\label{eq:P_gamma_4}
\end{equation}
where $P_{\rm int}(\sigma,W)$ is a smooth interpolating function.

The resulting model for $P_{\gamma}(\sigma)$ satisfies the constraints of Eq.~(\ref{eq:P_gamma_const}) and gives a good 
description of the total and diffraction dissociation photon--proton cross sections.
Indeed, at the typical $W=100$ GeV, 
it gives  $\sigma_{\gamma p}=\int d\sigma \sigma P_{\gamma}(\sigma)=135$ $\mu$b in agreement with the PDG value of $\sigma_{\gamma p}=146$ $\mu$b~\cite{Agashe:2014kda} and
$d\sigma_{\gamma p \to Xp}(t=0)/dt =\int d\sigma \sigma^2 P_{\gamma}(\sigma)/(16 \pi)=240$ $\mu$b/GeV$^2$ in agreement with 
the estimate
of $d\sigma_{\gamma p \to Xp}(t=0)/dt \approx 220$ $\mu$b/GeV$^2$, which is obtained by
integrating the data of~\cite{Chapin:1985mf} over the produced diffractive masses and extrapolating the resulting
cross section to the desired $W=100$ GeV.

In the virtual photon case, the distribution $P_{\gamma^{\ast}}(\sigma)$ can be constructed similarly to the model outlined above. Namely, for the small-$\sigma$ part one uses Eq.~(\ref{eq:P_gamma}), where the real photon wave function is substituted by the one of the virtual photon. In addition, the large-$\sigma$ contribution due to vector mesons receives a $Q^2$-dependent suppression factor
dictated by the vector meson dominance model,
\begin{equation}
P_{(\rho+\omega+\phi)/\gamma^{\ast}}(\sigma)=\frac{11}{9}\left(\frac{e}{f_{\rho}}\right)^2 \frac{m_{\rho}^2}{Q^2+m_{\rho}^2}
P_{\rho}(\sigma) \,.
\label{eq:P_gamma_3_virt}
\end{equation}
In the intermediate-$\sigma$ region, we use smooth interpolation as in Eq.~(\ref{eq:P_gamma_4}).

\begin{figure}[h]
\centering
 \includegraphics[width=0.75\textwidth]{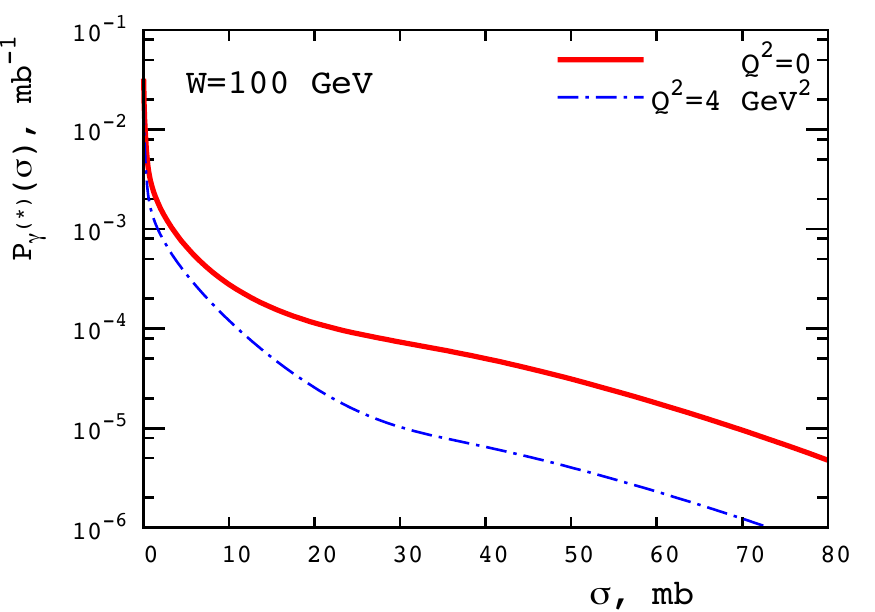}
\caption{The distributions $P_{\gamma}(\sigma)$
and $P_{\gamma^{\ast}}(\sigma)$ 
for the real and virtual photons at $W=100$ GeV.}
\label{fig:P_gamma}
\end{figure}

 The resulting distributions $P_{\gamma}(\sigma)$ for the real photon and $P_{\gamma^{\ast}}(\sigma)$ for the virtual photon at $Q^2=4$ GeV$^2$
 as a function of $\sigma$ at $W=100$ GeV are shown in Fig.~\ref{fig:P_gamma}. 
 Note that for small $\sigma$, the sensitivity to the choice of the quark mass $m_q$ is very weak.

Note that the distribution $P_{\gamma}(\sigma)$ can be probed in measurements of the total photon--nucleus cross sections at high energies,
which can done using, e.g., ultraperipheral heavy ion collisions in next run (Run 3) of the LHC operations.
Indeed, in this formalism, the photon--nucleus cross section can be written as a series in terms of moments of the distribution $\langle \sigma^k \rangle=\int d\sigma P_{\gamma}(\sigma) \sigma^k$, where each term corresponds to the simultaneous interaction with $1\leq k \leq A$ nucleons of the nuclear target; the terms with $k \geq 2$
contribute to the shadowing correction modifying the impulse approximation given by the $k=1$ term, see Sec.~\ref{sec:shadowing}.
Moreover, the effective number of $\langle \sigma^k \rangle$ terms, which one needs to take into account, increases with the nucleus mass number $A$.
Thus, the photon-nucleus cross section allows one to probe 
higher moments of the distribution $P_{\gamma}(\sigma)$ giving additional constraints on it compared to the free proton case.

It is expected that the effect of nuclear shadowing in this cross section is stronger than in the structure function $F_{2A}(x,Q^2)$
measured in lepton--nucleus DIS and has a significant $W$ dependence extending into a TeV range.

Another application of the distribution $P_{\gamma}(\sigma)$ is for the wounded nucleon model in the fixed-target case, see Sec. \ref{sec:diff_soft}.

\section{Inclusive coherent and incoherent diffraction in $eA$ DIS} 
\label{sec:shadowing}

%%%%%%%%%%%%%%%%%%%%%%%%%%%%%%%%%%%%%%%%%%%%%%%%%%%%%%%%%%%
\subsection{Connection of nuclear shadowing to diffraction}
\label{subsec:connection}

It is well known that at high energies, cross sections of hadron--nucleus scattering are smaller than the sum of 
individual hadron--nucleon cross sections. This phenomenon originates from destructive interference among the amplitudes for the interaction with one and two, tree, etc.~nucleons of the nuclear target. In the literature it
is called {\it nuclear shadowing} (NS) because it can be interpreted as a decrease of the effective number of nucleons of the nuclear target due to
their overlap or mutual geometric shadowing in the transverse 
plane~\cite{Glauber:1955qq,Glauber:1970jm,Bauer:1977iq,Gribov:1968jf}; the resulting theoretical approach is called the Gribov--Glauber model of nuclear shadowing. 

Note here that Glauber considered a quantum--mechanical potential model, which is different from the high-energy scattering situation, where the amplitude is predominantly imaginary. Moreover for this amplitude the diagrams considered by Glauber are cancelled out in the high energy limit. The reason for this disappearance is that the projectile does not have time to merge back to its initial state between the first and second interactions.
Gribov found the essential diagrams for hadron--nucleus scattering and, in the case of the interaction with two nucleons, expressed the correction to the total nuclear cross section through the diffractive cross section on the nucleon, see Eq.~(\ref{eq:GG}).
At the same time, for multiple scatterings, one needs modelling, which can be done by introducing the notion of cross section fluctuations, see discussion in Sec.~\ref{sec:fluctuations}. 

\begin{figure}[t]
\centering
 \includegraphics{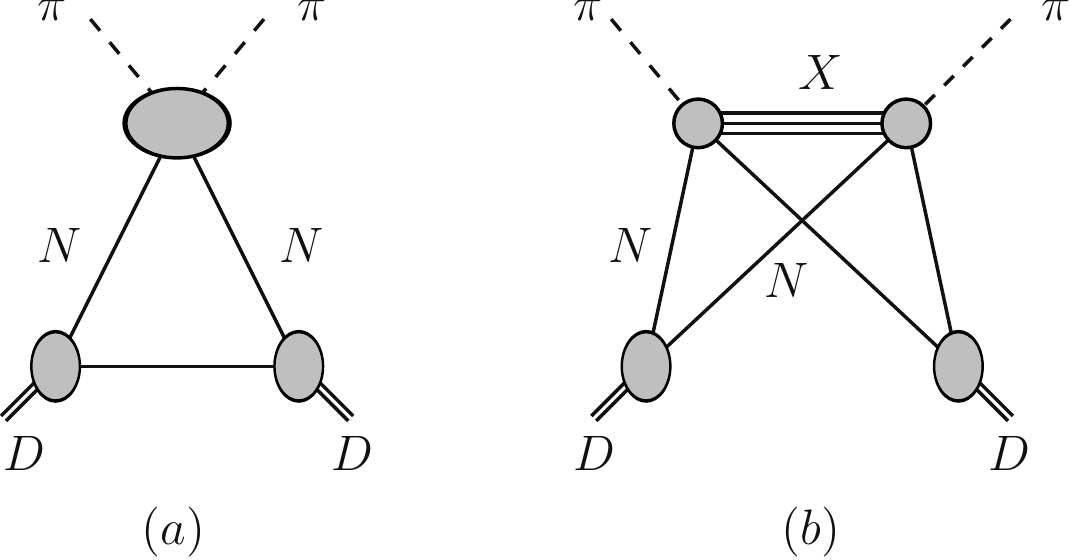}
\caption{The forward pion--deuteron scattering amplitude: graphs $a$ and $b$ correspond to the interaction with one and two nucleons of the nucleus and lead to the impulse approximation and the nuclear shadowing correction to $\sigma^{\pi D}_{\rm tot}(s)$,
respectively.}
\label{fig:GG_deuteron}
\end{figure}

A classic example of nuclear shadowing is the total pion--deuteron cross section $\sigma^{\pi D}_{\rm tot}(s)$. The corresponding 
forward pion--deuteron scattering amplitude is shown in Fig.~\ref{fig:GG_deuteron}, where graphs $a$ and $b$ correspond to the interaction with one and two nucleons of the target, respectively. An evaluation of these graphs gives~\cite{Gribov:1968jf}
\begin{equation}
\sigma^{\pi D}_{\rm tot}(s)\,=\,2 \sigma^{\pi N}_{\rm tot}(s)-2 \int dt\, \rho_D(4 t) \,\frac{d \sigma^{\pi N}_{\rm diff}(s,t)}{dt} \,,
\label{eq:GG}
\end{equation}
where the first term coming from graph $a$ is the impulse approximation given by twice the total pion--nucleon cross section 
$\sigma^{\pi N}_{\rm tot}(s)$ (we do not distinguish between protons and neutrons here) and
the second term coming from graph $b$ is the negative NS correction, which is expressed in terms of the pion--nucleon diffractive cross
section $d \sigma^{\pi N}_{\rm diff}(s,t)/dt$.
Note that $d \sigma^{\pi N}_{\rm diff}(s,t)/dt$ includes both the elastic and inelastic diffractive contributions; the former corresponds to the elastic intermediate state, which is implicitly included in graph $b$.   
 The distribution of nucleons is given by the deuteron form factor $\rho_D(4t)$ 
 evaluated at double the momentum transfer, which takes into account the effect of nuclear recoil;
 the integration runs over the invariant momentum transfer squared $t$. Finally, all involved cross sections depend
 on the invariant collision energy squared $s$. 
 
The diagrams corresponding to the sequential interaction of a projectile with target nucleons via elastic intermediate states, which lead to the Glauber shadowing correction, vanish in the high energy limit. At the same time, 
since the characteristic longitudinal distance over which the projectile--nucleus interaction develops (coherence length) 
increases with an increase of energy~\cite{Ioffe:1969kf,Gribov:1973jg},  the diagrams corresponding to the simultaneous interaction with all target nucleons contribute to this process; their contribution can be formulated in terms of diffractive scattering on a single nucleon.
 
 Note that the original derivation of Eq.~(\ref{eq:GG}) used the Pomeron trajectory with the intercept equal to unity, which resulted 
 in the vanishing real part of the scattering amplitude. The effect of the real part can be readily included using the 
 Gribov--Migdal relation~\cite{Gribov:1968uy}, see Eq.~(\ref{eq:gD}) below.

This connection between NS and diffraction  reflects the dominance of large longitudinal distances at high energies. It also holds in the case of the interaction of real and virtual photons with deuterons
and can be further generalized to heavier nuclei. In particular, the $\sigma^{\gamma^{(\ast)}D}(W)$ total photon--deuteron cross section at high energies can be written in the following form
\begin{equation}
\fl
\sigma^{\gamma^{(\ast)}D}(W)=2 \sigma^{\gamma^{(\ast)}N}(W)-2 \frac{1-\eta^2}{1+\eta^2} \int_{x_{\Pomeron,{\rm min}}}^{0.1} dx_{\Pomeron} d|\vec{k}_t|^2
 \rho_D\left(4t\right) \frac{d\sigma^{\gamma^{(\ast)}N}_{\rm diff}(W,t,x_{\Pomeron})}{dx_{\Pomeron} d|\vec{k}_t|^2} \,,
\label{eq:gD}
\end{equation}
where $\sigma^{\gamma^{(\ast)}N}(W)$ is the total photon--nucleon cross section; 
$d\sigma_{\gamma^{(\ast)}N}^{\rm diff}(W,t,x_{\Pomeron})/(dx_{\Pomeron} d|\vec{k}_t|^2)$ is the cross section for the 
$\gamma^{(\ast)}+ N \to X+N$ inclusive diffractive scattering on the nucleon; 
$x_{\Pomeron}=\xi$ is the longitudinal momentum fraction of the diffractive exchange, see Eq.~(\ref{eq:diff_def}); 
$\vec{k}_t$ is the transverse momentum transfer to the target nucleons so that $-t=(x_{\Pomeron}m_N)^2+|\vec{k}_t|^2$.

The lower $x_{\Pomeron}$ integration limit is defined by the lowest possible diffractively  produced mass corresponding to the $\rho$ vector  meson, 
\begin{equation}
x_{\Pomeron,{\rm min}}=\frac{M_{\rho}^2+Q^2}{W^2+Q^2} \,.
\label{eq:xpom_min}
\end{equation}
In the case of DIS with $Q^2={\cal O}(\rm few \ GeV^2$), $x_{\Pomeron,{\rm min}}=Q^2/(2 p \cdot q)=x$. 
The upper integration limit is defined by the usual condition on diffraction $M_X^2/W^2 \leq 0.1$. 
In fact the cutoff is stronger due to the presence of the nuclear form factor. Note that, as we discussed in Sec.~\ref{sec:inclusive_diffraction}, the diffractive exchange with the vacuum quantum numbers dies out for 
$x_{\Pomeron}>0.03$ and non-vacuum sub-leading contributions 
give a dominant contribution.

In Eq.~(\ref{eq:gD}), $\eta$ is the ratio of the real and imaginary parts of the $\gamma^{(\ast)}+ N \to X+N$ amplitude, which can be 
estimated using the asymptotic energy behavior of the corresponding cross section~\cite{Gribov:1968uy}
 \begin{equation}
 \eta=\frac{\Re e {\cal A}_{\gamma^{(\ast)}+ N \to X+N}}{\Im m {\cal A}_{\gamma^{(\ast)}+ N \to X+N}} \approx \frac{\pi}{2} 
 \frac{\partial \ln \Im m {\cal A}_{\gamma^{(\ast)}+ N \to X+N}}{\partial \ln s}
 =\frac{\pi}{2} \left(\alpha_{\Pomeron}(0)-1 \right) \,,
 \label{eq:eta}
 \end{equation}
where $\alpha_{\Pomeron}(0)$ determines the energy dependence of the imaginary part of the amplitude.  It corresponds to the intercept of the Pomeron exchange, which gives the dominant contribution at high $W$ and $M_X^2/W^2 \le 0.03$.

Equation~(\ref{eq:gD}) presents a general connection between the nuclear shadowing effect originating from the interaction with two nucleons of a nuclear target and diffraction on the nucleon. It is based on such general properties of 
scattering amplitudes as unitarity, which can also be formulated in form of the Abramovsky--Gribov--Kancheli (AGK) cutting rules~\cite{Abramovsky:1973fm}, and analyticity. It is important to emphasize that it does not require the decomposition over twists, i.e., it is 
valid for both leading-twist and higher-twist contributions and in the entire range of the photon virtualities including the
$Q^2=0$ photoproduction limit.

Equation~(\ref{eq:gD}) can be readily cast into the expression for the deuteron
structure function (SF) $F_{2D}(x,Q^2)$ probed in inclusive lepton--deuteron deep inelastic scattering (DIS) at small $x$,
\begin{equation}
\fl
F_{2D}(x,Q^2)=2 F_{2N}(x,Q^2)-2 \frac{1-\eta^2}{1+\eta^2} \int_x^{0.1} dx_{\Pomeron} \, d|\vec{k}_t|^2
 \rho_D\left(4t\right) F_{2N}^{D(4)}(\beta,Q^2,x_{\Pomeron},t) \;,
\label{eq:F_2D}
\end{equation}
where $F_{2N}(x,Q^2)$ and $F_{2N}^{D(4)}$ are the usual and diffractive nucleon structure functions (SFs) in 
inclusive $l+ N \to l^{\prime}+X$ and diffractive $l+ N \to l^{\prime}+X+N^{\prime}$ lepton--nucleon DIS, respectively (see Sec.~\ref{sec:inclusive_diffraction}).
Various approaches using Eq.~(\ref{eq:F_2D}), which describe the available data on the deuteron SFs in a wide range of photon 
virtualities $Q^2$ including the $Q^2=0$ photoproduction limit, differ mainly in modeling and parametrization
of the nucleon diffractive SF.
In particular, the phenomenological parameterizations included the scaling, quark--antiquark parton model contribution to 
the Pomeron part of the diffractive structure function and the contribution of $\rho$, $\omega$, and $\phi$ vector mesons to low-mass diffraction~\cite{Badelek:1991qa,Badelek:1994qg};
the Pomeron, vector meson, and meson
exchange contributions~\cite{Melnitchouk:1992eu,Melnitchouk:1995am}; the vector meson and quark--antiquark continuum 
contributions to the nuclear structure function~\cite{Piller:1995kh};
the aligned-jet model~\cite{Frankfurt:1988nt}; 
the leading-twist contribution based on next-to-leading order 
QCD analysis of hard diffraction in $ep$ DIS at HERA~\cite{Frankfurt:2003jf}.
In all the cases, the effect of NS is a few percent correction to the total lepton--deuteron cross section. 

The effect of NS is much larger for heavy nuclei and increases with an increase of the atomic mass number $A$.
However, a generalization of Eqs.~(\ref{eq:gD}) and (\ref{eq:F_2D}) to heavier nuclear targets requires modeling of interactions with
three and more nucleons of the target. This can be done using the eikonal or quasi-eikonal approximations \cite{Frankfurt:2002kd}, which assume that the
intermediate diffractive state elastically scatters on $ N \geq 3$ nucleons of the nuclear target with the 
effective cross section $\sigma_{XN}$
\begin{equation}
\sigma_{XN}=\frac{1}{16 \pi}  \frac{1}{(1+\eta^2) \sigma^{\gamma^{(\ast)}N}(W)} \int_{x}^{0.1} dx_{\Pomeron}\frac{d\sigma^{\gamma^{(\ast)}N}_{\rm diff}(W,t,x_{\Pomeron})}{dx_{\Pomeron} d|\vec{k}_t|^2}\bigg|_{t=0} \; .
\label{eq:sigmaXN}
\end{equation}

\begin{figure}[t]
\centering
 \includegraphics[width=1.0\textwidth]{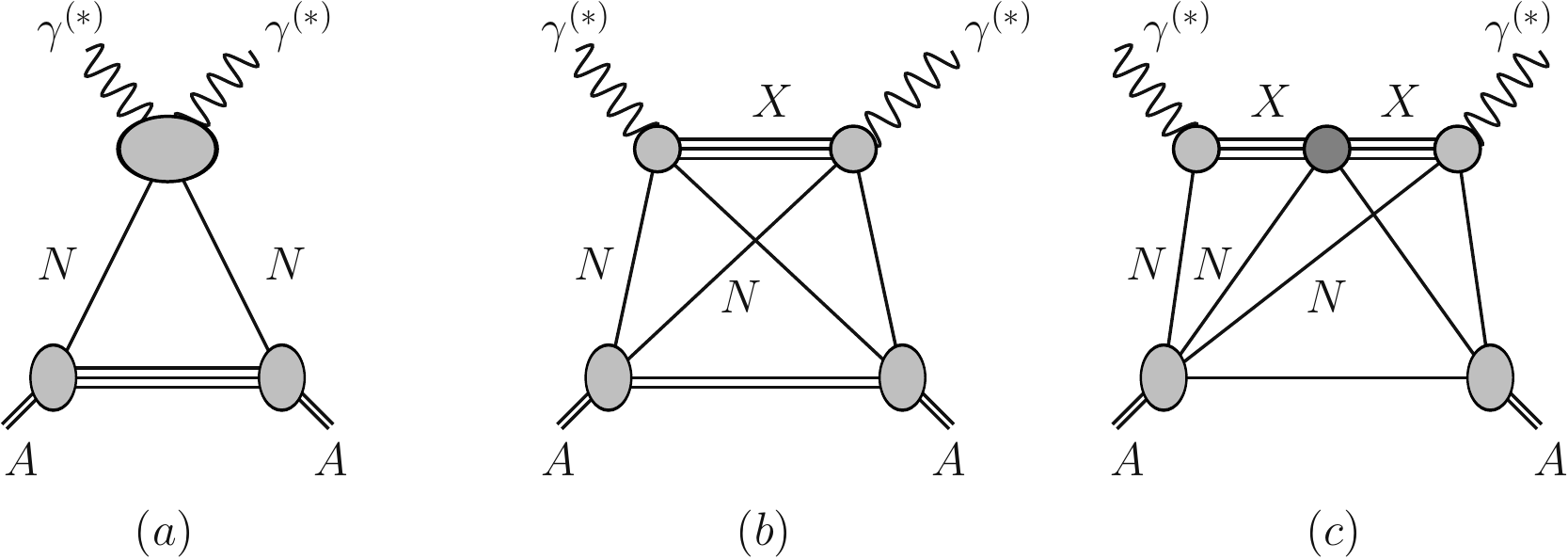}
\caption{The multiple scattering series for the forward (virtual) photon-nucleus scattering amplitude: graphs $a$, $b$, $c$ correspond to the interaction with one, two, and three nucleons of the target; the interaction with $A \geq N > 3$ nucleons is not shown, but implied.}
\label{fig:GG_A}
\end{figure}

Figure~\ref{fig:GG_A} shows the forward photon--nucleus scattering amplitude: graphs $a$, $b$, and $c$ present first three terms
of the multiple scattering series corresponding to the photon interaction with one, two, and three nucleons of the nucleus target;
the graphs for the interaction with $A \geq N > 3$ nucleons are not explicitly shown, but included in the final answer.
In graphs $b$ and $c$, $X$ denotes the intermediate diffractively produced state $X$; in
 graph $c$, the middle darker vertex denotes the interaction with the cross section $\sigma_{XN}$.
An evaluation and summation of these graphs gives the following expression for the (virtual) photon--nucleus cross section~\cite{Frankfurt:1988nt,Piller:1995kh,Piller:1997ny,Piller:1999wx,Adeluyi:2006xy}, 
\begin{eqnarray}
\fl
 \sigma^{\gamma^{(\ast)}A}(W) = A\sigma^{\gamma^{(\ast)}N}(s) -8 \pi A(A-1) \Re e   \frac{(1-i \eta)^2}{1+\eta^2} \int_{x}^{0.1} dx_{\Pomeron} \frac{d\sigma^{\gamma^{(\ast)}N}_{\rm diff}(W,t,x_{\Pomeron})}{dx_{\Pomeron} d|\vec{k}_t|^2}\bigg|_{t=0}  \nonumber\\
 \fl
 \times  \int d^2 \vec{b} \int^{\infty}_{-\infty} dz_1  \int^{\infty}_{z_1} dz_2 \rho_A(\vec{b},z_1) \rho_A(\vec{b},z_2)
e^{i (z_1-z_2) x_{\Pomeron}m_N} e^{-\frac{A}{2} (1-i\eta)\sigma_{XN} \int_{z_1}^{z_2} dz^{\prime}
\rho_A(\vec{b},z^{\prime})} \,,
\label{eq:gA}
\end{eqnarray}
where $\vec{b}$ and $z$ refer to the transverse and longitudinal coordinates of the involved nucleons, respectively. 
The derivation of Eq.~(\ref{eq:gA}) assumes independent nucleons in the nuclear ground state, where  $\rho_A(\vec{b},z)$ is the nuclear
density calculated using the Woods--Saxon (two-parameter Fermi model) parametrization~\cite{DeJager:1987qc}, 
and neglects the slope of the $\gamma^{(\ast)}+ N \to X+N^{\prime}$ diffractive cross section compared to that of the nuclear form factor. The latter means that the diffractive cross section on the nucleon is evaluated at $t=0$ and all involved nucleons
are positioned at the same transverse distance (impact parameter) $\vec{b}$.

The main difference between various approaches based on the Gribov--Glauber model lies in different parametrizations of the 
diffractive cross section on the nucleon (see the discussion above) and
different implementations of 
the interaction with $N \geq 3$ nucleons of the nuclear target. For instance, these interactions can be written in the form of 
summed fan diagrams~\cite{Capella:1997yv,Armesto:2003fi,Armesto:2010kr,Armesto:2006ph}, which leads to a different form of the expression for $\sigma^{\gamma^{(\ast)}A}(W)$.

The approaches based on the connection of nuclear shadowing to diffraction provide a good description and 
interpretation of data  on inclusive cross section in  lepton--nucleus and photon--nucleus scattering at high energies.  The useful feature of
 Eq.~(\ref{eq:gA}) is that it is applicable in a wide range of photon virtualities including the $Q^2=0$ limit, 
 which allows one to study the interplay of different mechanisms of NS and also higher-twist effects in NS.

In the following we will consider hard scattering off nuclei and an approach allowing one to generalize Eq.~(\ref{eq:gA}) to nuclear parton distribution functions (nPDFs). Since in this kinematics the involved structure functions can be expressed in 
terms of leading twist parton distributions, this approach is called the leading twist nuclear shadowing model~\cite{Frankfurt:1998ym,Frankfurt:2003zd,Frankfurt:2011cs}.

%%%%%%%%%%%%%%%%%%%%%%%%%%%%%%%%%%%%%%
\subsection{The model of leading twist nuclear shadowing}
\label{subsec:LT}

The model of leading twist nuclear shadowing is based on the combination of the following key elements:
 \renewcommand{\labelenumii}{\roman{enumii}}
\begin{enumerate}
\item 
The Gribov--Glauber model of nuclear shadowing allowing one to evaluate the effect of NS in hadron--nucleus scattering;\\
\item
 The QCD collinear factorization theorems for inclusive~\cite{Brock:1993sz} and diffractive DIS~\cite{Collins:1997sr} 
 and QCD analyses of inclusive diffraction in lepton--proton DIS at HERA~\cite{Aktas:2006hy,Aktas:2006hx,Chekanov:2009aa}, see Sec.~\ref{sec:inclusive_diffraction},
 allowing one to make predictions for parton distribution functions (PDFs) of different flavors $i$ in nuclei;\\
\item
A model for the hadronic structure of a hard probe to calculate the interaction with $N \geq 3$ nucleons of the target, see Sec.~\ref{sec:fluctuations}.
\end{enumerate}
Note that this approach employs the equivalence of pictures of hard scattering in the nuclear rest frame and in the frame, where the nucleus is fast.

\begin{figure}[t]
\centering
 \includegraphics[width=1.0\textwidth]{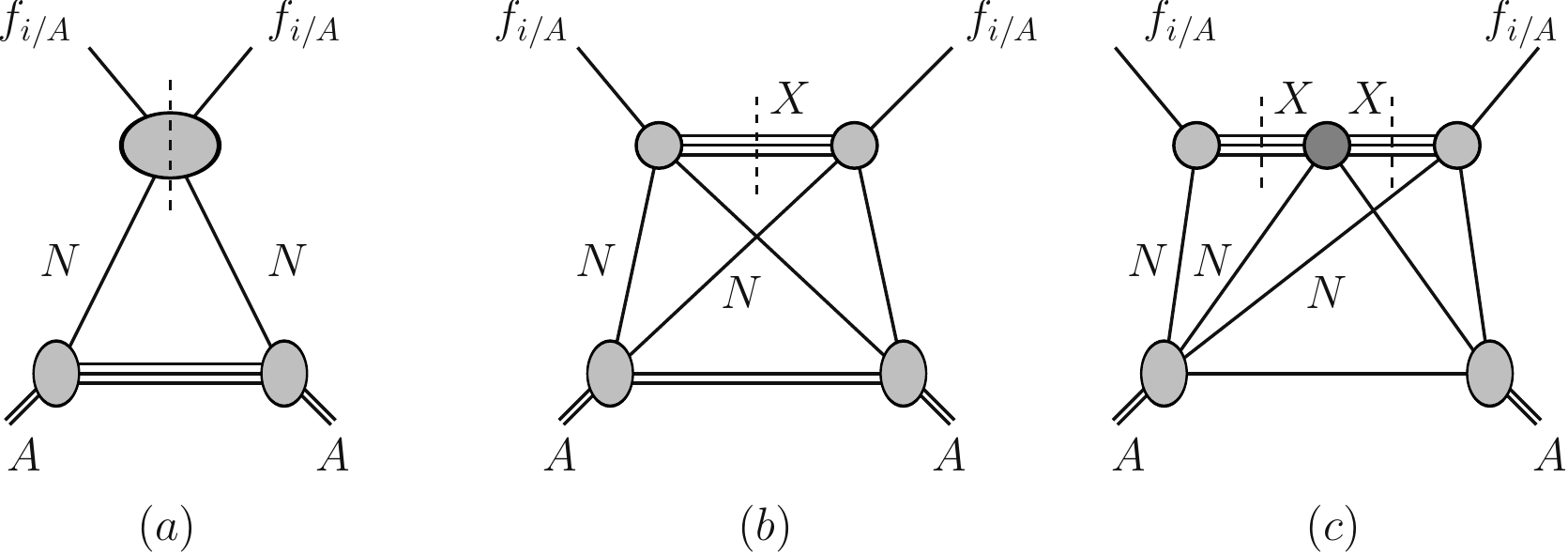}
\caption{The multiple scattering series for nPDFs $f_{i/A}(x,\mu^2)$: graphs $a$, $b$, $c$ correspond to the interaction with one, two, and three nucleons of the target; the interaction with $A \geq N > 3$ nucleons is not shown, but implied.
The dashed vertical lines stand for the imaginary part; $X$ denotes the diffractively produced intermediate state.
}
\label{fig:LTA_A}
\end{figure}

The multiple scattering series for nuclear PDFs $f_{i/A}(x,Q^2)$ is presented in Fig.~\ref{fig:LTA_A} (compare to Fig.~\ref{fig:GG_A}), 
where graphs $a$, $b$, $c$ correspond to the interaction with one, two, and three nucleons of the nuclear target.
The graphs for the interaction with $N > 3$ nucleons are implied and included in the resulting expression for 
$xf_{i/A}(x,Q_0^2)$, see Eq.~(\ref{eq:f_i/A}).
The dashed vertical lines denote the imaginary part of the corresponding graphs and correspond to the total parton--nucleon cross section in graph $a$ and to the so-called diffractive cut in graphs $b$ and $c$, where the outer vertices correspond to the
parton--nucleon diffractive scattering.
The middle darker vertex in graph $c$ stands for the interaction with $N > 3$ nucleons, which is modeled using the 
$\sigma_{\rm soft}^i(x,Q_0^2)$ effective cross section, see Eq.~(\ref{eq:sigma_soft}).

Applying the steps listed above to Eq.~(\ref{eq:gA}) and evaluating the graphs in  Fig.~\ref{fig:LTA_A}, nPDFs at small $x$  can be presented in the following form~\cite{Frankfurt:2011cs}
[compare to Eq.~(\ref{eq:gA})] 
\begin{eqnarray}
\fl
xf_{i/A}(x,Q_0^2) = A xf_{i/N}(x,Q_0^2)-8 \pi A(A-1) \Re e \frac{(1-i \eta)^2}{1+\eta^2} \int_{x}^{0.1} dx_{\Pomeron}
\beta f_{i}^{D(4)}(\beta,Q_0^2,x_{\Pomeron},t=0) \nonumber\\
\fl 
\times \int d^2 \vec{b} \int^{\infty}_{-\infty} dz_1  \int^{\infty}_{z_1} dz_2 \rho_A(\vec{b},z_1) \rho_A(\vec{b},z_2)
e^{i (z_1-z_2) x_{\Pomeron}m_N} \frac{\langle \sigma^2 e^{-\frac{A}{2} (1-i\eta)\sigma \int_{z_1}^{z_2} dz^{\prime}
\rho_A(\vec{b},z^{\prime})} \rangle_{i}}{\langle \sigma^2 \rangle_{i}} \;,
\label{eq:f_i/A}
\end{eqnarray}
where $i$ is the parton flavor; $f_{i}^{D(4)}$  are diffractive PDFs of the proton with $\beta=x/x_{\Pomeron}$;
$Q_0$ is the input, factorization scale chosen such that higher-twist effects in the proton diffractive PDFs are small. 
Note that the exponential factor $e^{i (z_1-z_2) x_{\Pomeron}m_N}$ automatically suppresses the shadowing term at large $x$. 

In our analysis, 
 $Q_0=2$ GeV. At smallest $x$ there is an indication that the higher twists effects 
in inclusive diffraction in $ep$ DIS at HERA
 may not be negligible up to $Q_0=3 \, \rm GeV$, see Sect.~\ref{sec:inclusive_diffraction} and \ref{sec:dipole_diff_structure_function_ht}.
 This effect is masked to some extent by the parametrization of proton diffractive PDFs $f_{i}^{D(4)}$ extracted from the fits to the 
 diffractive data.
 
The $\langle \dots \rangle_i$ term in Eq.~(\ref{eq:f_i/A}) sums the contributions corresponding to the interaction with 
$N \geq 3$ nucleons of the nuclear target and is expressed in terms of moments of the distribution $P_{i}(\sigma)$,
which takes into account the composite hadronic structure of a hard probe, see Sec.~\ref{sec:fluctuations},
\begin{equation}
\langle \sigma^k \rangle _{i} \equiv \int_0^{\infty} d\sigma  P_i(\sigma) \sigma^k \,,
\end{equation}
where we explicitly indicated the dependence on parton flavor $i$.
It is important to note that the $\langle \dots \rangle_i$ term and the entire expression in Eq.~(\ref{eq:f_i/A}) does not depend
on the type of hard probe and is valid for virtual photons, gauge bosons, neutrinos, etc. Hence, the resulting nPDFs obey
the QCD collinear factorization.

While the interaction with $N > 3$ target nucleons needs to be modeled in Eq.~(\ref{eq:f_i/A}),
the interaction with two target nucleons is unambiguously expressed in terms of usual and diffractive PDFs of the protons [compare to Eq.~(\ref{eq:sigmaXN})]
\begin{equation}
\sigma_{2,i}(x)=\frac{\langle  \sigma^2 \rangle _{i}}{\langle  \sigma \rangle _{i}} =\frac{1}{16 \pi} \frac{\int_{x}^{0.1} dx_{\Pomeron}f_{i}^{D(4)}(\beta,Q_0^2,x_{\Pomeron},t=0) }{(1+\eta^2) f_{i}(x,Q_0^2)} \,,
\label{eq:sigma2} 
\end{equation}
where  $f_{i}(x,Q_0^2)$ are usual PDFs of the proton.

\begin{figure}[t]
\includegraphics[width=1.0\textwidth]{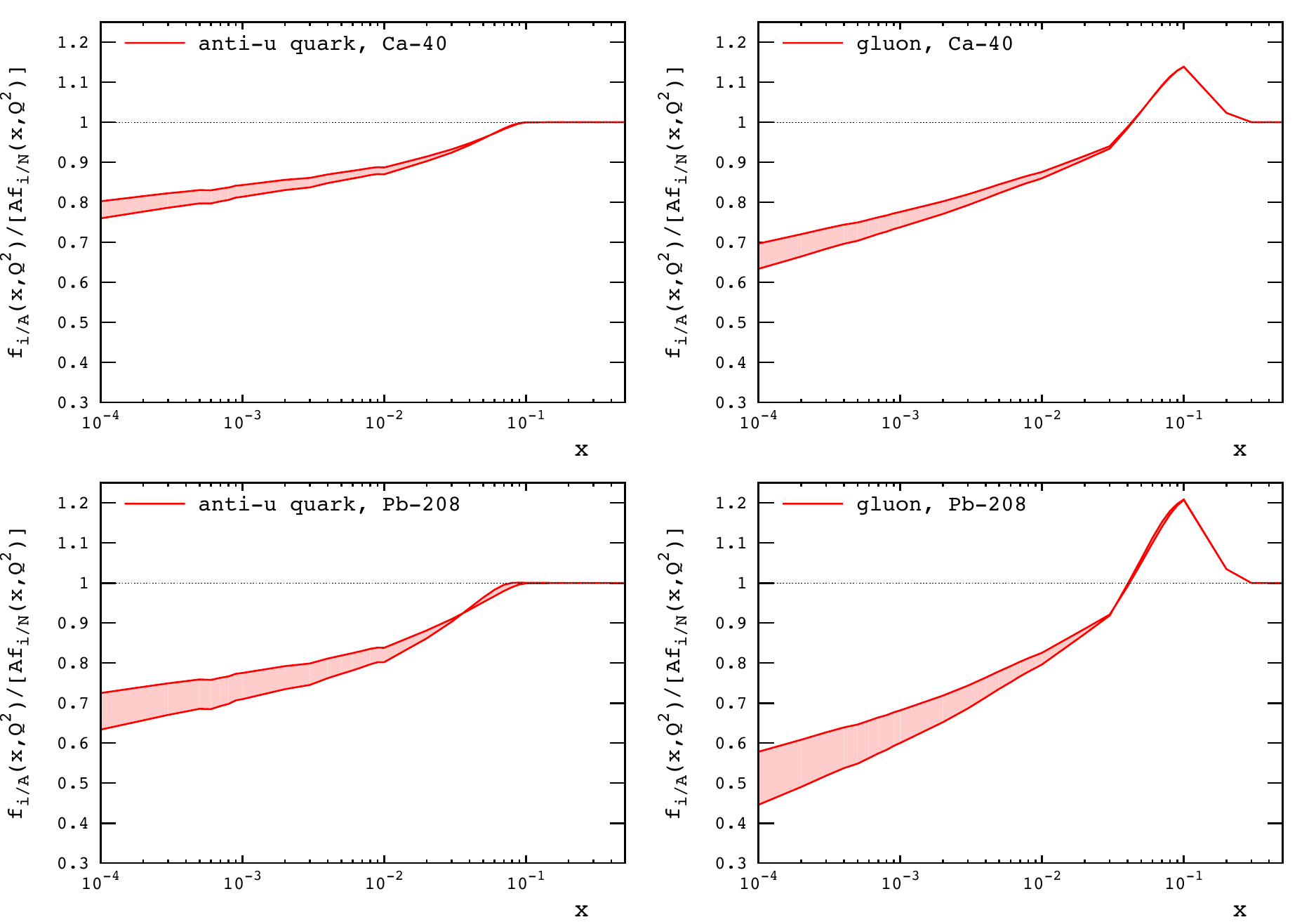}
\caption{The leading twist nuclear shadowing model predictions for the $f_{i/A}(x,Q_0^2)/[A f_{i/N}(x,Q_0^2)]$ ratios of the nuclear and nucleon PDFs for $\bar{u}$ quarks and gluons as a function of $x$ at the input scale $Q_0=2$ GeV.
The upper and lower panels correspond to Ca-40 and Pb-208, respectively.}
\label{fig:LT_pdfs}
\end{figure}

In the following, for definiteness we assume that the nuclear PDFs are probed by virtual photons and, hence, $P_i(\sigma)$ describes the hadronic structure of the virtual photon.
While in general hadronic fluctuations of small and large transverse sizes (small and large $\sigma$) are present in 
virtual photons, the structure of Eq.~(\ref{eq:f_i/A}) indicates that the $\langle \dots \rangle_i$ term is dominated by 
large-$\sigma$ fluctuations. Hence, it can be evaluated in terms of a single cross section
\begin{equation}
\sigma_{\rm soft}^i(x,Q_0^2)=\frac{\langle  \sigma^3 \rangle _{i}}{\langle  \sigma^2 \rangle _{i}} \,,
\label{eq:sigma_soft}
\end{equation}
which can be modeled using soft, non-perturbative part of the $P_i(\sigma)$ distribution. In the present approach, 
this is done by assuming that $P_i(\sigma)$ can be either calculated using a particular version
of the color dipole model~\cite{McDermott:1999fa}, i.e., $P_i(\sigma)=P_{\gamma^{\ast}}^{\rm dipole}(\sigma)$, or can be identified with that of the pion~\cite{Blaettel:1993ah}, i.e. $P_i(\sigma)=P_{\pi}(\sigma)$. For details, see Sec.~\ref{sec:fluctuations}.
Further, since the dispersion of $P_i(\sigma)$ for the calculation of the $\langle \dots \rangle_i$ term in Eq.~(\ref{eq:f_i/A})
can be safely neglected, i.e., 
\begin{equation}
\frac{\langle  \sigma^k \rangle _{i}}{\langle  \sigma^2 \rangle_{i}}=[\sigma_{\rm soft}^i(x,Q_0^2)]^{k-2} \,,
\label{eq:sigma_k}
\end{equation}
one obtains the following final expression for the nPDF $xf_{i/A}(x,Q_0^2)$ 
\begin{eqnarray}
\fl
xf_{i/A}(x,Q_0^2) = A xf_{i/N}(x,Q_0^2)-8 \pi A(A-1) \Re e \frac{(1-i \eta)^2}{1+\eta^2} \int_{x}^{0.1} dx_{\Pomeron}
\beta f_{i}^{D(4)}(\beta,Q_0^2,x_{\Pomeron},t=0) \nonumber\\
\fl  
\times \int d^2 \vec{b} \int^{\infty}_{-\infty} dz_1  \int^{\infty}_{z_1} dz_2 \rho_A(\vec{b},z_1) \rho_A(\vec{b},z_2)
e^{i (z_1-z_2) x_{\Pomeron}m_N} e^{-\frac{A}{2} (1-i\eta)\sigma_{\rm soft}^i(x,Q_0^2) \int_{z_1}^{z_2} dz^{\prime}
\rho_A(\vec{b},z^{\prime})} \,.
\label{eq:f_i/A_2}
\end{eqnarray}
The use of two plausible models for $\sigma_{\rm soft}^i(x,Q_0^2)$  
results in a spread of predictions for $xf_{i/A}(x,Q_0^2)$, 
which quantifies the theoretical uncertainty of the present approach. This is illustrated in Fig.~\ref{fig:LT_pdfs} showing 
predictions for the $f_{i/A}(x,Q_0^2)/[A f_{i/N}(x,Q_0^2)]$  ratios for $\bar{u}$ quarks and gluons as a function of $x$.
The upper row of panels corresponds to the nucleus of Ca-40, while the lower panels are for Pb-208. In the calculations, we used 
the 2006 H1 (Fit B) diffractive PDFs~\cite{Aktas:2006hy}.
Note that this set of diffractive PDFs is considered to be standard in the literature since it not only provides a good description of the cross sections of inclusive diffraction in DIS measured at HERA, but also gives an adequate description of diffractive dijet production, for reviews, see~\cite{Newman:2013ada,Butterworth:2005aq}.
Other analyses of diffractive PDFs, notably, the 2009 ZEUS Fit SJ~\cite{ZEUS:2009uxs} and that based on the xFitter framework~\cite{Goharipour:2018yov}, give comparable
results for the quark and gluon diffractive PDFs on the proton.
Note that the 2006 H1 Fit A is characterized by a very large gluon density at large momentum fractions leading to a poor agreement with the HERA diffractive jet data. Thus, the uncertainties associated with diffractive PDFs are rather modest and are smaller than the theoretical uncertainties of the leading twist nuclear shadowing model associated with modeling of the interaction with $N \geq 3$ nucleons, see Fig.~\ref{fig:LT_pdfs}.

Note that Eq.~(\ref{eq:f_i/A_2}) is applied to calculate the effect of nuclear shadowing for sea quark and gluon nuclear PDFs for $x \leq 0.1$.
For larger $x$, it is assumed that the sea quark distributions in nuclei are not modified at the input scale. For the gluon distribution, one additionally adds the effect of nuclear antishadowing in the interval $0.03 \leq x \leq 0.2$ by requiring that in the momentum sum rule for nPDFs, it compensates the suppression of nPDFs due to nuclear shadowing, see Ref.~\cite{Frankfurt:2011cs}. A similar dynamical model of the gluon antishadowing was proposed in~\cite{Frankfurt:2016qca}, where the mechanism of antishadowing was related to merging of two parton ladders
attached to two different nucleons of the nuclear target.
Alternatively, the enhancement of quark nuclear PDFs around $x=0.1$ was proposed to arise because of the real part 
the quark--nucleon scattering amplitude driven
by the Reggeon exchange~\cite{Brodsky:1989qz}.
However, this model contradicts the experimental observation 
that there is no enhancement of sea quarks in nuclei in this interval of $x$.

Figure~\ref{fig:LT_pdfs} presents the important feature of the leading twist nuclear shadowing model that the magnitude of
 NS is significant and nuclear shadowing for nuclear gluons is larger than that for quarks. This is a direct consequence of 
 the used connection between NS and diffraction on the proton and QCD factorization for inclusive diffraction in lepton--proton DIS.
  In particular, since the gluon diffractive PDF 
 is found to be large~\cite{Aktas:2006hy,Aktas:2006hx,Chekanov:2009aa}, this naturally leads to a large suppression of $g_{A}(x,Q_0^2)/[A g_{N}(x,Q_0^2)]$ for small $x$.
 
 In addition, it is predicted that the effect of NS increases with a decrease of $x$, see Fig.~\ref{fig:LT_pdfs}.
 This is different from predictions of global fits of nuclear PDFs~\cite{Hirai:2007sx,Kovarik:2015cma,Eskola:2016oht,Khanpour:2016pph}, which assume that the
  $f_{i/A}(x,Q_0^2)/[A f_{j/N}(x,Q_0^2)]$ ratios become $x$-independent at small $x$.

 One can see from Fig.~\ref{fig:LT_pdfs} that the sensitivity to modeling of the interaction with $N \geq 3$ nucleons of the nuclear
 target by the effective cross section $\sigma_{\rm soft}^i(x,Q_0^2)$ becomes reduced with an increase of $x$. In particular, at $x=10^{-3}$, the uncertainties in the predictions for $xf_{i/A}(x,Q_0^2)$ are at the
 level of 10\% for sea quarks, 15\% for gluons, and smaller for larger $x$.
 
 Note that unlike the quark channel, where the magnitude of nuclear shadowing for quark distributions can be inferred from the behavior of the ratio
 $F_{2A}(x,Q^2)/[A F_{2N}(x,Q^2)]$ of the structure functions,
 the nuclear gluon distribution at small $x$ can be predicted only using the collinear factorization for inclusive and diffractive DIS.

A good indication of the magnitude of the gluon nuclear shadowing is given by the ratio of the longitudinal nuclear and nucleon structure functions, $F_{LA}(x,Q^2)/[A F_{LN}(x,Q^2)]$. Thus, future measurements of this quantity at  EIC will rather directly constrain $xg_A(x,Q^2)$ at small $x$ and test predictions of the  model of the leading twist nuclear shadowing.

The leading twist model of nuclear shadowing naturally makes predictions for the impact parameter $\vec{b}$
 dependence of nuclear PDFs, 
 \begin{eqnarray}
 \fl
  xf_{i/A}(x,b,Q_0^2) \; = \nonumber \\
  \fl  A xf_{i/N}(x,Q_0^2)T_A(\vec{b})- 
   8 \pi A(A-1) \Re e \frac{(1-i \eta)^2}{1+\eta^2} \int_{x}^{0.1} dx_{\Pomeron}
\beta f_{i}^{D(4)}(\beta,Q_0^2,x_{\Pomeron},t=0) \nonumber\\
\fl
 \times \int^{\infty}_{-\infty} dz_1  \int^{\infty}_{z_1} dz_2 \rho_A(\vec{b},z_1) \rho_A(\vec{b},z_2)
e^{i (z_1-z_2) x_{\Pomeron}m_N} e^{-\frac{A}{2} (1-i\eta)\sigma_{\rm soft}^i(x,Q_0^2) \int_{z_1}^{z_2} dz^{\prime}
\rho_A(\vec{b},z^{\prime})} \,,
\label{eq:f_i/A_b}
\end{eqnarray}  
where $T_A(\vec{b})=\int dz \rho_A(\vec{b},z)$ is the so-called nuclear optical density.
Equation~(\ref{eq:f_i/A_b}) encodes the dynamical picture of NS resulting from an overlap of 
target nucleons in the transverse plane and naturally leads to an increase of NS with a decrease of $|\vec{b}|$,
where the nuclear density is larger. This leads to correlations of $x$ and $\vec{b}$ in the impact-parameter dependent
nPDFs $f_{i/A}(x,b,Q_0^2)$. 

\begin{figure}[t]
 \includegraphics[width=1.0\textwidth]{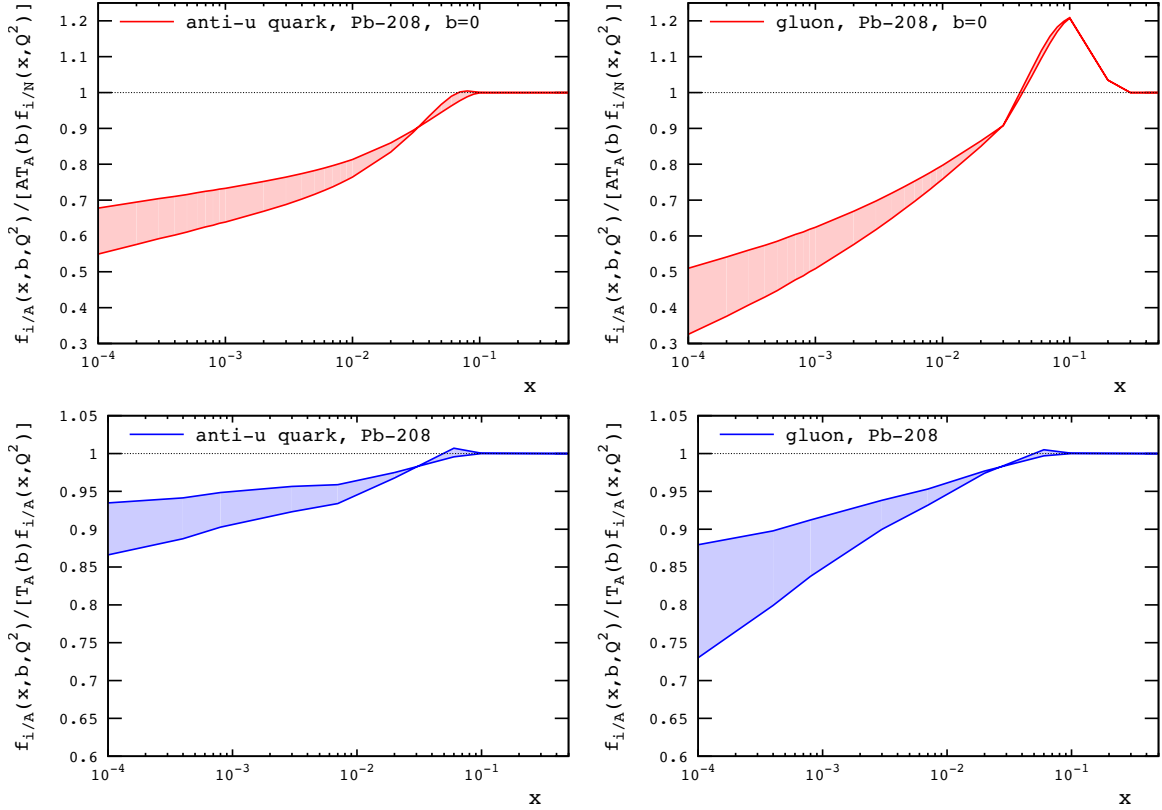}
\caption{The leading twist nuclear shadowing model predictions for the $\vec{b}$-dependence of nuclear PDFs.
Upper panels:
The $f_{i/A}(x,b,Q_0^2)/[A T_A(b)f_{j/N}(x,Q_0^2)]$ ratios for $\bar{u}$ quarks (left) and gluons
(right) in Pb-208  as a function of $x$ at $b=0$ and $Q_0^2=4$ GeV$^2$.
Lower panels: The $f_{i/A}(x,b,Q_0^2)/[{T_A(b) f_{i/A}(x,Q_0^2)}]$ ratios, see Eq.~(\ref{eq:impact_ratio}), for $\bar{u}$ quarks (left) and gluons (right) 
in Pb-208 as a function of $x$. }
\label{fig:LT_impact}
\end{figure}

The upper panels of Fig.~\ref{fig:LT_impact}
show the $f_{i/A}(x,b,Q_0^2)/[A T_A(b) f_{i/N}(x,Q_0^2)]$ ratios for $\bar{u}$ quarks (left) and gluons
(right) in Pb-208  as a function of $x$ at $b=0$ and $Q_0^2=4$ GeV$^2$.
The lower panels compare the predictions for $f_{i/A}(x,b,Q_0^2)$ to those of the $\vec{b}$-integrated $f_{i/A}(x,Q_0^2)$ and
give the ratios
\begin{equation}
 \frac{f_{i/A}(x,b,Q_0^2)}{A T_A(b) f_{i/N}(x,Q_0^2)}\Big/ \frac{f_{i/A}(x,Q_0^2)}{A f_{i/N}(x,Q_0^2)}=\frac{f_{i/A}(x,b,Q_0^2)}{T_A(b) f_{i/A}(x,Q_0^2)}
   \label{eq:impact_ratio}
 \end{equation}
for $\bar{u}$ quarks (left) and gluons (right) as a function of $x$. 
The suppression of these ratios indicates that the effect of NS is larger at $b=0$ than in the $\vec{b}$-integrated case.

 The global fits of nuclear PDFs can also in principle model the dependence of nPDFs on the impact parameter $b$, 
see, e.g.~\cite{Helenius:2012wd}. However, it depends on the assumed parametrization 
of the $b$ dependence and requires data with several nuclei to fix the additional free parameters.
  
\subsection{Nuclear diffractive structure functions and PDFs} 

\begin{figure}[t]
\includegraphics[width=1.0\textwidth]{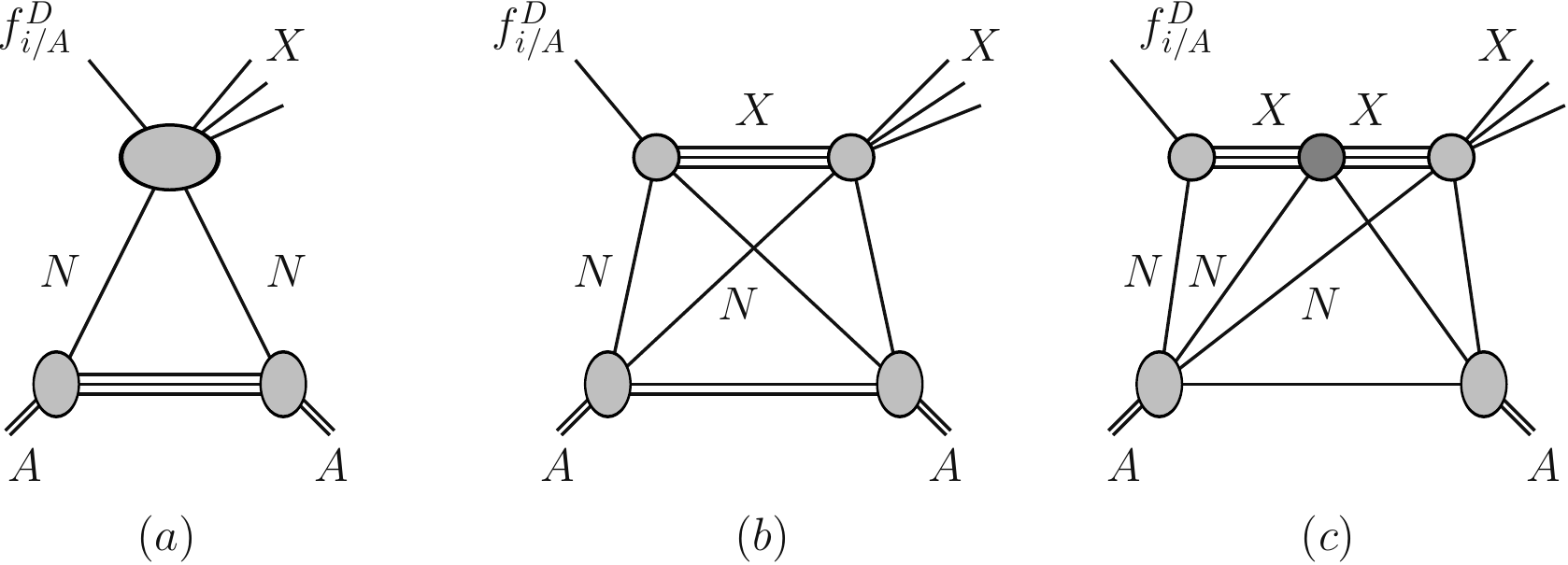}
\caption{The multiple scattering series for nuclear diffractive PDFs $f_{i/A}^{D}(\beta,\mu^2,x_{\Pomeron})$ at the amplitude level: graphs $a$, $b$, $c$ correspond to the interaction with one, two, and three nucleons of the target; the interaction with $A \geq N > 3$ nucleons is not shown, but implied. 
The diffractively produced final state and the intermediate states are denoted by $X$.
}
\label{fig:LTA_diff_A}
\end{figure}

Using the framework of the leading twist nuclear shadowing model, one can readily and without additional 
assumptions make predictions for nuclear diffractive structure functions and nuclear diffractive PDFs. These quantities can be probed, for instance, in 
$e+A \to e^{\prime}+X+A$ inclusive coherent diffraction in DIS with nuclei.
As in the case of usual nPDFs, in the leading twist nuclear shadowing model nuclear diffractive PDFs $f_{i/A}^{D}$ at small $x$
are obtained by summing the multiple scattering series for hard probe interaction with one, two, $\cdots$, $A$ nucleons of the nuclear target; first three terms of this series are shown in Fig.~\ref{fig:LTA_diff_A}.
 Summing these graphs, one obtains the following expression for
nuclear diffractive PDFs $f_{i/A}^{D(3)}(\beta,Q^2,x_{\Pomeron})$
\begin{eqnarray}
\fl
\beta f_{i/A}^{D(3)}(\beta,Q_0^2,x_{\Pomeron}) &=&
4 \pi A^2 \beta f_{i/N}^{D(4)}(\beta,Q_0^2,x_{\Pomeron},t_{\rm min}) \nonumber\\
& \times &  \int d^2 \vec{b} \left|\int^{\infty}_{-\infty} dz \rho_A(\vec{b},z)
e^{i z x_{\Pomeron}m_N} e^{-\frac{A}{2} (1-i\eta)\sigma_{\rm soft}^i(x,Q_0^2) \int_{z}^{\infty} dz^{\prime}
\rho_A(\vec{b},z^{\prime})}\right|^2 \,.
\label{eq:f_i/A_diff}
\end{eqnarray}

The structure of final states in diffraction with nuclei is richer than that  in the proton case and, in addition to coherent diffraction, 
 there is a class of diffractive events corresponding to disintegration of the target nucleus, which is called incoherent diffraction (without meson production).   Using completeness of final nuclear states $A^{\prime} \neq A$, one obtains for the nuclear structure function
 characterizing $e+A \to e^{\prime}+X+A^{\prime}$ incoherent diffraction in DIS with nuclei
\begin{eqnarray}
\fl
F_{2A,{\rm incoh}}^{D(3)}(\beta,Q_0^2,x_{\Pomeron})  &=&  A F_{2N}^{D(3)}(\beta,Q_0^2,x_{\Pomeron}) 
\int d^2 \vec{b} dz \rho_A(b,z) e^{-A \sigma_{\rm soft}^{\rm inel}(x,Q_0^2) \int_{z}^{\infty} dz^{\prime}
\rho_A(\vec{b},z^{\prime})} \nonumber\\
&-& 4 \pi A^2 \frac{\sigma_{\rm soft}}{8 \pi B_X} F_{2N}^{D(4)}(\beta,Q_0^2,x_{\Pomeron},t_{\rm min}) 
\int d^2 \vec{b} \Big|\int^{\infty}_{-\infty} dz \rho_A(\vec{b},z) \nonumber\\
&\times & 
e^{i z x_{\Pomeron}m_N} e^{-\frac{A}{2} (1-i\eta)\sigma_{\rm soft}(x,Q_0^2) \int_{z}^{\infty} dz^{\prime} \rho_A(\vec{b},z^{\prime})}\Big|^2 \,,
\label{eq:f_i/A_diff_inc}
\end{eqnarray}
where $\sigma_{\rm soft}(x,Q_0^2)$ is identified with $\sigma_{\rm soft}^i(x,Q_0^2)$ for quarks, and 
\begin{equation}
\sigma_{\rm soft}^{\rm inel}(x,Q_0^2)=\sigma_{\rm soft}(x,Q_0^2)-(1+\eta^2)[\sigma_{\rm soft}(x,Q_0^2)]^2/(16 \pi B_X) \; ,
\end{equation}
with $B_X=6-7$ GeV$^{-2}$.

We would like to emphasize that as in the inclusive case, in Eqs.~(\ref{eq:f_i/A_diff}) and (\ref{eq:f_i/A_diff_inc}),  the contributions due to the interaction with two target 
nucleons are model-independent. For the $N \geq 3$ terms, we made a simplifying assumption that all configurations 
contributing to $\sigma_{\rm soft}^i(x,Q_0^2)$ have similar cross sections. This is natural since 
$\sigma_{\rm soft}^i(x,Q_0^2)$ is already large.

\begin{figure}[t]
 \includegraphics[width=1.0\textwidth]{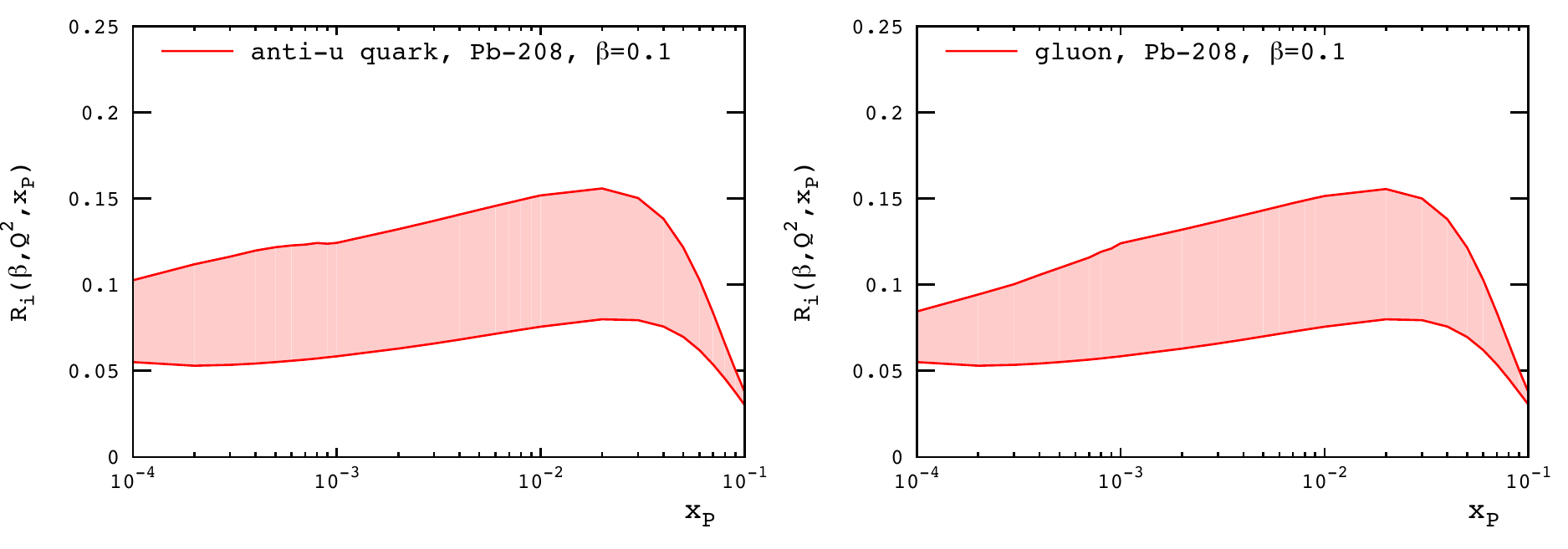}
\caption{The suppression factor of $R_i(\beta,Q_0^2,x_{\Pomeron})$ for nuclear diffractive PDFs as a function of $x_{\Pomeron}$ at $\beta=0.1$ and 
$Q_0=4$ GeV$^2$ for Pb-208. The left and right panels correspond to $\bar{u}$ quarks and gluons, respectively.
}
\label{fig:LT_diff}
\end{figure}

Multiple interactions with target nucleons encoded in Eqs.~(\ref{eq:f_i/A_diff}) and (\ref{eq:f_i/A_diff_inc}) lead to a significant
suppression of the predicted nuclear diffractive PDFs and structure functions at small $x$ due to the leading twist nuclear shadowing.
A convenient way to illustrate its magnitude is to introduce the nuclear suppression factor of $R_i$ with respect to the impulse
approximation (IA):
\begin{eqnarray}
\fl
f_{i/A}^{D(3)}(\beta,Q_0^2,x_{\Pomeron}) &=&R_i(\beta,Q_0^2,x_{\Pomeron}) 4 \pi A^2 \beta f_{i/N}^{D(4)}(\beta,Q_0^2,x_{\Pomeron},t_{\rm min}) \int d^2 b \, (T_A(b))^2 \nonumber\\
\fl
&=&R_i(\beta,Q_0^2,x_{\Pomeron}) A^2 \beta f_{i/N}^{D(4)}(\beta,Q_0^2,x_{\Pomeron},t_{\rm min})\int dt |F_A(t)|^2 \,,
\label{eq:R_i}
\end{eqnarray}
where $F_A(t)$ is the nuclear form factor.
Figure~\ref{fig:LT_diff} shows predictions for $R_i(\beta,Q_0^2,x_{\Pomeron})$ for Pb-208 as a function of $x_{\Pomeron}$ at 
$\beta=0.1$ and $Q_0=4$ GeV$^2$.
The left and right panels correspond to $\bar{u}$ quarks and gluons, respectively. One can see from the figure that the 
predicted nuclear suppression factor depends weakly on the parton flavor and $x_{\Pomeron}$ in the studied kinematic range.

One should note that the uncertainty bands for the predicted 
nuclear diffractive PDFs are much larger than those for the inclusive nPDFs, compare Figs.~\ref{fig:LT_pdfs}
and \ref{fig:LT_diff}. Mostly this is a consequence of the fact that in the case of coherent diffraction, the nuclear diffractive PDFs involve squares of the corresponding scattering amplitudes, which essentially doubles the resulting uncertainties.

Predictions of the leading twist nuclear shadowing model for nuclear structure functions and nuclear
diffractive structure functions can be best tested at the future Electron-Ion Collider~\cite{Boer:2011fh,Accardi:2012qut}.
Morevover, 
in the EIC kinematics with $Q^2 \ge 2$ GeV$^2$ and $x \ge 10^{-3}$, predictions of the presented approach have small uncertainties 
and cover a wide range of nuclei from deuteron to lead.

Note that the strong nuclear shadowing of nuclear diffractive PDFs and structure functions leads to the ratio of the diffractive to total cross sections on nuclei, which is similar to that on the proton \cite{Frankfurt:2011cs}.
Thus, the LT nuclear shadowing makes nuclei more transparent to high-energy probes and, hence, delays the onset of the black disk limit.

%%%%%%%%%%%%%%%%%%%%%%%%%%%%%%%%%
\section{Hard coherent diffraction in photoproduction of quarkonia and dijets on nuclei}
\label{sec:UPC}

\subsection{Exclusive photoproduction of charmonia on nuclei in Pb-Pb UPCs}
\label{sec:UPC_Jpsi}

Further progress in constraining nPDFs at small $x$ relies on studies of 
high energy hard processes with nuclei at collider energies, notably, in proton--nucleus ($pA$)
scattering at the LHC~\cite{Salgado:2011wc} and lepton--nucleus ($eA$) scattering at the future
EIC~\cite{Accardi:2012qut} and LHeC~\cite{AbelleiraFernandez:2012cc}.
However, the QCD analyses of the data on various hard processes in $pA$ scattering at the LHC
during Runs 1 and 2~\cite{Eskola:2013aya,Helenius:2014qla,Armesto:2015lrg,Kusina:2016fxy,Kusina:2017gkz,Eskola:2019dui,Eskola:2019bgf,Kusina:2020lyz} showed
that while they provide additional restrictions on nPDFs, the remaining uncertainties at small $x$ are still significant.
Therefore, it is important to explore the potential of 
complementary probes of small-$x$ nPDFs at the LHC.

It has been realized that collisions of ultrarelativistic ions at large impact parameters, when
the strong interaction is suppressed and the ions interact electromagnetically via the emission
of quasi-real photons in the so-called {\it ultraperipheral collisions} (UPCs), give an opportunity to
study photon--photon, photon--proton and photon--nucleus scattering at unprecedentedly 
high energies~\cite{Baltz:2007kq}. 

This program was realized during Runs 1 and 2 at the LHC 
by measuring exclusive photoproduction of charmonia ($J/\psi$ and $\psi(2S)$ vector mesons) in
Pb-Pb UPCs at $\sqrt{s_{NN}}=2.76$ TeV~\cite{Abbas:2013oua,Abelev:2012ba,Adam:2015sia,Khachatryan:2016qhq} and $\sqrt{s_{NN}}=5.02$ TeV~\cite{Acharya:2019vlb,LHCb:2018ofh,Acharya:2021ugn}.
Further high statistics studies are planned during Runs 3 and 4 at the LHC.

\begin{figure}[t]
\centering
 \includegraphics[width=0.65\textwidth]{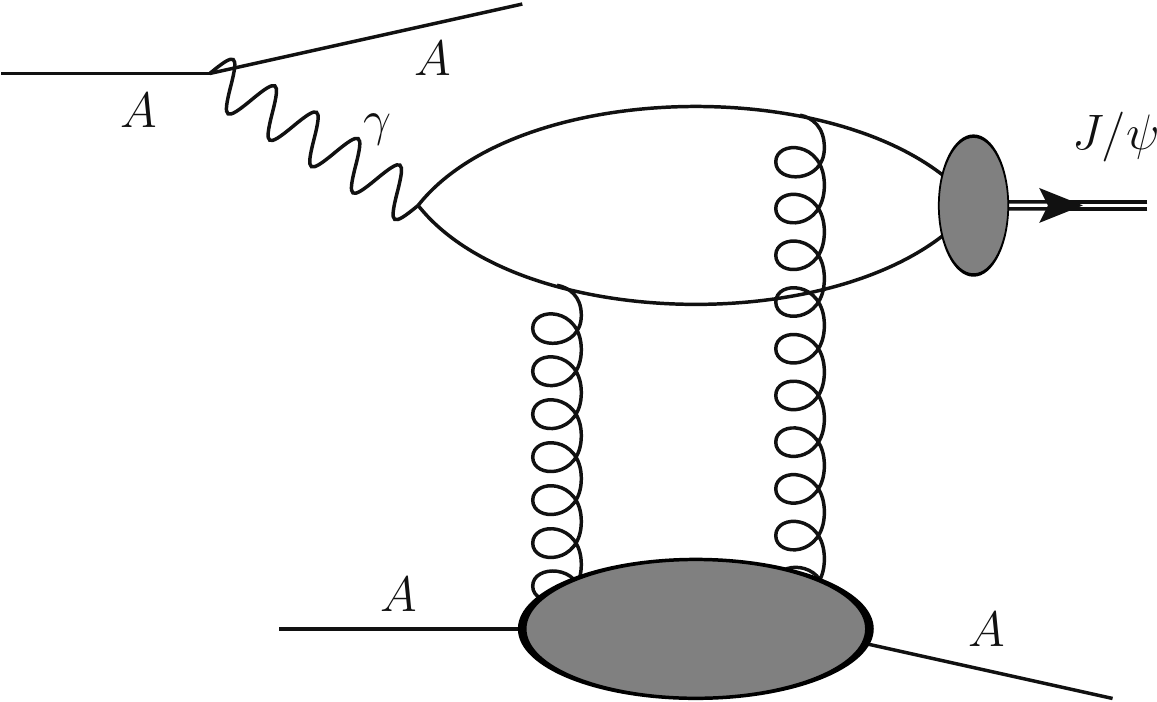}
\caption{Coherent $J/\psi$ photoproduction on nuclei in UPCs of ions $A$. The cross section is given by a sum of two terms, where each ion serves either as a source of equivalent photons or as a target.}
\label{fig:Jpsi_upc}
\end{figure}

A diagram for coherent $J/\psi$ photoproduction on nuclei in UPCs of ions $A$ is shown in Fig.~\ref{fig:Jpsi_upc}. 
Using the method of equivalent photons ~\cite{Budnev:1974de,Vidovic:1992ik},
the cross section of 
this process in nucleus--nucleus  UPCs
can be written in the following form~\cite{Baltz:2007kq}:
\begin{equation}
\frac {d\sigma_{AA\to AAJ/\psi}(y)} {dy}
=N_{\gamma/A}(y)\sigma_{\gamma A\to AJ/\psi}(y)+
N_{\gamma/A}(-y)\sigma_{\gamma A\to AJ/\psi}(-y) \,,
\label{csupc}
\end{equation}
where 
$N_{\gamma/A}(y) \equiv \omega dN_{\gamma/A}(\omega)/d \omega$ is the photon flux;
$y = \ln(2\omega/M_{J/\psi})=\ln(W^{2}_{\gamma p}/(2\gamma_{L}m_{N}M_{J/\psi}))$ is the $J/\psi$ rapidity, 
where $\omega$ is the photon 
energy in the laboratory frame, $W_{\gamma p}$ is the invariant photon--proton center-of-mass energy,
 $M_{J/\psi}$ is the mass of $J/\psi$ and $m_N$ is the nucleon mass.
 Note that interference between the two terms in Eq.~(\ref{csupc}) is sizable only at very small values of
the $J/\psi$ transverse momentum~\cite{Bertulani:2005ru,Klein:1999gv} and at central rapidities and, hence, can be safely neglected in the $t$-integrated UPC cross section.
 The presence of two terms in Eq.~(\ref{csupc}) is a reflection of the fact that in UPCs, both colliding nuclei serve as a source of quasi-real photons and a target.  For non-central rapidities $y \neq 0$, this leads to ambiguity in the value of probed $W_{\gamma p}$,
 \begin{equation}
W_{\gamma p}^{\pm}=\sqrt{2 E_N M_{J/\psi}}\,e^{\pm y/2} \,, 
\label{eq:W}
\end{equation}
where $E_N$ is the nuclear beam energy per nucleon.

The photon flux is calculated as convolution of the flux of quasireal 
photons emitted by an ultrarelativistic charged ion $N_{\gamma/A}(\omega,\vec{b})$~\cite{Budnev:1974de,Vidovic:1992ik} with the probability 
not to have inelastic strong ion--ion interactions $\Gamma_{AA}(\vec{b})=\exp(-\sigma_{NN} \int d^2 \vec{b}_1 T_A(\vec{b}_1)
T_A(\vec{b}-\vec{b}_1))$,
\begin{equation}
N_{\gamma/A}(W_{\gamma p})=\int d^2 \vec{b}\, N_{\gamma/A}(\omega,\vec{b}) \Gamma_{AA}(\vec{b}) \,,
\label{eq:flux}
 \end{equation}
where $\sigma_{NN}$ is the total nucleon--nucleon cross section and  
$T_A(\vec{b})$ is the nuclear optical density.

As we discussed in Sec.~\ref{sec:dipole_model},
coherent photoproduction of $J/\psi$ probes the gluon density of the target, see Eqs.~(\ref{eq:cs_photo}) and (\ref{eq:cs_photo_2}).  Thus, in the case of photoproduction on nuclei, this process allows one to access the $xg_A(x,\mu^2)$ nuclear gluon distribution at small $x$~\cite{Ryskin:1992ui}, 
\begin{equation}
\frac{d\sigma_{\gamma A \to J/\psi A}(W_{\gamma p},t=0)}{dt} \, \propto \, [x g_A(x,\mu^2)]^2 \,.
\label{eq:rel1}
\end{equation}
In Sec.~\ref{sec:dipole_model}, we discussed and quantified 
within the MFGS dipole model corrections to Eqs.~(\ref{eq:cs_photo}) and (\ref{eq:cs_photo_2}), hence, to the relation in Eq.~(\ref{eq:rel1}).
One should also mention additional corrections, which include
 the effects of the transverse momentum in the gluon loop and the charmonium wave function~\cite{Ryskin:1995hz},
 relativistic corrections to the charmonium wave function~\cite{Hoodbhoy:1996zg,Frankfurt:1995jw,Frankfurt:1997fj,Krelina:2019egg,Lappi:2020ufv}, and 
 next-to-leading order (NLO) QCD radiative corrections in the framework of collinear factorization for hard exclusive processes and generalized parton distribution functions (GPDs)~\cite{Ivanov:2004vd,Jones:2015nna}.

Keeping in mind all these corrections and using Eq.~(\ref{eq:rel1}), one
can express the cross section of coherent $J/\psi$ photoproduction on nuclei integrated over $t$ in the following form~\cite{Guzey:2013xba,Guzey:2013qza,Guzey:2016qwo}
\begin{eqnarray}
\fl
\sigma_{\gamma A \to J/\psi A}(W_{\gamma p}) &= \kappa_{A/N}^2 \frac{d\sigma_{\gamma p \to J/\psi p}(W_{\gamma p},t=0)}{dt} \int_{|t_{\rm min}|}^{\infty} dt
\left[\frac{ xg_A(x,t,\mu^2)}{A xg_N(x,\mu^2)}\right]^2   \nonumber\\
& \approx  \kappa_{A/N}^2 \frac{d\sigma_{\gamma p \to J/\psi p}(W_{\gamma p},t=0)}{dt}
\left[\frac{xg_A(x,\mu^2)}{A xg_N(x,\mu^2)}\right]^2 \int_{|t_{\rm min}|}^{\infty} dt  |F_A(t)|^2  \,,
\label{eq:rel2}
\end{eqnarray}
where $|t_{\rm min}|=x^2 m_N^2$ is the minimal momentum transfer squared.
In Eq.~(\ref{eq:rel2}), $xg_A(x,t,\mu^2)$ is the nuclear gluon GPD in the limit, 
when both gluon lines carry the same light-momentum fraction $x$, i.e., it is a two-gluon form factor of the target.
Fourier transform of this quantity relates it to the $b$-dependent nuclear gluon distribution, which we discussed in Sect.~\ref{subsec:LT},
\begin{equation}
g_A(x,t,\mu^2)=\int d^2 \vec{b} \, e^{i \vec{q}_{\perp} \cdot \vec{b}} g_A(x,b,\mu^2) \,,
\label{eq:rel2b}
\end{equation}
where $q_{\perp}$ is the transverse component of the momentum transfer, $t \approx -q_{\perp}^2$.
In the second line of Eq~(\ref{eq:rel2}), we used the commonly used factorized form $g_A(x,t,\mu^2)=F_A(t) g_A(x,\mu^2)$, where 
$F_A(t)$ is the nuclear form factor. This approximation works sufficiently well for the $t$-integrated cross section.
At the same time, for the differential cross section, the 
impact parameter dependence of the leading twist gluon nuclear shadowing modifies the shape of the $t$ distribution~\cite{Guzey:2016qwo}, see also the discussion below.

The factor of $\kappa_{A/N}^2=(1+\eta_A^2) R_{g,A}^2/[(1+\eta_p^2) R_{g,p}^2]$ takes into account the slightly different $x$ dependence of the nuclear and proton gluon distributions, where $\eta$ is the ratio of the real to the imaginary parts of the $\gamma T \to J/\psi T$ amplitude, and $R_{g}$ is a phenomenological enhancement factor relating the usual gluon density to the gluon generalized parton distribution, see Eq.~(\ref{eq:Rg}).
In the case, when the gluon distributions in a nucleus and the proton have a similar small-$x$ behavior, i.e., 
$R_g(x,\mu^2)=xg_A(x,\mu^2)/[A xg_N(x,\mu^2)]$ is a slow function of $x$, $\kappa_{A/N} \approx 1$ with a good precision.

To quantify nuclear modifications of the $J/\psi$ photoproduction cross section and minimize the effects 
affecting Eqs.~(\ref{eq:rel1}) and (\ref{eq:rel2}), which we mentioned above, it is useful to introduce the nuclear cross 
section in the impulse approximation (IA)
\begin{equation}
\sigma_{\gamma A \to J/\psi A}^{\rm IA}(W_{\gamma p})= \frac{d\sigma_{\gamma N \to J/\psi N}(W_{\gamma p},t=0)}{dt} 
\int_{|t_{\rm min}|}^{\infty} dt  |F_A(t)|^2  \,.
\label{eq:sigma_IA_tint}
\end{equation}
It is important to point out that  the theoretically-defined $\sigma_{\gamma A \to J/\psi A}^{\rm IA}(W_{\gamma p})$
is practically  model-independent since its estimate is based on the experimental data.

Then, taking the square root of the ratio of the cross sections in Eqs.~(\ref{eq:rel2}) and (\ref{eq:sigma_IA_tint}), 
one introduces the nuclear suppression factor of $S_{Pb}(x)$~\cite{Guzey:2013xba,Guzey:2013qza,Guzey:2020ntc}
\begin{equation}
S_{Pb}(x)=\sqrt{\frac{\sigma_{\gamma A \to J/\psi A}(W_{\gamma p})}{\sigma_{\gamma A \to J/\psi A}^{\rm IA}(W_{\gamma p})}}=
\kappa_{A/N} \frac{xg_A(x,\mu^2)}{A xg_N(x,\mu^2)} \equiv
\kappa_{A/N} R_g(x,\mu^2) \,.
\label{eq:rel3}
\end{equation}
It is expected that almost all kinematic factors and mentioned corrections cancel in the ratio of the nuclear and IA( proton) cross sections. Thus, Eq.~(\ref{eq:rel3}) establishes a direct correspondence between the suppression factor of $S_{Pb}(x)$ and the ratio of the nuclear and nucleon gluon distributions $R_g(x,\mu^2)$. Further, since at central rapidities $|y| \approx 0$, 
the $d\sigma_{AA\to AAJ/\psi}(y)/dy$ cross section is unambiguously related to the $\sigma_{\gamma A \to J/\psi A}(W_{\gamma p})$
photoproduction cross section at the definite value of $W_{\gamma p}=\sqrt{2 E_N M_{J/\psi}}$, Eq.~(\ref{eq:rel3}) gives a one-to-one correspondence between the measured UPC cross section at central rapidities and $R_g(x,\mu^2)$ at $x=M_{J/\psi}/(2 E_N)$.

\begin{figure}[t]
\centering
\includegraphics[width=0.80\textwidth]{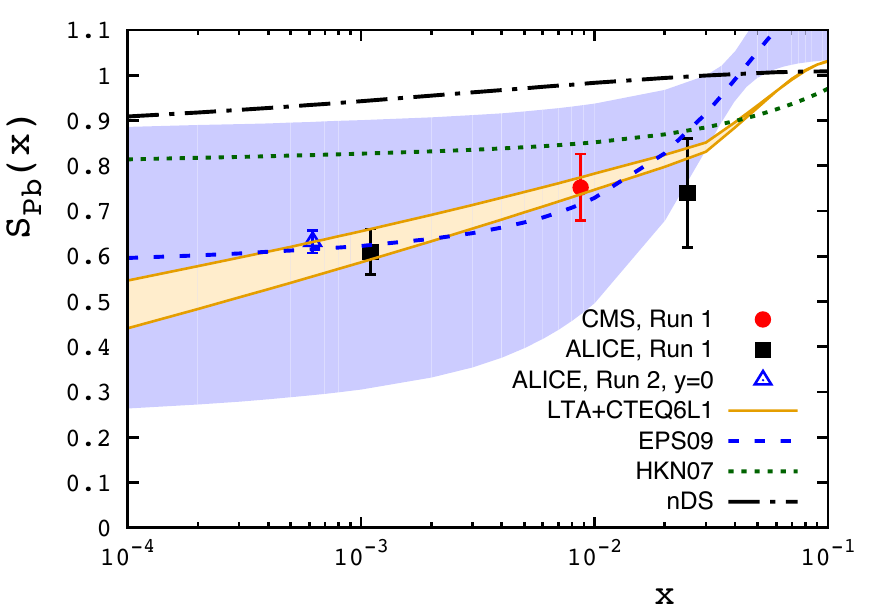}
\caption{The nuclear suppression factor of $S_{Pb}(x)$ as a function of the gluon momentum fraction of $x$:
the values extracted from the Run 1~\cite{Abbas:2013oua,Abelev:2012ba,Khachatryan:2016qhq} and the central rapidity 
Run 2~\cite{Acharya:2021ugn} UPC data on coherent $J/\psi$ photoproduction in Pb-Pb UPCs vs.~predictions of the leading twist model of nuclear shadowing and global fits of nPDFs. The bands indicate the uncertainties for the LTA model (yellow) and EPS09 parametrization (blue). }
\label{fig:S_pb208_2016}
\end{figure}

Figure~\ref{fig:S_pb208_2016} shows a comparison of the values of $S_{Pb}(x)$ extracted from the Run 1~\cite{Abbas:2013oua,Abelev:2012ba,Khachatryan:2016qhq} and the central rapidity 
Run 2~\cite{Acharya:2021ugn} UPC data on coherent $J/\psi$ photoproduction in Pb-Pb UPCs with $R_g(x,\mu^2)$ predicted in the leading twist model of nuclear shadowing and
global QCD fits of nPDFs. Note that
following the analysis of Ref.~\cite{Guzey:2013qza}, we take advantage of the ambiguity in the exact values of the scale $\mu$ and
take $\mu^2=3$ GeV$^2$ to best reproduce the available HERA and LHCb data on the $W_{\gamma p}$ dependence of the cross section
of exclusive $J/\psi$ photoproduction on the proton.
The good agreement with the predictions of the leading twist nuclear shadowing model and the EPS09 nPDFs, which however have much larger uncertainties, gives 
direct and weakly model-dependent evidence of large nuclear gluon shadowing at small $x$,
\begin{equation}
R_g(x=6 \times 10^{-4}-10^{-3},\mu^2=3 \ {\rm GeV}^2) \approx 0.6 \,.
\label{eq:Rg_shad}
\end{equation}

Note that the analysis of Ref.~\cite{Guzey:2020ntc} extracted the nuclear suppression factor of $S_{Pb}(x)$ in a wide range of $x$,
$10^{-5} \leq x \leq 0.04$ using all available Run 1 and 2 data on coherent $J/\psi$ photoproduction in Pb-Pb UPCs.
However, due to the ambiguity of the two terms in Eq.~(\ref{csupc}), such a procedure is in general model dependent and leads to
significant uncertainties in $S_{Pb}(x)$ for $x < 6 \times 10^{-4}$ and $x > 0.01$.
In this respect one should also mention the analysis of 
\cite{Contreras:2016pkc}, where $S_{Pb}(x)$ was extracted from measurements of coherent $J/\psi$ photoproduction in ultraperipheral and peripheral Pb-Pb collisions at the LHC at 2.76 TeV. The results of that analysis broadly agree with 
the trend of the nuclear suppression presented in 
Fig.~\ref{fig:S_pb208_2016}.

\begin{figure}[t]
\centering
\includegraphics[width=0.75\textwidth]{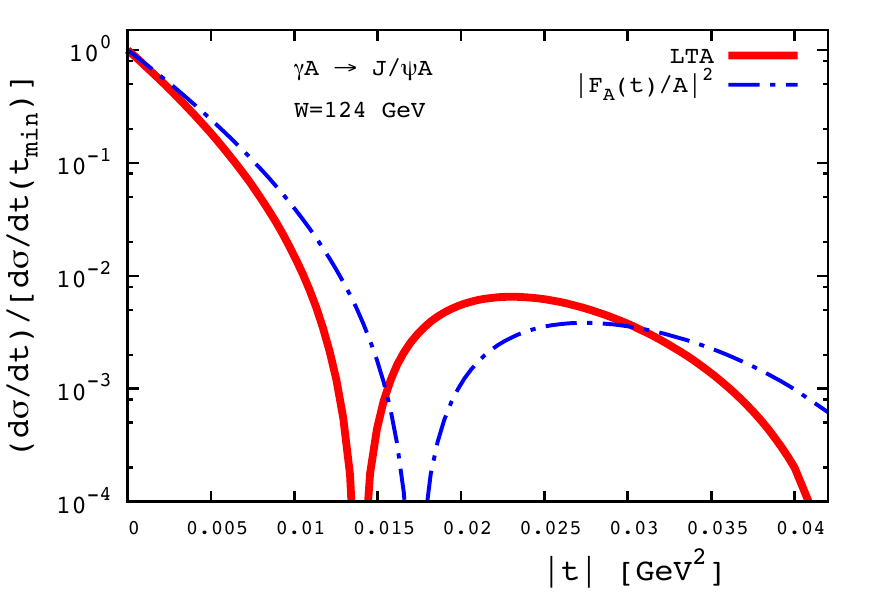}
\caption{The $d\sigma_{\gamma A \to J/\psi A}(W_{\gamma p},t)/dt$ 
cross section normalized to its value at $|t|=t_{\rm min}$ as a function of $t$ at $W=124$ GeV: predictions of the leading twist model of nuclear shadowing (red solid curve) vs. the factorized approximation (blue dot-dashed curve).
The figure is from~\cite{Guzey:2016qwo}, {\tt https://doi.org/10.1103/PhysRevC.95.025204}.}
\label{fig:Jpsi_tdep}
\end{figure}

The significant leading twist gluon nuclear shadowing also
affects the differential cross section of coherent $J/\psi$ photoproduction on nuclei,
\begin{equation}
\frac{d\sigma_{\gamma A \to J/\psi A}(W_{\gamma p},t)}{dt} = \kappa_{A/N}^2 \frac{d\sigma_{\gamma p \to J/\psi p}(W_{\gamma p},t=0)}{dt}
\left[\frac{xg_A(x,t,\mu^2)}{A xg_N(x,\mu^2)}\right]^2 \,.
\label{eq:cs_Jpsi_tdep}
\end{equation}
Figure~\ref{fig:Jpsi_tdep} shows the $d\sigma_{\gamma A \to J/\psi A}(W_{\gamma p},t)/dt$ 
cross section normalized to its value at $|t|=t_{\rm min}$ as a function of $t$ at $W=124$ GeV. This value corresponds to Pb-Pb UPCs during Run 2 at the LHC with $\sqrt{s_{NN}}=5.02$ TeV and the central rapidity $y=0$. The red solid curve is the prediction of Eq.~(\ref{eq:cs_Jpsi_tdep}), where for 
$xg_A(x,t,\mu^2)$ and $xg_A(x,b,\mu^2)$, see Eq.~(\ref{eq:rel2b}), we used predictions of the leading twist nuclear shadowing model for the impact parameter dependent nuclear PDFs, see Sec.~\ref{sec:shadowing}.
The blue dot-dashed curve gives the $t$ dependence of the nuclear form factor squared 
$[F_A(t)/A]^2$. One can see from the figure that the impact parameter dependence of the leading twist nuclear shadowing, i.e., the correlation between $b$ and $x$ in $xg_A(x,b,\mu^2)$, noticeably shifts the minimum of the $t$ distribution toward lower values of $t$. This can  be interpreted as broadening in impact parameter space of the small-$x$ gluon distribution in nuclei as a consequence of the fact that nuclear shadowing increases with a decrease of $b$ (increase of the nuclear density).

The predictions for the shift of the $t$ dependence of the
$d\sigma_{\gamma A \to J/\psi A}(W_{\gamma p},t)/dt$ 
cross section shown in Fig.~\ref{fig:Jpsi_tdep} have been nicely confirmed by the recent ALICE measurements~\cite{Acharya:2021bnz}.

The program of UPC measurements will continue with Runs 3 and 4 at the LHC with the increased collision energy and luminosity~\cite{Citron:2018lsq}, which  will enable one to probe and constrain the nuclear gluon distribution down to $x \approx 10^{-4}$.
An additional opportunity to access small-$x$ nuclear gluon distributions is offered by $J/\psi$ photoproduction accompanied by neutron emission due to electromagnetic excitation of one or both colliding nuclei, which allows one 
in principle to separate the contributions of two terms in Eq.~(\ref{csupc}), see \cite{Guzey:2013jaa}.

%%%%%%%%%%%%%%%%%%%%%%%%%%%%%%%%%%%%%%%%%%%%%%%%%%%%%%%%%%%%%%%%%%%%%%%%%%%%%%
\subsection{Diffractive dijet photoproduction in UPCs and at the EIC}

The program of UPC measurements also includes photoproduction of jets on the proton and nuclei, which provides information on the proton and nucleus structure in QCD, which is complementary to that obtained in DIS. As discussed in Sec.~\ref{sec:diff_dijets}, measurements of dijets in the presence of a rapidity gap allow one to access diffractive parton distributions of the target, test their universality, and get a new handle of factorization breaking in this process.

\begin{figure}[ht]
\begin{center}
\includegraphics[width=0.9\textwidth]{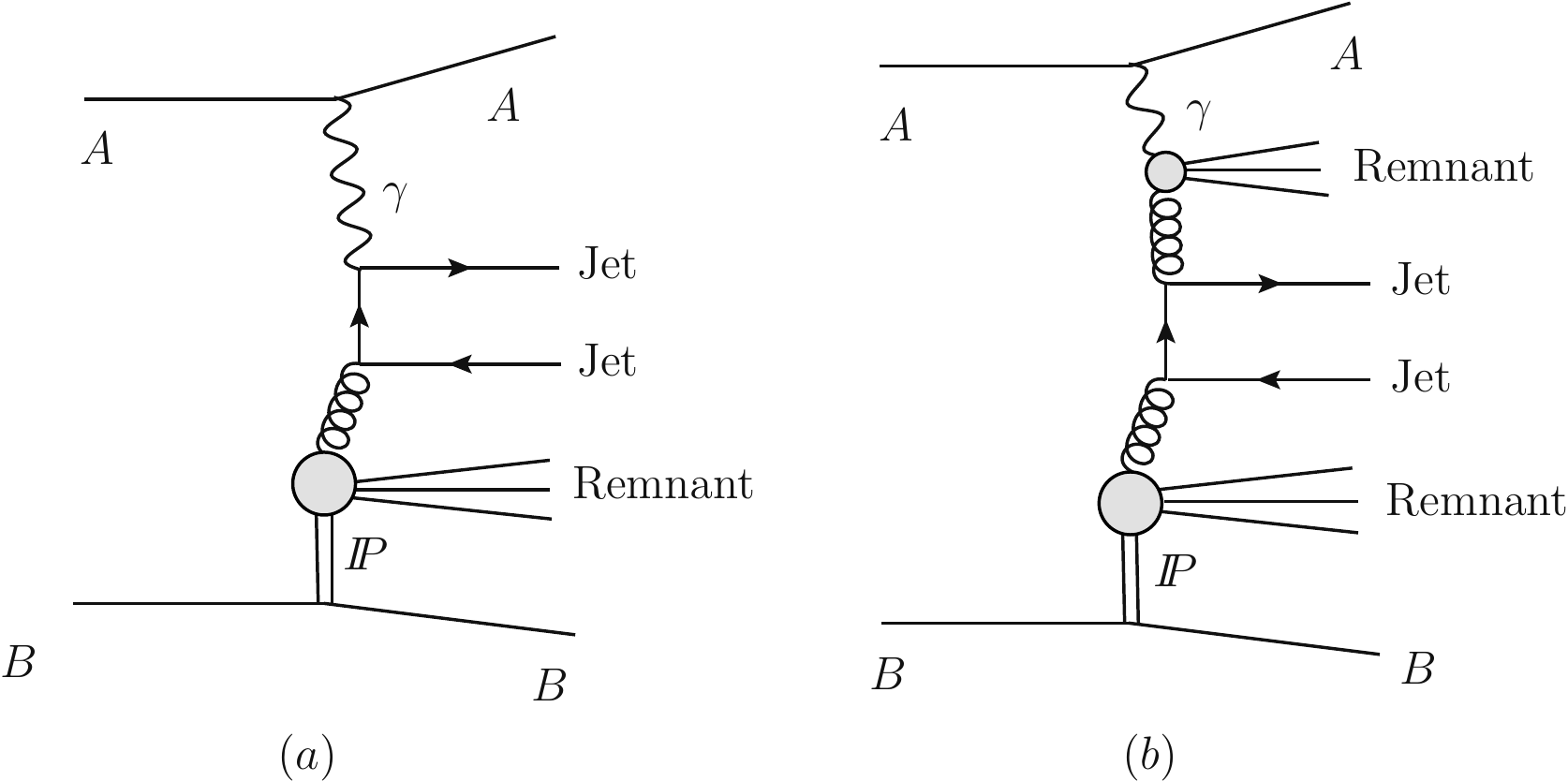}
\caption{Typical Feynman graphs for diffractive dijet photoproduction in UPCs.
Graphs $(a)$ and $(b)$ correspond to the direct and resolved photon contributions, respectively.}
\label{fig:photo_dijets_upc}
\end{center}
\end{figure}

Figure~\ref{fig:photo_dijets_upc} shows typical Feynman graphs for (quark) dijet photoproduction in the UPCs of ions $A$ and $B$,
where graphs $a$ and $b$ correspond to the direct and the resolved photon contributions, respectively. They are completely analogous to graphs in Fig.~\ref{fig:photo_dijets_ep} with the only difference that the electron and the proton are replaced by nuclei.

The corresponding cross section for symmetric heavy ion UPCs reads
\begin{eqnarray}
\fl
d \sigma(AA \to A+2\,{\rm jets}+X^{\prime}+A) &=& d\sigma(AA\to A+2\,{\rm jets}+X^{\prime}+A)^{(+)} \nonumber\\
&+& d \sigma(AA\to A+2\,{\rm jets}+X^{\prime}+A)^{(-)} \,,
\label{eq:AA}
\end{eqnarray}
where $X^{\prime}$ stands for the produced diffractive final state $X$ after removing two jets.
The two terms in Eq.~(\ref{eq:AA}) have the same meaning as in Eq.~(\ref{csupc}) and correspond to two possible directions of the diffracting nucleus. They are related to each other  by inverting the sign of the $\eta_1$ and $\eta_2$ jet rapidities:
\begin{equation}
\fl
d \sigma(AA\to A+2\,{\rm jets}+X^{\prime}+A)^{(-)}=d \sigma(AA\to A+2\,{\rm jets}+X^{\prime}+A)^{(+)}_{|\eta_1 \to -\eta_1,\, \eta_2 \to -\eta_2} \,.
\label{eq:reverse_rule}
\end{equation}

The cross section of dijet photoproduction in $AA$ UPCs  is given by the following convolution in analogy to the photoproduction in DIS collisions, see Eq.~(\ref{eq:fact_diff_dijet_photo}) in Sec.~\ref{sec:diff_dijets},
\begin{eqnarray}
\fl
d \sigma(AA\to A &+&2\,{\rm jets}+X^{\prime}+A)^{(+)}
=\sum_{a,b} \int dt \int  dx_{\Pomeron}
\int_0^1 dz_{\Pomeron} \int  dy \int dx_{\gamma}
\nonumber\\
&\times & f_{\gamma/A}(y)  f_{a/\gamma}(x_{\gamma},\mu^2) f^{D(4)}_{b/A}(x_{\Pomeron},z_{\Pomeron},t,\mu^2)
d \hat{\sigma}_{ab \to {\rm jets}}^{(n)} \,,
\label{eq:AA_2}
\end{eqnarray}
where $a$ and $b$ are parton flavors including the case, when $a$ stands for the photon for the direct photon contribution;
$y$, $x_{\gamma}$,  $x_{\Pomeron}$, and $z_{\Pomeron}$ are longitudinal momentum fractions carried by photons, partons in the photon, the diffractive exchange in the target nucleus, and partons in the diffractive exchange, respectively;
$f_{\gamma/A}(y)$ is the photon flux calculated in the equivalent photon approximation;
$f_{a/\gamma}(x_{\gamma},\mu^2)$ and $f^{D(4)}_{b/A}(x_{\Pomeron},z_{\Pomeron},t,\mu^2)$
are PDFs of the photon in the resolved photon case and the diffractive PDFs of the target nucleus,
respectively; $d \hat{\sigma}_{ab \to {\rm jets}}^{(n)}$ is the cross section for production of 
an $n$-parton final state from two initial partons $a$ and $b$.

As we mentioned above, Eq.~(\ref{eq:AA_2}) should be in principle complemented by the effect of factorization breaking in diffractive dijet photoproduction. 
In connection to this, one of the best observables to distinguish various schemes of the factorization breaking is provided by the 
observed photon momentum fraction of $x_{\gamma}^{\rm jets}$,
\begin{equation}
x_{\gamma}^{\rm jets} = \frac{m_{\rm jets}}{y \sqrt{s_{NN}}}e^{y_{\rm jets}} \,,
\label{eq:variables_jets}
\end{equation}
where 
\begin{equation}
\fl
m_{\rm jets} = \left[(E_1+E_2)^2-(\vec{p}_1+\vec{p}_2)^2 \right]^{1/2} \,, \quad
y_{\rm jets} =\frac{1}{2} \ln \frac{E_1+E_2+p_{1z}+p_{2z}}{E_1+E_2-p_{1z}-p_{2z}} \,.
\label{eq:variables_jets2}
\end{equation}
In Eq.~(\ref{eq:variables_jets2}), $E_{1,2}$, $\vec{p}_{1,2}$ and $p_{1,2z}$ are the energies, momentum vectors, 
and the longitudinal component of the momentum of two most energetic jets.

\begin{figure}[t]
\centering
\includegraphics[width=0.8\textwidth]{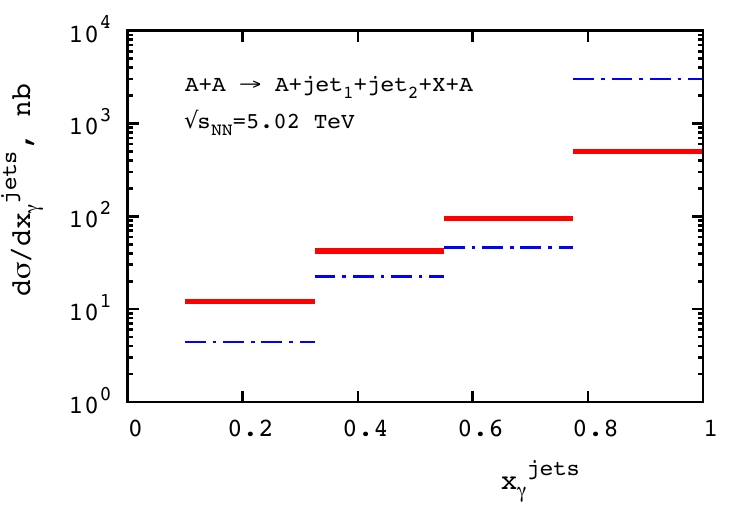}
 \caption{The cross section of diffractive dijet photoproduction in Pb-Pb UPCs at
 $\sqrt{s_{NN}}=5.02$ TeV as a function of $x_{\gamma}^{\rm jets}$ for two scenarios of diffractive QCD factorization breaking:
 the red solid curve corresponds to the global suppression factor of $R_{\rm glob}=0.1$, while
the blue dot-dashed curve assumes the suppression only for the resolved photon contribution by the factor of $R_{\rm res}=0.04$. 
These predictions also include the effect of
 leading twist nuclear shadowing in nuclear diffractive PDFs suppressing the cross section by approximately 
a factor  of seven.
}
 \label{fig:Summary_rffi}
\end{figure}

Using the framework presented above, one can make predictions for various distributions in 
diffractive dijet photoproduction in UPC in the LHC kinematics~\cite{Guzey:2016tek}.
This analysis used the $k_T$-cluster algorithm with the jet radius $R=1$ and assumed that $E_1 \geq 20$ GeV for the leading jet and $E_2 \geq 18$ GeV for the subleading one.
An example is presented in Fig.~\ref{fig:Summary_rffi}, where the cross section of dijet photoproduction in 
Pb-Pb UPCs at $\sqrt{s_{NN}}=5.02$ TeV is shown as a function $x_{\gamma}^{\rm jets}$. The results are shown 
for two schemes of diffractive QCD
factorization breaking. Namely, the red solid curve corresponds to the global suppression factor of $R_{\rm glob}=0.1$, while
the blue dot-dashed curve assumes the suppression only for the resolved photon contribution by the factor of $R_{\rm res}=0.04$.

Thus, studies of this process on nuclei may shed some light on the mechanism of QCD
factorization breaking in diffractive photoproduction and, for the first time, give access to
nuclear diffractive PDFs and test their models. For instance, if one neglected the strong
suppression of nuclear diffractive PDFs by the leading twist nuclear
shadowing, see Fig.~\ref{fig:LT_diff}, the predicted cross section shown in
Fig.~\ref{fig:Summary_rffi} would increase by approximately 
a factor  of seven.

Similarly, one can study diffractive dijet photoproduction in lepton--proton and lepton--nucleus scattering 
at the future Electron-Ion Collider (EIC) in the US.
Using the same framework of NLO pQCD,
the cross section for the reaction $e+A \rightarrow e+2~{\rm jets}+X'+A$ can be written as follows~\cite{Guzey:2020gkk}
\begin{equation}
\fl
d\sigma =
  \sum_{a,b} \int dy \int dx_{\gamma} \int dt \int dx_{\Pomeron} \int dz_{\Pomeron}
  f_{\gamma/e}(y) f_{a/\gamma }(x_\gamma,\mu^2) f^{D(4)}_{b/A}(x_{\Pomeron},z_{\Pomeron},t,\mu^2)  d\hat{\sigma}_{ab}^{(n)} \,.
  \label{eq:gk_1}
\end{equation}
The only difference between Eqs.~(\ref{eq:gk_1})
 and (\ref{eq:fact_diff_dijet_photo}) is the use 
 of the nuclear diffractive PDFs instead of those of the proton.
In the case of lepton--nucleus (lepton--nucleus) scattering, the longitudinal momentum fraction $x_\gamma^{\rm obs}$ 
is usually experimentally determined from the two observed leading jets through
\begin{equation}
x_{\gamma}^{\rm obs} = {p_{T1}\,e^{-\eta_1}+p_{T2}\,e^{-\eta_2}\over 2yE_e}  \,,
\label{eq:gk_2}
\end{equation}
where $p_{T}$ and $\eta$ is the transverse momentum and pseudorapidity of jet 1 or 2; $E_e$ is the electron beam energy.

\begin{figure}[t]
\centering
\includegraphics[width=0.8\textwidth]{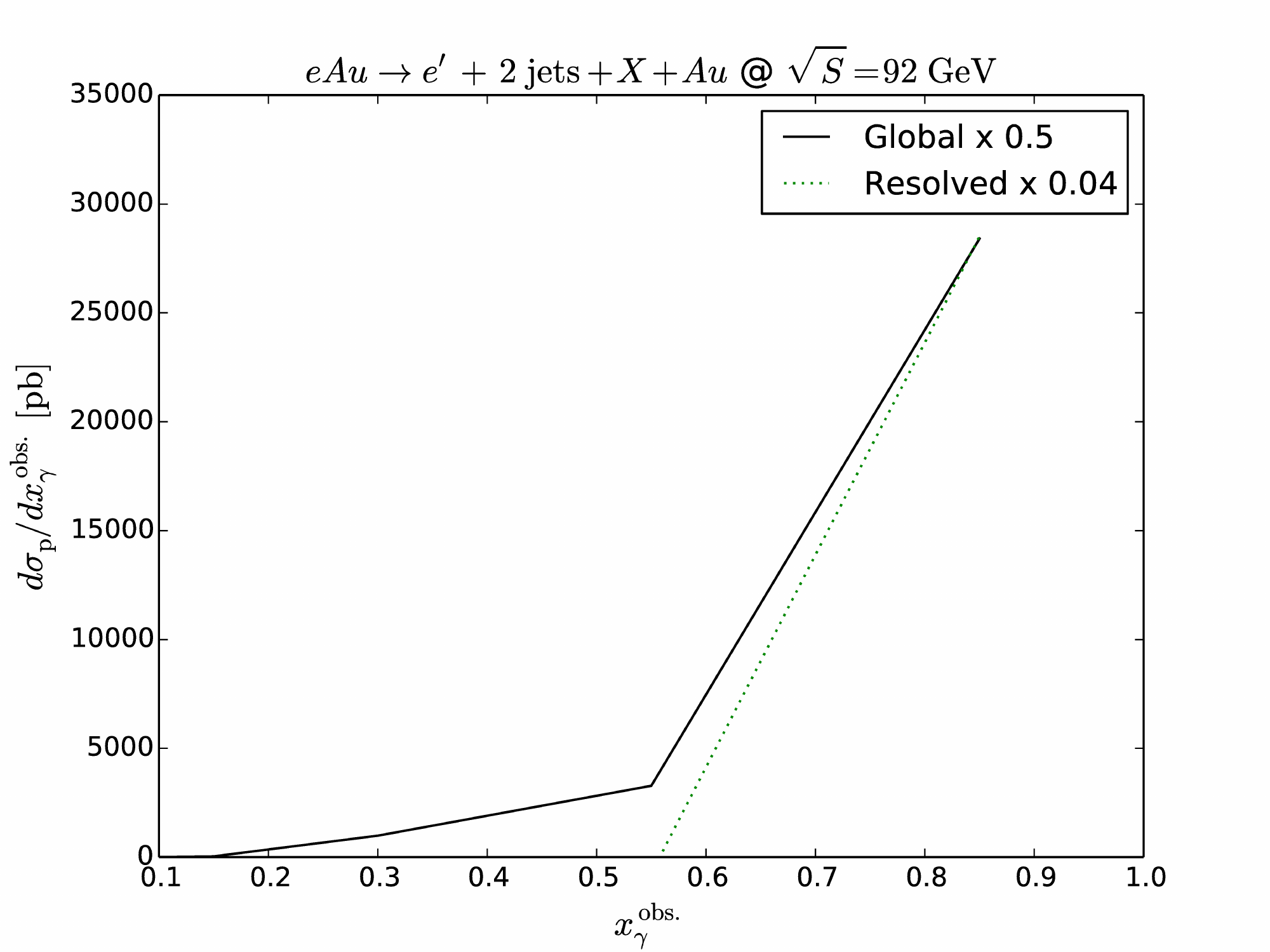}
\caption{NLO QCD cross section for coherent diffractive dijet photoproduction 
   $e+Au \to e^{\prime}+2\ {\rm jets}+ X+Au$ as a function of $x_{\gamma}^{\rm obs}$
   at $\sqrt{S}=92$ GeV. The effect of factorization breaking is included either by the 
    global suppression by a factor of $R_{\rm glob.}=0.5$ (full black) or by the suppression of only
   the resolved photon contribution by a factor of $R_{\rm dir.}=0.04$ (dotted green curves).
   The figure is from \cite{Guzey:2020gkk}, {\tt DOI:10.1007/JHEP05(2020)074}.
}
\label{fig:gk}
\end{figure}

Using the formalism outlined above, one can make predictions for diffractive dijet photoproduction in the EIC 
kinematics as a function of  
the jet average transverse momentum $\bar{p}_T$, the observed longitudinal momentum fractions $x_{\gamma}^{\rm obs}$
and $z_{\Pomeron}^{\rm obs}$, the proton longitudinal momentum transfer $x_{\Pomeron}$, and
the jet pseudorapidity difference $\Delta \eta$~\cite{Guzey:2020gkk}.
Given experience from HERA, one can define jets with the anti-$k_T$ algorithm with the distance parameter $R=1$
and assume that the detectors
can identify jets above relatively low transverse energies of $p_{T1}>5$ GeV
(leading jet) and $p_{T2}>4.5$ GeV (subleading jet).
This is because the underlying event in the photon fragmentation region is much smaller than in the studied kinematics of $pA$ collisions at the LHC.
An example of the resulting predictions is presented in Fig.~\ref{fig:gk}, which shows the dijet cross section as a function of $x_{\gamma}^{\rm obs}$.
 We compare two different schemes of factorization breaking, namely, global suppression by a factor of 0.5 (full black)
   with only resolved-photon suppression by a factor of 0.04 (dotted green curves).

Note that the discussion above assumed that diffractive dijet photoproduction is dominated by the electromagnetic (UPC) mechanism of the photon--Pomeron scattering. 
It was shown in \cite{Basso:2017mue} that this is indeed the case for heavy ion UPCs. Furthermore, as one goes from heavy to light nuclei, the Pomeron-Pomeron contribution becomes more competitive and the two contributions become approximately equal for oxygen beams.

\section{Coherent and incoherent photoproduction of light vector mesons in heavy ion UPCs} 
\label{sec:UPC_rho}

In addition to photoproduction of quarkonia and dijets in UPCs, which address the hadron structure and 
nuclear shadowing in perturbative QCD, it is important to study these phenomena in soft photon--nucleus scattering at high energies. 
In particular, coherent and incoherent photoproduction of light vector mesons ($\rho$, $\omega$, $\phi$)
in heavy ion UPCs gives complementary information on the non-perturbative hadron structure of real photons and the magnitude and energy-dependence of the inelastic (Gribov) shadowing correction to the photon--nucleus cross sections.

Recalling the discussion in Sec.~\ref{sec:UPC_Jpsi}, 
the cross section of coherent and incoherent photoproduction
of light 
vector mesons (for definiteness, we consider the case of $\rho$ mesons) can be written in the following form
\begin{equation}
\frac{d \sigma_{AA \to \rho AA^{\prime}}}{dy}\; =\;N_{\gamma/A}(y) \sigma_{\gamma A \to \rho A^{\prime}}(y)
+N_{\gamma/A}(-y) \sigma_{\gamma A \to \rho A^{\prime}}(-y) \;.
\label{eq:cs_upc}
\end{equation}
where $A^{\prime}$ denotes the final state of the target nucleus and includes both the coherent case $A^{\prime}=A$ 
and the case of nuclear break-up $A^{\prime} \neq A$.

\subsection{Hadron structure of real photons and 
coherent and incoherent cross sections of photoproduction on nuclei}

In Sec.~\ref{sec:fluctuations}, we discussed that at high energies, hadrons and photons interact with hadronic targets by means of their hadronic fluctuations. This can be quantified by introducing the distribution $P(\sigma)$ giving the probability for the projectile to interact with the target hadron with the cross section $\sigma$. The resulting formalism enables for transparent calculation and 
interpretation of coherent and incoherent (quasi-elastic) cross sections of hadron--nucleus and photon--nucleus scattering at high energies in  the Gribov--Glauber framework.

Applying the notion of cross section fluctuations to the $\gamma \to \rho$ transition,
one readily obtains the cross section of coherent $\rho$ photoproduction on nuclei~\cite{Frankfurt:2015cwa}
\begin{equation}
\sigma_{\gamma A \to \rho A}=\left(\frac{e}{f_{\rho}}\right)^2 \int d^2 {\bf b} \left|\int d\sigma P_{\rho}(\sigma)
\left(1-e^{-\frac{1}{2} \sigma T_A(b)} \right)  \right|^2 \,,
\label{eq:sigma_rhoA}
\end{equation}
where $f^2_{\rho}/( 4 \pi) = 2.01 \pm 0.1$ is determined from the $\rho \to e^{+}e^{-}$ decay~\cite{Bauer:1977iq};
$P_{\rho}(\sigma)$ is the distribution extensively discussed in Sec.~\ref{sec:fluctuations}, 
see Fig.~\ref{fig:P_rho}.
Equation~(\ref{eq:sigma_rhoA}) has a clear physics interpretation: long before the target, the photon fluctuates into a coherent superposition of eigenstates of the scattering operator; each state interacts with the nucleus according to the Gribov--Glauber approach; the result is summed over all possible fluctuations with the probability distribution $P_{\rho}(\sigma)$
corresponding to photoproduction of $\rho$ in the final state.
Since fluctuations corresponding to different values of $\sigma$ are present in the $\gamma -\rho$ transition, Eq.~(\ref{eq:sigma_rhoA}) naturally takes into account the inelastic diffractive intermediate states leading to the inelastic (Gribov) shadowing correction.

In the absence of cross section fluctuations, one obtains the standard Glauber model expression for the cross section of coherent 
$\rho$ photoproduction on nuclei,
\begin{equation}
\sigma_{\gamma A \to \rho A}=\left(\frac{e}{f_{\rho}}\right)^2 \int d^2 {\bf b} 
\left(1-e^{-\frac{1}{2} \sigma_{\rho N} T_A(b)} \right)^2 \,,
\label{eq:sigma_rhoA_Glauber}
\end{equation}
where $\sigma_{\rho N}$ is the total $\rho$ meson-nucleon cross section. In this case, nuclear shadowing is determined by multiple elastic rescattering with the $\sigma_{\rho N}$ cross section, which leads to the standard Glauber nuclear shadowing correction.

In the incoherent case, using the completeness (closure) of the nuclear final states $A^{\prime}$, one obtains the following
expression for the cross section of incoherent (quasi-elastic) $\rho$ photoproduction on nuclei~\cite{Guzey:2020pkq}
\begin{equation}
\fl
\sigma_{\gamma A \to \rho A^{\prime}} 
= \left(\frac{e}{f_{\rho}}\right)^2 \int d^2 {\bf b}\, T_A(b) \left(\int d\sigma P_{\rho}(\sigma) 
\frac{\sigma}{\sqrt{16\pi B}}\exp\left[-\frac{\sigma^{\rm in}}{2}T_A(b)\right]\right)^2\,,
\label{eq:inc7}
\end{equation}
where $B$ is the slope of the $t$ dependence of the $\gamma p \to \rho p$ cross section;
$\sigma^{\rm in}=\sigma-\sigma^{\rm el}$ with $\sigma^{\rm el}=\sigma^2/(16 \pi B)$, where $\sigma^{\rm in}$ and $\sigma_{\rm el}$ are the inelastic and elastic cross sections, respectively.
Neglecting hadronic fluctuations of the photon, i.e., replacing $P_{\rho}(\sigma)$ by the  $\delta$-function in Eq.~(\ref{eq:inc7}),
one obtains the Glauber model expression for $\sigma_{\gamma A \to \rho A^{\prime}}$,
\begin{equation}
\fl
\sigma_{\gamma A \to \rho A^{\prime}} =\left(\frac{e}{f_{\rho}}\right)^2 \frac{\sigma_{\rho N}^2}{16\pi B}
\int d^2 {\bf b}\, T_A(b) e^{-\sigma^{\rm in}T_A(b)} =\sigma_{\gamma p \to \rho p}
\int d^2 {\bf b}\, T_A(b) e^{-\sigma_{\rho N}^{\rm in}T_A(b)} \,,
\label{eq:rho_inc_Glauber}
\end{equation}
where $\sigma_{\rho N}^{\rm in}$ is the inelastic $\rho$ meson-nucleon cross section.

Equations (\ref{eq:inc7}) and (\ref{eq:rho_inc_Glauber}) have a clear physical meaning and interpretation: elastic scattering of states $|\sigma \rangle$ (elastic photoproductuon of $\rho$ mesons) takes place on any of $A$ nucleons of the target, whose distribution in the transverse plane is given by $T_A(b)$; 
these states further interact with the rest of target nucleons, which leads to the attenuation (nuclear shadowing) of the nuclear cross section. While elastic interactions are allowed, the inelastic rescattering would destroy the final-state elastic state ($\rho$ meson) and, hence, should be rejected; the probability not to have inelastic scattering is given by 
$\exp[-\sigma^{\rm in}T_A(b)]$.

The incoherent (quasi-elastic) cross section in Eq.~(\ref{eq:inc7}) does not include the contribution of 
$\rho$ photoproduction with nucleon dissociation, $\gamma N \to \rho Y$, 
where $Y$ denotes the hadronic system with mass $M_Y$. 
If this contribution is not rejected experimentally, it will
increase the incoherent cross section. This effect can be taken into 
account by combining the standard Glauber technique with cross section fluctuations of target nucleons; the resulting 
approach was developed for incoherent $J/\psi$ photoproduction on nuclei~\cite{Guzey:2018tlk}. 
Applying this method to the case of $\rho$ photoproduction, the cross section of incoherent 
$\rho$ photoproduction on nuclei, which includes both elastic and 
nucleon-dissociative photoproduction on target nucleons, can be written 
in the following form [compare to Eq.~(\ref{eq:inc7})]
\begin{equation}
\fl
\sigma_{\gamma A \to \rho A^{\prime}+Y} =\left(1+\frac{\sigma_{\gamma p \to \rho Y}}{\sigma_{\gamma p \to \rho p}}\right)
 \left(\frac{e}{f_{\rho}}\right)^2 \int d^2 {\bf b}\, T_A(b) \left(\int d\sigma P(\sigma) 
\frac{\sigma}{\sqrt{16\pi B}}\exp\left[-\frac{\sigma^{\rm in}}{2}T_A(b)\right]\right)^2 \,,
\label{eq:inc7_dd}
\end{equation}
where $\sigma_{\gamma p \to \rho p}$ and $\sigma_{\gamma p \to \rho Y}$ are 
the $t$-integrated cross sections of 
elastic and nucleon-dissociative $\rho$ photoproduction on the proton, respectively.
Using the ZEUS analysis of elastic and proton-dissociative $\rho^0$ photoproduction at HERA~\cite{Breitweg:1997ed}
that found
$\sigma_{\gamma p \to \rho p}/\sigma_{\gamma p \to \rho Y}=2.0 \pm 0.2 ({\rm stat}.) \pm 0.7 (\rm syst.)$
in kinematic domain $M_Y < 0.1W_{\gamma p}^2$ and $|t|<0.5$ GeV$^2$,
it was estimated in Ref.~\cite{Guzey:2020pkq} that
the nucleon dissociation may increase the cross section of one-step
incoherent $\rho$ photoproduction by as much as 50\%.
The exact magnitude of this contribution depends on such data selection 
criteria as the mass of the produced state $Y$ and
the range of the momentum transfer $t$.

Note that in the derivation of Eq.~(\ref{eq:inc7_dd}), one used that 
all but one amplitude describing the inelastic interaction correspond to small $t$, which allows for the application of 
the formalism of cross section (color) fluctuations.

\subsection{Predictions for the cross section of $\rho$ photoproduction in heavy ion UPCs and comparison to RHIC and LHC data}

\begin{figure}[tp]
\centering
 \includegraphics[width=0.75\textwidth]{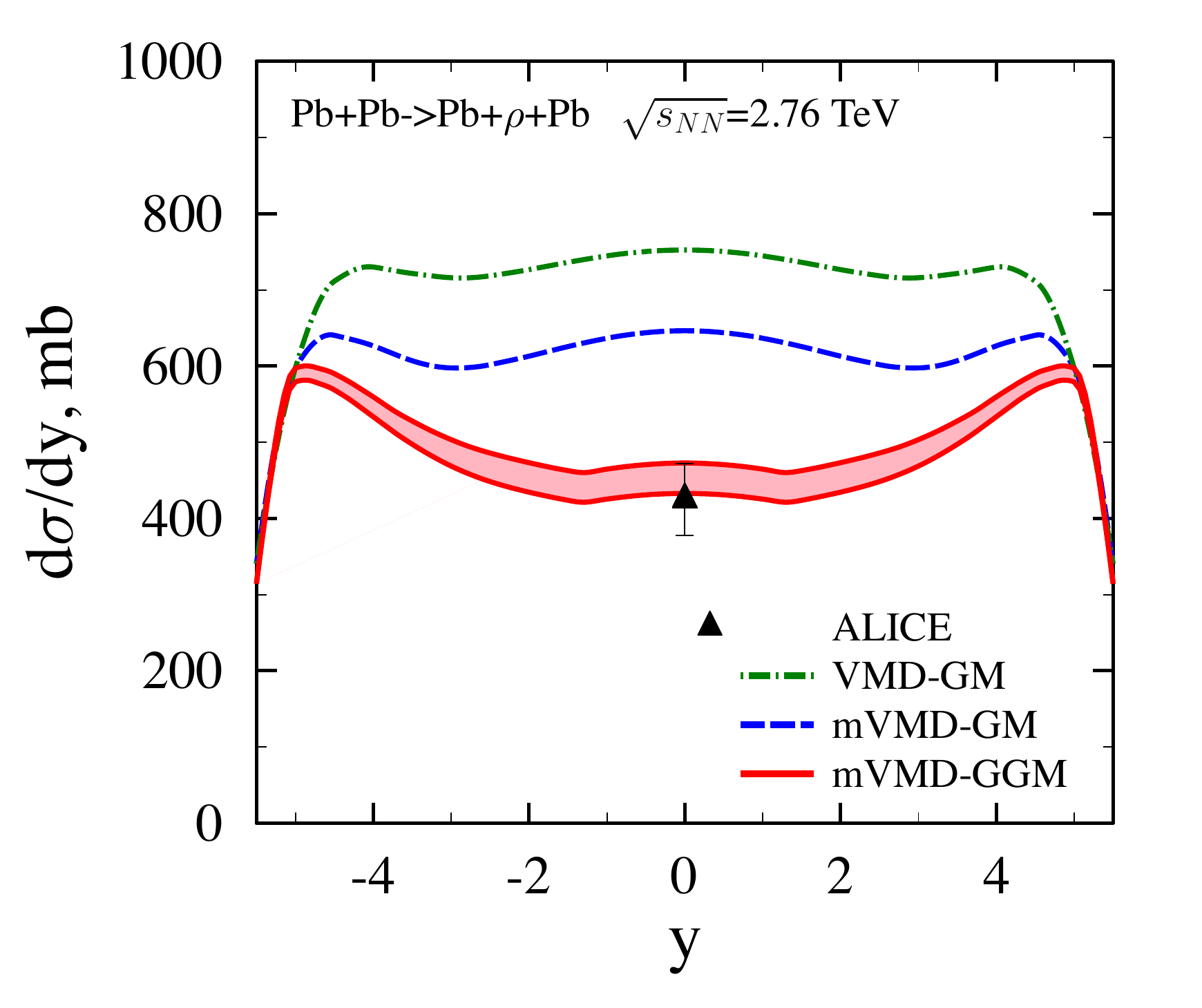}
\caption{The cross section of coherent $\rho$ photoproduction
in Pb-Pb UPCs at $\sqrt{s_{NN}}=2.76$ TeV as a function of the rapidity $y$: predictions of the Gribov--Glauber model
(red solid curves)
and the Glauber model (blue dashed and green dot-dashed curves)
 are compared to the ALICE data~\cite{Adam:2015gsa}.  The shaded band gives the theoretical uncertainty due to modeling of the $P_{\rho}(\sigma)$ distribution.
The figure is from \cite{Frankfurt:2015cwa}, {\tt https://doi.org/10.1016/j.physletb.2015.11.012}.}
\label{fig:rho_coh1}
\end{figure}

Using the framework described above, one can readily make predictions for the cross section of $\rho$ photoproduction in heavy ion UPCs 
and compare them to the available RHIC and LHC data. 
Figure~\ref{fig:rho_coh1} shows the cross section of coherent $\rho$ photoproduction
in Pb-Pb UPCs at $\sqrt{s_{NN}}=2.76$ TeV as a function of the
$\rho$ meson rapidity $y$.
Three sets of theoretical predictions are compared to the ALICE measurement~\cite{Adam:2015gsa}. 
The best description of the ALICE data is provided by the mVMD-GGM
calculation, which is obtained using Eq.~(\ref{eq:sigma_rhoA}) 
and which includes the effects of the photon hadronic structure and inelastic (Gribov) nuclear shadowing 
The acronym mVMD-GGM stands for the modified Vector Meson Dominance model combined with the Gribov--Glauber model.
It is presented by the
red solid curves with the shaded area showing the uncertainty due to the variation of the fluctuation strength. The latter is modeled by varying the parameter $\omega_{\sigma}^{\rho}$ in Eqs.~(\ref{eq:photon_DD}) and (\ref{eq:photon_DD_4}) 
due to the variation of the parameter $\beta$ in Eq.~(\ref{eq:omega_fit})
in the interval $\beta=0.25-0.35$.

In contrast, the mVMD-GM (blue dashed curve) and the VMD-GM (green dot-dashed curve) calculations are based on 
Eq.~(\ref{eq:sigma_rhoA_Glauber}), which includes only elastic intermediate states in the calculation of nuclear shadowing
(Glauber model) and significantly overestimate the data (the calculations are based on different parametrizations of the 
$d\sigma_{\gamma p \to \rho p}(t=0)/dt$ on the proton, see Ref.~\cite{Frankfurt:2015cwa}).

 Note also that in all these calculations, one neglects a small additional effect due to the leading twist shadowing  for configurations with small $\sigma \leq 10$ mb, 
 see Sec.~\ref{sec:UPC},
 which would lead to a small decrease of $d\sigma_{\gamma A \to \rho A}/dy$.

\begin{figure}[t]
\centering
 \includegraphics[width=0.85\textwidth,trim=0 50 0 120]{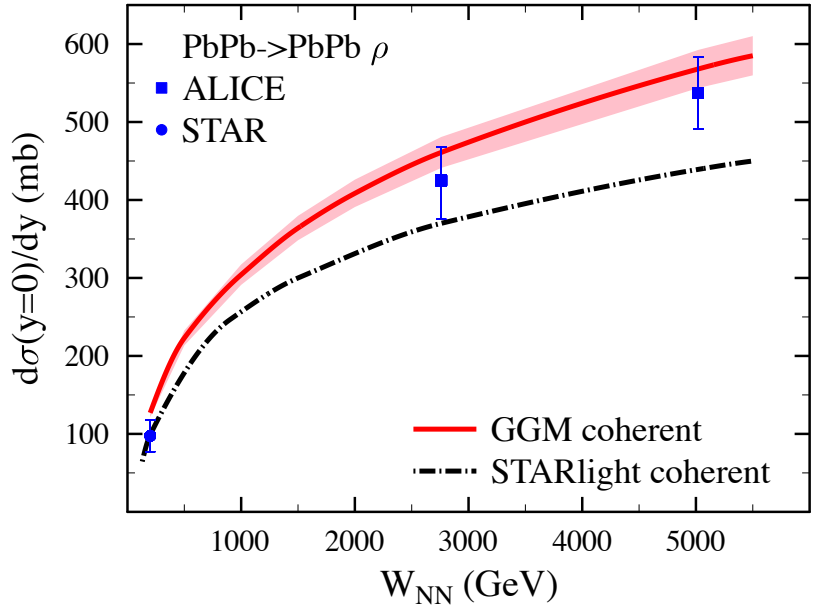}
\caption{The cross section of coherent $\rho$ photoproduction
in Pb-Pb UPCs as a function of $W_{NN}=\sqrt{s_{NN}}$ at $y=0$. Predictions of 
the Gribov--Glauber model (red solid curve with the shaded band) 
and STARlight Monte Carlo (black dot-dashed curve) are compared to
the scaled STAR data at $\sqrt{s_{NN}}=200$ GeV~\cite{Abelev:2007nb} and
the ALICE data at $\sqrt{s_{NN}}=2.76$ TeV~\cite{Adam:2015gsa} and $\sqrt{s_{NN}}=5.02$ TeV~\cite{Acharya:2020sbc}.
The figure is from \cite{Guzey:2020pkq}, {\tt https://doi.org/10.1103/PhysRevC.102.015208}.
}
\label{fig:rho_coh2}
\end{figure}

Figure~\ref{fig:rho_coh2} presents the energy dependence of the cross section of coherent $\rho$ photoproduction
in Pb-Pb UPCs at $y=0$. Theoretical predictions of 
the Gribov--Glauber model (red solid curve with the shaded band) 
and STARlight Monte Carlo (black dot-dashed curve) are compared to
the scaled STAR data at $\sqrt{s_{NN}}=200$ GeV~\cite{Abelev:2007nb} and
the ALICE data at $\sqrt{s_{NN}}=2.76$ TeV~\cite{Adam:2015gsa} and $\sqrt{s_{NN}}=5.02$ TeV~\cite{Acharya:2020sbc}.
One can see from the figure that the approach based on the Gribov--Glauber model of nuclear shadowing describes the normalization and the energy dependence of the cross section very well.

In contrast, the models for the photon--nucleus interaction implemented in the STARlight Monte Carlo~\cite{Klein:2016yzr}, which is often used in processing and analysis of UPC data, 
tends to underestimate the cross section at LHC energies.
This is the result of the assumption that the $t$ dependence of
$d\sigma_{\gamma A \to \rho A}/dt$ is given by the nuclear form factor squared and identification of the $\sigma_{\rho A}^{\rm in}$ inelastic nuclear cross section with the total one, see detailed discussion in~\cite{Guzey:2020pkq}.

\begin{figure}[t]
\centering
 \includegraphics[width=0.85\textwidth,,trim=0 20 0 150]{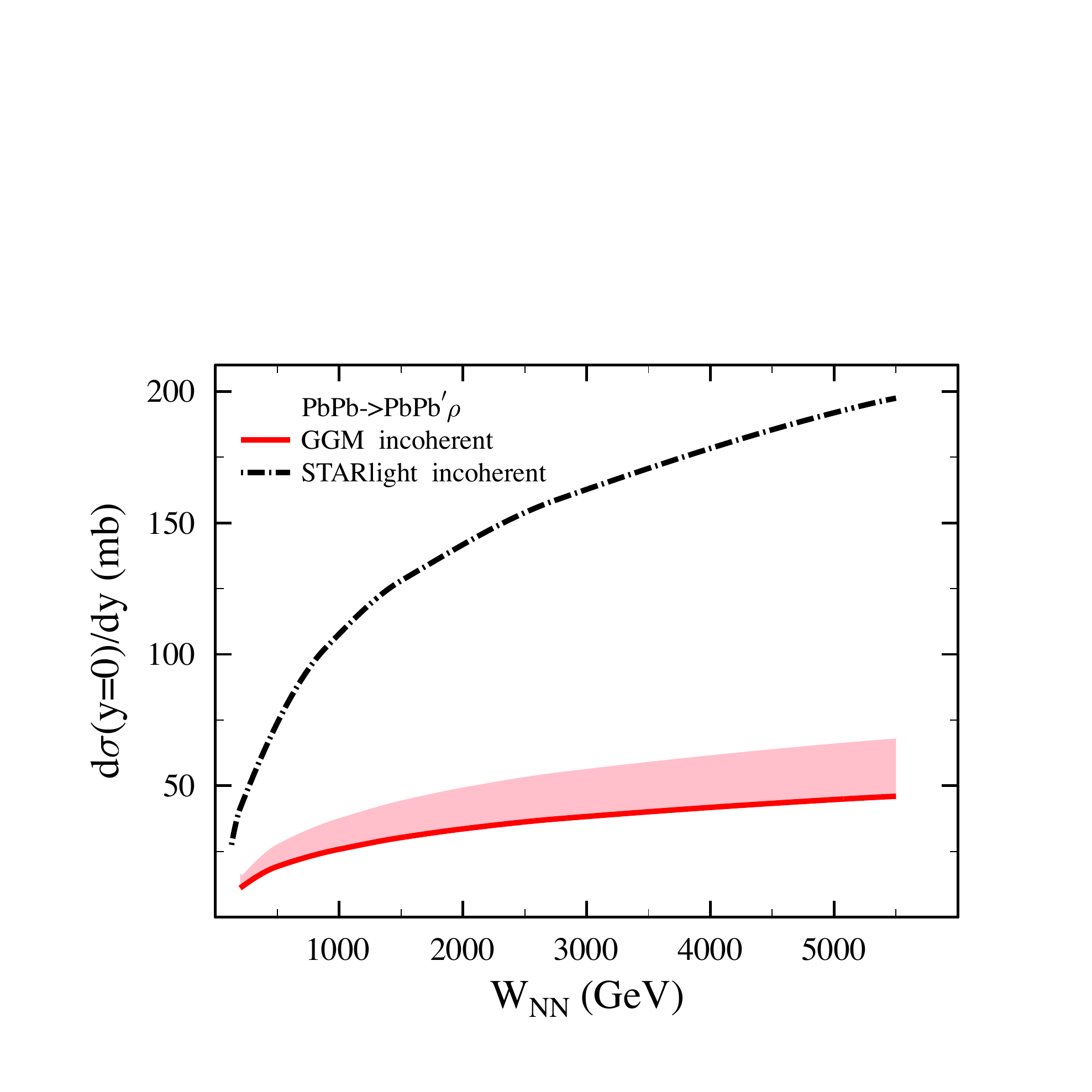}
\caption{The cross section of incoherent $\rho$ photoproduction in Pb-Pb UPCs as a function of the collision energy at $y=0$: predictions of the Gribov--Glauber (red solid) and STARlight (black dot-dashed) approaches. The red shaded band shows the range of predictions due to the inclusion of both elastic and nucleon-dissociative photoproduction on target nucleons.
The figure is from~\cite{Guzey:2020pkq}, {\tt https://doi.org/10.1103/PhysRevC.102.015208}.
}
\label{fig:rho_incoh}
\end{figure}

Finally, predictions for the cross section of 
incoherent $\rho$ photoproduction in Pb-Pb UPCs as a function of the collision energy at $y=0$ are shown in Fig.~\ref{fig:rho_incoh}. As in Fig.~\ref{fig:rho_coh2},
the result of the Gribov--Glauber model is contrasted with the
STARlight prediction. As one can see from the figure, the STARlight predictions exceed severalfold those of the Gribov--Glauber model. This is the result of the STARlight framework assumption that the cross section of the incoherent photoproduction of vector mesons on nuclear targets is proportional to the ratio of the inelastic $\rho A$ and $\rho N$ cross sections. The latter is in conflict with the  Glauber expression for the quasi-elastic $\gamma A \to \rho A^{\prime}$ cross section.

Figures~\ref{fig:rho_coh1}, \ref{fig:rho_coh2}, and \ref{fig:rho_incoh} clearly demonstrate that it is important to properly take into account the effects of both elastic and inelastic nuclear shadowing, which dramatically suppress the cross sections of light vector meson photoproduction on heavy nuclei.

It is also instructive to compare the magnitudes of nuclear suppression in the case of coherent $\rho$ and $J/\psi$ photoproduction on heavy nuclei. It can be quantified by the factor of $S_{Pb}^2(W)$ [compare to Eq.~(\ref{eq:rel3})]
\begin{equation}
S_{Pb}^2(W)=\frac{\sigma_{\gamma A \to \rho A}}{\sigma^{\rm IA}_{\gamma A \to \rho A}} \,,
\label{eq:S2_Pb}
\end{equation}
where $\sigma^{\rm IA}_{\gamma A \to \rho A}$ is the cross section calculated in the impulse approximation (IA),
\begin{equation}
{\sigma^{\rm IA}_{\gamma A \to \rho A}}=\left(\frac{e}{f_{\rho}}\right)^2 \frac{\sigma_{\rho N}^2}{4} \int d^2 {\bf b} [T_A(b)]^2 \,. 
\label{eq:S2_Pb_IA}
\end{equation}

\begin{figure}[t]
\centering
 \includegraphics[width=0.75\textwidth]{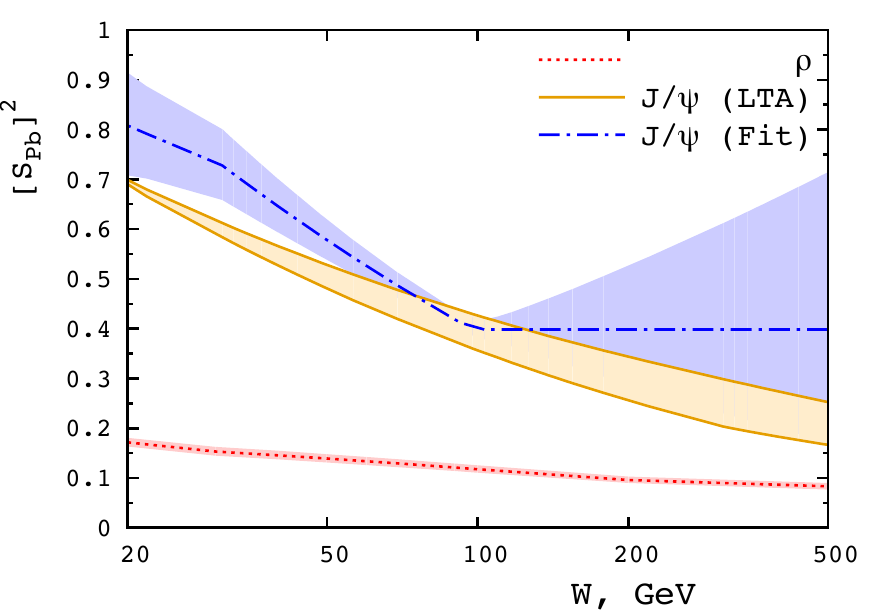}
\caption{The nuclear suppression factor of $S_{Pb}^2(W)$ as a function of $W$ for coherent photoproduction of $\rho$ and $J/\psi$ mesons on Pb. See text for explanations.}
\label{fig:S2_rho_2021}
\end{figure}

Figure~\ref{fig:S2_rho_2021} shows $S_{Pb}^2(W)$ as a function of $W$ for coherent photoproduction of $\rho$ on the heavy nucleus of lead (Pb). The uncertainty band corresponds to the variation of the parameter $\beta =0.3 \pm 0.05$, see Eq.~(\ref{eq:omega_fit}).
For comparison, we also show $S_{Pb}^2(W)$ for coherent 
$J/\psi$ photoproduction. Predictions of the leading twist model of nuclear shadowing are represented by the orange band and are obtained by squaring the corresponding values in Fig.~\ref{fig:S_pb208_2016} and keeping in mind that $x=M^2_{J/\psi}/W^2$. In addition, we show the result of 
the analysis of Ref.~\cite{Guzey:2020ntc}, where the nuclear suppression factor for coherent $J/\psi$ photoproduction was extracted from the fit to all available LHC data (Runs 1 and 2, central and forward rapidities) 
on coherent $J/\psi$ photoproduction in Pb-Pb UPCs
(the dot-dashed blue curve and the associated error band that fans out toward large $W$ or small $x$).

\begin{figure}[t]
\centering
\includegraphics[width=1.0\textwidth]{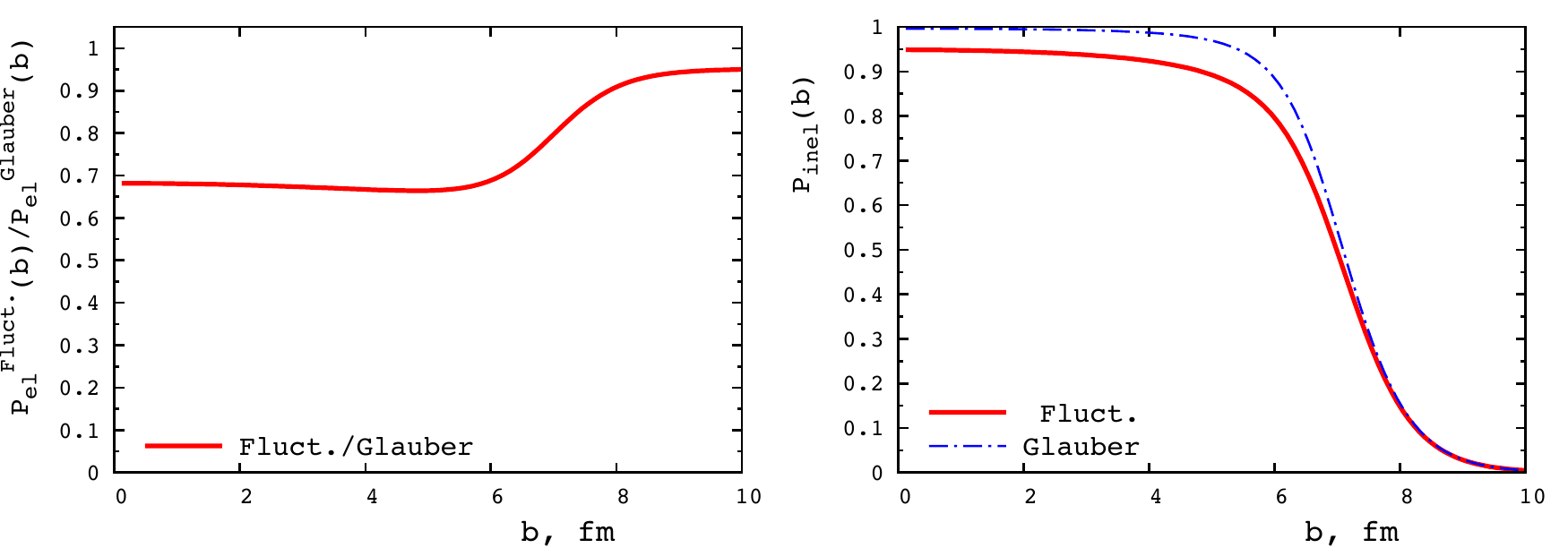}
\caption{The ratio of the nuclear profile functions squared~(\ref{eq:ratio_fluct_glauber}) with and without cross section fluctuations (left) and the probability of inelastic interactions $P_{\rm inel}$~(\ref{eq:pinel}) (right) as a function of $|{\bf b}|$ at $W=100$ GeV.}
\label{fig:BD_rho_2021}
\end{figure}

Note that cross section fluctuations, which lead to inelastic nuclear shadowing in elastic photoproduction of $\rho$ mesons on nuclei, significantly suppress the contributions to the elastic cross section for given impact parameter compared to the expectation of the Glauber model. This is illustrated in the left panel of Fig.~\ref{fig:BD_rho_2021} showing the ratio
\begin{equation}
\frac{P_{\rm el}^{\rm Fluct.}(b)}{P_{\rm el}^{\rm Glauber}(b)}=\frac{\left[\int d\sigma P_{\rho}(\sigma)
\left(1-e^{-\frac{1}{2} \sigma T_A(b)}\right)\right]^2}{\left(1-e^{-\frac{1}{2} \sigma_{\rho N} T_A(b)}\right)^2} \,,
\label{eq:ratio_fluct_glauber}
\end{equation}
as a function of $|{\bf b}|$ at $W=100$ GeV.
As in the case of nuclear diffractive structure functions and PDFs, inelastic nuclear shadowing strongly suppresses the coherent nuclear cross section, makes it more transparent and delays the onset of the black disk limit compared to the expectations of the Glauber model, where the interactions are nearly black (fully absorptive) for $b=0$. This is further illustrated in the right panel of Fig.~\ref{fig:BD_rho_2021}
presenting the probability of inelastic interactions 
$P_{\rm inel}$,
\begin{eqnarray}
&& P_{\rm inel}^{\rm Fluct.}(b)=1-\int d\sigma P_{\rho}(\sigma)
\left(1 -\Gamma_A(b,\sigma)\right)^2 \,, \nonumber\\
&& P_{\rm inel}^{\rm Glauber}(b)=1-
(1 -\Gamma_A(b,\sigma_{\rho N}))^2\,,
\label{eq:pinel}
\end{eqnarray}
as a function of $b$. Here 
$\Gamma_A(b,\sigma)=1-\exp[-\sigma T_A(b)/2]$ is the nuclear profile function (scattering amplitude in impact parameter space).
This figure quantifies the effect of inelastic nuclear shadowing on the onset of the black disk limit
in coherent photoproduction of $\rho$ on heavy nuclei of Pb.

Note that the elastic cross section is much more sensitive to deviations from black disk regime than the inelastic one.

  \section{Soft diffraction in hadron--nucleus and photon--nucleus collisions and related phenomena} 
\label{sec:diff_soft}

The presence of cross section fluctuations in energetic hadrons and photons that we introduced and discussed in Sec.~\ref{sec:fluctuations} leads to a number of characteristic effects in soft diffraction on nuclei. 

\subsection{Coherent diffractive dissociation off nuclei}

The classic application of the Good--Walker formalism deals with diffraction of high-energy projectiles on nuclear targets~\cite{Good:1960ba}. Considering diffractive dissociation of an incoming hadron in the state $|h \rangle$ into a complete set of states $|n \rangle$, the cross section of coherent diffractive dissociation on a nuclear target can be written in the following form,
\begin{equation}
\fl
\sigma_{\rm diff}^{hA}=\int d^2 {\bf b} \Bigg(\int d \sigma P_{h}(\sigma)\sum_{n} |\langle h| F(\sigma,b)|n \rangle|^2-\Big(\int d \sigma P_{h}(\sigma) |\langle h| F(\sigma,b)|h \rangle| \Big)^2 \Bigg) \, ,
\label{Adiff}
\end{equation} 
where ${\bf b}$ is the impact parameter; 
$F(\sigma,b)=1-e^{-\sigma T(b) /2}$ is the Glauber profile function, which is the state $|n \rangle$--nucleus scattering amplitude in  the impact parameter space.
Note that Eq.~(\ref{Adiff}) combines the discrete and continuous versions of the formalism of cross section fluctuations and implies that states $|n \rangle$ interact with nuclear target nucleons with the cross section $\sigma$;
the corresponding distribution is given by $P_h(\sigma)$.

Using completeness of the states $|n \rangle$, the $\sigma_{\rm diff}^{hA}$
cross section can be presented in the following final form
\begin{equation}
\fl
\sigma_{\rm diff}^{hA}=\int d^2 {\bf b} \Bigg(\int d \sigma P_{h}(\sigma) |\langle h| F^{2}(\sigma,b)|h \rangle|-\Big(\int d \sigma P(\sigma) |\langle h| F(\sigma,b)|h \rangle| \Big)^2 \Bigg) \, .
\label{csdiff}
\end{equation}
Equation~(\ref{csdiff}) is applicable for sufficiently heavy nuclei $A\ge 10$, where the essential momentum transfer squared $t$ in the rescatterings is small due to  the suppression by the nucleus form factor. 
For the proton beam, 
it predicts that $\sigma_{\rm diff}^{pA} \propto A^{0.8}$ for $A \approx 16$ and $\sigma_{\rm diff}^{pA}  \propto A^{0.4}$ for $A \approx 200$. 
This is consistent with the $A$ dependence observed in semi-inclusive $n + A \to p \pi^{-} + A$ data for the diffractive mass interval $1.35 \leq M \leq 1.45$ GeV \cite{Zielinski:1983ty} and with the absolute cross section of coherent inelastic diffraction off emulsion targets~\cite{Boos:1978hr}. For a review, see~\cite{Frankfurt:2000tya}.

In the case of scattering off light nuclei, one needs to take into account the
$t$ dependence of the  elementary diffractive amplitude. The results of the theoretical analysis including this effect~\cite{Strikman:1995jf} agrees well with the data on proton
coherent diffractive dissociation on He-4~\cite{Bujak:1981mp}. 

At collider energies and in a wide range of impact parameters, the interaction becomes practically completely black leading to a strong suppression of the inelastic diffraction and the dominance of the excitation of the proton via the Coulomb photon exchange
\cite{Guzey:2005tk}. 
The corresponding cross section is given by convolution of the flux of the equivalent photons $N_{\gamma/A}(\omega)$ with the total photon-proton cross section $\sigma^{\gamma p}_{\rm tot}$
\begin{equation}
\sigma^{pA}_{e.m.}=\int \frac{d\omega}{\omega} N_{\gamma/A}(\omega) \sigma^{\gamma p}_{\rm tot}(s) \,,
\end{equation}
where $\omega$ is the photon energy
Since the Coulomb contribution can be calculated 
with a high precision, hadronic diffraction can be measured up to the energies, where the ratio of two contributions approaches unity. 

\begin{figure}[t]
\centering
 \includegraphics[width=0.75\textwidth]{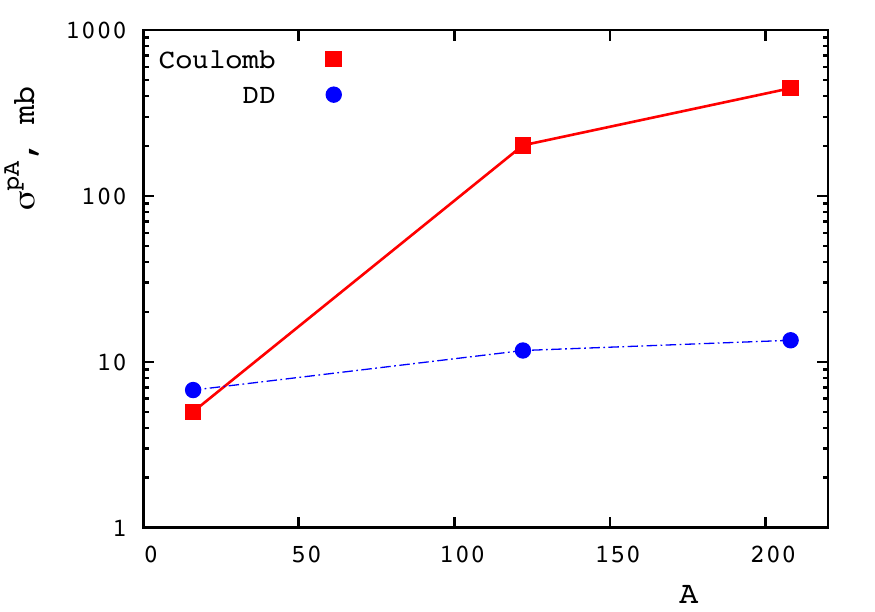}
\caption{The $A$ dependence of the Coulomb and hadronic contributions to the coherent diffractive proton--nucleus cross section at the LHC.}
\label{test2}
\end{figure}

The $A$ dependence of the Coulomb and hadronic diffractive contributions in the LHC energy range is presented in Fig.~\ref{test2} for a sample of nuclei (Pb, Xe, O). One can see that at the LHC, measurements of the hadronic contribution is possible for $A \leq 16$.
At the same time, the hadronic contribution to coherent diffraction on heavy nuclei begins to compete with the e.m. mechanism only for $\sqrt{s_{NN}} \le 100$ GeV, see the discussion in \cite{Guzey:2005tk}. 

Note that measurements of coherent diffraction in $pp$, $p D$, and $p ^{4}He$ scattering were performed with internal jet targets at FNAL. 
Therefore, it is maybe possible to perform similar studies using the gas targets within the ALICE and LHCb gas target projects.

\subsection{Distribution over the number of wounded nucleons}

To determine the inelastic cross section $\sigma_{\nu}$ for the
proton to interact with $\nu$ nucleons in proton--nucleus collisions, the standard Glauber formalism~\cite{Bertocchi:1976bq} at high energies can be generalized to include the effect of cross section fluctuations~\cite{Heiselberg:1991is}. When the impact
parameters in nucleon--nucleon ($NN$) interactions are small compared
to the typical distance between neighboring nucleons, one obtains \cite{Blaettel:1993rd}
\begin{equation} 
\fl
\sigma_{\nu} =  \int d\sigma P_{p}(\sigma)  
\left(\begin{array}{c}
A\\
\nu \end{array} \right)
\int
    d^2\vec{b}\, \left[{\sigma_{\rm in}(\sigma) T(b)\over A}\right]^{\nu}
    \left[1-{\sigma_{\rm in}(\sigma) T(b)\over A}\right]^{A-\nu} \,, 
    \label{eq:wounded}
  \end{equation}
where 
$\sigma_{\rm in}(\sigma)$ is the inelastic
cross-section for a configuration with the given total cross-section,
which following Refs.~\cite{Blaettel:1993rd,Alvioli:2014sba} is taken
to be a fixed fraction of $\sigma$. In the limit,
when the effect of cross section fluctuations is neglected,
$P_p(\sigma)=\delta(\sigma-\left<\sigma\right>)$ and  Eq.~(\ref{eq:wounded}) 
reduces to the Glauber model expression. The distribution over $\nu$
can be calculated with a Monte Carlo Glauber procedure, which includes
$NN$ correlations and finite-size effects~\cite{Alvioli:2013vk}.

We explained previously in Sec.~\ref{sec:fluctuations} that there is a $t \neq 0$ component of the inelastic diffraction, which is not  related  to the fluctuations of the cross section. In principle,  this contribution  has to be included  in 
$\sigma_{in}$ and in the modeling of the final states corresponding  to the interaction with $\nu$ nucleons. 
However, it appears that the current uncertainties in the strength of the diffraction at the LHC and, in particular, in the value of $\omega_\sigma$ are significantly larger than this effect.

There can be several sources of fluctuations of the cross section.  Among them is 
the strength of the interaction of a quark--gluon configuration,   which is likely to depend on $x$ of the selected parton.
This could be tested in the processes, which involve production of jets in the very forward region in proton--nucleus collisions with an additional constraint on the transverse energy in the backward region.
In particular, large $x \ge 0.5$ are likely to be built of more compact configurations of  valence partons resulting in  fewer $q\bar q $ pairs. One may expect that the average cross section $\langle \sigma(x) \rangle$ for
 such   configurations  should be smaller than $\sigma_{\rm in}$. This would result in a gross violation of the geometrical estimate of the distribution over $\nu$  (even including fluctuations around the average value given by $\sigma_{in}$). Such a violation was observed at the LHC \cite{ATLAS:2014cpa,Chatrchyan:2014hqa} and RHIC \cite{Adare:2015gla} in $pA$ and $dA$ collisions,  where the production of  jets originating from large $x$ quarks was studied as a function of the centrality of the collision. This pattern was explained in the model,  where
 $\lambda(x)=\left<\sigma(x)\right>/\sigma_{\rm in}$ drops with an increase of $x$ reaching $\lambda(0.5) \sim 0.6$, \cite{Alvioli:2014eda,Alvioli:2017wou}.
 A moderate difference of $\lambda(x)$ at RHIC and the LHC was also explained as a consequence of the energy dependence of $P_p(\sigma)$.
 No alternative explanations of the discussed effect have   been suggested yet, primarily because the integrated inclusive cross section does not show any nuclear effects.

  \section{Hard diffraction in proton--proton scattering}
\label{sec:pp_hard}

\subsection{Rapidity gap survival probability}

We have discussed in Sec.~\ref{sec:inclusive_diffraction} that the approximation of the 
leading twist factorization works well in diffractive DIS. 
This implies that if a hard probe interacts with a parton belonging to the diffractive fracture function of the proton, 
the final state would include a proton carrying the momentum close to that of the initial proton ($x_{\Pomeron} \le 0.01$ )  and a diffractive state with the mass squared equal to  $M_X^2=W^2 x_{\Pomeron}$ (for $Q^2,|t| \ll W^2$, see Eq.~(\ref{eq:diff_def})).

\begin{figure}[t]
\centering
\includegraphics[width=5cm,trim=0 120 0 120]{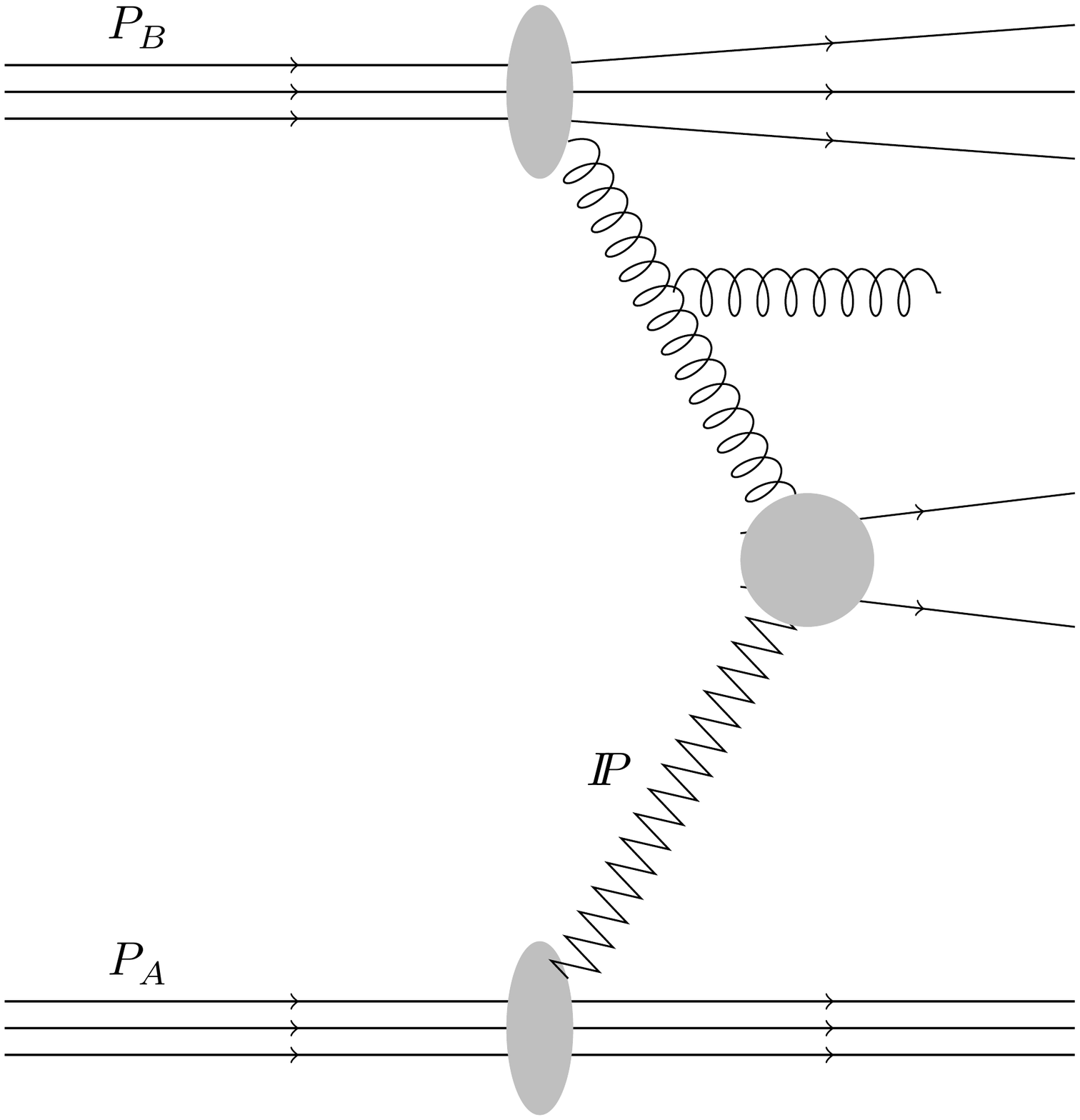}
\caption{Single diffraction process in hadron--hadron collision with production of two jets. Zigzag line denotes schematically an exchange of the Pomeron responsible for the existence of the rapidity gap.} \label{fig:single_pomeron}
\end{figure}

In the case of the diffractive proton--proton ($pp$) scattering with a hard subprocess like production of two jets, $pp \to p +X$, where $X= 2\ {\rm jets} + Y$, see Fig.~\ref{fig:single_pomeron}, the cross section can be 
written
as a product of the cross section calculated in the impulse approximation and the rapidity gap survival factor $S^2 < 1$,
\begin{equation}
\fl
 {d\sigma (pp\to 2 \, {\rm jets} + Y + p) \over dx_{\Pomeron} \, d\Omega} \; = \; S^2  \sum_{ij}\,\int dx_1 d\beta \, f_i (x_1, \mu^2) \,  f^{\rm D}_j(\beta, x_{\Pomeron},t,\mu^2) \, \frac{\hat{\sigma}^{\rm hard}_{ij}}{d\Omega} \,.
 \label{eq:pp_1}
\end{equation}
Here $f_i (x_1, \mu^2)$ is the PDF of the non-diffracting nucleon; $f^{\rm D}_j(\beta, 
x_{\Pomeron},t,\mu^2) $ is the diffractive fracture function and $\hat{\sigma}^{\rm hard}_{ij}$ is the hard  scattering cross section, $\Omega$ is the phase space for produced partons which form the dijet system and the sum goes over different partons.

The presence of the suppression factor $S^2$ reflects  the high probability that the constituents  of the diffractive nucleon  not involved in the hard collision may interact with the second nucleon. Subsequently this  will typically lead to the hadron production in the gap region.
The suppression factor may be estimated using 
generalized parton distribution functions (GPDs) as follows.
At high energies, $s \gg |t|$ and the angular momentum conservation in the c.m.~frame implies that the scattering
amplitude is effectively diagonal in the impact parameter of the
colliding $pp$ system. Hence, it is convenient to represent it as a Fourier
integral over the transverse coordinate variable ${\bf b}$,
\begin{equation}
A_{\rm el} (s, t = -\Delta_{\perp}^2) 
= \frac{i s}{4\pi} \int d^2 {\bf b} e^{-i (\Delta_{\perp} {\bf b})}
 \Gamma (s,b) ,
\label{Gamma_def}
\end{equation}
where $\Gamma (s, b)$  is the (dimensionless) profile function
 of the $pp$ elastic
amplitude in the conventions of Ref.~\cite{Frankfurt:2003td}.
The probability of inelastic interaction to occur  at a given impact parameter is then equal to
\begin{equation}  
P_{\rm inel} (b) = 1- |1- \Gamma (s, b)|^2 \,.
\end{equation}
Hence, the  probability that no inelastic interactions would occur in the collision with a hard subprocess  at a given impact parameter can be written as $ |1- \Gamma(s, b)|^2$
(in the absence of correlation between soft and hard interactions).
For numerical estimates, we employ a simple analytic parametrization of 
the profile function, which satisfies unitarity and reflects the 
approach to the black-disk regime ($\Gamma(s,b)\rightarrow 1$) 
at small $b$:
\begin{equation}
\Gamma(s, b)=\Gamma_0 \, \exp[-b^2/ (2 B(s))]\,,
\label{Gamma_gaussian}
\end{equation}
where $\Gamma_0 = 1$ and the slope parameter is given in terms
of the total cross section as $B(s) = \sigma_{\rm tot}(s)/(4\pi)$;
the inelastic cross section for this profile is
$\sigma_{\rm in}(s) = 3\pi B(s)$.

The transverse spatial distribution of gluons involved in the hard diffractive collision
is given by a Fourier transform of the gluon GPD,
\begin{equation}
F_g(x,t = -\Delta_{\perp}^2; Q^2) = 
\int d^2 {\vec{\rho}}  \, e^{-i ({\vec{\Delta}_{\perp}} \cdot  {\vec{\rho})}} F_g(x,\rho; Q^2) \,,
\label{rhoprof_def}
\end{equation}
where $\rho$ is the transverse coordinate variable.
The distribution of the interaction point over $b$, which is the transverse distance between the c.m. of two interacting nucleons, is given by the convolution of transverse distributions of the diffractive  and inclusive  GPDs 

\begin{equation}
\fl
P_{\rm hard} (b) \equiv \int d^2 \vec{\rho}_1 \int d^2 \vec{\rho}_2 
\delta^{(2)} (\vec{b} + \vec{\rho}_1 - \vec{\rho}_2 ) 
\times
\frac{F^2_g (\rho_1)}{[\int d^2 \rho_1' F^2_{g} (\rho_1')]}
\; \frac{F^2_{\rm diff} (\rho_2)}{[\int d^2 \rho_2' \; F^2_{\rm diff} (\rho_2')]} \,,
\label{P_hard}
\end{equation}
which satisfies
\begin{equation}
\int d^2 \vec{b} \; P_{\rm hard}(b) = 1 \,.
\label{P_hard_normalization}
\end{equation}
In terms of this distribution, the suppression factor can be   
is expressed as
\begin{equation}
S^2 = \int d^2 b \; P_{\rm hard}(b) |1 - \Gamma(b)|^2 .
\label{survb}
\end{equation}

Experimentally,  the $t$ slope of the gluon GPD squared is $B_g (Q_0^2,  x\sim 10^{-3 }) \sim 4$ GeV$^{-2}$; 
 the $t$ slope of the diffractive hard  (fracture PDF) cross section $B_{\rm diff} \sim 6$  GeV$^{-2}$, and slope of the elastic $pp$ cross section at $\sqrt{s} = 8$ TeV is $B=20$ GeV$^{-2}$. 
 Using Eqs.~(\ref{Gamma_gaussian}) and (\ref{survb}), one finds
 \begin{equation}
 S^2={2B^2_{av} \over (B+B_{av}) (B+2B_{av})} \,,
 \label{sup}
 \end{equation}
 where $B_{av}=(2B_{\rm diff} B_g)/(B_{\rm diff} + B_g)$.

 Equation~(\ref{sup}) leads to a strong suppression of hard diffraction in $pp$ scattering at high energies. 
 It reflects a much smaller transverse range of hard processes than that of the soft interactions~\cite{Frankfurt:2003td}. 
 
  Numerically one finds that $S^2=0.07 \cdot (1\pm 0.2)$  in the kinematics of the CMS--TOTEM experiment \cite{Sirunyan:2020ifc}, which reported $\left< S^2\right> = 0.074  \cdot ^{+15\%}_{-14\%}$.
However, one should look at the agreement of the estimate with the data with a grain of salt.

Note that $b$ defined as a distance between the c.m. of two nucleons in the case of  hard collisions (Eq.~(\ref{P_hard}))   is somewhat different from $b$ defined for elastic $pp$ scattering.  
However, this effect is likely to be significant only if $x$ is close to unity.

 On the theoretical side, we have made an implicit assumption that $b$ defined  as the distance between the transverse c.m. of the interacting nucleons (\ref{P_hard}) is close to $b$ defined from a Fourier transform of the elastic $pp$ amplitude. Such an assumption is natural for $b=0$, where the interaction is practically black, but  maybe less accurate for large $b$, for partons at large $x$, where the transverse distance between the partons is sensitive to their momentum fractions.
 Also inelastic soft diffraction of the diffracting nucleon off the intact nucleon does not lead to filling of the gap. 
 Such a process constitutes  $\sim 5- 10\%$ of the inelastic cross section and leads to increase of $S^2$. On the other hand, multiparton interactions involving the parton of diffracting nucleon produced in the splitting in the DGLAP evolution lead to hard collisions filling the gap. Such interactions are not included in the mean field 
 approximation of Eq.~(\ref{survb}) and lead to a reduction of $S^2$.

 On the experimental side, the interval of $x_{\Pomeron}$ included values up to $x_{\Pomeron} =0.1$
 for which  non-Pomeron contributions have to be included. Also the averaging  over $x$ of the parton in the diffracting nucleon effectively corresponds to averaging over  $B_g$, which increases with a decrease of $x$ resulting in an increase of $S^2$.

 In the future it would be highly desirable to study the rate of diffraction as a function of $x$  preferably up to $x \sim 0.5$. Indeed, at large $x$ there is potentially an interesting connection with centrality dependence of the jet production mentioned in Sec.~\ref{sec:diff_soft}, which may indicate that configurations in nucleon with large $x$ have a smaller interaction cross section than the average ones (a factor of two at  $x\sim 0.5$). A possible interpretation of these data is that at large $x$, one selects configurations in the nucleon smaller than average ones. If so, it corresponds to $pp$ interaction in more transparent configurations and hence to larger $S^2$.

  Note in passing that information about $F_g(x, \rho)$ extracted from exclusive $J/\psi$ production allows one to calculate the distribution of hard collisions in $pp$ scattering over impact parameters and establish that hard parton--parton interactions occur on average at significantly smaller impact parameters than the soft inelastic interactions 
 leading to correlations of dijet production and underlying hadron multiplicity
 \cite{Frankfurt:2003td}. 
Also the analysis of  the double parton interactions using information on $F_g(x, \rho)$ 
indicates that at  virtualities $\sim 1$  GeV$^2$, small-$x$ partons are weakly correlated in the transverse plane, and that the observed high rate of  double parton high-$p_T$ interactions originates from  
 pQCD induced correlations, for detailed discussion, see  \cite{Blok:2017alw}.

 \subsection{Probing correlations of partons near the nucleon edge}

Studies of the multiparton interaction (MPI) in $pp$ scattering indicate that correlations between the partons involved in MPI are rather weak at low $Q^2$  and grow with an increase of virtuality due to the QCD evolution and subsequent
$1 \to 2$ parton splittings.

Studies of the diffractive and double diffractive hard processes allow one to examine possible correlations of partons closer to the nucleon periphery as well as the multiparton structure of the Pomeron ladders.
In particular, one could consider 
single and double diffraction with production of two and four jets,
\begin{equation}
 pp  \to p + X (2 jets +Y, 4 jets +Y)
 \label{SD}
 \end{equation}
 \begin{equation}
pp  \to p p + X (2 jets +Y, 4 jets +Y).
\label{DD}
 \end{equation} 

\begin{figure}[t]
\centering
\includegraphics[width=0.4\textwidth,angle=90,trim=200 0 140 250]{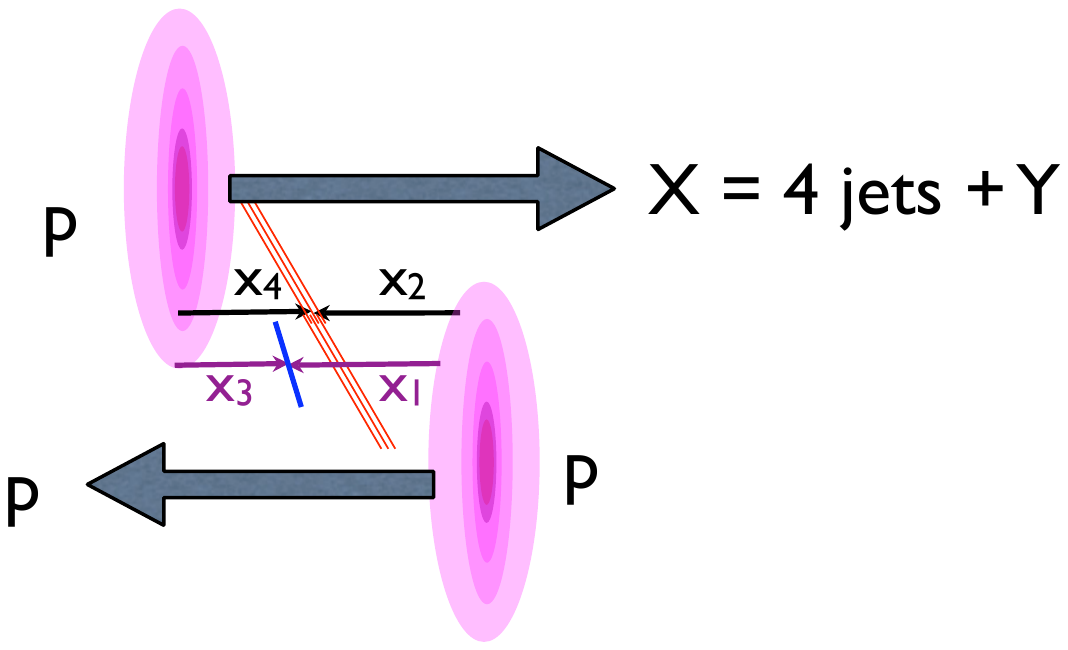}
\caption{Single diffraction process with production of four jets.}
\label{fig:sd}
\end{figure}

 In the case of single diffraction with production of four jets
  depicted in Fig.~\ref{fig:sd}, one can study (i) the rate of such events: the smaller the transverse size of the "Pomeron" exchange, the larger the cross section; (ii) factorization of
  the dependence on $x_1, x_2$ into the product of single parton distributions as measured in the single diffraction with production of two jets; (iii) dependence of the spectrum in $x_1+x_2$ on $t$: the larger $t$, the closer interaction to the perturbative regime, and hence the harder the spectrum. In particular, for large $-t$, one could look for the peak near $x_1+x_2=x_{\Pomeron}$.
  It is important to study also the dependence of the cross section on $x_3, x_4$
 in production of both two and four jets. Large $x_3$ correspond to partons which are likely to be closer to the center of the nucleon than small-$x$ partons, leading to a decrease of the probability of the gap survival with an increase of $x_3,x_4$.

 Correlations between the partons should also enhance the cross section of the exclusive 
channel of four-jet production in the double diffraction, when the light-cone momentum fractions carried by two of the interacting partons of both  
nucleons are close to maximal: $ (x_1+x_2) / x_{\Pomeron} \sim 1$. Such a contribution 
should be enhanced, if $-t_1,-t_2$ are large enough (few GeV$^2$) to squeeze 
the transverse sizes of the exchanged ladders (see~\cite{Rogers:2009ke}).

\begin{figure}[t]
\centering
\includegraphics[scale=.45]{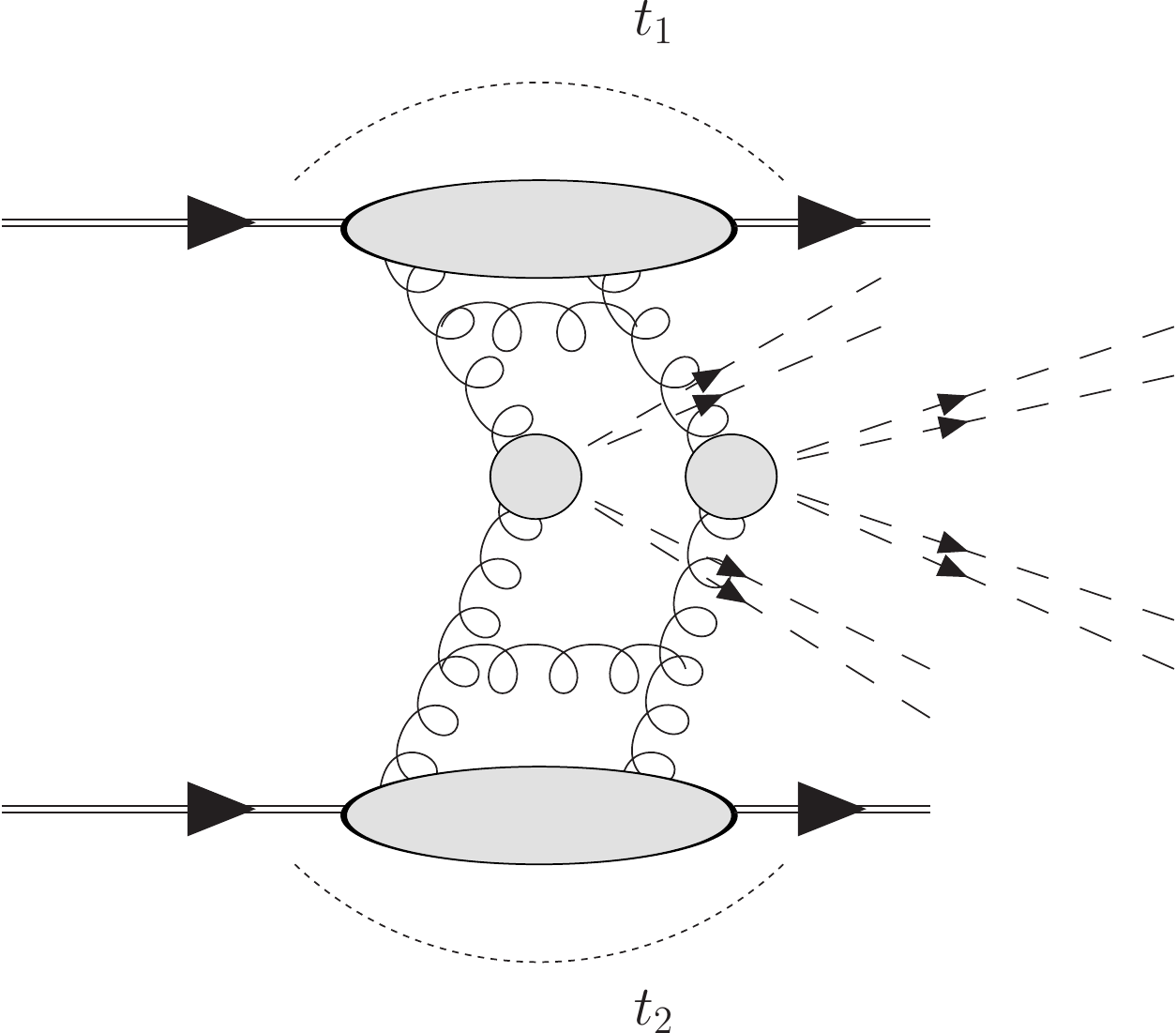}
\caption{Double Pomeron process with production of two pairs of dijets.}
\label{fig:diffraction}
\end{figure}

  \section{Summary and outlook}
\label{sec:summary}

Diffractive processes provide unique tools for accessing the nucleon 3D structure and fluctuations of the gluon density and the interaction strength. They are perfectly suited for probing the transition between non-perturbative and perturbative regimes as a function of virtuality in exclusive and semi-inclusive  processes, as well as testing the approach  to the  nonlinear dynamics at high energies.

Inclusive diffraction at HERA at large $Q^2$ is successfully described by the factorization theorem with diffractive PDFs,
with extracted $\alpha_{\pom}(t=0)\simeq 1.1$ close to that in soft processes. There is possibly an indication of the limitation of the collinear factorization at low $Q^2< 5\, \rm GeV^2$ and low $x\lesssim 0.001$.
The mechanism of breakdown of factorization in diffractive  dijet photoproduction needs further investigation, 
and studies of this process in UPCs at the LHC might help.

 The studies of hard exclusive processes ($Q^2 > 15 \; \rm GeV^2$ for light vector mesons) provide unique information on the transverse distribution of gluons and sea quarks as a function of $x$, obtaining as a result 3D single parton image of the nucleon.
The  information 
%vg about
on
the gluon GPDs made it possible to start building a QCD theory of multiparton hadron--hadron  collisions.
While inelastic diffraction at $t \sim 0 $ provides unique information about  the fluctuations of interaction strength,
%vg (though 
this is not the case for finite $t$.
Studies of 
%vg the fluctuations of the strength of interaction 
these fluctuation
provide information 
%vg about 
on the hadron structure, hence, probing the  properties of hadrons beyond the single particle (GPD) level.

There is 
%vg also 
an
important connection between diffraction in the scattering off a nucleon and the leading twist shadowing of nuclear PDFs. 
In the large $Q^2$ regime, 
%vg the confirmation 
the use 
of the QCD factorization theorem for hard inclusive diffraction at HERA allowed one to make predictions for the significant leading twist  nuclear shadowing  consistent with the current data on $J/\psi$ coherent photoproduction measured 
 in  ultraperipheral lead--lead collisions at the LHC.

A comparison of  the coherent  diffractive production of light and heavy vector mesons in  UPC at  the LHC  provides  valuable insights in the transition from predominantly soft to hard dynamics. Namely, the dipole--nucleus interaction  in the case of $\rho$ meson production is close to the black disk regime.  It was also found that  the cross sections fluctuations slow down the approach to the black disk regime.
Note also that shadowing is parametrically larger in the leading twist approach that in the eikonal model, where it is a higher twist effect.

 Information on inclusive diffraction obtained at HERA allows one to make quantitative progress in the understanding of hard diffraction dynamics and gap survival probability in $pp$ collisions at the LHC.

In the coming decade more information will come from UPC at the LHC to be followed by high precision data from EIC for $x \ge 0.001$. 
In particular,   studies  of the inelastic diffraction in the rapidity gap kinematics in UPCs  would allow one to probe the BFKL dynamics under conditions of minimal interference with  the DGLAP effects as well as to probe the new regime of proximity of the interaction to the black regime.

Complementary information on hard diffraction may come from the planned very forward detector upgrades at the
LHC~\cite{Citron:2018lsq,ALICE:2020mso}.

The  EIC will be able to  provide detailed information about the 3D nucleon structure and will allow one to separate leading  and higher twist effects in a wide range of diffraction and inclusive processes. In particular, it has the potential to test the limits of collinear and soft factorization in diffraction.

  \section*{Acknowledgments}

We thank
 M. Alvioli, N. Armesto,  B. Blok, M. Deak,  M. Klasen,  P. Kotko,  E. Kryshen,  L. Motyka, P. Newman,  W. Slominski, C. Weiss, and M. Zhalov for numerous discussions and collaborations. Our special thanks goes to J.D. Bjorken, V.N. Gribov and A.H. Mueller for numerous discussions for many decades. This research was supported by the U.S. Department of Energy  grants DE-FG02-93ER40771, DE-SC-0002145, National Science Centre in Poland, grant 2019/33/B/ST2/02588,  and by the Binational Science Foundation United States-Israel grant 202115.
 
 \vspace*{1cm}
% \bibliography{mybib}
\bibliography{main}

\begin{thebibliography}{100}

\bibitem{Feinberg}
E.~L. Feinberg and I.~Y. Pomeranchuk.
\newblock {\em Doklady Akad. Nauk SSSR}, 93:439, 1953.

\bibitem{Feinberg2}
E.~L. Feinberg and I.~Y. Pomeranchuk.
\newblock {\em Suppl. Nuovo Cimento III}, serie X:652, 1956.

\bibitem{Good:1960ba}
M.~L. Good and W.~D. Walker.
\newblock {Diffraction disssociation of beam particles}.
\newblock {\em Phys. Rev.}, 120:1857--1860, 1960.

\bibitem{Gribov:1968jf}
V.~N. Gribov.
\newblock {Glauber corrections and the interaction between high-energy hadrons
  and nuclei}.
\newblock {\em Sov. Phys. JETP}, 29:483--487, 1969.

\bibitem{Baltz:2007kq}
A.J. Baltz.
\newblock {The Physics of Ultraperipheral Collisions at the LHC}.
\newblock {\em Phys. Rept.}, 458:1--171, 2008.

\bibitem{Accardi:2012qut}
A.~Accardi et~al.
\newblock {Electron Ion Collider: The Next QCD Frontier}.
\newblock {\em Eur. Phys. J.}, A52(9):268, 2016.

\bibitem{LHeCStudyGroup:2012zhm}
J.~L. Abelleira~Fernandez et~al.
\newblock {A Large Hadron Electron Collider at CERN: Report on the Physics and
  Design Concepts for Machine and Detector}.
\newblock {\em J. Phys. G}, 39:075001, 2012.

\bibitem{Bruning:2019scy}
Oliver Br\"uning and Max Klein.
\newblock {Exploring the energy frontier with deep inelastic scattering at the
  LHC}.
\newblock {\em J. Phys. G}, 46(12):123001, 2019.

\bibitem{FCC:2018byv}
A.~Abada et~al.
\newblock {FCC Physics Opportunities}: {Future Circular Collider Conceptual
  Design Report Volume 1}.
\newblock {\em Eur. Phys. J. C}, 79(6):474, 2019.

\bibitem{Derrick:1993xh}
M.~Derrick et~al.
\newblock {Observation of events with a large rapidity gap in deep inelastic
  scattering at HERA}.
\newblock {\em Phys. Lett. B}, 315:481--493, 1993.

\bibitem{Ahmed:1994nw}
T.~Ahmed et~al.
\newblock {Deep inelastic scattering events with a large rapidity gap at HERA}.
\newblock {\em Nucl. Phys. B}, 429:477--502, 1994.

\bibitem{Ahmed:1995ns}
T.~Ahmed et~al.
\newblock {First measurement of the deep inelastic structure of proton
  diffraction}.
\newblock {\em Phys. Lett. B}, 348:681--696, 1995.

\bibitem{Breitweg:1997rg}
J.~Breitweg et~al.
\newblock {Measurement of elastic J/psi photoproduction at HERA}.
\newblock {\em Z. Phys. C}, 75:215--228, 1997.

\bibitem{Trentadue:1993ka}
L.~Trentadue and G.~Veneziano.
\newblock {Fracture functions: An Improved description of inclusive hard
  processes in QCD}.
\newblock {\em Phys. Lett. B}, 323:201--211, 1994.

\bibitem{Collins:1997sr}
John~C. Collins.
\newblock {Proof of factorization for diffractive hard scattering}.
\newblock {\em Phys. Rev. D}, 57:3051--3056, 1998.
\newblock [Erratum: Phys.Rev.D 61, 019902 (2000)].

\bibitem{Berera:1995fj}
Arjun Berera and Davison~E. Soper.
\newblock {Behavior of diffractive parton distribution functions}.
\newblock {\em Phys. Rev. D}, 53:6162--6179, 1996.

\bibitem{Armesto:2019gxy}
N.~Armesto, P.~R. Newman, W.~Slominski, and A.~M. Stasto.
\newblock {Inclusive diffraction in future electron-proton and electron-ion
  colliders}.
\newblock {\em Phys. Rev. D}, 100(7):074022, 2019.

\bibitem{Collins:1989gx}
John~C. Collins, Davison~E. Soper, and George~F. Sterman.
\newblock {Factorization of Hard Processes in QCD}.
\newblock {\em Adv. Ser. Direct. High Energy Phys.}, 5:1--91, 1989.

\bibitem{Abe:1997rg}
F.~Abe et~al.
\newblock {Measurement of diffractive dijet production at the Tevatron}.
\newblock {\em Phys. Rev. Lett.}, 79:2636--2641, 1997.

\bibitem{Affolder:2000hd}
T.~Affolder et~al.
\newblock {Dijet production by double pomeron exchange at the Fermilab
  Tevatron}.
\newblock {\em Phys. Rev. Lett.}, 85:4215--4220, 2000.

\bibitem{Affolder:2000vb}
T.~Affolder et~al.
\newblock {Diffractive dijets with a leading antiproton in $\bar{p}p$
  collisions at $\sqrt{s} = 1800$ GeV}.
\newblock {\em Phys. Rev. Lett.}, 84:5043--5048, 2000.

\bibitem{Gribov:1972ri}
V.~N. Gribov and L.~N. Lipatov.
\newblock {Deep inelastic e p scattering in perturbation theory}.
\newblock {\em Sov. J. Nucl. Phys.}, 15:438--450, 1972.
\newblock [Yad. Fiz.15,781(1972)].

\bibitem{Gribov:1972rt}
V.~N. Gribov and L.~N. Lipatov.
\newblock {e+ e- pair annihilation and deep inelastic e p scattering in
  perturbation theory}.
\newblock {\em Sov. J. Nucl. Phys.}, 15:675--684, 1972.
\newblock [Yad. Fiz.15,1218(1972)].

\bibitem{Altarelli:1977zs}
Guido Altarelli and G.~Parisi.
\newblock {Asymptotic Freedom in Parton Language}.
\newblock {\em Nucl. Phys.}, B126:298--318, 1977.

\bibitem{Dokshitzer:1977sg}
Yuri~L. Dokshitzer.
\newblock {Calculation of the Structure Functions for Deep Inelastic Scattering
  and e+ e- Annihilation by Perturbation Theory in Quantum Chromodynamics.}
\newblock {\em Sov. Phys. JETP}, 46:641--653, 1977.
\newblock [Zh. Eksp. Teor. Fiz.73,1216(1977)].

\bibitem{Aktas:2006hy}
A.~Aktas et~al.
\newblock {Measurement and QCD analysis of the diffractive deep-inelastic
  scattering cross-section at HERA}.
\newblock {\em Eur.Phys.J.}, C48:715--748, 2006.

\bibitem{Chekanov:2009aa}
S.~Chekanov et~al.
\newblock {A QCD analysis of ZEUS diffractive data}.
\newblock {\em Nucl.Phys.}, B831:1--25, 2010.

\bibitem{Ingelman:1984ns}
G.~Ingelman and P.~E. Schlein.
\newblock {Jet Structure in High Mass Diffractive Scattering}.
\newblock {\em Phys. Lett. B}, 152:256--260, 1985.

\bibitem{Balitsky:1978ic}
I.~I. Balitsky and L.~N. Lipatov.
\newblock {The Pomeranchuk Singularity in Quantum Chromodynamics}.
\newblock {\em Sov. J. Nucl. Phys.}, 28:822--829, 1978.
\newblock [Yad. Fiz.28,1597(1978)].

\bibitem{Kuraev:1977fs}
E.~A. Kuraev, L.~N. Lipatov, and V.~S. Fadin.
\newblock {The Pomeranchuk Singularity in Nonabelian Gauge Theories}.
\newblock {\em Sov. Phys. JETP}, 45:199--204, 1977.
\newblock [Zh. Eksp. Teor. Fiz.72,377(1977)].

\bibitem{Lipatov:1985uk}
L.~N. Lipatov.
\newblock {The Bare Pomeron in Quantum Chromodynamics}.
\newblock {\em Sov. Phys. JETP}, 63:904--912, 1986.
\newblock [Zh. Eksp. Teor. Fiz.90,1536(1986)].

\bibitem{Goharipour:2018yov}
Muhammad Goharipour, Hamzeh Khanpour, and Vadim Guzey.
\newblock {First global next-to-leading order determination of diffractive
  parton distribution functions and their uncertainties within the xFitter
  framework}.
\newblock {\em Eur. Phys. J. C}, 78(4):309, 2018.

\bibitem{Khanpour:2019pzq}
Hamzeh Khanpour.
\newblock {Phenomenology of diffractive DIS in the framework of fracture
  functions and determination of diffractive parton distribution functions}.
\newblock {\em Phys. Rev. D}, 99(5):054007, 2019.

\bibitem{Zlebcik:2019tiu}
Radek Zlebcik.
\newblock {Diffractive PDF determination from HERA inclusive and jet data at
  NNLO QCD}.
\newblock {\em PoS}, DIS2019:059, 2019.

\bibitem{Chekanov:2007tv}
S.~Chekanov et~al.
\newblock {Leading neutron energy and pT distributions in deep inelastic
  scattering and photoproduction at HERA}.
\newblock {\em Nucl. Phys. B}, 776:1--37, 2007.

\bibitem{Chekanov:2008tn}
S.~Chekanov et~al.
\newblock {Leading proton production in deep inelastic scattering at HERA}.
\newblock {\em JHEP}, 06:074, 2009.

\bibitem{Owens:1984zj}
J.~F. Owens.
\newblock {Q**2 Dependent Parametrizations of Pion Parton Distribution
  Functions}.
\newblock {\em Phys. Rev. D}, 30:943, 1984.

\bibitem{Gluck:1991ey}
M.~Gluck, E.~Reya, and A.~Vogt.
\newblock {Pionic parton distributions}.
\newblock {\em Z. Phys. C}, 53:651--656, 1992.

\bibitem{Aaron:2012ad}
F.D. Aaron et~al.
\newblock {Inclusive Measurement of Diffractive Deep-Inelastic Scattering at
  HERA}.
\newblock {\em Eur. Phys. J. C}, 72:2074, 2012.

\bibitem{Derrick:1995ue}
M.~Derrick et~al.
\newblock {Measurement of the cross-section for the reaction $\gamma p \to
  J/\psi p$ with the ZEUS detector at HERA}.
\newblock {\em Phys. Lett. B}, 350:120--134, 1995.

\bibitem{Aid:1996dn}
S.~Aid et~al.
\newblock {Elastic and inelastic photoproduction of $J/\psi$ mesons at HERA}.
\newblock {\em Nucl. Phys. B}, 472:3--31, 1996.

\bibitem{Adloff:2001rw}
C.~Adloff et~al.
\newblock {On the rise of the proton structure function F(2) towards low x}.
\newblock {\em Phys. Lett. B}, 520:183--190, 2001.

\bibitem{Barone:2002cv}
Vincenzo Barone and Enrico Predazzi.
\newblock {\em {High-Energy Particle Diffraction}}, volume v.565 of {\em Texts
  and Monographs in Physics}.
\newblock Springer-Verlag, Berlin Heidelberg, 2002.

\bibitem{Aktas:2006hx}
A.~Aktas et~al.
\newblock {Diffractive deep-inelastic scattering with a leading proton at
  HERA}.
\newblock {\em Eur. Phys. J. C}, 48:749--766, 2006.

\bibitem{Aaron:2010aa}
F.D. Aaron et~al.
\newblock {Measurement of the cross section for diffractive deep-inelastic
  scattering with a leading proton at HERA}.
\newblock {\em Eur. Phys. J. C}, 71:1578, 2011.

\bibitem{Chekanov:2008fh}
S.~Chekanov et~al.
\newblock {Deep inelastic scattering with leading protons or large rapidity
  gaps at HERA}.
\newblock {\em Nucl. Phys. B}, 816:1--61, 2009.

\bibitem{Motyka:2012ty}
L.~Motyka, M.~Sadzikowski, and W.~Slominski.
\newblock {Evidence of strong higher twist effects in diffractive DIS at HERA
  at moderate $Q^2$}.
\newblock {\em Phys. Rev. D}, 86:111501, 2012.

\bibitem{Maktoubian:2019ppi}
Atefeh Maktoubian, Hossein Mehraban, Hamzeh Khanpour, and Muhammad Goharipour.
\newblock {Role of higher twist effects in diffractive DIS and determination of
  diffractive parton distribution functions}.
\newblock {\em Phys. Rev. D}, 100(5):054020, 2019.

\bibitem{Klasen:2002xb}
Michael Klasen.
\newblock {Theory of hard photoproduction}.
\newblock {\em Rev. Mod. Phys.}, 74:1221--1282, 2002.

\bibitem{Khoze:2000wk}
Valery~A. Khoze, Alan~D. Martin, and M.~G. Ryskin.
\newblock {Soft diffraction and the elastic slope at Tevatron and LHC energies:
  A MultiPomeron approach}.
\newblock {\em Eur. Phys. J. C}, 18:167--179, 2000.

\bibitem{Klasen:2010vk}
Michael Klasen and Gustav Kramer.
\newblock {Suppression factors in diffractive photoproduction of dijets}.
\newblock {\em Eur. Phys. J. C}, 70:91--106, 2010.

\bibitem{Guzey:2016awf}
V.~Guzey and M.~Klasen.
\newblock {A fresh look at factorization breaking in diffractive
  photoproduction of dijets at HERA at next-to-leading order QCD}.
\newblock {\em Eur. Phys. J. C}, 76(8):467, 2016.

\bibitem{Klasen:1994bj}
M.~Klasen, G.~Kramer, and S.~G. Salesch.
\newblock {Photoproduction of jets at HERA: Comparison of next-to-leading order
  calculation with ZEUS data}.
\newblock {\em Z. Phys. C}, 68:113--120, 1995.

\bibitem{Klasen:2013cba}
M.~Klasen, G.~Kramer, and M.~Michael.
\newblock {Next-to-next-to-leading order contributions to jet photoproduction
  and determination of $\alpha_s$}.
\newblock {\em Phys. Rev. D}, 89(7):074032, 2014.

\bibitem{Klasen:1995ab}
M.~Klasen and G.~Kramer.
\newblock {Inclusive dijet production at HERA: Direct photon cross-sections in
  next-to-leading order QCD}.
\newblock {\em Z. Phys. C}, 72:107--122, 1996.

\bibitem{Klasen:1996it}
M.~Klasen and G.~Kramer.
\newblock {Inclusive two jet production at HERA: Direct and resolved
  cross-sections in next-to-leading order QCD}.
\newblock {\em Z. Phys. C}, 76:67--74, 1997.

\bibitem{Klasen:1997br}
M.~Klasen, T.~Kleinwort, and G.~Kramer.
\newblock {Inclusive Jet Production in $\gamma p$ and $\gamma\gamma$ Processes:
  Direct and Resolved Photon Cross Sections in Next-To-Leading Order QCD}.
\newblock {\em Eur. Phys. J. direct}, 1(1):1, 1998.

\bibitem{Budnev:1974de}
V.M. Budnev, I.F. Ginzburg, G.V. Meledin, and V.G. Serbo.
\newblock {The Two photon particle production mechanism. Physical problems.
  Applications. Equivalent photon approximation}.
\newblock {\em Phys. Rept.}, 15:181--281, 1975.

\bibitem{Vidovic:1992ik}
M.~Vidovic, M.~Greiner, C.~Best, and G.~Soff.
\newblock {Impact parameter dependence of the electromagnetic particle
  production in ultrarelativistic heavy ion collisions}.
\newblock {\em Phys. Rev. C}, 47:2308--2319, 1993.

\bibitem{Chekanov:2007aa}
S.~Chekanov et~al.
\newblock {Dijet production in diffractive deep inelastic scattering at HERA}.
\newblock {\em Eur. Phys. J. C}, 52:813--832, 2007.

\bibitem{Chekanov:2007rh}
Sergei Chekanov et~al.
\newblock {Diffractive photoproduction of dijets in ep collisions at HERA}.
\newblock {\em Eur. Phys. J. C}, 55:177--191, 2008.

\bibitem{Aktas:2007bv}
A.~Aktas et~al.
\newblock {Dijet Cross Sections and Parton Densities in Diffractive DIS at
  HERA}.
\newblock {\em JHEP}, 10:042, 2007.

\bibitem{Aktas:2007hn}
A.~Aktas et~al.
\newblock {Tests of QCD factorisation in the diffractive production of dijets
  in deep-inelastic scattering and photoproduction at HERA}.
\newblock {\em Eur. Phys. J.}, C51:549--568, 2007.

\bibitem{Aaron:2010su}
F.D. Aaron et~al.
\newblock {Diffractive Dijet Photoproduction in ep Collisions at HERA}.
\newblock {\em Eur. Phys. J. C}, 70:15--37, 2010.

\bibitem{Aaron:2011mp}
F.D. Aaron et~al.
\newblock {Measurement of Dijet Production in Diffractive Deep-Inelastic
  Scattering with a Leading Proton at HERA}.
\newblock {\em Eur. Phys. J. C}, 72:1970, 2012.

\bibitem{Andreev:2014yra}
V.~Andreev et~al.
\newblock {Measurement of Dijet Production in Diffractive Deep-Inelastic ep
  Scattering at HERA}.
\newblock {\em JHEP}, 03:092, 2015.

\bibitem{Andreev:2015cwa}
V.~Andreev et~al.
\newblock {Diffractive Dijet Production with a Leading Proton in $ep$
  Collisions at HERA}.
\newblock {\em JHEP}, 05:056, 2015.

\bibitem{Zlebcik:2011kq}
Radek Zlebcik, Karel Cerny, and Alice Valkarova.
\newblock {Factorisation breaking in diffractive dijet photoproduction at
  HERA?}
\newblock {\em Eur. Phys. J. C}, 71:1741, 2011.

\bibitem{Aktas:2006up}
A.~Aktas et~al.
\newblock {Diffractive open charm production in deep-inelastic scattering and
  photoproduction at HERA}.
\newblock {\em Eur. Phys. J. C}, 50:1--20, 2007.

\bibitem{Chekanov:2007pm}
S.~Chekanov et~al.
\newblock {Diffractive photoproduction of D*+-(2010) at HERA}.
\newblock {\em Eur. Phys. J. C}, 51:301--315, 2007.

\bibitem{H1:2017bnb}
V.~Andreev et~al.
\newblock {Measurement of $D^{*}$ production in diffractive deep inelastic
  scattering at HERA}.
\newblock {\em Eur. Phys. J. C}, 77(5):340, 2017.

\bibitem{Agostini:2020fmq}
P.~Agostini et~al.
\newblock {The Large Hadron-Electron Collider at the HL-LHC}.
\newblock 7 2020.

\bibitem{AbdulKhalek:2021gbh}
R.~Abdul~Khalek et~al.
\newblock {Science Requirements and Detector Concepts for the Electron-Ion
  Collider: EIC Yellow Report}.
\newblock 3 2021.

\bibitem{Armesto:2021fws}
Nestor Armesto, Paul~R. Newman, Wojciech Slominski, and Anna~M. Stasto.
\newblock {Diffractive longitudinal structure function at the Electron Ion
  Collider}.
\newblock 12 2021.

\bibitem{Dainton:2006wd}
J.~B. Dainton, M.~Klein, P.~Newman, E.~Perez, and F.~Willeke.
\newblock {Deep inelastic electron-nucleon scattering at the LHC}.
\newblock {\em JINST}, 1:P10001, 2006.

\bibitem{AbelleiraFernandez:2012cc}
J.~L. Abelleira~Fernandez et~al.
\newblock {A Large Hadron Electron Collider at CERN: Report on the Physics and
  Design Concepts for Machine and Detector}.
\newblock {\em J. Phys.}, G39:075001, 2012.

\bibitem{Klein:2018rhq}
Max Klein.
\newblock {Future Deep Inelastic Scattering with the LHeC}.
\newblock In A.~Levy, S.~Forte, and G.~Ridolfi, editors, {\em From My Vast
  Repertoire ...: Guido Altarelli's Legacy}, pages 303--347. 2019.

\bibitem{Bordry:2018gri}
Frederick Bordry, Michael Benedikt, Oliver Bruning, John Jowett, Lucio Rossi,
  Daniel Schulte, Steinar Stapnes, and Frank Zimmermann.
\newblock {Machine Parameters and Projected Luminosity Performance of Proposed
  Future Colliders at CERN}.
\newblock 2018.

\bibitem{LHeClumi}
Oliver Bruning, John Jowett, Max Klein, Daniel Pellegrini, Daniel Schulte, and
  Frank Zimmermann.
\newblock {Future Circular Collider Study FCC-eh Baseline Parameters}.
\newblock 2017.

\bibitem{FCC_CDRv1}
Edited by~M. Mangano~et al.
\newblock {Future Circular Collider Study. Volume 1: Physics Opportunities.
  Conceptual Design Report}.
\newblock 2018.

\bibitem{FCC_CDRv3}
Edited by~M. Benedikt~et al.
\newblock {Future Circular Collider Study. Volume 3: The Hadron Collider
  (FCC-hh) . Conceptual Design Report}.
\newblock 2018.

\bibitem{Aaron:2012hua}
F.D. Aaron et~al.
\newblock {Combined inclusive diffractive cross sections measured with forward
  proton spectrometers in deep inelastic $ep$ scattering at HERA}.
\newblock {\em Eur.Phys.J.}, C72:2175, 2012.

\bibitem{Guzey:2020gkk}
Vadim Guzey and Michael Klasen.
\newblock {Diffractive dijet photoproduction at the EIC}.
\newblock {\em JHEP}, 05:074, 2020.

\bibitem{Mueller:1989st}
Alfred~H. Mueller.
\newblock {Small x Behavior and Parton Saturation: A QCD Model}.
\newblock {\em Nucl. Phys. B}, 335:115--137, 1990.

\bibitem{Gribov:1968gs}
V.~N. Gribov.
\newblock {Interaction of gamma quanta and electrons with nuclei at
  high-energies}.
\newblock {\em Zh. Eksp. Teor. Fiz.}, 57:1306--1323, 1969.

\bibitem{Bjorken:1973gc}
J.~D. Bjorken and John~B. Kogut.
\newblock {Correspondence Arguments for High-Energy Collisions}.
\newblock {\em Phys. Rev. D}, 8:1341, 1973.

\bibitem{Frankfurt:1988nt}
L.~L. Frankfurt and M.~I. Strikman.
\newblock {Hard Nuclear Processes and Microscopic Nuclear Structure}.
\newblock {\em Phys. Rept.}, 160:235--427, 1988.

\bibitem{Ioffe:1969kf}
B.~L. Ioffe.
\newblock {Space-time picture of photon and neutrino scattering and
  electroproduction cross-section asymptotics}.
\newblock {\em Phys. Lett. B}, 30:123--125, 1969.

\bibitem{Buchmuller:1996xw}
W.~Buchmuller, M.~F. McDermott, and Arthur Hebecker.
\newblock {Gluon radiation in diffractive electroproduction}.
\newblock {\em Nucl. Phys. B}, 487:283--310, 1997.
\newblock [Erratum: Nucl.Phys.B 500, 621--622 (1997)].

\bibitem{Buchmuller:1998jv}
W.~Buchmuller, T.~Gehrmann, and Arthur Hebecker.
\newblock {Inclusive and diffractive structure functions at small x}.
\newblock {\em Nucl. Phys. B}, 537:477--500, 1999.

\bibitem{Wusthoff:1997fz}
M.~Wusthoff.
\newblock {Large rapidity gap events in deep inelastic scattering}.
\newblock {\em Phys. Rev. D}, 56:4311--4321, 1997.

\bibitem{Kowalski:2008sa}
H.~Kowalski, T.~Lappi, C.~Marquet, and R.~Venugopalan.
\newblock {Nuclear enhancement and suppression of diffractive structure
  functions at high energies}.
\newblock {\em Phys. Rev. C}, 78:045201, 2008.

\bibitem{Golec-Biernat:2001gyl}
Krzysztof~J. Golec-Biernat and M.~Wusthoff.
\newblock {Diffractive parton distributions from the saturation model}.
\newblock {\em Eur. Phys. J. C}, 20:313--321, 2001.

\bibitem{GolecBiernat:2008gk}
K.~Golec-Biernat and A.~Luszczak.
\newblock {Dipole model analysis of the newest diffractive deep inelastic
  scattering data}.
\newblock {\em Phys. Rev. D}, 79:114010, 2009.

\bibitem{Blaettel:1993rd}
B.~Blaettel, G.~Baym, L.~L. Frankfurt, and M.~Strikman.
\newblock {How transparent are hadrons to pions?}
\newblock {\em Phys. Rev. Lett.}, 70:896--899, 1993.

\bibitem{Frankfurt:1993it}
L.~Frankfurt, G.~A. Miller, and M.~Strikman.
\newblock {Coherent nuclear diffractive production of mini - jets: Illuminating
  color transparency}.
\newblock {\em Phys. Lett. B}, 304:1--7, 1993.

\bibitem{Nikolaev:1994ce}
Nikolai~N. Nikolaev and B.~G. Zakharov.
\newblock {On determination of the large 1/x gluon distribution at HERA}.
\newblock {\em Phys. Lett. B}, 332:184--190, 1994.

\bibitem{Frankfurt:1995jw}
Leonid Frankfurt, Werner Koepf, and Mark Strikman.
\newblock {Hard diffractive electroproduction of vector mesons in QCD}.
\newblock {\em Phys. Rev. D}, 54:3194--3215, 1996.

\bibitem{Gribov:1984tu}
L.~V. Gribov, E.~M. Levin, and M.~G. Ryskin.
\newblock {Semihard Processes in QCD}.
\newblock {\em Phys. Rept.}, 100:1--150, 1983.

\bibitem{Mueller:1985wy}
Alfred~H. Mueller and Jian-wei Qiu.
\newblock {Gluon Recombination and Shadowing at Small Values of x}.
\newblock {\em Nucl. Phys. B}, 268:427--452, 1986.

\bibitem{Kovchegov:2012mbw}
Yuri~V. Kovchegov and Eugene Levin.
\newblock {\em {Quantum chromodynamics at high energy}}, volume~33.
\newblock Cambridge University Press, 8 2012.

\bibitem{McLerran:1993ka}
Larry~D. McLerran and Raju Venugopalan.
\newblock {Gluon distribution functions for very large nuclei at small
  transverse momentum}.
\newblock {\em Phys. Rev. D}, 49:3352--3355, 1994.

\bibitem{McLerran:1993ni}
Larry~D. McLerran and Raju Venugopalan.
\newblock {Computing quark and gluon distribution functions for very large
  nuclei}.
\newblock {\em Phys. Rev. D}, 49:2233--2241, 1994.

\bibitem{McLerran:1994vd}
Larry~D. McLerran and Raju Venugopalan.
\newblock {Green's functions in the color field of a large nucleus}.
\newblock {\em Phys. Rev. D}, 50:2225--2233, 1994.

\bibitem{Iancu:2000hn}
Edmond Iancu, Andrei Leonidov, and Larry~D. McLerran.
\newblock {Nonlinear gluon evolution in the color glass condensate. 1.}
\newblock {\em Nucl. Phys. A}, 692:583--645, 2001.

\bibitem{Iancu:2001ad}
Edmond Iancu, Andrei Leonidov, and Larry~D. McLerran.
\newblock {The Renormalization group equation for the color glass condensate}.
\newblock {\em Phys. Lett. B}, 510:133--144, 2001.

\bibitem{Ferreiro:2001qy}
Elena Ferreiro, Edmond Iancu, Andrei Leonidov, and Larry McLerran.
\newblock {Nonlinear gluon evolution in the color glass condensate. 2.}
\newblock {\em Nucl. Phys. A}, 703:489--538, 2002.

\bibitem{JalilianMarian:1996xn}
Jamal Jalilian-Marian, Alex Kovner, Larry~D. McLerran, and Heribert Weigert.
\newblock {The Intrinsic glue distribution at very small x}.
\newblock {\em Phys. Rev. D}, 55:5414--5428, 1997.

\bibitem{JalilianMarian:1997dw}
Jamal Jalilian-Marian, Alex Kovner, and Heribert Weigert.
\newblock {The Wilson renormalization group for low x physics: Gluon evolution
  at finite parton density}.
\newblock {\em Phys. Rev. D}, 59:014015, 1998.

\bibitem{JalilianMarian:1997gr}
Jamal Jalilian-Marian, Alex Kovner, Andrei Leonidov, and Heribert Weigert.
\newblock {The Wilson renormalization group for low x physics: Towards the high
  density regime}.
\newblock {\em Phys. Rev. D}, 59:014014, 1998.

\bibitem{JalilianMarian:1997jx}
Jamal Jalilian-Marian, Alex Kovner, Andrei Leonidov, and Heribert Weigert.
\newblock {The BFKL equation from the Wilson renormalization group}.
\newblock {\em Nucl. Phys. B}, 504:415--431, 1997.

\bibitem{Balitsky:1995ub}
I.~Balitsky.
\newblock {Operator expansion for high-energy scattering}.
\newblock {\em Nucl. Phys. B}, 463:99--160, 1996.

\bibitem{Balitsky:1998ya}
Ian Balitsky.
\newblock {Factorization and high-energy effective action}.
\newblock {\em Phys. Rev. D}, 60:014020, 1999.

\bibitem{Kovchegov:1999ua}
Yuri~V. Kovchegov.
\newblock {Unitarization of the BFKL pomeron on a nucleus}.
\newblock {\em Phys. Rev. D}, 61:074018, 2000.

\bibitem{Kovchegov:1999yj}
Yuri~V. Kovchegov.
\newblock {Small x F(2) structure function of a nucleus including multiple
  pomeron exchanges}.
\newblock {\em Phys. Rev. D}, 60:034008, 1999.

\bibitem{Albacete:2010sy}
Javier~L. Albacete, Nestor Armesto, Jose~Guilherme Milhano, Paloma
  Quiroga-Arias, and Carlos~A. Salgado.
\newblock {AAMQS: A non-linear QCD analysis of new HERA data at small-x
  including heavy quarks}.
\newblock {\em Eur. Phys. J. C}, 71:1705, 2011.

\bibitem{Iancu:2015joa}
E.~Iancu, J.~D. Madrigal, A.~H. Mueller, G.~Soyez, and D.~N.
  Triantafyllopoulos.
\newblock {Collinearly-improved BK evolution meets the HERA data}.
\newblock {\em Phys. Lett. B}, 750:643--652, 2015.

\bibitem{Ducloue:2019jmy}
B.~Duclou\'e, E.~Iancu, G.~Soyez, and D.~N. Triantafyllopoulos.
\newblock {HERA data and collinearly-improved BK dynamics}.
\newblock {\em Phys. Lett. B}, 803:135305, 2020.

\bibitem{Beuf:2020dxl}
G.~Beuf, H.~H\"anninen, T.~Lappi, and H.~M\"antysaari.
\newblock {Color Glass Condensate at next-to-leading order meets HERA data}.
\newblock {\em Phys. Rev. D}, 102:074028, 2020.

\bibitem{Bartels:2002cj}
J.~Bartels, Krzysztof~J. Golec-Biernat, and H.~Kowalski.
\newblock {A modification of the saturation model: DGLAP evolution}.
\newblock {\em Phys. Rev. D}, 66:014001, 2002.

\bibitem{GolecBiernat:1998js}
Krzysztof~J. Golec-Biernat and M.~Wusthoff.
\newblock {Saturation effects in deep inelastic scattering at low Q**2 and its
  implications on diffraction}.
\newblock {\em Phys. Rev. D}, 59:014017, 1998.

\bibitem{GolecBiernat:1999qd}
Krzysztof~J. Golec-Biernat and M.~Wusthoff.
\newblock {Saturation in diffractive deep inelastic scattering}.
\newblock {\em Phys. Rev. D}, 60:114023, 1999.

\bibitem{Golec-Biernat:2017lfv}
Krzysztof Golec-Biernat and Sebastian Sapeta.
\newblock {Saturation model of DIS : an update}.
\newblock {\em JHEP}, 03:102, 2018.

\bibitem{Iancu:2003ge}
E.~Iancu, K.~Itakura, and S.~Munier.
\newblock {Saturation and BFKL dynamics in the HERA data at small x}.
\newblock {\em Phys. Lett. B}, 590:199--208, 2004.

\bibitem{Marquet:2007nf}
Cyrille Marquet.
\newblock {A Unified description of diffractive deep inelastic scattering with
  saturation}.
\newblock {\em Phys. Rev. D}, 76:094017, 2007.

\bibitem{Soyez:2007kg}
G.~Soyez.
\newblock {Saturation QCD predictions with heavy quarks at HERA}.
\newblock {\em Phys. Lett. B}, 655:32--38, 2007.

\bibitem{McDermott:1999fa}
M.~McDermott, L.~Frankfurt, V.~Guzey, and M.~Strikman.
\newblock {Unitarity and the QCD improved dipole picture}.
\newblock {\em Eur. Phys. J. C}, 16:641--656, 2000.

\bibitem{Glauber:1955qq}
R.~J. Glauber.
\newblock {Cross-sections in deuterium at high-energies}.
\newblock {\em Phys. Rev.}, 100:242--248, 1955.

\bibitem{Franco:1965wi}
V.~Franco and R.~J. Glauber.
\newblock {High-energy deuteron cross-sections}.
\newblock {\em Phys. Rev.}, 142:1195--1214, 1966.

\bibitem{Glauber:1970jm}
R.~J. Glauber and G.~Matthiae.
\newblock {High-energy scattering of protons by nuclei}.
\newblock {\em Nucl. Phys. B}, 21:135--157, 1970.

\bibitem{Frankfurt:2002kd}
L.~Frankfurt, V.~Guzey, M.~McDermott, and M.~Strikman.
\newblock {Nuclear shadowing in deep inelastic scattering on nuclei: Leading
  twist versus eikonal approaches}.
\newblock {\em JHEP}, 02:027, 2002.

\bibitem{Bjorken:1970ah}
J.~D. Bjorken, John~B. Kogut, and Davison~E. Soper.
\newblock {Quantum Electrodynamics at Infinite Momentum: Scattering from an
  External Field}.
\newblock {\em Phys. Rev. D}, 3:1382, 1971.

\bibitem{Nikolaev:1990ja}
Nikolai~N. Nikolaev and B.~G. Zakharov.
\newblock {Color transparency and scaling properties of nuclear shadowing in
  deep inelastic scattering}.
\newblock {\em Z. Phys. C}, 49:607--618, 1991.

\bibitem{Nikolaev:1991et}
Nikolai Nikolaev and Bronislav~G. Zakharov.
\newblock {Pomeron structure function and diffraction dissociation of virtual
  photons in perturbative QCD}.
\newblock {\em Z. Phys. C}, 53:331--346, 1992.

\bibitem{Kovchegov:1999kx}
Yuri~V. Kovchegov and Larry~D. McLerran.
\newblock {Diffractive structure function in a quasiclassical approximation}.
\newblock {\em Phys. Rev. D}, 60:054025, 1999.
\newblock [Erratum: Phys.Rev.D 62, 019901 (2000)].

\bibitem{Hatta:2006hs}
Y.~Hatta, E.~Iancu, C.~Marquet, G.~Soyez, and D.~N. Triantafyllopoulos.
\newblock {Diffusive scaling and the high-energy limit of deep inelastic
  scattering in QCD at large N(c)}.
\newblock {\em Nucl. Phys. A}, 773:95--155, 2006.

\bibitem{Munier:2003zb}
Stephane Munier and Arif Shoshi.
\newblock {Diffractive photon dissociation in the saturation regime from the
  Good and Walker picture}.
\newblock {\em Phys. Rev. D}, 69:074022, 2004.

\bibitem{Bartels:1993du}
Jochen Bartels and H.~Lotter.
\newblock {A Note on the BFKL pomeron and the 'hot spot' cross-section}.
\newblock {\em Phys. Lett. B}, 309:400--408, 1993.

\bibitem{Bojak:1997me}
I.~Bojak and M.~Ernst.
\newblock {Limitations of small x resummation methods from F2 data}.
\newblock {\em Nucl. Phys. B}, 508:731--752, 1997.

\bibitem{Fadin:1998py}
R.~S. Fadin and L.~N. Lipatov.
\newblock {BFKL pomeron in the next-to-leading approximation}.
\newblock {\em Phys. Lett.}, B429:127--134, 1998.

\bibitem{Ciafaloni:1998gs}
M.~Ciafaloni and G.~Camici.
\newblock {Energy scale(s) and next-to-leading BFKL equation}.
\newblock {\em Phys. Lett.}, B430:349--354, 1998.

\bibitem{Salam:1998tj}
G.~P. Salam.
\newblock {A Resummation of large subleading corrections at small x}.
\newblock {\em JHEP}, 07:019, 1998.

\bibitem{Salam:1999cn}
G.~P. Salam.
\newblock {An Introduction to leading and next-to-leading BFKL}.
\newblock {\em Acta Phys. Polon.}, B30:3679--3705, 1999.

\bibitem{Kuraev:1973sj}
E.~A. Kuraev and L.~N. Lipatov.
\newblock {Electron and muonic production in e- e- and e+ e- colliding beams}.
\newblock {\em Yad. Fiz.}, 16:1060--1077, 1972.

\bibitem{Ciafaloni:2003ek}
M.~Ciafaloni, D.~Colferai, D.~Colferai, G.~P. Salam, and A.~M. Stasto.
\newblock {Extending QCD perturbation theory to higher energies}.
\newblock {\em Phys. Lett.}, B576:143--151, 2003.

\bibitem{Ciafaloni:2003rd}
M.~Ciafaloni, D.~Colferai, G.~P. Salam, and A.~M. Stasto.
\newblock {Renormalization group improved small x Green's function}.
\newblock {\em Phys. Rev.}, D68:114003, 2003.

\bibitem{Andersson:1995ju}
Bo~Andersson, G.~Gustafson, and J.~Samuelsson.
\newblock {The Linked dipole chain model for DIS}.
\newblock {\em Nucl. Phys. B}, 467:443--478, 1996.

\bibitem{Kwiecinski:1996td}
J.~Kwiecinski, A.~D. Martin, and P.~J. Sutton.
\newblock {Constraints on gluon evolution at small x}.
\newblock {\em Z. Phys.}, C71:585--594, 1996.

\bibitem{Deak:2019wms}
Michal Deak, Krzysztof Kutak, Wanchen Li, and Anna~M. Sta\'sto.
\newblock {On the different forms of the kinematical constraint in BFKL}.
\newblock {\em Eur. Phys. J. C}, 79(8):647, 2019.

\bibitem{Altarelli:2000mh}
G.~Altarelli, R.~D. Ball, and S.~Forte.
\newblock {Small x resummation and HERA structure function data}.
\newblock {\em Nucl. Phys.}, B599:383--423, 2001.

\bibitem{Altarelli:2001ji}
G.~Altarelli, R.~D. Ball, and S.~Forte.
\newblock {Factorization and resummation of small x scaling violations with
  running coupling}.
\newblock {\em Nucl. Phys.}, B621:359--387, 2002.

\bibitem{Altarelli:2003hk}
G.~Altarelli, R.~D. Ball, and S.~Forte.
\newblock {An Anomalous dimension for small x evolution}.
\newblock {\em Nucl. Phys.}, B674:459--483, 2003.

\bibitem{Altarelli:2008aj}
G.~Altarelli, R.~D. Ball, and S.~Forte.
\newblock {Small x Resummation with Quarks: Deep-Inelastic Scattering}.
\newblock {\em Nucl. Phys.}, B799:199--240, 2008.

\bibitem{Ciafaloni:1999yw}
M.~Ciafaloni, D.~Colferai, and G.~P. Salam.
\newblock {Renormalization group improved small x equation}.
\newblock {\em Phys. Rev.}, D60:114036, 1999.

\bibitem{Ciafaloni:1999au}
M.~Ciafaloni, D.~Colferai, and G.~P. Salam.
\newblock {A collinear model for small x physics}.
\newblock {\em JHEP}, 10:017, 1999.

\bibitem{Ciafaloni:2003kd}
M.~Ciafaloni, D.~Colferai, G.~P. Salam, and A.~M. Stasto.
\newblock {The Gluon splitting function at moderately small x}.
\newblock {\em Phys. Lett.}, B587:87--94, 2004.

\bibitem{Ciafaloni:2007gf}
M.~Ciafaloni, D.~Colferai, G.~P. Salam, and A.~M. Stasto.
\newblock {A Matrix formulation for small-x singlet evolution}.
\newblock {\em JHEP}, 08:046, 2007.

\bibitem{Thorne:2001nr}
R.~S. Thorne.
\newblock {The Running coupling BFKL anomalous dimensions and splitting
  functions}.
\newblock {\em Phys. Rev.}, D64:074005, 2001.

\bibitem{Vera:2005jt}
A.~Sabio~Vera.
\newblock {An 'All-poles' approximation to collinear resummations in the Regge
  limit of perturbative QCD}.
\newblock {\em Nucl. Phys.}, B722:65--80, 2005.

\bibitem{Bonvini:2016wki}
M.~Bonvini, S.~Marzani, and T.~Peraro.
\newblock {Small-$x$ resummation from HELL}.
\newblock {\em Eur. Phys. J.}, C76(11):597, 2016.

\bibitem{Kwiecinski:1997ee}
J.~Kwiecinski, A.~D. Martin, and A.~M. Stasto.
\newblock {A Unified BFKL and GLAP description of F2 data}.
\newblock {\em Phys. Rev.}, D56:3991--4006, 1997.

\bibitem{Caron-Huot:2016tzz}
Simon Caron-Huot and Matti Herranen.
\newblock {High-energy evolution to three loops}.
\newblock {\em JHEP}, 02:058, 2018.

\bibitem{Deak:2020zay}
Michal Deak, Leonid Frankfurt, Mark Strikman, and Anna~M. Sta\'sto.
\newblock {Taming of preasymptotic small $x$ evolution within resummation
  framework}.
\newblock {\em Eur. Phys. J. C}, 80(4):315, 2020.

\bibitem{Ball:2017otu}
R.~D. Ball, V.~Bertone, M.~Bonvini, S.~Marzani, J.~Rojo, and L.~Rottoli.
\newblock {Parton distributions with small-x resummation: evidence for BFKL
  dynamics in HERA data}.
\newblock {\em Eur. Phys. J.}, C78(4):321, 2018.

\bibitem{Martin:1999rn}
Alan~D. Martin, M.~G. Ryskin, and T.~Teubner.
\newblock {$\upsilon$ photoproduction at HERA compared to estimates of
  perturbative QCD}.
\newblock {\em Phys. Lett. B}, 454:339--345, 1999.

\bibitem{Kutak:2012rf}
Krzysztof Kutak and Sebastian Sapeta.
\newblock {Gluon saturation in dijet production in p-Pb collisions at Large
  Hadron Collider}.
\newblock {\em Phys. Rev. D}, 86:094043, 2012.

\bibitem{Hentschinski:2013id}
Martin Hentschinski, Agustin Sabio~Vera, and Clara Salas.
\newblock {$F_2$ and $F_L$ at small $x$ using a collinearly improved BFKL
  resummation}.
\newblock {\em Phys. Rev. D}, 87(7):076005, 2013.

\bibitem{Abdolmaleki:2018jln}
Hamed Abdolmaleki et~al.
\newblock {Impact of low-$x$ resummation on QCD analysis of HERA data}.
\newblock {\em Eur. Phys. J. C}, 78(8):621, 2018.

\bibitem{Frankfurt:2010ea}
L.~Frankfurt, M.~Strikman, and C.~Weiss.
\newblock {Transverse nucleon structure and diagnostics of hard parton-parton
  processes at LHC}.
\newblock {\em Phys. Rev. D}, 83:054012, 2011.

\bibitem{Frankfurt:2001nt}
L.~Frankfurt, V.~Guzey, M.~McDermott, and M.~Strikman.
\newblock {Revealing the black body regime of small x DIS through final state
  signals}.
\newblock {\em Phys. Rev. Lett.}, 87:192301, 2001.

\bibitem{Frankfurt:2001av}
L.~Frankfurt, V.~Guzey, M.~McDermott, and M.~Strikman.
\newblock {Electron nucleus collisions at THERA}.
\newblock 4 2001.

\bibitem{Gelis:2008zzc}
Francois Gelis.
\newblock {Gluon saturation from DIS to nucleus-nucleus collisions}.
\newblock {\em Acta Phys. Polon. B}, 39:2419--2454, 2008.

\bibitem{Mueller:1993rr}
Alfred~H. Mueller.
\newblock {Soft gluons in the infinite momentum wave function and the BFKL
  pomeron}.
\newblock {\em Nucl. Phys. B}, 415:373--385, 1994.

\bibitem{GolecBiernat:2003ym}
Krzysztof~J. Golec-Biernat and A.~M. Stasto.
\newblock {On solutions of the Balitsky-Kovchegov equation with impact
  parameter}.
\newblock {\em Nucl. Phys. B}, 668:345--363, 2003.

\bibitem{Berger:2010sh}
Jeffrey Berger and Anna Stasto.
\newblock {Numerical solution of the nonlinear evolution equation at small x
  with impact parameter and beyond the LL approximation}.
\newblock {\em Phys. Rev. D}, 83:034015, 2011.

\bibitem{Balitsky:2008zza}
I.~Balitsky and G.~A. Chirilli.
\newblock {Next-to-leading order evolution of color dipoles}.
\newblock {\em Phys. Rev.}, D77:014019, 2008.

\bibitem{Kovner:2014lca}
A.~Kovner, M.~Lublinsky, and Y.~Mulian.
\newblock {NLO JIMWLK evolution unabridged}.
\newblock {\em JHEP}, 08:114, 2014.

\bibitem{Lublinsky:2016meo}
M.~Lublinsky and Y.~Mulian.
\newblock {High Energy QCD at NLO: from light-cone wave function to JIMWLK
  evolution}.
\newblock {\em JHEP}, 05:097, 2017.

\bibitem{Motyka:2009gi}
L.~Motyka and A.~M. Stasto.
\newblock {Exact kinematics in the small x evolution of the color dipole and
  gluon cascade}.
\newblock {\em Phys. Rev.}, D79:085016, 2009.

\bibitem{Beuf:2014uia}
G.~Beuf.
\newblock {Improving the kinematics for low-$x$ QCD evolution equations in
  coordinate space}.
\newblock {\em Phys. Rev.}, D89(7):074039, 2014.

\bibitem{Iancu:2015vea}
E.~Iancu, J.~D. Madrigal, A.~H. Mueller, G.~Soyez, and D.~N.
  Triantafyllopoulos.
\newblock {Resumming double logarithms in the QCD evolution of color dipoles}.
\newblock {\em Phys. Lett.}, B744:293--302, 2015.

\bibitem{Hatta:2016ujq}
Yoshitaka Hatta and Edmond Iancu.
\newblock {Collinearly improved JIMWLK evolution in Langevin form}.
\newblock {\em JHEP}, 08:083, 2016.

\bibitem{Kovchegov:1999ji}
Yuri~V. Kovchegov and Eugene Levin.
\newblock {Diffractive dissociation including multiple pomeron exchanges in
  high parton density QCD}.
\newblock {\em Nucl. Phys. B}, 577:221--239, 2000.

\bibitem{Levin:2001pr}
E.~Levin and M.~Lublinsky.
\newblock {Diffractive dissociation and saturation scale from nonlinear
  evolution in high-energy DIS}.
\newblock {\em Eur. Phys. J. C}, 22:647--654, 2002.

\bibitem{Levin:2002fj}
E.~Levin and M.~Lublinsky.
\newblock {Diffractive dissociation from nonlinear evolution in DIS on nuclei}.
\newblock {\em Nucl. Phys. A}, 712:95--109, 2002.

\bibitem{Levy:2007fb}
Aharon Levy.
\newblock {Exclusive vector meson electroproduction at HERA}.
\newblock In {\em {12th International Conference on Elastic and Diffractive
  Scattering: Forward Physics and QCD}}, pages 2--9, 11 2007.

\bibitem{Newman:2013ada}
Paul Newman and Matthew Wing.
\newblock {The Hadronic Final State at HERA}.
\newblock {\em Rev. Mod. Phys.}, 86(3):1037, 2014.

\bibitem{Collins:1996fb}
John~C. Collins, Leonid Frankfurt, and Mark Strikman.
\newblock {Factorization for hard exclusive electroproduction of mesons in
  QCD}.
\newblock {\em Phys. Rev. D}, 56:2982--3006, 1997.

\bibitem{Brodsky:1994kf}
Stanley~J. Brodsky, L.~Frankfurt, J.~F. Gunion, Alfred~H. Mueller, and
  M.~Strikman.
\newblock {Diffractive leptoproduction of vector mesons in QCD}.
\newblock {\em Phys. Rev. D}, 50:3134--3144, 1994.

\bibitem{Ryskin:1992ui}
M.G. Ryskin.
\newblock {Diffractive J/psi electroproduction in LLA QCD}.
\newblock {\em Z. Phys. C}, 57:89--92, 1993.

\bibitem{Frankfurt:2000ez}
L.~Frankfurt, M.~McDermott, and M.~Strikman.
\newblock {A Fresh look at diffractive J / psi photoproduction at HERA, with
  predictions for THERA}.
\newblock {\em JHEP}, 03:045, 2001.

\bibitem{Frankfurt:1997fj}
Leonid Frankfurt, Werner Koepf, and Mark Strikman.
\newblock {Diffractive heavy quarkonium photoproduction and electroproduction
  in QCD}.
\newblock {\em Phys. Rev. D}, 57:512--526, 1998.

\bibitem{Mueller:1998fv}
Dieter M\"uller, D.~Robaschik, B.~Geyer, F.~M. Dittes, and J.~Ho\v{r}ej\v{s}i.
\newblock {Wave functions, evolution equations and evolution kernels from light
  ray operators of QCD}.
\newblock {\em Fortsch. Phys.}, 42:101--141, 1994.

\bibitem{Radyushkin:1997ki}
A.~V. Radyushkin.
\newblock {Nonforward parton distributions}.
\newblock {\em Phys. Rev. D}, 56:5524--5557, 1997.

\bibitem{Ji:1996nm}
Xiang-Dong Ji.
\newblock {Deeply virtual Compton scattering}.
\newblock {\em Phys. Rev. D}, 55:7114--7125, 1997.

\bibitem{Goeke:2001tz}
K.~Goeke, Maxim~V. Polyakov, and M.~Vanderhaeghen.
\newblock {Hard exclusive reactions and the structure of hadrons}.
\newblock {\em Prog. Part. Nucl. Phys.}, 47:401--515, 2001.

\bibitem{Belitsky:2001ns}
Andrei~V. Belitsky, Dieter Mueller, and A.~Kirchner.
\newblock {Theory of deeply virtual Compton scattering on the nucleon}.
\newblock {\em Nucl. Phys. B}, 629:323--392, 2002.

\bibitem{Diehl:2003ny}
M.~Diehl.
\newblock {Generalized parton distributions}.
\newblock {\em Phys. Rept.}, 388:41--277, 2003.

\bibitem{Belitsky:2005qn}
A.~V. Belitsky and A.~V. Radyushkin.
\newblock {Unraveling hadron structure with generalized parton distributions}.
\newblock {\em Phys. Rept.}, 418:1--387, 2005.

\bibitem{Brock:1993sz}
Raymond Brock et~al.
\newblock {Handbook of perturbative QCD: Version 1.0}.
\newblock {\em Rev. Mod. Phys.}, 67:157--248, 1995.

\bibitem{Frankfurt:1997ha}
L.~Frankfurt, A.~Freund, V.~Guzey, and M.~Strikman.
\newblock {Nondiagonal parton distribution in the leading logarithmic
  approximation}.
\newblock {\em Phys. Lett. B}, 418:345--354, 1998.
\newblock [Erratum: Phys.Lett.B 429, 414 (1998)].

\bibitem{Shuvaev:1999fm}
A.~Shuvaev.
\newblock {Solution of the off forward leading logarithmic evolution equation
  based on the Gegenbauer moments inversion}.
\newblock {\em Phys. Rev. D}, 60:116005, 1999.

\bibitem{Shuvaev:1999ce}
A.~G. Shuvaev, Krzysztof~J. Golec-Biernat, Alan~D. Martin, and M.~G. Ryskin.
\newblock {Off diagonal distributions fixed by diagonal partons at small x and
  xi}.
\newblock {\em Phys. Rev. D}, 60:014015, 1999.

\bibitem{Martin:2007sb}
A.~D. Martin, C.~Nockles, Mikhail~G. Ryskin, and Thomas Teubner.
\newblock {Small x gluon from exclusive J/psi production}.
\newblock {\em Phys. Lett. B}, 662:252--258, 2008.

\bibitem{Harland-Lang:2013xba}
L.~A. Harland-Lang.
\newblock {Simple form for the low-x generalized parton distributions in the
  skewed regime}.
\newblock {\em Phys. Rev. D}, 88(3):034029, 2013.

\bibitem{Frankfurt:1998yf}
L.~L. Frankfurt, M.~F. McDermott, and M.~Strikman.
\newblock {Diffractive photoproduction of $\upsilon$ at HERA}.
\newblock {\em JHEP}, 02:002, 1999.

\bibitem{Kowalski:2006hc}
H.~Kowalski, L.~Motyka, and G.~Watt.
\newblock {Exclusive diffractive processes at HERA within the dipole picture}.
\newblock {\em Phys. Rev. D}, 74:074016, 2006.

\bibitem{Munier:2001nr}
S.~Munier, A.~M. Stasto, and Alfred~H. Mueller.
\newblock {Impact parameter dependent S matrix for dipole proton scattering
  from diffractive meson electroproduction}.
\newblock {\em Nucl. Phys. B}, 603:427--445, 2001.

\bibitem{Guzey:2013qza}
V.~Guzey and M.~Zhalov.
\newblock {Exclusive $J/{\psi}$ production in ultraperipheral collisions at the
  LHC: constrains on the gluon distributions in the proton and nuclei}.
\newblock {\em JHEP}, 10:207, 2013.

\bibitem{Adloff:2000vm}
C.~Adloff et~al.
\newblock {Elastic photoproduction of J / psi and Upsilon mesons at HERA}.
\newblock {\em Phys. Lett. B}, 483:23--35, 2000.

\bibitem{Alexa:2013xxa}
C.~Alexa et~al.
\newblock {Elastic and Proton-Dissociative Photoproduction of J/psi Mesons at
  HERA}.
\newblock {\em Eur. Phys. J. C}, 73(6):2466, 2013.

\bibitem{Chekanov:2002xi}
S.~Chekanov et~al.
\newblock {Exclusive photoproduction of J / psi mesons at HERA}.
\newblock {\em Eur. Phys. J. C}, 24:345--360, 2002.

\bibitem{Aktas:2005xu}
A.~Aktas et~al.
\newblock {Elastic J/psi production at HERA}.
\newblock {\em Eur. Phys. J. C}, 46:585--603, 2006.

\bibitem{Aaij:2014iea}
Roel Aaij et~al.
\newblock {Updated measurements of exclusive $J/\psi$ and $\psi$(2S) production
  cross-sections in pp collisions at $\sqrt{s}=7$ TeV}.
\newblock {\em J. Phys. G}, 41:055002, 2014.

\bibitem{Aaij:2018arx}
Roel Aaij et~al.
\newblock {Central exclusive production of $J/\psi$ and $\psi(2S)$ mesons in
  $pp$ collisions at $\sqrt{s}=13~$TeV}.
\newblock {\em JHEP}, 10:167, 2018.

\bibitem{TheALICE:2014dwa}
Betty~Bezverkhny Abelev et~al.
\newblock {Exclusive $\mathrm{J/}\psi$ photoproduction off protons in
  ultra-peripheral p-Pb collisions at $\sqrt{s_{\rm NN}}=5.02$ TeV}.
\newblock {\em Phys. Rev. Lett.}, 113(23):232504, 2014.

\bibitem{Pumplin:2002vw}
J.~Pumplin, D.~R. Stump, J.~Huston, H.~L. Lai, Pavel~M. Nadolsky, and W.~K.
  Tung.
\newblock {New generation of parton distributions with uncertainties from
  global QCD analysis}.
\newblock {\em JHEP}, 07:012, 2002.

\bibitem{Jones:2015nna}
S.~P. Jones, A.~D. Martin, M.~G. Ryskin, and T.~Teubner.
\newblock {Exclusive $J/\psi$ and $\Upsilon$ photoproduction and the low $x$
  gluon}.
\newblock {\em J. Phys. G}, 43(3):035002, 2016.

\bibitem{Flett:2020duk}
C.~A. Flett, A.~D. Martin, M.~G. Ryskin, and T.~Teubner.
\newblock {Very low $x$ gluon density determined by LHCb exclusive $J/\psi$
  data}.
\newblock {\em Phys. Rev. D}, 102:114021, 2020.

\bibitem{Garcia:2019tne}
A.~Arroyo~Garcia, M.~Hentschinski, and K.~Kutak.
\newblock {QCD evolution based evidence for the onset of gluon saturation in
  exclusive photo-production of vector mesons}.
\newblock {\em Phys. Lett. B}, 795:569--575, 2019.

\bibitem{Hentschinski:2020yfm}
Martin Hentschinski and Emilio Padr\'on~Molina.
\newblock {Exclusive $J/\Psi$ and $\Psi(2s)$ photo-production as a probe of QCD
  low $x$ evolution equations}.
\newblock {\em Phys. Rev. D}, 103(7):074008, 2021.

\bibitem{Rogers:2003vi}
T.~Rogers, V.~Guzey, M.~Strikman, and X.~Zu.
\newblock {Determining the proximity of gamma* N scattering to the black body
  limit using DIS and J / psi production}.
\newblock {\em Phys. Rev. D}, 69:074011, 2004.

\bibitem{Mantysaari:2016jaz}
Heikki M\"antysaari and Bj\"orn Schenke.
\newblock {Revealing proton shape fluctuations with incoherent diffraction at
  high energy}.
\newblock {\em Phys. Rev. D}, 94(3):034042, 2016.

\bibitem{Mantysaari:2016ykx}
Heikki M\"antysaari and Bj\"orn Schenke.
\newblock {Evidence of strong proton shape fluctuations from incoherent
  diffraction}.
\newblock {\em Phys. Rev. Lett.}, 117(5):052301, 2016.

\bibitem{Mantysaari:2019jhh}
Heikki M\"antysaari and Bj\"orn Schenke.
\newblock {Accessing the gluonic structure of light nuclei at a future
  electron-ion collider}.
\newblock {\em Phys. Rev. C}, 101(1):015203, 2020.

\bibitem{Krelina:2019gee}
M.~Krelina, V.~P. Goncalves, and J.~Cepila.
\newblock {Coherent and incoherent vector meson electroproduction in the future
  electron-ion colliders: the hot-spot predictions}.
\newblock {\em Nucl. Phys. A}, 989:187--200, 2019.

\bibitem{Aaron:2009xp}
F.~D. Aaron et~al.
\newblock {Diffractive Electroproduction of rho and phi Mesons at HERA}.
\newblock {\em JHEP}, 05:032, 2010.

\bibitem{Patrignani:2016xqp}
C.~Patrignani et~al.
\newblock {Review of Particle Physics}.
\newblock {\em Chin. Phys. C}, 40(10):100001, 2016.

\bibitem{Chekanov:2004mw}
S.~Chekanov et~al.
\newblock {Exclusive electroproduction of J/psi mesons at HERA}.
\newblock {\em Nucl. Phys. B}, 695:3--37, 2004.

\bibitem{Chekanov:2007zr}
S.~Chekanov et~al.
\newblock {Exclusive rho0 production in deep inelastic scattering at HERA}.
\newblock {\em PMC Phys. A}, 1:6, 2007.

\bibitem{Mankiewicz:1999tt}
L.~Mankiewicz and G.~Piller.
\newblock {Comments on exclusive electroproduction of transversely polarized
  vector mesons}.
\newblock {\em Phys. Rev. D}, 61:074013, 2000.

\bibitem{Escobedo:2019bxn}
Miguel~\'Angel Escobedo and Tuomas Lappi.
\newblock {Dipole picture and the nonrelativistic expansion}.
\newblock {\em Phys. Rev. D}, 101(3):034030, 2020.

\bibitem{Lappi:2020ufv}
Tuomas Lappi, Heikki M\"antysaari, and Jani Penttala.
\newblock {Relativistic corrections to the vector meson light front wave
  function}.
\newblock {\em Phys. Rev. D}, 102(5):054020, 2020.

\bibitem{Frankfurt:2008vi}
L.~Frankfurt, M.~Strikman, D.~Treleani, and C.~Weiss.
\newblock {Evidence for color fluctuations in the nucleon in high-energy
  scattering}.
\newblock {\em Phys. Rev. Lett.}, 101:202003, 2008.

\bibitem{Blok:2017alw}
B.~Blok and M.~Strikman.
\newblock {Multiparton pp and pA Collisions: From Geometry to
  Parton\textendash{}Parton Correlations}.
\newblock {\em Adv. Ser. Direct. High Energy Phys.}, 29:63--99, 2018.

\bibitem{Janssen:2010zz}
Xavier Janssen.
\newblock {Diffractive electroproduction of rho and phi mesons at H1}.
\newblock {\em PoS}, DIS2010:071, 2010.

\bibitem{Bartels:1996fs}
J.~Bartels, J.~R. Forshaw, H.~Lotter, and M.~Wusthoff.
\newblock {Diffractive production of vector mesons at large t}.
\newblock {\em Phys. Lett.}, B375:301--309, 1996.

\bibitem{Ginzburg:1996vq}
I.F. Ginzburg and D.Yu. Ivanov.
\newblock {The Q**2 dependence of the hard diffractive photoproduction of
  vector meson or photon and the range of pQCD validity}.
\newblock {\em Phys. Rev. D}, 54:5523--5535, 1996.

\bibitem{Chekanov:2002rm}
S.~Chekanov et~al.
\newblock {Measurement of proton dissociative diffractive photoproduction of
  vector mesons at large momentum transfer at HERA}.
\newblock {\em Eur. Phys. J. C}, 26:389--409, 2003.

\bibitem{Aktas:2003zi}
A.~Aktas et~al.
\newblock {Diffractive photoproduction of J/psi mesons with large momentum
  transfer at HERA}.
\newblock {\em Phys. Lett. B}, 568:205--218, 2003.

\bibitem{Kotko:2019kma}
P.~Kotko, L.~Motyka, M.~Sadzikowski, and A.~M. Stasto.
\newblock {BFKL Pomeron loop contribution in diffractive photoproduction and
  inclusive hadroproduction of J/psi and Upsilon}.
\newblock {\em JHEP}, 07:129, 2019.

\bibitem{Frankfurt:2008er}
L.~Frankfurt, M.~Strikman, and M.~Zhalov.
\newblock {Large t diffractive J/psi photoproduction with proton dissociation
  in ultraperipheral pA collisions at LHC}.
\newblock {\em Phys. Lett. B}, 670:32--36, 2008.

\bibitem{Motyka:2001zh}
L.~Motyka, A.~D. Martin, and M.~G. Ryskin.
\newblock {The Nonforward BFKL amplitude and rapidity gap physics}.
\newblock {\em Phys. Lett.}, B524:107--114, 2002.

\bibitem{Forshaw:2001pf}
J.~R. Forshaw and G.~Poludniowski.
\newblock {Vector meson photoproduction at high t and comparison to HERA data}.
\newblock {\em Eur. Phys. J.}, C26:411--415, 2003.

\bibitem{Frankfurt:2006tp}
L.~Frankfurt, M.~Strikman, and M.~Zhalov.
\newblock {Elastic and large t rapidity gap vector meson production in
  ultraperipheral proton-ion collisions}.
\newblock {\em Phys. Lett. B}, 640:162--169, 2006.

\bibitem{Alberi:1981af}
G.~Alberi and G.~Goggi.
\newblock {Diffraction of Subnuclear Waves}.
\newblock {\em Phys. Rept.}, 74:1--207, 1981.

\bibitem{Miettinen:1978jb}
Hannu~I. Miettinen and Jon Pumplin.
\newblock {Diffraction Scattering and the Parton Structure of Hadrons}.
\newblock {\em Phys. Rev. D}, 18:1696, 1978.

\bibitem{Blaettel:1993ah}
B.~Blaettel, G.~Baym, L.~L. Frankfurt, H.~Heiselberg, and M.~Strikman.
\newblock {Hadronic cross-section fluctuations}.
\newblock {\em Phys. Rev. D}, 47:2761--2772, 1993.

\bibitem{Frankfurt:1996ri}
L.~Frankfurt, A.~Radyushkin, and M.~Strikman.
\newblock {Interaction of small size wave packet with hadron target}.
\newblock {\em Phys. Rev. D}, 55:98--104, 1997.

\bibitem{Guzey:2005tk}
V.~Guzey and M.~Strikman.
\newblock {Proton-nucleus scattering and cross section fluctuations at RHIC and
  LHC}.
\newblock {\em Phys. Lett. B}, 633:245--252, 2006.

\bibitem{Levin:1965mi}
E.~M. Levin and L.~L. Frankfurt.
\newblock {The Quark hypothesis and relations between cross-sections at
  high-energies}.
\newblock {\em JETP Lett.}, 2:65--70, 1965.

\bibitem{Derrick:1995vq}
M.~Derrick et~al.
\newblock {Measurement of elastic $\rho^0$ photoproduction at HERA}.
\newblock {\em Z. Phys. C}, 69:39--54, 1995.

\bibitem{Derrick:1996vw}
M.~Derrick et~al.
\newblock {Study of elastic $\rho^0$ photoproduction at HERA using the ZEUS
  leading proton spectrometer}.
\newblock {\em Z. Phys. C}, 73:253--268, 1997.

\bibitem{Breitweg:1997ed}
J.~Breitweg et~al.
\newblock {Elastic and proton dissociative $\rho^0$ photoproduction at HERA}.
\newblock {\em Eur. Phys. J. C}, 2:247--267, 1998.

\bibitem{Weber:2006di}
Ronald~M. Weber.
\newblock {\em {Diffractive $\rho^0$ photoproduction at HERA}}.
\newblock PhD thesis, Zurich, ETH, 2006.

\bibitem{Frankfurt:2015cwa}
L.~Frankfurt, V.~Guzey, M.~Strikman, and M.~Zhalov.
\newblock {Nuclear shadowing in photoproduction of \ensuremath{\rho} mesons in
  ultraperipheral nucleus collisions at RHIC and the LHC}.
\newblock {\em Phys. Lett. B}, 752:51--58, 2016.

\bibitem{Chapin:1985mf}
T.~J. Chapin, R.~L. Cool, Konstantin~A. Goulianos, K.~A. Jenkins, J.~P.
  Silverman, G.~R. Snow, H.~Sticker, Sebastian~N. White, and Yue-Hua Chou.
\newblock {DIFFRACTION DISSOCIATION OF PHOTONS ON HYDROGEN}.
\newblock {\em Phys. Rev. D}, 31:17--30, 1985.

\bibitem{Bauer:1977iq}
T.~H. Bauer, R.~D. Spital, D.~R. Yennie, and F.~M. Pipkin.
\newblock {The Hadronic Properties of the Photon in High-Energy Interactions}.
\newblock {\em Rev. Mod. Phys.}, 50:261, 1978.
\newblock [Erratum: Rev.Mod.Phys. 51, 407 (1979)].

\bibitem{Bjorken:1972uk}
J.~D. Bjorken.
\newblock {Final state hadrons in deep inelastic processes and colliding
  beams}.
\newblock {\em Conf. Proc. C}, 710823:281--297, 1971.

\bibitem{Alvioli:2016gfo}
M.~Alvioli, L.~Frankfurt, V.~Guzey, M.~Strikman, and M.~Zhalov.
\newblock {Mapping color fluctuations in the photon in ultraperipheral heavy
  ion collisions at the Large Hadron Collider}.
\newblock {\em Phys. Lett. B}, 767:450--457, 2017.

\bibitem{Agashe:2014kda}
K.~A. Olive et~al.
\newblock {Review of Particle Physics}.
\newblock {\em Chin. Phys. C}, 38:090001, 2014.

\bibitem{Gribov:1973jg}
V.~N. Gribov.
\newblock {Space-time description of hadron interactions at high-energies}.
\newblock 1973.

\bibitem{Gribov:1968uy}
V.~N. Gribov and Alexander~A. Migdal.
\newblock {Properties of the pomeranchuk pole and the branch cuts related to it
  at low momentum transfer}.
\newblock {\em Sov. J. Nucl. Phys.}, 8:583--590, 1969.

\bibitem{Abramovsky:1973fm}
V.~A. Abramovsky, V.~N. Gribov, and O.~V. Kancheli.
\newblock {Character of Inclusive Spectra and Fluctuations Produced in
  Inelastic Processes by Multi - Pomeron Exchange}.
\newblock {\em Yad. Fiz.}, 18:595--616, 1973.

\bibitem{Badelek:1991qa}
Barbara~Maria Badelek and Jan Kwiecinski.
\newblock {Shadowing in inelastic lepton - deuteron scattering}.
\newblock {\em Nucl. Phys. B}, 370:278--298, 1992.

\bibitem{Badelek:1994qg}
B.~Badelek and J.~Kwiecinski.
\newblock {Shadowing in the deuteron and the new f2(n) / f2(p) measurements}.
\newblock {\em Phys. Rev. D}, 50:4--8, 1994.

\bibitem{Melnitchouk:1992eu}
W.~Melnitchouk and Anthony~William Thomas.
\newblock {Shadowing in deuterium}.
\newblock {\em Phys. Rev. D}, 47:3783--3793, 1993.

\bibitem{Melnitchouk:1995am}
W.~Melnitchouk and Anthony~William Thomas.
\newblock {Q**2 dependence of nuclear shadowing}.
\newblock {\em Phys. Rev. C}, 52:3373--3377, 1995.

\bibitem{Piller:1995kh}
G.~Piller, W.~Ratzka, and W.~Weise.
\newblock {Phenomenology of nuclear shadowing in deep inelastic scattering}.
\newblock {\em Z. Phys. A}, 352:427--439, 1995.

\bibitem{Frankfurt:2003jf}
L.~Frankfurt, V.~Guzey, and M.~Strikman.
\newblock {Nuclear shadowing and extraction of F2(p) - F2(n) at small x from
  deuteron collider data}.
\newblock {\em Phys. Rev. Lett.}, 91:202001, 2003.

\bibitem{Piller:1997ny}
G.~Piller, G.~Niesler, and W.~Weise.
\newblock {Diffractive phenomena and shadowing in deep inelastic scattering}.
\newblock {\em Z. Phys. A}, 358:407--413, 1997.

\bibitem{Piller:1999wx}
Gunther Piller and Wolfram Weise.
\newblock {Nuclear deep inelastic lepton scattering and coherence phenomena}.
\newblock {\em Phys. Rept.}, 330:1--94, 2000.

\bibitem{Adeluyi:2006xy}
A.~Adeluyi and G.~Fai.
\newblock {Mass dependence of nuclear shadowing at small Bjorken-x from
  diffractive scattering}.
\newblock {\em Phys. Rev. C}, 74:054904, 2006.

\bibitem{DeJager:1987qc}
H.~De~Vries, C.~W. De~Jager, and C.~De~Vries.
\newblock {Nuclear charge and magnetization density distribution parameters
  from elastic electron scattering}.
\newblock {\em Atom. Data Nucl. Data Tabl.}, 36:495--536, 1987.

\bibitem{Capella:1997yv}
A.~Capella, A.~Kaidalov, C.~Merino, D.~Pertermann, and J.~Tran Thanh~Van.
\newblock {Structure functions of nuclei at small x and diffraction at HERA}.
\newblock {\em Eur. Phys. J. C}, 5:111--117, 1998.

\bibitem{Armesto:2003fi}
N.~Armesto, A.~Capella, A.~B. Kaidalov, J.~Lopez-Albacete, and C.~A. Salgado.
\newblock {Nuclear structure functions at small x from inelastic shadowing and
  diffraction}.
\newblock {\em Eur. Phys. J. C}, 29:531--540, 2003.

\bibitem{Armesto:2010kr}
Nestor Armesto, Alexei~B. Kaidalov, Carlos~A. Salgado, and Konrad Tywoniuk.
\newblock {Nuclear shadowing in Glauber-Gribov theory with Q2-evolution}.
\newblock {\em Eur. Phys. J. C}, 68:447--457, 2010.

\bibitem{Armesto:2006ph}
Nestor Armesto.
\newblock {Nuclear shadowing}.
\newblock {\em J. Phys. G}, 32:R367--R394, 2006.

\bibitem{Frankfurt:1998ym}
L.~Frankfurt and M.~Strikman.
\newblock {Diffraction at HERA, color opacity and nuclear shadowing}.
\newblock {\em Eur. Phys. J. A}, 5:293--306, 1999.

\bibitem{Frankfurt:2003zd}
L.~Frankfurt, V.~Guzey, and M.~Strikman.
\newblock {Leading twist nuclear shadowing: Uncertainties, comparison to
  experiments, and higher twist effects}.
\newblock {\em Phys. Rev. D}, 71:054001, 2005.

\bibitem{Frankfurt:2011cs}
L.~Frankfurt, V.~Guzey, and M.~Strikman.
\newblock {Leading Twist Nuclear Shadowing Phenomena in Hard Processes with
  Nuclei}.
\newblock {\em Phys. Rept.}, 512:255--393, 2012.

\bibitem{Butterworth:2005aq}
J.~M. Butterworth and M.~Wing.
\newblock {High energy photoproduction}.
\newblock {\em Rept. Prog. Phys.}, 68:2773--2828, 2005.

\bibitem{ZEUS:2009uxs}
S.~Chekanov et~al.
\newblock {A QCD analysis of ZEUS diffractive data}.
\newblock {\em Nucl. Phys. B}, 831:1--25, 2010.

\bibitem{Frankfurt:2016qca}
L.~Frankfurt, V.~Guzey, and M.~Strikman.
\newblock {Dynamical model of antishadowing of the nuclear gluon distribution}.
\newblock {\em Phys. Rev. C}, 95(5):055208, 2017.

\bibitem{Brodsky:1989qz}
Stanley~J. Brodsky and Hung~Jung Lu.
\newblock {Shadowing and Antishadowing of Nuclear Structure Functions}.
\newblock {\em Phys. Rev. Lett.}, 64:1342, 1990.

\bibitem{Hirai:2007sx}
M.~Hirai, S.~Kumano, and T.~H. Nagai.
\newblock {Determination of nuclear parton distribution functions and their
  uncertainties in next-to-leading order}.
\newblock {\em Phys. Rev. C}, 76:065207, 2007.

\bibitem{Kovarik:2015cma}
K.~Kovarik et~al.
\newblock {nCTEQ15 - Global analysis of nuclear parton distributions with
  uncertainties in the CTEQ framework}.
\newblock {\em Phys. Rev. D}, 93(8):085037, 2016.

\bibitem{Eskola:2016oht}
Kari~J. Eskola, Petja Paakkinen, Hannu Paukkunen, and Carlos~A. Salgado.
\newblock {EPPS16: Nuclear parton distributions with LHC data}.
\newblock {\em Eur. Phys. J. C}, 77(3):163, 2017.

\bibitem{Khanpour:2016pph}
Hamzeh Khanpour and S.~Atashbar~Tehrani.
\newblock {Global Analysis of Nuclear Parton Distribution Functions and Their
  Uncertainties at Next-to-Next-to-Leading Order}.
\newblock {\em Phys. Rev. D}, 93(1):014026, 2016.

\bibitem{Helenius:2012wd}
Ilkka Helenius, Kari~J. Eskola, Heli Honkanen, and Carlos~A. Salgado.
\newblock {Impact-Parameter Dependent Nuclear Parton Distribution Functions:
  EPS09s and EKS98s and Their Applications in Nuclear Hard Processes}.
\newblock {\em JHEP}, 07:073, 2012.

\bibitem{Boer:2011fh}
Daniel Boer et~al.
\newblock {Gluons and the quark sea at high energies: Distributions,
  polarization, tomography}.
\newblock 8 2011.

\bibitem{Salgado:2011wc}
C.~A. Salgado et~al.
\newblock {Proton-Nucleus Collisions at the LHC: Scientific Opportunities and
  Requirements}.
\newblock {\em J. Phys. G}, 39:015010, 2012.

\bibitem{Eskola:2013aya}
Kari~J. Eskola, Hannu Paukkunen, and Carlos~A. Salgado.
\newblock {A perturbative QCD study of dijets in p+Pb collisions at the LHC}.
\newblock {\em JHEP}, 10:213, 2013.

\bibitem{Helenius:2014qla}
Ilkka Helenius, Kari~J. Eskola, and Hannu Paukkunen.
\newblock {Probing the small-$x$ nuclear gluon distributions with isolated
  photons at forward rapidities in p+Pb collisions at the LHC}.
\newblock {\em JHEP}, 09:138, 2014.

\bibitem{Armesto:2015lrg}
N\'estor Armesto, Hannu Paukkunen, Jos\'e~Manuel Pen\'\i{}n, Carlos~A. Salgado,
  and P\'\i{}a Zurita.
\newblock {An analysis of the impact of LHC Run I proton\textendash{}lead data
  on nuclear parton densities}.
\newblock {\em Eur. Phys. J. C}, 76(4):218, 2016.

\bibitem{Kusina:2016fxy}
A.~Kusina, F.~Lyonnet, D.~B. Clark, E.~Godat, T.~Jezo, K.~Kovarik, F.~I.
  Olness, I.~Schienbein, and J.~Y. Yu.
\newblock {Vector boson production in pPb and PbPb collisions at the LHC and
  its impact on nCTEQ15 PDFs}.
\newblock {\em Eur. Phys. J. C}, 77(7):488, 2017.

\bibitem{Kusina:2017gkz}
Aleksander Kusina, Jean-Philippe Lansberg, Ingo Schienbein, and Hua-Sheng Shao.
\newblock {Gluon Shadowing in Heavy-Flavor Production at the LHC}.
\newblock {\em Phys. Rev. Lett.}, 121(5):052004, 2018.

\bibitem{Eskola:2019dui}
Kari~J. Eskola, Petja Paakkinen, and Hannu Paukkunen.
\newblock {Non-quadratic improved Hessian PDF reweighting and application to
  CMS dijet measurements at 5.02 TeV}.
\newblock {\em Eur. Phys. J. C}, 79(6):511, 2019.

\bibitem{Eskola:2019bgf}
Kari~J. Eskola, Ilkka Helenius, Petja Paakkinen, and Hannu Paukkunen.
\newblock {A QCD analysis of LHCb D-meson data in p+Pb collisions}.
\newblock {\em JHEP}, 05:037, 2020.

\bibitem{Kusina:2020lyz}
A.~Kusina et~al.
\newblock {Impact of LHC vector boson production in heavy ion collisions on
  strange PDFs}.
\newblock {\em Eur. Phys. J. C}, 80(10):968, 2020.

\bibitem{Abbas:2013oua}
E.~Abbas et~al.
\newblock {Charmonium and $e^+e^-$ pair photoproduction at mid-rapidity in
  ultra-peripheral Pb-Pb collisions at $\sqrt{s_{\rm NN}}$=2.76 TeV}.
\newblock {\em Eur. Phys. J. C}, 73(11):2617, 2013.

\bibitem{Abelev:2012ba}
Betty Abelev et~al.
\newblock {Coherent $J/\psi$ photoproduction in ultra-peripheral Pb-Pb
  collisions at $\sqrt{s_{NN}} = 2.76$ TeV}.
\newblock {\em Phys. Lett. B}, 718:1273--1283, 2013.

\bibitem{Adam:2015sia}
Jaroslav Adam et~al.
\newblock {Coherent $\psi$(2S) photo-production in ultra-peripheral Pb Pb
  collisions at $\sqrt{s}_{\rm NN}$ = 2.76 TeV}.
\newblock {\em Phys. Lett. B}, 751:358--370, 2015.

\bibitem{Khachatryan:2016qhq}
Vardan Khachatryan et~al.
\newblock {Coherent $J/\psi$ photoproduction in ultra-peripheral PbPb
  collisions at $\sqrt {s_{NN}} =$ 2.76 TeV with the CMS experiment}.
\newblock {\em Phys. Lett. B}, 772:489--511, 2017.

\bibitem{Acharya:2019vlb}
Shreyasi Acharya et~al.
\newblock {Coherent J/$\psi$ photoproduction at forward rapidity in
  ultra-peripheral Pb-Pb collisions at $\sqrt{s_{\rm{NN}}}=5.02$ TeV}.
\newblock {\em Phys. Lett. B}, 798:134926, 2019.

\bibitem{LHCb:2018ofh}
A.~Bursche.
\newblock {Study of coherent $J/\psi$ production in lead-lead collisions at
  $\sqrt{s_{\rm NN}} =5\ \rm{TeV}$ with the LHCb experiment}.
\newblock {\em Nucl. Phys. A}, 982:247--250, 2019.

\bibitem{Acharya:2021ugn}
Shreyasi Acharya et~al.
\newblock {Coherent $\rm{J/\psi}$ and $\rm{\psi'}$ photoproduction at
  midrapidity in ultra-peripheral Pb-Pb collisions at
  $\sqrt{s_{\rm{NN}}}~=~5.02$ TeV}.
\newblock 1 2021.

\bibitem{Bertulani:2005ru}
Carlos~A. Bertulani, Spencer~R. Klein, and Joakim Nystrand.
\newblock {Physics of ultra-peripheral nuclear collisions}.
\newblock {\em Ann. Rev. Nucl. Part. Sci.}, 55:271--310, 2005.

\bibitem{Klein:1999gv}
Spencer~R. Klein and Joakim Nystrand.
\newblock {Interference in exclusive vector meson production in heavy ion
  collisions}.
\newblock {\em Phys. Rev. Lett.}, 84:2330--2333, 2000.

\bibitem{Ryskin:1995hz}
M.~G. Ryskin, R.~G. Roberts, Alan~D. Martin, and E.~M. Levin.
\newblock {Diffractive J / psi photoproduction as a probe of the gluon
  density}.
\newblock {\em Z. Phys. C}, 76:231--239, 1997.

\bibitem{Hoodbhoy:1996zg}
Pervez Hoodbhoy.
\newblock {Wave function corrections and off forward gluon distributions in
  diffractive J / psi electroproduction}.
\newblock {\em Phys. Rev. D}, 56:388--393, 1997.

\bibitem{Krelina:2019egg}
Michal Krelina, Jan Nemchik, and Roman Pasechnik.
\newblock {$D$-wave effects in diffractive electroproduction of heavy quarkonia
  from the photon-like $V\rightarrow Q{\bar{Q}}$ transition}.
\newblock {\em Eur. Phys. J. C}, 80(2):92, 2020.

\bibitem{Ivanov:2004vd}
D.~Yu. Ivanov, A.~Schafer, L.~Szymanowski, and G.~Krasnikov.
\newblock {Exclusive photoproduction of a heavy vector meson in QCD}.
\newblock {\em Eur. Phys. J. C}, 34(3):297--316, 2004.
\newblock [Erratum: Eur.Phys.J.C 75, 75 (2015)].

\bibitem{Guzey:2013xba}
V.~Guzey, E.~Kryshen, M.~Strikman, and M.~Zhalov.
\newblock {Evidence for nuclear gluon shadowing from the ALICE measurements of
  PbPb ultraperipheral exclusive $J/{\psi}$ production}.
\newblock {\em Phys. Lett. B}, 726:290--295, 2013.

\bibitem{Guzey:2016qwo}
V.~Guzey, M.~Strikman, and M.~Zhalov.
\newblock {Accessing transverse nucleon and gluon distributions in heavy nuclei
  using coherent vector meson photoproduction at high energies in ion
  ultraperipheral collisions}.
\newblock {\em Phys. Rev. C}, 95(2):025204, 2017.

\bibitem{Guzey:2020ntc}
V.~Guzey, E.~Kryshen, M.~Strikman, and M.~Zhalov.
\newblock {Nuclear suppression from coherent $J /\psi$ photoproduction at the
  Large Hadron Collider}.
\newblock {\em Phys. Lett. B}, 816:136202, 2021.

\bibitem{Contreras:2016pkc}
J.~G. Contreras.
\newblock {Gluon shadowing at small $x$ from coherent $\mathrm{J/}\psi$
  photoproduction data at energies available at the CERN Large Hadron
  Collider}.
\newblock {\em Phys. Rev. C}, 96(1):015203, 2017.

\bibitem{Acharya:2021bnz}
Shreyasi Acharya et~al.
\newblock {First measurement of the |$t$|-dependence of coherent $J/\psi$
  photonuclear production}.
\newblock {\em Phys. Lett. B}, 817:136280, 2021.

\bibitem{Citron:2018lsq}
Z.~Citron et~al.
\newblock {Report from Working Group 5}: {Future physics opportunities for
  high-density QCD at the LHC with heavy-ion and proton beams}.
\newblock {\em CERN Yellow Rep. Monogr.}, 7:1159--1410, 2019.

\bibitem{Guzey:2013jaa}
V.~Guzey, M.~Strikman, and M.~Zhalov.
\newblock {Disentangling coherent and incoherent quasielastic $J/\psi$
  photoproduction on nuclei by neutron tagging in ultraperipheral ion
  collisions at the LHC}.
\newblock {\em Eur. Phys. J. C}, 74(7):2942, 2014.

\bibitem{Guzey:2016tek}
V.~Guzey and M.~Klasen.
\newblock {Diffractive dijet photoproduction in ultraperipheral collisions at
  the LHC in next-to-leading order QCD}.
\newblock {\em JHEP}, 04:158, 2016.

\bibitem{Basso:2017mue}
E.~Basso, V.~P. Goncalves, A.~K. Kohara, and M.~S. Rangel.
\newblock {Photon and Pomeron -- induced production of Dijets in $pp$, $pA$ and
  $AA$ collisions}.
\newblock {\em Eur. Phys. J. C}, 77(9):600, 2017.

\bibitem{Guzey:2020pkq}
V.~Guzey, E.~Kryshen, and M.~Zhalov.
\newblock {Incoherent \ensuremath{\rho} meson photoproduction in
  ultraperipheral nuclear collisions at the CERN Large Hadron Collider}.
\newblock {\em Phys. Rev. C}, 102(1):015208, 2020.

\bibitem{Guzey:2018tlk}
V.~Guzey, M.~Strikman, and M.~Zhalov.
\newblock {Nucleon dissociation and incoherent $J/\psi$ photoproduction on
  nuclei in ion ultraperipheral collisions at the Large Hadron Collider}.
\newblock {\em Phys. Rev. C}, 99(1):015201, 2019.

\bibitem{Adam:2015gsa}
Jaroslav Adam et~al.
\newblock {Coherent \ensuremath{\rho}$^{0}$ photoproduction in ultra-peripheral
  Pb-Pb collisions at $ \sqrt{s_{\mathrm{NN}}}=2.76 $ TeV}.
\newblock {\em JHEP}, 09:095, 2015.

\bibitem{Abelev:2007nb}
B.~I. Abelev et~al.
\newblock {$\rho^0$ photoproduction in ultraperipheral relativistic heavy ion
  collisions at $\sqrt{s_{NN}}$ = 200 GeV}.
\newblock {\em Phys. Rev. C}, 77:034910, 2008.

\bibitem{Acharya:2020sbc}
Shreyasi Acharya et~al.
\newblock {Coherent photoproduction of $\rho^{0}$ vector mesons in
  ultra-peripheral Pb-Pb collisions at $ \sqrt{{\mathrm{s}}_{\rm{NN}}} $ = 5.02
  TeV}.
\newblock {\em JHEP}, 06:035, 2020.

\bibitem{Klein:2016yzr}
Spencer~R. Klein, Joakim Nystrand, Janet Seger, Yuri Gorbunov, and Joey
  Butterworth.
\newblock {STARlight: A Monte Carlo simulation program for ultra-peripheral
  collisions of relativistic ions}.
\newblock {\em Comput. Phys. Commun.}, 212:258--268, 2017.

\bibitem{Zielinski:1983ty}
M.~Zielinski et~al.
\newblock {THREE PION PRODUCTION ON NUCLEI AT 200-GEV}.
\newblock {\em Z. Phys. C}, 16:197--204, 1983.

\bibitem{Boos:1978hr}
E.~G. Boos et~al.
\newblock {DIFFRACTIVE COHERENT PRODUCTION IN INTERACTIONS OF 400-GEV/C PROTONS
  ON EMULSION NUCLEI}.
\newblock {\em Nucl. Phys. B}, 137:37--45, 1978.

\bibitem{Frankfurt:2000tya}
Leonid Frankfurt, Vadim Guzey, and Mark Strikman.
\newblock {Color coherent phenomena on nuclei and the QCD evolution equation}.
\newblock {\em J. Phys. G}, 27:R23--146, 2001.

\bibitem{Strikman:1995jf}
M.~Strikman and V.~Guzey.
\newblock {Hadronic cross-section fluctuations and proton coherent diffractive
  dissociation on helium}.
\newblock {\em Phys. Rev. C}, 52:R1189--R1192, 1995.

\bibitem{Bujak:1981mp}
A.~Bujak et~al.
\newblock {Coherent Proton Diffraction Dissociation on Helium From 46-{GeV} to
  400-{GeV}}.
\newblock {\em Phys. Rev. D}, 23:1911, 1981.

\bibitem{Bertocchi:1976bq}
L.~Bertocchi and D.~Treleani.
\newblock {Glauber Theory, Unitarity, and the AGK Cancellation}.
\newblock {\em J. Phys. G}, 3:147, 1977.

\bibitem{Heiselberg:1991is}
H.~Heiselberg, G.~Baym, B.~Blaettel, L.~L. Frankfurt, and M.~Strikman.
\newblock {Color transparency, color opacity, and fluctuations in nuclear
  collisions}.
\newblock {\em Phys. Rev. Lett.}, 67:2946--2949, 1991.

\bibitem{Alvioli:2014sba}
M.~Alvioli, L.~Frankfurt, V.~Guzey, and M.~Strikman.
\newblock {Revealing \textquotedblleft{}flickering\textquotedblright{} of the
  interaction strength in pA collisions at the CERN LHC}.
\newblock {\em Phys. Rev. C}, 90:034914, 2014.

\bibitem{Alvioli:2013vk}
M.~Alvioli and M.~Strikman.
\newblock {Color fluctuation effects in proton-nucleus collisions}.
\newblock {\em Phys. Lett. B}, 722:347--354, 2013.

\bibitem{ATLAS:2014cpa}
Georges Aad et~al.
\newblock {Centrality and rapidity dependence of inclusive jet production in
  $\sqrt{s_{\rm{NN}}} = 5.02$ TeV proton-lead collisions with the ATLAS
  detector}.
\newblock {\em Phys. Lett. B}, 748:392--413, 2015.

\bibitem{Chatrchyan:2014hqa}
Serguei Chatrchyan et~al.
\newblock {Studies of dijet transverse momentum balance and pseudorapidity
  distributions in pPb collisions at $\sqrt{s_{\rm{NN}}} = 5.02$ TeV}.
\newblock {\em Eur. Phys. J. C}, 74(7):2951, 2014.

\bibitem{Adare:2015gla}
A.~Adare et~al.
\newblock {Centrality-dependent modification of jet-production rates in
  deuteron-gold collisions at $\sqrt{s_{NN}}$=200 GeV}.
\newblock {\em Phys. Rev. Lett.}, 116(12):122301, 2016.

\bibitem{Alvioli:2014eda}
Massimiliano Alvioli, Brian~A. Cole, Leonid Frankfurt, D.~V. Perepelitsa, and
  Mark Strikman.
\newblock {Evidence for $x$-dependent proton color fluctuations in pA
  collisions at the CERN Large Hadron Collider}.
\newblock {\em Phys. Rev. C}, 93(1):011902, 2016.

\bibitem{Alvioli:2017wou}
Massimiliano Alvioli, Leonid Frankfurt, Dennis Perepelitsa, and Mark Strikman.
\newblock {Global analysis of color fluctuation effects in proton\textendash{}
  and deuteron\textendash{}nucleus collisions at RHIC and the LHC}.
\newblock {\em Phys. Rev. D}, 98(7):071502, 2018.

\bibitem{Frankfurt:2003td}
L.~Frankfurt, M.~Strikman, and C.~Weiss.
\newblock {Dijet production as a centrality trigger for $p p$ collisions at
  CERN LHC}.
\newblock {\em Phys. Rev. D}, 69:114010, 2004.

\bibitem{Sirunyan:2020ifc}
Albert~M Sirunyan et~al.
\newblock {Measurement of single-diffractive dijet production in proton-proton
  collisions at $\sqrt{s} =$ 8 TeV with the CMS and TOTEM experiments}.
\newblock {\em Eur. Phys. J. C}, 80(12):1164, 2020.
\newblock [Erratum: Eur.Phys.J.C 81, 383 (2021)].

\bibitem{Rogers:2009ke}
T.~C. Rogers and M.~Strikman.
\newblock {Multiple Hard Partonic Collisions with Correlations in Proton-Proton
  Scattering}.
\newblock {\em Phys. Rev. D}, 81:016013, 2010.

\bibitem{ALICE:2020mso}
{Letter of Intent: A Forward Calorimeter (FoCal) in the ALICE experiment}.
\newblock 6 2020.

\end{thebibliography}
 \bibliographystyle{unsrt}

\end{document}